\ifdefined\DoubleSided
  \documentclass[twoside,openright,a4paper]{nusthesis}
\else
  \documentclass[a4paper]{nusthesis}
\fi

\dsp % pseudo double spacing

\usepackage{indentfirst} % indent the first paragraph of a section

\usepackage{microtype} % Better typography
\usepackage{tabulary} % Automatic table sizing
\usepackage{hhline}

\usepackage{enumitem}

\usepackage{hyperref}

\usepackage{listings}

\usepackage{amsthm}
\theoremstyle{definition}
 
\theoremstyle{plain}

\usepackage{mathtools}

\usepackage{interval}
\intervalconfig{soft open fences}

\usepackage[linesnumbered,ruled,vlined]{algorithm2e}
\SetAlgorithmName{Algorithm}{Algorithm}{Algorithm}
\IncMargin{0.5em}
\SetCommentSty{textnormal}
\SetNlSty{}{}{:}
\SetAlgoNlRelativeSize{0}
\SetKwInput{KwGlobal}{Global}
\SetKwInput{KwPrecondition}{Precondition}
\SetKwProg{Proc}{Procedure}{:}{}
\SetKwProg{Func}{Function}{:}{}
\SetKw{And}{and}
\SetKw{Or}{or}
\SetKw{To}{to}
\SetKw{DownTo}{downto}
\SetKw{Break}{break}
\SetKw{Continue}{continue}
\SetKw{SuchThat}{\textit{s.t.}}
\SetKw{WithRespectTo}{\textit{wrt}}
\SetKw{Iff}{\textit{iff.}}
\SetKw{MaxOf}{\textit{max of}}
\SetKw{MinOf}{\textit{min of}}
\SetKwBlock{Match}{match}{}{}

\usepackage{array}
\newcolumntype{L}[1]{>{\raggedright\let\newline\\\arraybackslash\hspace{0pt}}m{#1}}
\newcolumntype{C}[1]{>{\centering\let\newline\\\arraybackslash\hspace{0pt}}m{#1}}
\newcolumntype{R}[1]{>{\raggedleft\let\newline\\\arraybackslash\hspace{0pt}}m{#1}}

\usepackage{color}
\usepackage[dvipsnames]{xcolor}

\usepackage{soul}
\soulregister\cite7
\soulregister\ref7
\soulregister\pageref7
\soulregister\autoref7
\soulregister\eqref7

\renewcommand{\eqref}{Equation~\ref}

\newcommand*{\RootPicDir}{pic}
\newcommand*{\PicDir}{\RootPicDir}

\newcommand*{\SetPicSubDir}[1]{\renewcommand*{\PicDir}{\RootPicDir /#1}}

\newcommand*{\RootExpDir}{exp}
\newcommand*{\ExpDir}{\RootExpDir}

\newcommand*{\SetExpSubDir}[1]{\renewcommand*{\ExpDir}{\RootExpDir /#1}}

\usepackage{lipsum} % for generating dummy text
\usepackage{wasysym}
\usepackage{amssymb,mathtools,amsmath}
\usepackage{physics}
\usepackage{siunitx}
\usepackage{mathrsfs}
\usepackage{amssymb}
\usepackage{graphicx}
\usepackage{float}
\usepackage[caption=false]{subfig}
\usepackage{dcolumn}
\usepackage{appendix}
\usepackage{tabularx}
\usepackage{enumitem}
\usepackage{caption}
\usepackage{epstopdf}
\usepackage[normalem]{ulem}
\usepackage{verbatim}
\usepackage{bbold}
\usepackage{bm}
\usepackage{empheq}
\usepackage{rotating}
\usepackage{multirow}
\usepackage{hhline}
\usepackage{nicematrix,tikz}
\PassOptionsToPackage{table,xcdraw}{xcolor}
\usetikzlibrary{fit}
\usepackage{hyperref}
\hypersetup{
	colorlinks,
	citecolor=teal,
	linkcolor=red,
	urlcolor=magenta}

\begin{document}

\title{TOPOLECTRICAL CIRCUIT THEORY AND REALIZATIONS OF TOPOLOGICAL, NON-LINEAR, AND NON-HERMITIAN PHENOMENA}

\author{Haydar Sahin}
%\prevdegrees{%
%  B.S., Culinary University}
\degree{Doctor of Philosophy}
\field{Electrical and Computer Engineering}
\degreeyear{2024}
\supervisor{%
Associate Professor Mansoor Bin Abdul Jalil - Main Advisor \\
Assistant Professor Lee Ching Hua - Co-Advisor\\
Dr. Kong Jian Feng - Co-Advisor}

% only involve the examiners in the final submission
\examiners{%
Adjunct Professor Liang Gengchiau\\
Associate Professor Andrivo Rusydi}

\maketitle

%\declaresign{} % remove this if you prefer to sign physically
\declaredate{24 January 2024}
\declarationpage

\begin{frontmatter}
  \dedicate{To my father}
  \begin{acknowledgments}

First and foremost, I am profoundly grateful to my esteemed supervisors, Prof. Mansoor Bin Abdul Jalil, Prof. Ching Hua Lee, and Dr. Jian Feng Kong, for the extraordinary privilege of studying under their mentorship. Throughout the four years of my PhD journey, their unwavering support, exceptional guidance, and nurturing mentorship have been the cornerstones of my academic growth.

From the moment I embarked on this journey under the guidance of Prof. Mansoor Bin Abdul Jalil, I have been continually inspired and supported in my development as a scientist. His immense scientific knowledge, coupled with his patient and thoughtful approach, has not only bolstered my research but also set a stellar example for my future endeavors.

Prof. Ching Hua Lee's profound mentorship has significantly shaped my perspective on scientific research. His insightful advice on research directions has been crucial in steering my studies. His versatility across various research fields is not only admirable but has also provided me with a unique opportunity to delve into pioneering areas of study. His clarity, patience, and motivational guidance have been invaluable, and they are qualities I strive to emulate in my own professional journey.

Dr. Jian Feng Kong's role in my academic development has been instrumental, providing me with deeper insights into our research. Our discussions, extending beyond research to a myriad of topics, have been both enlightening and enjoyable. His advice and suggestions have not only enhanced my research but also enriched my experience in Singapore.

Words alone cannot fully express my gratitude towards each of my supervisors. Their collective mentorship has been fundamental in shaping me into the scientist I am today. I am eternally appreciative of their generous support and the profound impact they have had on my personal and professional growth.

I would like to extend my sincere gratitude to Prof. Fatih \"Ozaydın and Dr. Can Yeşilyurt for their pivotal role in initiating and supporting my doctoral studies in Singapore. Their unwavering assistance from the very beginning of my application process, and throughout my PhD journey, has been invaluable. I am deeply appreciative of their guidance and support.

I am deeply grateful to be part of an outstanding research group with Dr. Zhuo Bin Siu, Dr. S.M. Rafi-Ul-Islam, and my colleague Md Saddam Hossain Razo. Our engaging discussions, both in formal meetings and casual settings, have been invaluable. A special note of thanks goes to Zhuo Bin, who has become a role model for me in the realm of scientific inquiry. His enthusiasm during our discussions, his innovative approaches to new problems, and his exceptional motivation are aspects I deeply admire. The knowledge I have gained from Zhuo Bin, especially through his engaging presentations at our group meetings, which I always attended with rapt attention, has been invaluable. I want to express my gratitude to Rafi, whose valuable input has greatly helped me to improve my understanding of our field, especially in my early years. His ongoing encouragement has been particularly influential. I also extend my thanks to Saddam for his dedicated efforts and valuable contributions to our group's success. Their presence during my PhD journey has been more than just a privilege; it has been an integral part of my academic experience. Without their contributions, the research presented in this thesis would not have reached the high standards it currently embodies.

I am equally grateful to my dear friends Can Berk Saner, Garen Haddeler, İlayda Canyakmaz, Kuluhan Bilici, Merve Esmebaşı, and Cem Ataman, whose companionship and support have enriched my life immensely during these four years. Our shared moments have been a source of joy and comfort. I also wish to thank my labmate He Yihan for the fruitful discussions that have broadened my perspective in various physics fields. Additionally, my sincere thanks go to Shuhan Yang for his substantial contributions, tremendous efforts, and expertise in our experimental studies, which have been immensely valuable.

I extend my gratitude to the Agency for Science, Technology and Research (A*STAR) for its invaluable support through the SINGA fellowship program. This support was instrumental in enabling me to pursue my PhD at the National University of Singapore. Additionally, I am deeply appreciative of the warm hospitality extended by the people of Singapore. Their friendliness and kindness have greatly alleviated my homesickness and made Singapore feel like a second home.

Last but certainly not least, I extend my deepest love and gratitude to my family, whose endless support has been a cornerstone in my academic journey and personal development. Their encouragement has not only fueled my studies but also inspired me to strive for greatness in all aspects of life. I would like to express my deepest gratitude to my beloved Hazal, who has shared in both my happiest moments and times of sorrow, and whose steadfast support has been invaluable in overcoming the challenges faced during my PhD journey.

I dedicate this thesis to the memory of my father. His dedication to ensuring our education has left a lasting impact on me. This achievement is a tribute to his enduring influence and the values he instilled in me.

The essence of this thesis is fully realized through its contribution to my lovely country and its people. Inspired by the vision of Mustafa Kemal Atatürk, the founder of the modern Turkish Republic, this work aims to embody his enduring and inspiring principles indefinitely.

%The more we learn about the world, and the deeper our learning, the more conscious, specific, and articulate will be our knowledge of what we do not know. --Karl Popper

\end{acknowledgments}

  \tableofcontents
  \clearpage
\chapter*{Publications during PhD Study}  % important: chapter*, not section*
\markboth{}{}

%\section*{Publications during PhD Study}
%\addcontentsline{toc}{chapter}{Publications during PhD Study}
%\markboth{}{}

%\vspace{1em}
\textbf{\large Journal Publications}

\begin{enumerate}[label={[\arabic*]}]
  \item H. Sahin, Z. B. Siu, S. M. Rafi-Ul-Islam, J. F. Kong, M. B. A. Jalil, and C. H. Lee, ``Impedance responses and size-dependent resonances in topolectrical circuits via the method of images'', \textit{Physical Review B}, vol. 107, no. 24, p. 245114, 2023. \url{https://link.aps.org/doi/10.1103/PhysRevB.107.245114}
  \item X. Zhang, B. Zhang, H. Sahin, Z. B. Siu, S. M. Rafi-Ul-Islam, J. F. Kong, B. Shen, M. B. A. Jalil, R. Thomale, and C. H. Lee, ``Anomalous fractal scaling in two-dimensional electric networks'', \textit{Communications Physics}, vol. 6, no. 1, p. 151, 2023. \url{https://www.nature.com/articles/s42005-023-01266-1}
  \item S. M. Rafi-Ul-Islam, H. Sahin, Z. B. Siu, and M. B. A. Jalil, ``Interfacial skin modes at a non-Hermitian heterojunction'', \textit{Physical Review Research}, vol. 4, no. 4, p. 043021, 2022. \url{https://link.aps.org/doi/10.1103/PhysRevResearch.4.043021}
  \item S. M. Rafi-Ul-Islam, Z. B. Siu, H. Sahin, C. H. Lee, and M. B. A. Jalil, ``Critical hybridization of skin modes in coupled non-Hermitian chains'', \textit{Physical Review Research}, vol. 4, no. 1, p. 013243, 2022. \url{https://link.aps.org/doi/10.1103/PhysRevResearch.4.013243}
  \item S. M. Rafi-Ul-Islam, Z. B. Siu, H. Sahin, C. H. Lee, and M. B. A. Jalil, ``System size dependent topological zero modes in coupled topolectrical chains'', \textit{Physical Review B}, vol. 106, no. 7, p. 075158, 2022. \url{https://link.aps.org/doi/10.1103/PhysRevB.106.075158}
  \item S. M. Rafi-Ul-Islam, Z. B. Siu, H. Sahin, C. H. Lee, and M. B. A. Jalil, ``Unconventional skin modes in generalized topolectrical circuits with multiple asymmetric couplings'', \textit{Physical Review Research}, vol. 4, no. 4, p. 043108, 2022. \url{https://link.aps.org/doi/10.1103/PhysRevResearch.4.043108}
  \item S. M. Rafi-Ul-Islam, Z. B. Siu, H. Sahin, and M. B. A. Jalil, ``Valley and spin quantum Hall conductance of silicene coupled to a ferroelectric layer'', \textit{Frontiers in Physics}, vol. 10, p. 1021192, 2022. \url{https://www.frontiersin.org/articles/10.3389/fphy.2022.1021192/full}
  \item S. M. Rafi-Ul-Islam, Z. B. Siu, H. Sahin, and M. B. A. Jalil, ``Valley Hall effect and kink states in topolectrical circuits'', \textit{Physical Review Research}, vol. 5, no. 1, p. 013107, 2023. \url{https://link.aps.org/doi/10.1103/PhysRevResearch.5.013107}
  \item S. M. Rafi-Ul-Islam, Z. B. Siu, H. Sahin, and M. B. A. Jalil, ``Type-II corner modes in topolectrical circuits'', \textit{Physical Review B}, vol. 106, no. 24, p. 245128, 2022. \url{https://link.aps.org/doi/10.1103/PhysRevB.106.245128}
  \item S. M. Rafi-Ul-Islam, Z. B. Siu, H. Sahin, and M. B. A. Jalil, ``Conductance modulation and spin/valley polarized transmission in silicene coupled with ferroelectric layer'', \textit{Journal of Magnetism and Magnetic Materials}, vol. 571, p. 170559, 2023. \url{https://linkinghub.elsevier.com/retrieve/pii/S0304885323002081}
  \item S. M. Rafi-Ul-Islam, Z. B. Siu, H. Sahin, and M. B. A. Jalil, ``Chiral surface and hinge states in higher-order Weyl semimetallic circuits'', \textit{Physical Review B}, vol. 109, no. 8, p. 085430, 2024. \url{https://link.aps.org/doi/10.1103/PhysRevB.109.085430}
  \item S. M. Rafi-Ul-Islam, Z. B. Siu, H. Sahin, Razo, Md. Saddam Hossain, and M. B. A. Jalil, ``Twisted topology of non-Hermitian systems induced by long-range coupling'', \textit{Physical Review B}, vol. 109, no. 4, p. 045410, 2024. \url{https://link.aps.org/doi/10.1103/PhysRevB.109.045410}
\end{enumerate}

\textbf{\large Conferences}

\begin{enumerate}[label={[\arabic*]}]
     \item H. Sahin, Md S.H. Razo, S. Yang, Z. B. Siu, F. Ozaydin, S. M. Rafi-Ul-Islam, J. F. Kong, H. Yang, C. H. Lee., and M. B. A. Jalil ``PT-sensing through topological state morphing in a PT-symmetric topolectrical circuit'', \textit{2024 Annual Meeting of the Physical Society of Taiwan}, Pages 24, 2024.
     \item H. Sahin, H. Akgun, Z. B. Siu, S. M. Rafi-Ul-Islam, J. F. Kong, M. B. A. Jalil, and C. H. Lee. ``Topological Classification of Non-linear Chaotic Topolectrical Circuits Through Machine Learning'', \textit{IPS Meeting 2023 - Institute of Physics Singapore}, Pages 69, 2023.
    \item H. Sahin, S. M. Rafi-Ul-Islam, Z. B. Siu, J. F. Kong, C. H. Lee. and M. B. A. Jalil, ``N-th Root Topological Lattices'', \textit{The 2021 Around-the-Clock Around-the-Globe Magnetics Conference - IEEE Magnetics}, Pages 44, 2021.
    \item H. Sahin, S. M. Rafi-Ul-Islam, Z. B. Siu, C. H. Lee. and M. B. A. Jalil, ``Unconventional Hybridization of Skin Modes in Coupled non-Hermitian Chains'', \textit{APS March Meeting Abstracts}, Pages N70. 013, 2022.
     \item H. Sahin, S. M. Rafi-Ul-Islam, Z. B. Siu, C. H. Lee. and M. B. A. Jalil, ``Interfacial Corner Modes in a Topolectrical Heterojunction'', \textit{APS March Meeting Abstracts}, Pages N70. 011, 2022.
     \item S. M. Rafi-Ul-Islam, H. Sahin, Z. B. Siu, and M. B. A. Jalil, ``A Circuit Lattice Representation of a Second-Order Dirac Semimetal'', \textit{Joint MMM-Intermag Conference - AIP Publishing and the IEEE Magnetics Society}, Pages 35, 2022.
     \item S. M. Rafi-Ul-Islam, H. Sahin, C. Sun, Z. B. Siu, J. F. Kong, C. H. Lee and M. B. A. Jalil, ``AC Frequency Modulated Phase Transitions in Topolectrical Circuits'', \textit{Joint MMM-Intermag Conference - AIP Publishing and the IEEE Magnetics Society}, Pages 27, 2022.
    
\end{enumerate}

%\defbibnote{PubListPrenote}{\large Journal Publications}
%\defbibnote{PubListPostnote}{\large Conferences}

%\bookmarksetup{startatroot}
%\newrefcontext[sorting=none]
%\printbibliography[
%  heading=subbibliography,
%  title={Publications during PhD Study},
%  prenote=PubListPrenote,
%  postnote=PubListPostnote,
%] % optional to include your publication list
  \begin{abstract}

A wide array of physical phenomena, despite sharing common conceptual underpinnings, manifest themselves in diverse manner across various physical systems. From the domains of topology to non-Hermitian phenomena, metamaterial platforms lay the groundwork for implementing and observing the fundamental principles of our universe. Electrical circuits, a prime example, not only provide an ideal platform for studying a diverse range of physical phenomena but also mediate to unveil these fundamental principles. The advent of topolectrical (TE) circuits theory marks a groundbreaking development in exploring topological concepts, significantly expanding the application of electrical circuits across the entire field of physics. In this thesis, we will initially delve into the fundamental characteristics of TE circuits and subsequently extend our discussion to their applications in profound physical phenomena.

Our initial investigation delves into the voltage responses of TE circuits. While a simple one-dimensional circuit array might typically be expected to display a trivial voltage response to current injection, our research reveals a more complex dynamic. We will demonstrate that the fundamental response of the simplest electrical circuit array can significantly form the responses of more complex circuit networks. Moreover, while global structures of the electrical circuits, such as boundary conditions, categorize the voltage responses into unilateral and bilateral, the local features like the excitation position define the lateral extent of voltage existence. It shows a clear dependency on the parity of the total number of nodes in the circuit, as well as on the specific node where the current is introduced.

We will later extend our investigation to the two-point impedance characteristics of TE circuits. We reveal the size-dependent impedance responses in various dimensional LC circuits and TE circuits, providing comprehensive analytical methods and expressions for open boundary circuits. These impedance resonances challenge the current understanding of LC circuit resonance conditions, previously thought to depend only on the parameter space, and not the size of the LC network. Experimental verification for these size-dependent impedance responses will be provided, with both theoretical and experimental analyses revealing the fractal scaling of these responses due to the circuit size.

Building on these insights, we will present implementations of TE circuits that encompass non-linear dynamics. Whilst most TE implementations involve linear systems, we will push our exploration into the interplay of topology and non-linear systems in a topological Su–Schrieffer–Heeger circuit lattice with onsite chaotic Chua’s circuits. This includes introducing the concept of topological chaos and a novel method for studying non-linear topological systems.

Following this, we present a real-world application of TE circuits in a non-Hermitian parity-time (PT) symmetric circuit, proposing a sensitive sensor application. The sensitivity of our model relies on integrating profound concepts of topology, non-Hermitian physics, and PT symmetry, demonstrated through realistic circuit simulations. The major advantage of our proposed circuit is its ability to modulate the dynamics of topological modes through a single coupling, translating into measurable real-world applications.

Therefore, this thesis provides a comprehensive study on the fundamental properties of electrical circuits and their implementations in condensed matter physics phenomena.
\end{abstract}

  \listoffigures
  \listoftables
\end{frontmatter}
  
\SetPicSubDir{ch-Intro}

\chapter{Introduction to the Thesis}
\vspace{2em}

Since its introduction to the field in 2018, topolectrical (TE) circuits theory has been instrumental in realizing diverse phenomena in condensed matter physics, paving the way for a nuanced exploration of their fundamental characteristics, sophisticated behaviors, and potential for innovative real-world applications as will be explored in this thesis.

The first pioneering paper titled Topolectrical Circuit by Lee. et.al~\cite{lee_topolectrical_2018} includes an introductory review to the Laplacian formalism relating the circuit Laplacian to the condensed matter counterpart of Hamiltonian. The initial interest in TE circuits emerged due to their resemblance to quantum mechanics (QM), specifically the parallel between the circuit Laplacian in circuits and the Schr\"odinger equation of QM, in which the Hamiltonian is the fundamental element providing the mathematical framework for a physical system. This concept is crucial for understanding a wide range of materials, including graphene and its variants such as silicene and black phosphorus~\cite{rafi-ul-islam_valley_2022}. These materials are often modeled as lattices, representing atoms or, in more detailed models, the individual atomic orbitals. Such models focus on the movement of electrons between these atomic sites. The Hamiltonians that describe the electron transitions between adjacent atomic sites or orbitals are known as `tight-binding (TB)' Hamiltonians. On the other hand, while the positions of electrons are represented by lattice sites within a TB Hamiltonian, the voltage nodes are a direct analogy to this TB Hamiltonian in a TE circuit. Both the TB Hamiltonian and the Laplacian matrix define the connections between these points, yet they are distinct. The Hamiltonian is Hermitian (not always the case, but we will later discuss the non-Hermitian case), meaning it is equal to its own conjugate transpose ($H = H^\dagger$), a property not shared by the Laplacian. The Laplacian involves the anti-Hermitian combination ($J = -J^\dagger$), which can be transformed the Laplacian matrix into a Hermitian Laplacian by disregarding the imaginary unit $i$ since it is a common product to the Laplacian due to the admittance of the components such as capacitors and inductors. This simplification leads to using the Laplacian itself multiplied by $i$, or alternatively by $i\omega$, which is a common practice in our research field. This results in a modified Laplacian matrix that we refer to as the `normalized Laplacian' in which the entries are real, which is more suitable for analyzing the features of a TE system. However, although the eigenvalues of the normalized circuit Laplacian are real as satisfying the Hermicity, the eigenvalues of the Laplacian matrix itself are imaginary due to the common prefix of imaginary unit $i$. 
\begin{figure}
    \centering
    \includegraphics[width=14cm]{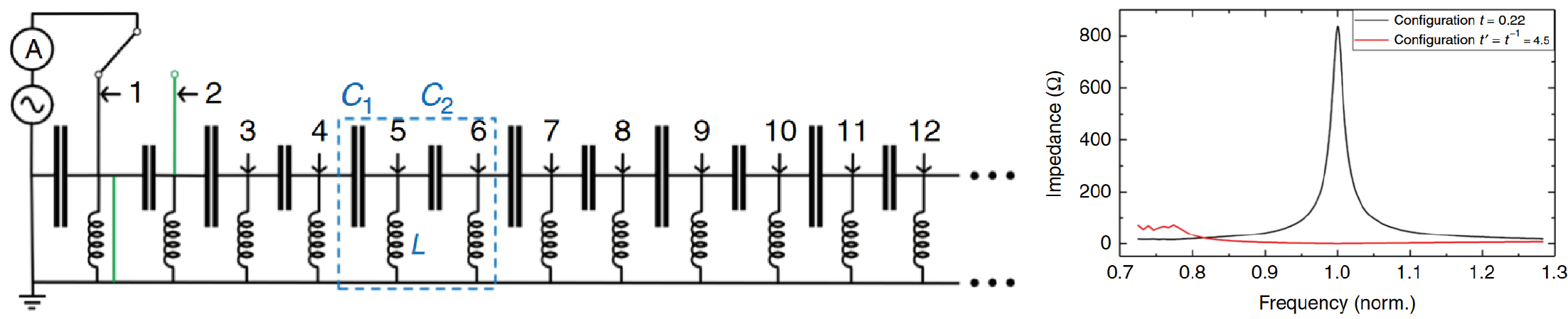}
    \caption{The figure presents the first topolectrical SSH model, comprised of commonly used components like capacitors and inductors. The right panel displays the inaugural experimental realization of the impedance response in topolectrical circuits. It demonstrates a high impedance in the topologically non-trivial setting, contrasting with the absence of impedance resonance in the trivial phase. This figure is adopted from Ref.~\cite{lee_topolectrical_2018}.}
    \label{fig:enter-label}
\end{figure}

In a TE circuit, the presence or absence of a zero eigenvalue in a normalized Laplacian matrix is of significant relevance. Considering the TB correspondence of the normalized circuit Laplacian ($L$), the time-independent Schrödinger equation can be formulated as $L|\psi\rangle=E|\psi\rangle$, where $|\psi\rangle$ is an eigenvector of $L$ with energy $E$. Notably, when the TE Laplacian matrix includes a zero eigenvalue, it results in $L|\psi_0\rangle=|0\rangle$ where $|0\rangle$ is a vector of 0 with the same dimension of $L$. This equation demonstrates that the application of the matrix $L$ to the eigenvector $|\psi_0\rangle$ yields a zero vector. This concept indicates that the right side of the equation signifies the absence of current sources at each node. Consequently, the eigenvector $|\psi_0\rangle$ represents a voltage distribution across the nodes that complies with Kirchhoff's current law (KCL), eliminating the need for external current sources. It is crucial to note that this scenario does not violate energy conservation principles, as the system must initially be energized to establish this voltage distribution. The remarkable output is the correspondence between the KCL and the eigenvector of $L$ corresponding to a TB Hamiltonian. This results in a voltage profile with the same as the spatial distribution of the eigenstate of $L$. Specifically, since the topological zero modes (TZMs) are present in the non-trivial topological phase and their eigenstate distribution exhibits an exponential localization at the boundaries, the voltage response of a topologically designed electrical circuit must comply with the eigenstate distribution of the TZMs. This output is one of the main hallmarks of the TE circuits. Another important aspect of the zero eigenvalue is its relationship with the current frequency. When considering the TE Laplacian, denoted as $L$ as a function of frequency $\omega$, the presence of a zero eigenvalue indicates a resonant frequency within the circuit. This resonant frequency is characterized by a notable increase in the circuit's impedance, as evidenced in Ref.~\cite{lee_topolectrical_2018}. 

\begin{figure}
    \centering
    \includegraphics[width=14cm]{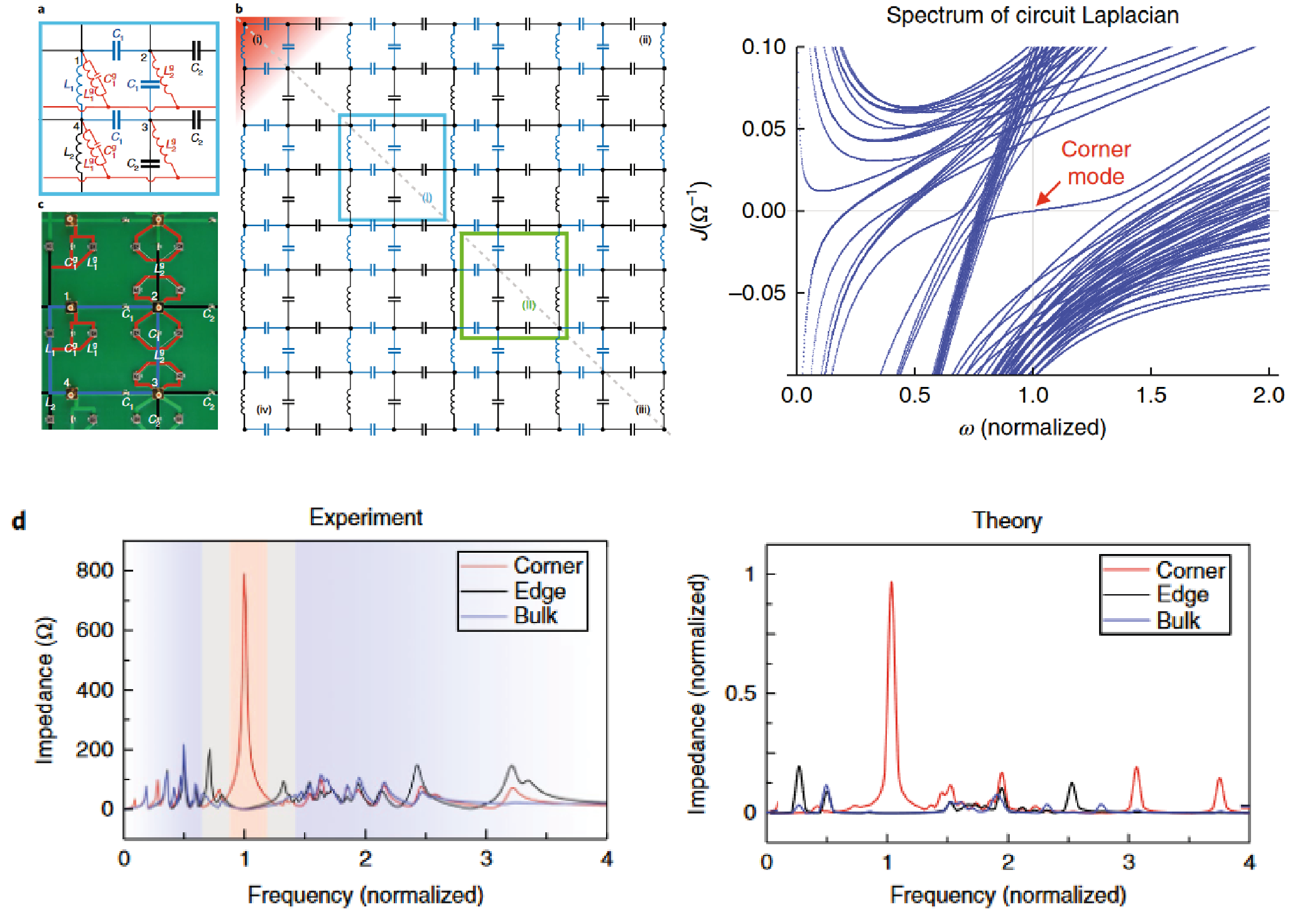}
    \caption{This figure showcases the first comprehensive realization of topological corner modes in topolectrical circuits. The figure is adopted from Ref.~\cite{imhof_topolectrical-circuit_2018}.}
    \label{fig:TEcornermodes}
\end{figure}

The establishment of this fundamental correspondence has led to various circuit implementations in the years following the introduction of the TE concept. For instance, one of the earliest realizations of topological boundary modes was the topological corner modes, as demonstrated by the high corner impedance in a two-dimensional (2D) TE circuit, as referenced in Imhof et al. (2018)~\cite{imhof_topolectrical-circuit_2018}. The studied 2D circuit exhibits high corner impedance when mirror symmetry is preserved alongside an appropriate open boundary configuration. The termination at the corners keeps the mirror symmetry invariant, which depends on the choice of the unit cell in the circuit. This initial experimental work showcased a quadrupole insulator with topologically protected corner states in a 2D TE circuit, characterized by the impedance resonance features of TE circuits (see Fig.~\ref{fig:TEcornermodes}). 

In the evolving landscape of condensed matter physics, a following work by the same leading group~\cite{helbig_band_2019} marks a pivotal advancement, which introduces an innovative approach for designing, engineering, and measuring band structures in synthetic crystals composed of electric circuit elements. Central to this methodology is the novel Laplacian formalism for synthetic matter, enabling the exploration of various TB models through Laplacian eigenmodes. Significantly, this study demonstrates this technique in a honeycomb circuit, revealing a Dirac cone admittance bulk dispersion and distinct flat band admittance edge modes. Furthermore, it delves into the analysis of topological phase transitions, exemplified by the measurement in a TE Su-Schrieffer-Heeger (SSH) circuit. This paper not only underscores the versatility and advantages of using electric circuits in synthetic matter studies but also opens new horizons for understanding complex physical phenomena in synthetic systems, highlighting the potential of electric circuit networks in unlocking fundamental insights into topological band structures and beyond. 

Alongside the TE realizations with the passive components, the paper by Hofmann et.al.~\cite{hofmann2019chiral} in 2019 introduces the concept of a topolectrical Chern circuit (TCC), which is characterized by topologically protected unidirectional voltage modes at its boundaries via active components. A key feature of the TCC is its use of negative impedance converters with current inversion (INICs), enabling the creation of a nonreciprocal, time-reversal symmetry-broken electronic network. The TCC is notable for its admittance bulk gap, which can be fully tuned using the resistors in the INICs, along with a chiral voltage boundary mode that reflects the Berry flux monopole observed in the admittance band structure. This study emphasizes the capability to calibrate active circuit elements within the TCC. The TCC is an innovative advancement in topological circuitry, showcasing a topological voltage Chern mode induced by the non-reciprocity of the INICs. 

This advancement of INICs has paved the way for further exploration and analysis of TEs with active components. A later experimental paper by Helbig et.al.~\cite{helbig_generalized_2020} represents a significant advancement in the context of non-Hermitian topological TEs. This study delves into the concept of bulk–boundary correspondence (BBC) in non-Hermitian systems, which is a fundamental principle that connects the surface states of a material to its bulk topological properties. Through a non-Hermitian TE circuit, the realization of generalized BBC, which constitutes a unique characteristic of non-Hermitian systems is reported via the incorporation of gain and loss elements, and the violation of reciprocity. These elements have been predicted to dramatically affect the traditional understanding of BBC. The realization of a generalized BBC contributes significantly to the broader understanding of topological phases in non-Hermitian systems, offering new perspectives and potential applications in the design and analysis of TE circuits.

\begin{figure}
    \centering
    \includegraphics[width=10cm]{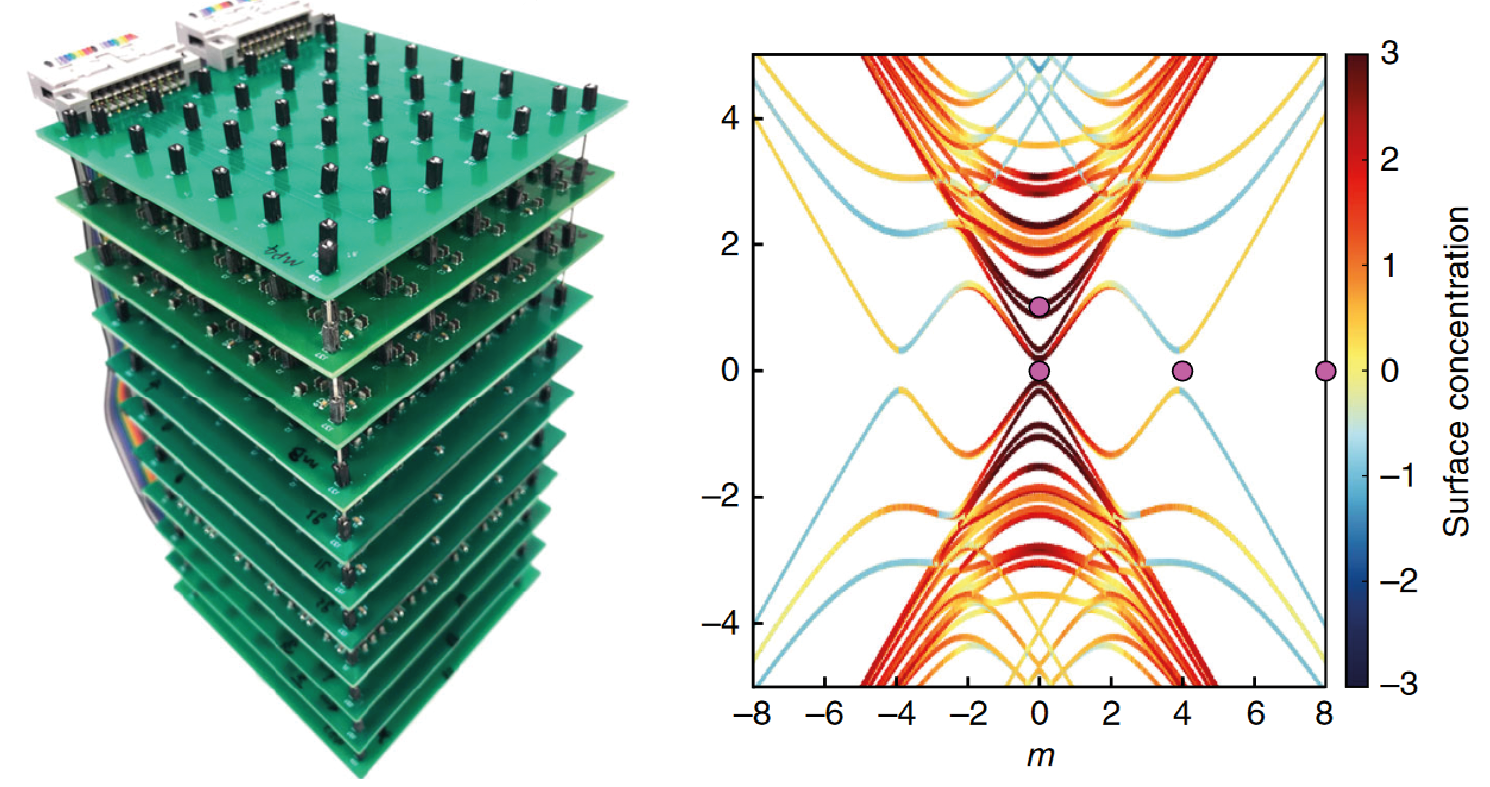}
    \caption{This figure depicts the realization of a four-dimensional topolectrical circuit that hosts surface states on its 3D surface. The figure is adopted from Ref.~\cite{wang_circuit_2020}.}
    \label{fig:TE4D}
\end{figure}

An important superiority of the TE circuits is their independence from the real space dimensions. TE circuits enable us to realize the synthetic dimensions through the proper node connections. A pioneering work by Wang et al.~\cite{wang_circuit_2020} presented a significant leap in the field of topological insulators and the exploration of high-dimensional topological models using TE circuits. The implementation of a four-dimensional Quantum Hall (4DQH) phase, a theoretical concept in the realm of topological insulators that has not been realizable in real materials due to the inherent three-dimensional nature of physical space. The implementation involves a 4D lattice model consisting of four sublattices with sites connected by real nearest neighbor hoppings. The experimental design includes capacitive and inductive connections, allowing the system to transition between a 4DQH phase and a conventional 4D band insulator phase. Impedance measurements equivalent to finding the local density of states (LDOS) were used to demonstrate that the 4DQH phase hosts surface states on the 3D surface (refer to Fig.~\ref{fig:TE4D}). This observation is unrealizable in other platforms, but easily accessible in TE circuits.

\begin{figure}
    \centering
    \includegraphics[width=14cm]{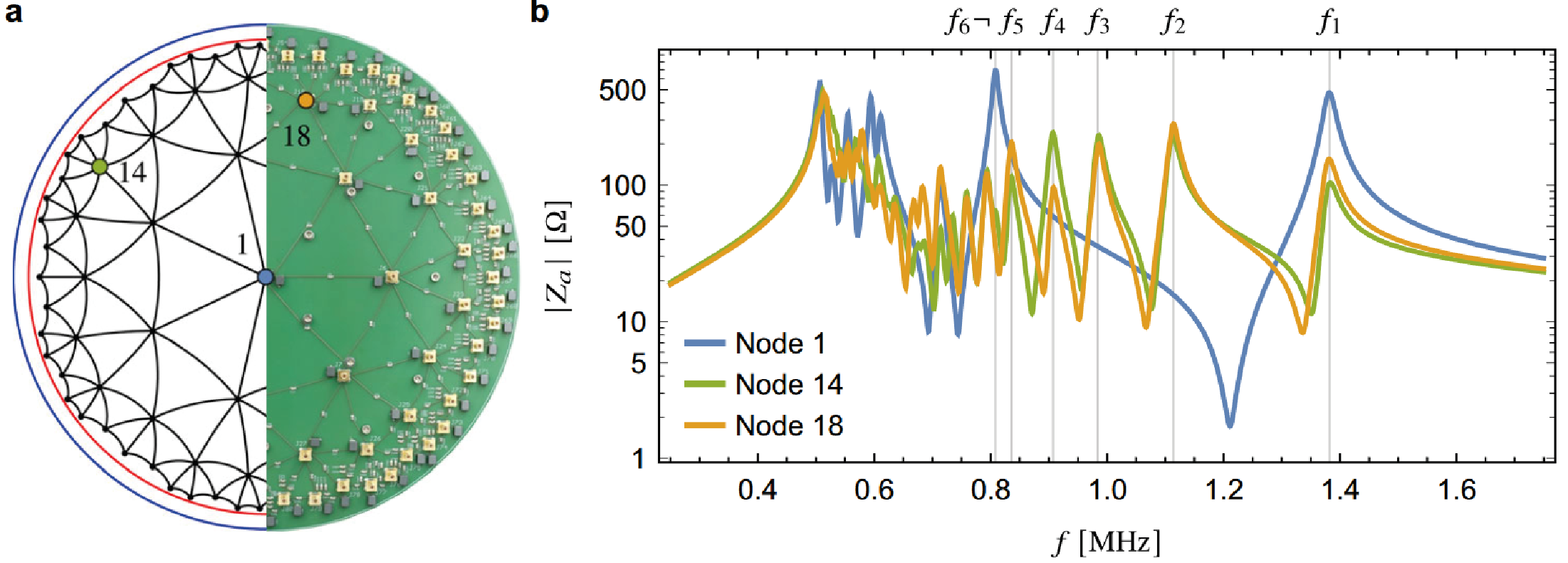}
    \caption{This figure presents the experimental realization of a hyperbolic lattice, implementing a two-dimensional surface with negative curvature. The figure is adopted from Ref.~\cite{lenggenhager2022simulating}.}
    \label{fig:TEhyperbolic}
\end{figure}

Following the exploration of high-dimensional topological models and four-dimensional topological insulators in previous studies, Lenggenhager et al~\cite{lenggenhager2022simulating} presented the innovative concept of emulating hyperbolic space (see Fig.~\ref{fig:TEhyperbolic}), characterized by negative curvature, using a TE circuit network~\cite{chen_hyperbolic_2023}. In this TE, the hyperbolic plane, a two-dimensional surface with negative curvature, is embedded in a $(2 + 1)$-dimensional Minkowski space as a hyperboloid sheet. The TE circuit network effectively simulates the eigenstates of a `hyperbolic drum', allowing the investigation of the eigenmodes and their behavior in this curved space. One of the significant findings of this study is the spectral ordering of Laplacian eigenstates in hyperbolic (negatively curved) and flat two-dimensional spaces, which has a universally different structure. This experiment also demonstrates that TE circuits can serve as a versatile platform to emulate hyperbolic lattices and provide a method to verify the effective hyperbolic metric.

Another flexibility of TE circuits is the ease in realizing long-range couplings. Such long-range couplings are crucial in constructing knot structures in the momentum space. The study presented by Lee et al.~\cite{lee_imaging_2020} is a remarkable example for the visualization of complex knot structures in momentum space. The design and measurement of three-dimensional knotted nodal structures are achieved through the simulations of nodal structures and drumhead state measurements under various boundary conditions, providing a comprehensive analysis of the complex structures in momentum space. The experimental results involve points in reciprocal space corresponding to different admittance eigenvalues, which show the visualization of various knot structures, such as the Hopf-link, and trefoil knot.

A further implementation of TE circuits is the non-linear topological systems. An earlier experimental example to non-linear implementation in TE circuits was presented by Kotwal et.al~\cite{kotwal_active_2021}, which introduces active topolectrical circuits (ATCs), combining the principles of topolectrical circuits with non-linear van der Pol (vdP) circuits. ATCs utilize nonlinear vdP circuits to create autonomous units that self-organize into collective, coherent dynamics, producing self-excited topologically protected signal patterns. The study confirms the emergence of self-organized protected edge oscillations in 1D and 2D ATCs, demonstrating that nonlinear ATCs are a robust platform for high-dimensional autonomous TE circuits with topologically protected functionalities. The experiments include the creation of self-organized topological wave patterns in active nonlinear electronic circuits, aligning closely with predictions from a generic mathematical model.

Non-linear systems have been extensively studied within photonic platforms, largely due to the intricate interactions between light and matter~\cite{berger1998nonlinear}. While earlier efforts in implementing TE circuits primarily focused on linear systems, recent discussions have showcased their potential in the exploration of non-linear topological phenomena~\cite{sahin_protected_2025}. Historically, classical electrical circuits have been employed to investigate complex non-linear systems. Notable examples include oscillatory circuits with components exhibiting non-linear current-voltage (IV) characteristics, such as the van der Pol (vdP) circuit and Chua's diode, explored in depth by Leon Chua over the past decade~\cite{leon_chua_ten_years}. This electronic component is unique due to its non-linear, asymmetric conductivity, which makes it an ideal subject for studying non-linear dynamics and chaos theory. Chua's diode has been a cornerstone in the exploration of chaotic circuits, contributing significantly to our understanding of non-linear electronic systems. An intriguing question arises when considering the intersection of topological elements with non-linear characteristics: what are the resulting behaviors, and how can topological signatures be identified within these non-linear systems? This thesis delves into this query, offering a comprehensive analysis that paves the way for a deeper understanding of non-linear topological systems, particularly through the lens of chaotic TE circuits.

In addition to the non-linear topology, the non-Hermitian phenomenon represents a recent direction in physics. It unveils a new avenue for understanding interactive systems that interact with their environments. The most intriguing characteristic of non-Hermitian systems is the non-Hermitian skin effect (NHSE), wherein all eigenstates display exponential boundary localization. Research on NHSE dynamics has yielded new insights in physics, such as the generalized BZ and spectral graph topology\cite{lin_topological_2023}. One of the earliest examples of a non-Hermitian (NH) Hamiltonian was studied through a PT-symmetric Hamiltonian, which described ring resonators. This Hamiltonian is represented as $H_\text{PT}=\begin{pmatrix} \omega+i\gamma & \kappa \\ \kappa & \omega-i\gamma \end{pmatrix}$, where $\pm\gamma$ are the gain and loss terms, respectively, $\omega$ is the natural frequency of both systems, and $\kappa$ is the coupling term. In this NH Hamiltonian, the eigenvalues are always complex when the polarities of both $\gamma$ in the diagonal elements are identical, indicating that both systems are either gainy or lossy. Conversely, when the polarities of $\gamma$ are opposite, a unique feature emerges. In such cases, the eigenvalues are expressed as $E=\omega\pm\sqrt{\kappa^2-\gamma^2}$. Despite being an NH Hamiltonian ($H_\text{PT}^\dagger\neq H_\text{PT}$), the eigenvalues remain real when $\kappa>\gamma$ (indicating a PT-symmetric phase) but become complex when $\kappa<\gamma$ (signifying a broken PT-symmetric phase). The scenario where $|\kappa|=|\gamma|$ leads to a particularly compelling feature known as an exceptional point (EP), at which the eigenfrequencies of the two systems become identical. This profound phenomenon has garnered considerable attention due to the system's heightened sensitivity to perturbations at the EP, exhibiting sensitivity as a function of the square root of the perturbation. For instance, a perturbation $\varepsilon_\kappa$ in the coupling between two ring resonators introduces an additional term to the eigenvalue spectrum, such as $\omega = \omega_0 \pm \sqrt{(\kappa+\varepsilon_\kappa)^2 - \gamma^2}$. This sensitivity is particularly advantageous for sensing purposes, including single-particle detection. Various implementations have been introduced across different platforms, such as electrical circuits, photonic crystals, and acoustic metamaterials. Alongside the EP characteristic of PT-symmetric systems, a significant observation involves defect engineering in lattices with alternating onsite gain and loss, as discussed in Ref.~\cite{stegmaier_topological_2021}. In this TE model, a 1D SSH chain with a defective node, where translational symmetry is broken, exhibits distinct topological and defect states depending on the PT, broken PT, and APT phases. Defect engineering based on PT symmetry reveals that the eigenvalue spectrum in the PT phase is entirely imaginary, becoming complex in the broken PT phase. Conversely, the eigenvalues shift to the real plane in the APT phase. Interestingly, while defect states appear in the PT and APT phases, they disappear in the broken PT phase. Notably, the topological edge states are present in all PT phases.

Particularly interesting is the Non-Hermitian Skin Effect resulting from predominantly non-reciprocal coupling mechanisms~\cite{sahin_unconventional_2022,sahin_interfacial_2022}. These non-reciprocal couplings lead to a non-unitary Hamiltonian, where eigenstates are amplified under open boundary conditions, yet this mechanism is absent under periodic boundary conditions. The dependency of NHSE on singular coupling underscores its sensitivity to boundary conditions, a feature proposed for sensing applications~\cite{budich_non-hermitian_2020}. In this thesis, we delve further into the sensitivity response of non-Hermitian systems through a proposal for a PT-symmetric, non-Hermitian, and topological circuit. Our circuit proposal integrates concepts from the aforementioned papers, offering a practical application owing to the versatility of TE circuits.

\subsection{Overview of thesis structure}

This thesis embarks on a comprehensive exploration of topolectrical circuits, aiming to bridge the gap between theoretical research and practical applications in this field. TE circuits theory has become a key instrument in interpreting various phenomena in condensed matter physics. This work delves into the subtleties of TE circuits, exploring their complex behaviors, essential features, and real-world applicability.

Chapter 2 initiates a detailed examination of parity-dependent voltage responses in TE circuits. By studying simple one-dimensional electrical circuits, analogous to the 1D free electron gas in condensed matter physics, the chapter illuminates intricate voltage behaviors in both topological and non-Hermitian circuits. This investigation not only uncovers the fundamental traits of these circuits but also highlights their exponential voltage localization properties, setting a foundation for understanding their operational dynamics in diverse conditions, essential for practical use.

Chapter 3 further investigates electrical circuits by examining their impedance responses. Challenging conventional beliefs, our research shows that circuit size can significantly impact impedance resonance. This insight is gleaned through two-point impedance measurements in various dimensional LC classical and topolectrical circuits, revealing the emergence of fractal structures in circuit size and parameter space, thus contributing new perspectives to circuit theory.

Chapter 4 connects to Chapter 3 by experimentally validating the size-dependent impedance resonances. The chapter focuses on impedance scaling in a 2D LC circuit, confirming the applicability and consistency of our theoretical insights.

In Chapter 5, the concept of topological chaos is introduced by incorporating non-linear dynamics into TE circuits. This chapter explores the interaction between chaos and topology, especially through merging the topological 1D SSH circuit with Chua's chaotic circuit. This combination results in topological chaos, broadening the scope of TE circuits beyond linear models and offering novel methods to analyze the dynamics of both topological and non-linear circuits.

Chapter 6 concludes the thesis by proposing a practical application for TE circuits. It demonstrates a single circuit model that integrates topological, non-Hermitian, and PT-symmetry elements, showing highly sensitive responses at PT-symmetric phase transition points. Simulations under realistic conditions validate the model's effectiveness and highlight the practicality of TE circuits in real-world scenarios.

Overall, this thesis has a dual goal: to deepen the theoretical understanding of TE circuits and to showcase their practical applications. By fusing theoretical analysis with experimental proof and practical implementation, this work contributes significantly to the fields of condensed matter physics and electrical engineering. It underscores the adaptability and potential of TE circuits in enhancing our comprehension of complex physical phenomena.

\begin{figure}
    \centering
    \includegraphics[width=15cm]{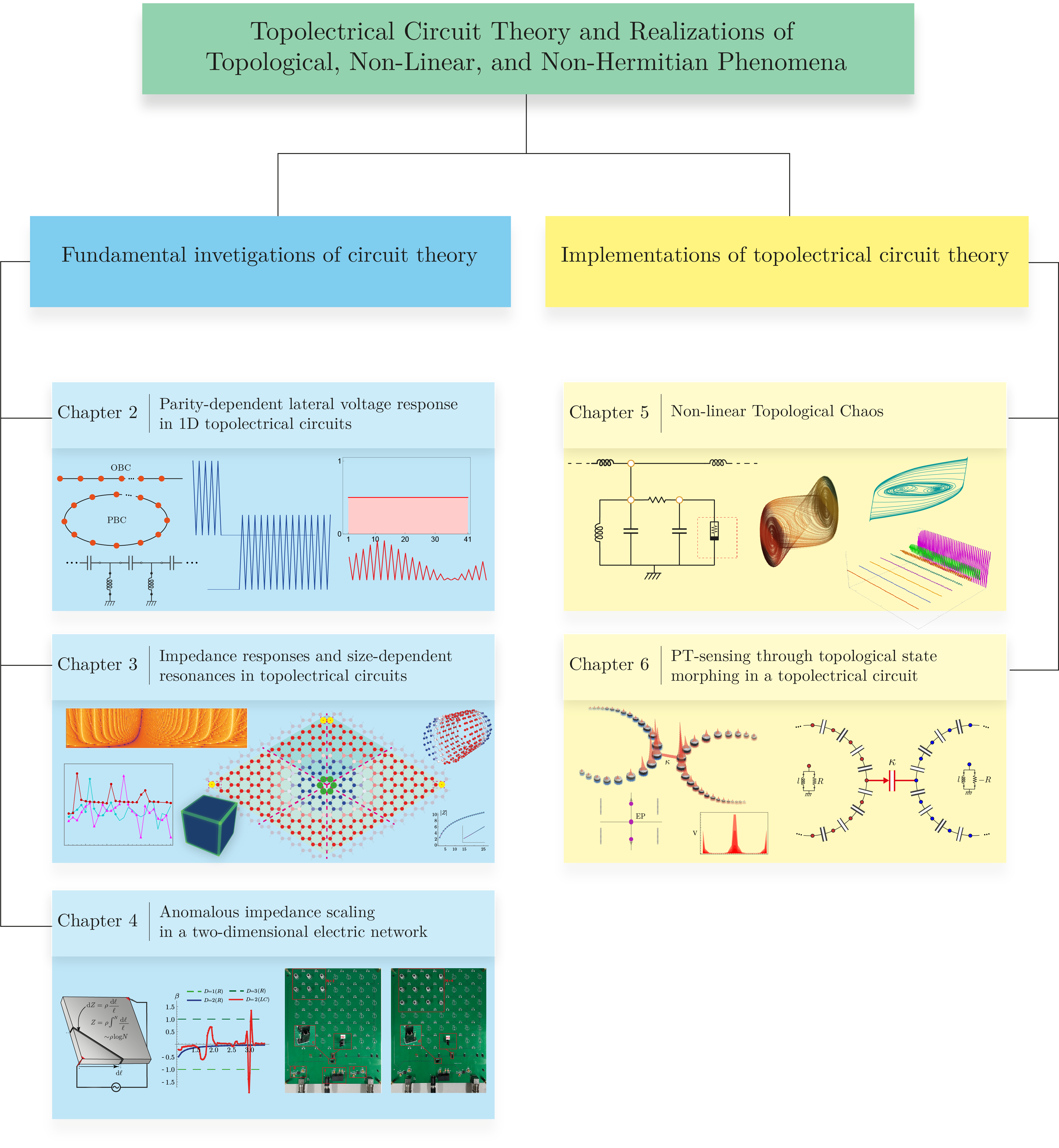}
    \caption*{An overview of the thesis content: Chapters 2, 3, and 4 focus on the fundamental properties of electrical circuits, such as voltage and impedance responses. Based on these voltage and impedance characteristics, Chapters 5 and 6 present the implementation of electrical circuits in condensed matter systems, covering topological, non-linear, and non-Hermitian phenomena.}
    \label{fig:introductionsummary}
\end{figure}
\SetPicSubDir{ch-lateralvoltage}
\SetExpSubDir{ch-lateralvoltage}

\chapter{Parity-dependent lateral voltage response in 1D topolectrical circuits}
\label{ch:lateralvoltage}
\vspace{2em}

In this chapter, we begin to examine the voltage response of electrical circuits, focusing on the parity of the circuit size and the current injection node, rather than on the bulk properties. Conventionally, drastic transitions in condensed matter systems are often attributed to the intriguing bulk properties. For instance, in topological insulators, a phase transition from trivial to non-trivial is determined by the bulk's degree of freedom. However, we unveil that even a simple electrical circuit comprising a single type of node, which is not inherently interesting in terms of its bulk properties, can exhibit a different voltage response upon inclusion or exclusion of a single node. We outline that while boundary conditions determine whether the voltage response is unilateral or bilateral, the parity of the total number of nodes determines the exact profiles of these two supergroups. We classify all possible voltage profiles through rank and determinant analysis of the circuit Laplacian and later examine the lateral voltage response in various circuit models, such as dimer and trimer Su-Schrieffer-Heeger (SSH) and non-Hermitian Hatano-Nelson (HN) circuits. The complex voltage response of these general circuits such as topological or non-Hermitian circuits can be directly understood from the primary classification of the simplest circuit corresponding to the one-dimensional (1D) free electron gas model. Just as diverse biological life forms exhibit complex functions within the framework of simple bilateral symmetry, our circuit, despite its simplicity, demonstrates varied voltage profiles, thereby offering insights into the complex voltage response of more advanced circuits. The lateral voltage response is particularly significant for real-world applications where voltage response is affected by various inevitable effects. Our classification and discussion of realistic circuits would be particularly helpful for experimentalists in analyzing and designing topolectrical (TE) circuits to realize phenomena of condensed matter physics.

Overall, chapter 2 explores the essential aspects of the fundamental characteristics of electrical circuit networks, beginning with an analysis of an elementary one-dimensional circuit and progressing to more complex TE circuits. The unique contribution of the chapter lies in characterizing the fundamental voltage profiles and unveiling their parity-dependent properties.

\subsection{Introduction}
Electrical circuits have proven to be a highly versatile metamaterial platform, covering a range of phenomena from topological phases~\cite{hasan_colloquium_2010,haldane_model_1988,su_solitons_1979,kane_z_2005,rafi-ul-islam_realization_2020,imhof_topolectrical-circuit_2018,helbig_generalized_2020,rafi-ul-islam_critical_2022,rafi-ul-islam_system_2022} to various other aspects of condensed matter systems. This versatility is mainly attributable to the direct correspondence between the tight-binding models and the circuit Laplacian~\cite{lee_topolectrical_2018,jia_time_2015}. A distinctive feature of the TE circuits is their impedance response~\cite{sahin_impedance_2023,zhang_anomalous_2023}, which serves as a hallmark distinguishing between topologically trivial and non-trivial phases. The spatial distribution of eigenmodes of non-trivial topological zero eigenvalues presents exponential localization at the boundaries of the circuit. These boundary modes have been observed in TE circuits as voltage localization across various dimensions, such as edge states in 1D systems and edge, corner and hinge states in higher-order topological systems~\cite{wang_circuit_2020}. In addition to realizing topological phenomena, electrical circuits have successfully demonstrated non-Hermitian~\cite{lin_topological_2023,gong_topological_2018,lee_anomalous_2016,lee_anatomy_2019,kawabata_symmetry_2019,rafi-ul-islam_unconventional_2022,rafi-ul-islam_interfacial_2022,yokomizo_non-bloch_2020} and non-linear phenomena~\cite{kotwal_active_2021,hohmann_observation_2022}, hyperbolic band theory~\cite{lenggenhager_simulating_2022}, and non-Abelian and Floquet mechanisms, thanks to their rich degree of freedom and independency from the real space dimensions. However, although the theory of topolectrical circuits is well-established and has seen wide application, several fundamental questions remain beyond current knowledge. For instance, in a finite 1D SSH circuit, while both eigenmodes corresponding to zero eigenvalues have identical spatial distributions, injecting current at different nodes with opposite parity results in voltage localization at either the left or right edge. Specifically, voltage exponentially localizes at the left edge when current is injected at any even node, and at the right edge for any odd node injection in an ideal SSH TE. While the exponential nature of the voltage is a topological feature, this parity-determined edge localization is related to the fundamental characteristics of electrical circuits, which is the main discussion of this chapter. Indeed, even a trivial 1D circuit consisting of a single-type-node exhibits distinctive voltage responses depending on both the parity of the total number of nodes and the parity of the node where the current is injected. For example, one might intuitively expect a symmetrical voltage profile on either side of this circuit with open boundary conditions (OBC) when injecting current at any node. Yet, what we observe is an asymmetrical voltage profile on either side relative to the current injection node in OBC circuits, a phenomenon we dub as 'unilateral voltage response'. In contrast, circuits with periodic boundary conditions (PBC) exhibit a symmetrical voltage profile on either side, denoted as 'bilateral voltage response'. Particularly, the unilateral voltage response inherent in OBC circuits provides insight into the underlying mechanism of voltage localization direction in the 1D Hermitian and non-Hermitian SSH circuits. Specifically, in all 1D OBC circuits with even sizes, the injected current excites only one side of the circuit: while the voltage on one side is entirely zero, the voltage at nodes with the opposite parity as the injection node is non-zero. In the case of 1D SSH TE circuit, the circuit's response to a current injection is similar, except that the voltage profile has an exponential characteristic.

The lateral voltage responses categorize the two main voltage characteristics based on the boundary conditions, but their exact profiles are determined by the parity of the circuit size. We have found that all possible voltage profiles in generalized 1D topolectrical circuits can be categorized through rank and determinant analysis of the most basic circuit. Intriguingly, despite the simplicity of the single-type-node trivial circuit, it exhibits a richness akin to the diversity seen in biological nature~\cite{wiltschko_lateralization_2002}, where basic symmetries such as radial and bilateral symmetries play a crucial role in forming complex body structures of organisms. Similar to biological systems, we observe diverse voltage responses originating from the simple symmetries of the circuit array. This culminates in two main categorical profiles based on lateral symmetry, and five subcategories based on the parity of the circuit size.

In this study, our initial focus is on ideal 1D circuits with pure reactance to establish a baseline understanding of the lateral voltage response. The analysis begins with the simplest 1D single-type-node circuit, considering both OBC and PBC. For each boundary condition, we explore the circuit's voltage response when a single current source is employed. Our examination specifically targets how voltage profiles vary with different circuit sizes and locations of current injection. Consequently, we categorize the circuits based on the oddness and evenness of their size and examine their responses under a current injection at different nodes. Because the voltage profile can be expressed in terms of the eigenspectrum composition of the circuit Laplacian, we explore the main features of the eigenmodes and eigenvalues for each configuration of the single-type-node circuit. Later, we discuss the lateral voltage response originating from the single-type-node circuit in topolectrical circuits such as dimer SSH, trimer SSH, and non-Hermitian HN circuits. We then extend our investigation to realistic circuits incorporating parasitic elements. This analysis is particularly vital as it may offer significant insights for experimental implementations. We observe that while resistances inherent in these circuits contribute to their stabilization, they also induce an exponentially localized voltage with a phase identical to the current. The real part of the complex voltage results in an additional exponential voltage component caused by the series resistances and significantly alter the desired voltage profile. Nevertheless, our categorization of lateral voltage responses, along with our exploration in more realistic circuits, lays the groundwork for a comprehensive understanding of voltage profiles in topolectrical circuits.

\subsection{Parity dependent lateral voltage response}
To examine the parity dependency of the lateral voltage response, we begin with the simplest trivial single-type-node circuit network where each node is connected by a capacitor with capacitance $c$. One might anticipate that the voltage response of this circuit, in the case of current injection at a single node at one time, would be superficial and identical due to its simple structure. However, it is neither trivial nor intuitive. The ideal non-dissipative circuits would exhibit dependency solely on the boundary conditions alongside the coupling strength. However, we shall demonstrate that the voltage response is also highly dependent on the parity of both the excitation node as well as the total number of nodes in the circuit.

To understand the origin of the parity-dependent lateral voltage response, we consider a circuit lattice with single-type $N$ nodes. Throughout this study, $N$ always stands for the total number of nodes. The OBC real-space circuit Laplacian for this circuit lattice is written as 

\begin{equation}
	L = j\omega\sum_{x=1}^{N-1} (c a_x^\dagger a_{x+1} + c^* a_{x+1}^\dagger a_x) - \sum_{x=1}^{N} \epsilon a_x^\dagger a_x,
	\label{LaplacianRealSpace}
\end{equation}
where $a_x^\dagger$ and $a_x$ are the creation and annihilation operators at site $x$, respectively. $j$ is the imaginary unit and $\omega = 2\pi f$ where $f$ is the frequency of the driving AC signal. $c$ represents the capacitance of the coupling capacitors. $\epsilon$ denotes the total node conductance, which is explicitly defined as $\epsilon = 2j\omega c + (j\omega l)^{-1} + \Delta$, where $l$ is the inductance of the common grounding inductors and $\Delta$ equals zero in the absence of additional onsite components. We assume the circuit operates at its resonant frequency, i.e., $f = f_r$, where $f_r$ is calculated by $(2\pi\sqrt{lc})^{-1}$. Given this condition and $\Delta = 0$, it follows that $\epsilon$ effectively vanishes. Note that the above Laplacian satisfies PBC with $a_{N+1}=a_{1}$ when the sum runs over all $N$. The circuit Laplacian above describes the voltage response of a circuit to a current injection, satisfying the relation $\mathbf{I} = L \mathbf{V}$ where $\mathbf{I}$ and $\mathbf{V}$ represent the current and voltage vectors, respectively. In this study, our objective is to investigate the voltage response based on the parity of both the circuit size and the current injection node. To this end, we employ a single current source set to the resonant frequency $f_r$, and characterized by a constant current magnitude, satisfying $I_i = I \delta_{i,j}$, where $I_i$ denotes the current of magnitude $I$ injected at node $i$. This implies that the current magnitude is $I$ exclusively at the injection node and zero elsewhere, in accordance with Kirchhoff's Current Law (KCL). When the magnitude of the driving AC current remains constant for each injection, implying that the current matrix is indeed an identity vector, then the inverse of the Laplacian matrix corresponds to the voltage matrix. This holds true when the Laplacian matrix can be diagonalized, meaning it is non-singular. To broaden our discussion and include singular cases, we explore the voltage response characteristics in terms of the eigenspace of the Laplacian matrix, as expressed by $L \Psi = \lambda \Psi$. Here, $\Psi$ and $\lambda$ represent the eigenvector and eigenvalue of $L$, respectively. In the case of a Hermitian Laplacian, the left and right eigenvectors form an orthonormal basis satisfying $\Psi_i^\ast \Psi_j = \delta_{i,j}$ where $\delta$ is the Kronecker delta. However, in the case of a non-Hermitian Laplacian matrix, the left and right eigenvectors are generally not the same and do not necessarily form an orthonormal set. Considering that the circuit we describe in \eqref{LaplacianRealSpace} is a trivial single-site circuit, its Laplacian adheres to Hermiticity. To determine the eigenspectrum of the Laplacian $L$, we solve the characteristic equation $(L-\lambda \mathbb{1})\Psi = 0$, where $\mathbb{1}$ denotes the identity matrix, corresponding to the dimension of $L$. Now, since $\mathbf{V} = L^{-1} \mathbf{I}$ where $\Psi^\dagger L^{-1} \Psi = \lambda^{-1}$, we can express the voltage in terms of the eigenspectrum of the circuit Laplacian as
\begin{equation}
	V'(i,x) = I_i \sum_{\alpha=1, \scriptstyle (\lambda_{\alpha} \notin 0)}^{N} \frac{1}{\lambda_\alpha} \Psi_{\alpha,i}^{\ast} \Psi_{\alpha,x},
	\label{voltage1}
\end{equation}
but for our analysis, we will employ an equivalent expression to calculate the node voltage. Instead of summing the $i$th and $x$th components of all eigenvectors, as in \eqref{voltage1}, we sum over their components. This alternative approach involves summing all the components of the $i$th and $x$th eigenvectors. This approach becomes applicable when we organize the eigensystem in ascending order with respect to the eigenvalues. Given that the circuit Laplacian in our initial analysis is a Hermitian matrix, this equivalent summation method simplifies our eigenspectrum analysis. Consequently, the node voltages can now be computed using
\begin{equation}
	V(i,x) = I_i \sum_{\beta=1, \scriptstyle (\lambda_{\beta} \notin 0)}^{N} \frac{1}{\lambda_\beta} \Psi_{i,\beta}^{\ast} \Psi_{x,\beta},
	\label{voltage2}
\end{equation}
where $\beta$ denotes the scalar coefficients of eigenvector $\Psi$ and $x$ indicates the position of the node at which the voltage $V$ is measured. We will now proceed to examine the characteristics of the eigenspectrum of the circuit Laplacian. Subsequently, we will analyze the resultant voltage profiles derived from the eigenspectrum analysis and \eqref{voltage2}.

\begin{figure*}[t!]
	\centering
	\includegraphics[width=\textwidth]{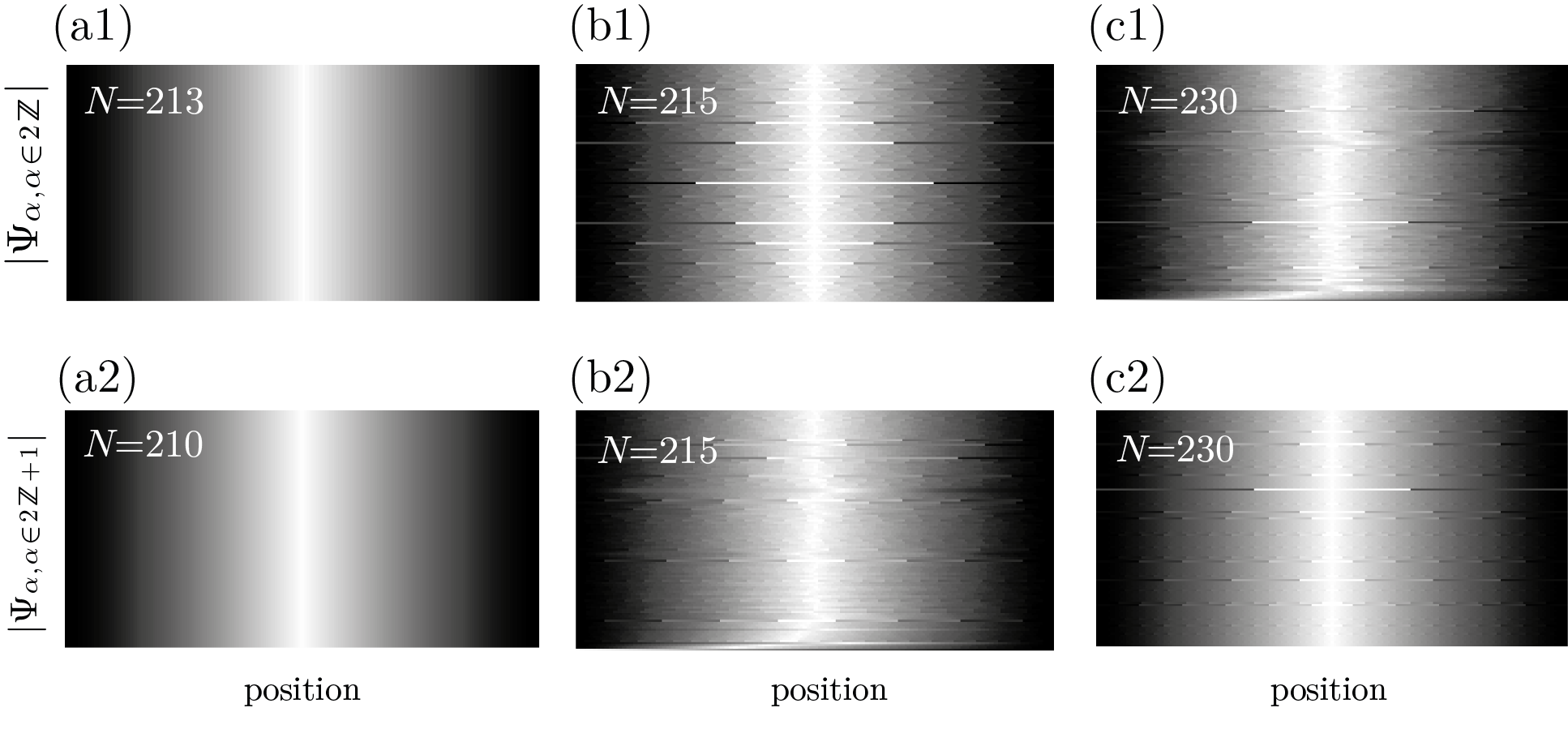}
	\caption{\textbf{Examples of even and odd labeled eigenvectors and emergent fractal structures in eigenvector space.} These plots display the component distributions of even and odd labeled eigenvectors for various sizes of the single-type-node OBC circuit. The eigenvectors are sorted in ascending order of their respective eigenvalues, and the components within each eigenvector are also sorted. \textbf{(a1) and (a2)} For $N=213$ and $N=210$, the even and odd eigenvectors, respectively, exhibit symmetrical component distributions. This symmetry occurs when the parity of the eigenvector label contrasts with the parity of the circuit size. \textbf{(b1) and (b2)} Intriguingly, in some circuit sizes, a fractal-like structure becomes apparent in the distribution of the even-labeled eigenvector components as exemplified in (b1). These fractal-like structures also emerge in other circuit sizes and PBC circuit. \textbf{(c1) and (c2)} When the circuit size is even, the component distribution of the odd-labeled eigenvectors is symmetric. This symmetric scaling is attributed to the paired nature of the eigenvector components.}
	\label{FigFractal}
\end{figure*}

\begin{figure*}[ht!]
	\centering
	\includegraphics[width=13.5cm]{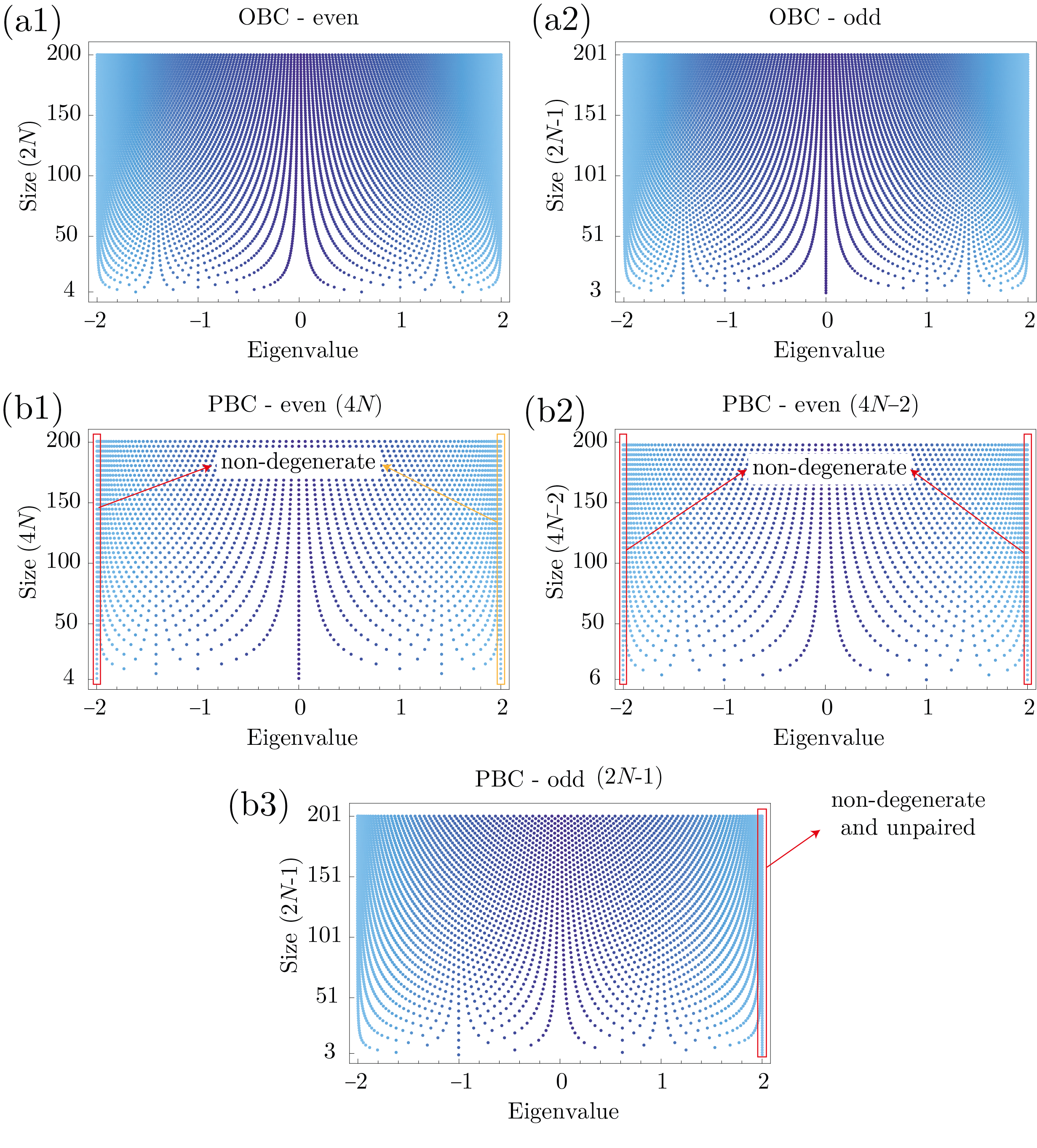}
	\caption{\textbf{Eigenvalue spectra of 1D OBC and PBC single-type-node circuits with even and odd sizes.} In all plots, points of the same color represent eigenvalues with the same magnitude. The parameters used are $c = 1\,\mu$F, $l = 1\,\mu$H. \textbf{(a1)} In even-sized OBC circuits, all eigenvalues form pairs with opposite polarities. \textbf{(a2)} In odd-sized OBC circuits, each eigenvalue pairs with another of opposite polarity. Unlike even-sized circuits, odd-sized circuits include a zero eigenvalue. \textbf{Panel b:} In PBC single-type-node circuits, all eigenvalues are degenerate except those marked by the red and yellow rectangles. \textbf{(b1)}  PBC circuits with sizes that are multiples of 4 exhibit two sets of doubly degenerate eigenvalues, including two zero eigenvalues. The eigenvectors corresponding to eigenvalues marked by the yellow rectangle have negative constant components. \textbf{(b2)} For PBC circuits with sizes that are multiples of $4N-2$, the eigenvalues form pairs. \textbf{(b3)} Odd-sized PBC circuits feature a unique unpaired eigenvalue. }
	\label{FigEigenvalue}
\end{figure*}

Initially, let us consider the circuit has OBC and $N$ is even. The eigenvalues of $L$ consist of pairs of the same magnitude but with opposite signs due to the chiral symmetry defined as $\mathcal{C}^{-1} L^\intercal \mathcal{C} = - L $ where $\mathcal{C}$ is a unitary matrix. The eigenvalues are denoted as $\lambda = \{-\lambda_1, +\lambda_1, -\lambda_2, +\lambda_2,\dots,-\lambda_{N/2},+\lambda_{N/2} \}$. Here, each number in subscript varying from 1 to $N/2$ indicates the eigenvalues of the same magnitude. We sort the eigenvalues and the corresponding eigenvectors such that $\lambda = \{-\lambda_{1}, -\lambda_{2}, \ldots, -\lambda_{N/2},+\lambda_{N/2},\ldots,+\lambda_2,+\lambda_1 \}$ and $\Psi = \{ \Psi_{-\lambda_1}, \Psi_{-\lambda_2}, \ldots, \Psi_{+\lambda_2}, \Psi_{+\lambda_1} \}$. Here, for the feasibility of the discussion, we replace the subscripts of each column vector with $\alpha$ varying from $1$ to $N$. For an odd $\alpha$, we have\\ $\Psi_\alpha = \{\psi_{(1,\pm)},\psi_{(1,\mp)},\psi_{(2,\pm)},\psi_{(2,\mp)},\dots,\psi_{(N/2,\pm)},\psi_{(N/2,\mp)}\}^{\intercal}$, and for an even $\alpha$, we have $\Psi_\alpha = \{\psi_{(1,\pm)},\psi_{(1,\pm)},\psi_{(2,\pm)},\psi_{(2,\pm)},\dots,\psi_{(N/2,\pm)},\psi_{(N/2,\pm)} \}^{\intercal}$. Here, the notation $(\cdot \,\ \cdot)$ in the subscripts assigns numbers to the components of $\Psi_\alpha$ with the same magnitude and labels the polarity combination of each pair. The three key features of eigenspectrum emerging from the sign of the elements of each pair can be summarized as
\begin{enumerate}
	\renewcommand{\labelenumi}{\roman{enumi})}
	\item $|\lambda_\alpha^\pm| = |\lambda_\alpha^\mp|$, where $|\cdot|$ denotes the cardinality. There are equal number of negative and positive eigenvalues each which has a pair with opposite sign, due to the chiral symmetry (see Fig.~\ref{FigEigenvalue}a1).
	\item $ |\psi_\beta^\pm| = |\psi_\beta^\mp| $ when $\alpha$ is odd. Each element of a pair labeled by $\beta$ has an opposite polarity. There are equal number of positive and negative scalar elements, which is always symmetric (Fig.~\ref{FigFractal}a2). This leads to $\sum_\beta \psi_{\beta} = 0$.
	\item When $\alpha$ is even, the elements of a pair have the same polarity. The total number of positive and negative scalar elements can be asymmetric (Fig.~\ref{FigFractal}c1). This leads to $\sum_\beta \psi_{\beta} \neq 0$.
\end{enumerate}
These three key behaviors always hold true for an even $N$ due to the linear algebraic features of the Laplacian matrix $L$. 

Now, let us examine the scenarios where $N$ is odd. In these cases, the eigenvalues form pairs with the same magnitudes but opposite polarities with the notable exception of a single zero eigenvalue (refer to Fig.~\ref{FigEigenvalue}a2). Similar to even $N$ cases, we sort the eigenvalues and their corresponding eigenvectors in the following manner: $\lambda = \{-\lambda_{1}, -\lambda_{2}, \ldots, -\lambda_{N/2},0,+\lambda_{N/2},\ldots,+\lambda_2,+\lambda_1 \}$ and $\Psi = \{ \Psi_{-\lambda_1}, \Psi_{-\lambda_2}, \ldots, \Psi_{+\lambda_2}, \Psi_{+\lambda_1} \}$. This arrangement helps in succinctly outlining the three key features of the eigenspectrum as follows:
\begin{enumerate}
	\renewcommand{\labelenumi}{\roman{enumi})}
	\item Out of the eigenvalues, $N-1$ of them fulfill the condition $|\lambda_\alpha^\pm| = |\lambda_\alpha^\mp|$, and one eigenvalue specifically satisfies $\lambda_{(N/2)+1} = 0$.
	\item When $\alpha$ is odd, there are $N-1$ pairs of scalar elements for which the condition $|\psi_\beta^\pm| = |\psi_\beta^\pm|$ holds. However, there is consistently one scalar element that remains unpaired (Fig.~\ref{FigFractal}b2). This results in $\sum_\beta \psi_{\beta} \neq 0$.
	\item When $\alpha$ is even, there are a total of $N-1$ pairs of scalar elements satisfying the condition $|\psi_\beta^\pm| = |\psi_\beta^\mp|$ (Fig.~\ref{FigFractal}b1). The remaining scalar element is always zero. Therefore, this leads to $\sum_\beta \psi_{\beta} = 0$.
\end{enumerate}

Intriguingly, the distribution of symmetric eigenvector pairs, depending on the parity of $\alpha$, leads to the formation of fractal patterns. These fractal-like distributions in the eigenvectors emerge when the parity of $N$ and $\alpha$ are opposite, as illustrated in Fig.\ref{FigFractal}b1. Conversely, in cases of asymmetric eigenvector pairs as discussed earlier, the distribution of eigenvector components appears noisy. Such scenarios occur when $N$ and $\alpha$ share the same parity, exemplified in Fig.\ref{FigFractal}b2 and c1. The distinction between symmetric and asymmetric pairs of eigenvector components and eigenvalues is crucial, as it underpins the voltage profile which we will discuss in more detail later.

We now expand the eigenspectrum analysis from the OBC single-type-node circuit to the PBC, applying the same methodology. Similarly, we begin by sorting the eigenvalues and their corresponding sets of eigenvectors. Following this, we relabel the eigenvectors as $\alpha$ and their components as $\beta$. This process leads us to identify several key features of the eigenspectrum under PBC, which are summarized as follows:
\begin{enumerate}
	\renewcommand{\labelenumi}{\roman{enumi})}
	\item For a circuit size of $4N$:
		\begin{enumerate}
		\item There are a total of $(N-1)/2$ eigenvalue pairs with opposite polarity, satisfying $|\lambda_\alpha^\pm| = |\lambda_\alpha^\mp|$. Excluding the two non-degenerate eigenvalues, as shown in Fig.~\ref{FigEigenvalue}b1, each of these pairs is degenerate. This results in a total of 4 eigenvalues for each pair, all having the same magnitude.
		\item For both even and odd values of $\alpha$, the components of $N-1$ eigenvectors typically form pairs with opposite polarity. This pairing leads to a sum of $\sum_\beta \psi_\beta=0$. The notable exception is the eigenvector corresponding to a single non-degenerate eigenvalue, which is highlighted in yellow in Fig.~\ref{FigEigenvalue}b1. The components of this particular eigenvector are consistently the same and exhibit a negative value.
		\end{enumerate}
	\item For a circuit size of $4N-2$:
		\begin{enumerate}
			\item There are in total $N/2$ eigenvalue pairs with opposite polarity, each satisfying $|\lambda_\alpha^\pm| = |\lambda_\alpha^\mp|$, and none of these pairs is zero. 
			\item In all cases of $\alpha$, be it even or odd, the eigenvector components align in pairs of contrasting polarity, leading to a total sum of $\sum_\beta \psi_\beta=0$. An exception exists for the eigenvector associated with the non-zero unpaired eigenvalue, wherein the sum of its components invariably remains non-zero.
		\end{enumerate}
	\item When $N$ is odd
		\begin{enumerate}
		\item While a total of $(N-1)/2$ eigenvalues form pairs with the same polarity, satisfying $|\lambda_\alpha^\pm| = |\lambda_\alpha^\pm|$, there exists one non-zero unpaired eigenvalue (see Fig.~\ref{FigEigenvalue}b3).
		\item Excluding the eigenvector associated with the non-zero unpaired eigenvalue, the summation of the components of all other eigenvectors consistently equals zero, denoted as $\sum_\beta \psi_\beta=0$. However, a notable behavior is that the unpaired components of exactly $(N+1)/2$ eigenvectors is non-zero, whereas the unpaired components of exactly $(N-1)/2$ eigenvectors are zero, i.e., $\Psi_{\alpha1\in\{1,\ldots,(N+1)/2\}} =\{ \psi_1, \psi_2, \ldots, \psi_{N-1}, \psi_N\neq0\} $ and $\Psi_{\alpha2\in\{1,2,\ldots,(N-1)/2\}} =\{ \psi_1, \psi_2,  \ldots, \psi_{N-1}, \psi_N=0\} $.
		\end{enumerate}
\end{enumerate}

\begin{table}[t!]
	\renewcommand{\arraystretch}{2}
	\centering
	\begin{tabular}{|c||cc|ccc|}
		\hline
		\textbf{Voltage}               & \multicolumn{2}{c|}{\textbf{Unilateral}} & \multicolumn{3}{c|}{\textbf{Bilateral}}                      \\ \hhline{|=||==|===|}
		\textbf{Boundary}              & \multicolumn{2}{c|}{OBC}                 & \multicolumn{3}{c|}{PBC}                                     \\ \hline
		\multirow{2}{*}{\textbf{Size}} & \multicolumn{1}{c|}{    even   }     & odd      & \multicolumn{2}{c|}{even}                             & odd  \\ \cline{2-6} 
		& \multicolumn{1}{c|}{\(2N\)}       & \(2N-1\)     & \multicolumn{1}{c|}{\(4N\)}   & \multicolumn{1}{c|}{\(4N-2\)} & \(2N-1\) \\ \hline
		\textbf{Rank}                  & \multicolumn{1}{c|}{\(2N\)}       & \(2N-2\)     & \multicolumn{1}{c|}{\(4N-2\)} & \multicolumn{1}{c|}{\(4N-2\)} & \(2N-1\) \\ \hline
		\textbf{Determinant}           & \multicolumn{1}{c|}{\(\pm1\)}        & \(0\)        & \multicolumn{1}{c|}{\(0\)}    & \multicolumn{1}{c|}{\(-4\)}   & \(2\)    \\ \hline
	\end{tabular}
	\caption{\textbf{The general classification of the lateral voltage response based on the rank and determinant analysis of the Laplacian matrix.} The five distinct voltage profiles under OBC and PBC, as illustrated in Fig.~\ref{figVoltageOBCandPBC}, can be categorized based on these analyses. Specifically, the rank of an even-sized circuit under OBC equals the circuit size, whereas it is one less in an odd-sized circuit. Under PBC, the rank of an odd-sized circuit matches the circuit size. In contrast, for even-sized circuits, two scenarios arise: the rank of circuits sized $4N-2$ remains the same, but for circuits sized $4N$, the rank is always two less than the circuit size. A zero determinant indicates the presence of at least one zero eigenvalue, while a non-zero determinant implies that all eigenvalues are non-zero. This distinction in zero or non-zero determinants provides insights into the stability of the circuit.}
	\label{tableRankDet}
\end{table}

As deduced from the key points summarized above for both even and odd $N$, the parity of $N$ significantly influences the eigenspectrum of $L$, even with the inclusion of just a single node in the circuit lattice under both OBC and PBC. These distinct behaviors due to the addition of a single node can be comprehended through the rank and determinant analysis of matrices of both even and odd sizes.

For even-sized square Hermitian matrices, the rank of the matrix is always equal to its size, and its determinant is non-zero. This indicates that the row vectors (or, equivalently, column vectors) of the matrix are linearly independent and there is a unique solution. Conversely, the rank of an odd-sized square matrix is unequal to the size of the matrix and the determinant of it is zero. This leads to an eigenvector which is linearly dependent on its rows or columns. These properties of even and odd-sized matrices outline whether $L$ is singular or non-singular. From the electrical circuit perspective, while a non-singular circuit Laplacian indicates a stability of the circuit, a singular Laplacian indicates a floating circuit. As we have concluded through the analysis of the eigenspectrum of circuit Laplacian with even and odd sizes, when $N$ is odd, there is one zero eigenvalue and an unpaired component in the eigenvectors. This parity dependency significantly alters the resultant voltage profile. Given that the eigenspectrum of $L$ is intrinsically linked to the linear algebraic properties of the Laplacian matrix, we present a classification of the voltage response of LC circuits based on the rank and determinant analysis of the circuit Laplacian. As outlined in Table~\ref{tableRankDet}, when $N$ is even under OBC, $\text{rank}(L_\text{OBC}) = \text{dim}(L_\text{OBC})$ and $\text{det}(L_\text{OBC}) \neq 0$. In contrast, when $N$ is odd, $\text{rank}(L_\text{OBC}) = \text{dim}(L_\text{OBC})-1$ and $\text{det}(L_\text{OBC}) = 0$, where det($\cdot$) and dim($\cdot$) refer to the determinant and dimension of $L$, respectively. This leads to the two distinct voltage responses which we dub as unilateral voltage response for even $N$ and semi-lateral voltage responses for odd $N$. As we will detail in the following sections,  when $N$ is even, the composition of the eigenvectors cancels each other due to the symmetrical pairs of scalar elements, resulting in a unilateral response. However, when $N$ is odd, the unpaired scalar components of the eigenvectors lead to a voltage at the other lateral of the circuit but with different profile. As for PBC, the rank of Laplacian matrices with odd $N$s is always equal to $N$ and the determinant is non-zero. This indicates the invertibility and stability of $L$. However, when $N$ is even, we have two distinct rank and determinant. While the sizes the multiples of $4$ gives rise to $\text{rank}(L_\text{PBC}) = \text{dim}(L_\text{PBC})-2$ and $\text{det}(L_\text{PBC}) = 0$, the sizes expressed as $4N-2$ gives rise to $\text{rank}(L_\text{PBC}) = \text{dim}(L_\text{PBC})$ and $\text{det}(L_\text{PBC}) = -4$. These three size classification leads to a bilateral voltage response in all cases that exists only under PBC. We will now discuss the lateral and bilateral voltage responses in the following sections by integrating the linear algebraic concept of $L$ and its eigenspectrum.

\begin{figure*}[ht!]
	\centering
	\includegraphics[width=\textwidth]{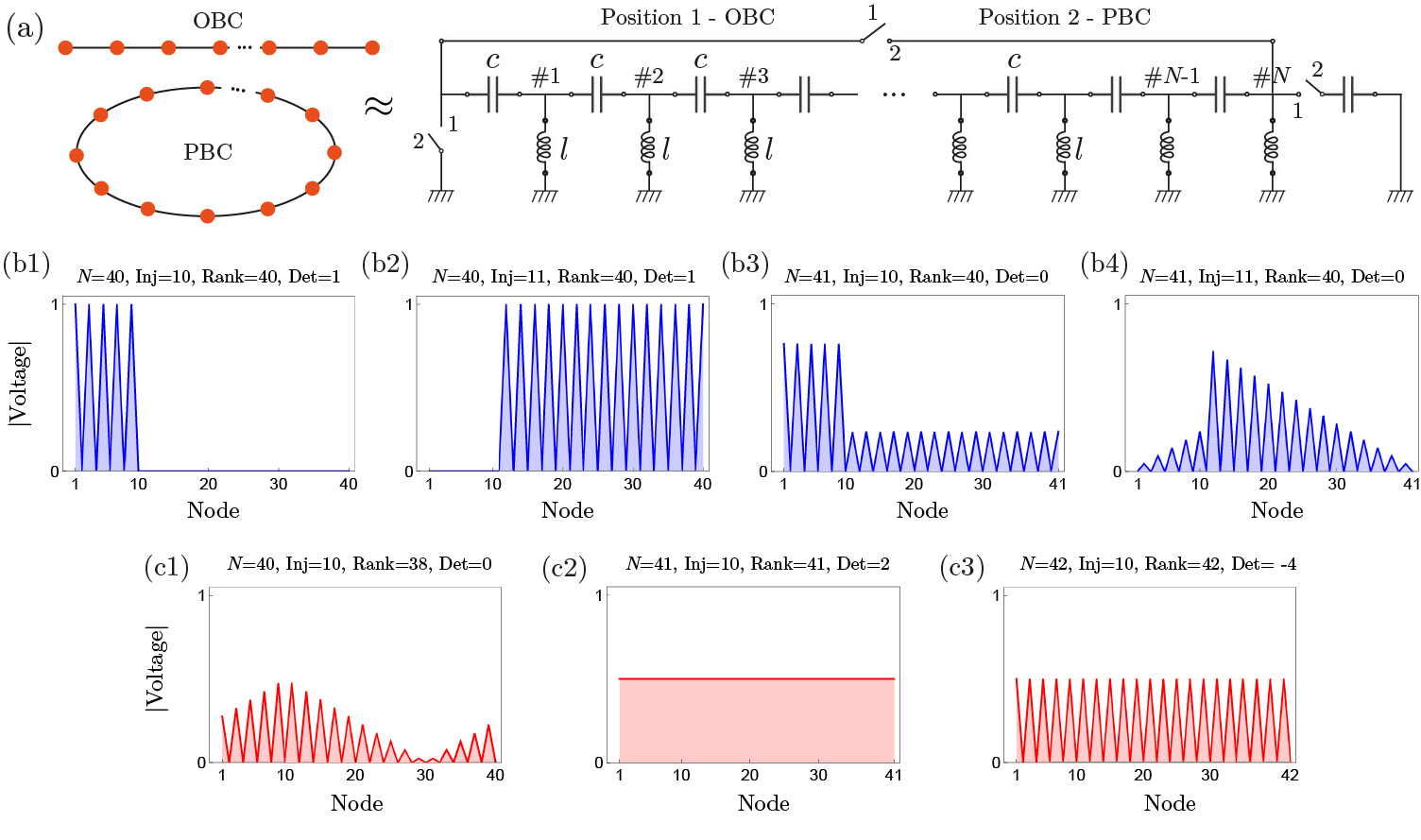}
	\caption{\textbf{The studied circuit schematic and its lateral voltage response under both configurations OBC and PBC with odd and even sizes.} \textbf{(a)} The circuit comprises a single type node corresponding to the 1D free electron gas model. Each node is connected by a capacitor with capacitance $c$, and the nodes are grounded through an inductor with inductance $l$. The three switches illustrate the boundary conditions: OBC at position 1 and PBC at position 2. \textbf{(b1-b4)} The voltage response of the circuit shown in (a) under OBC (blue plots). (b1) and (b2) illustrate the unilateral voltage response in a circuit with an even size of $N=40$, featuring current injections at nodes 10 and 11, respectively. (b3) and (b4) depict the semi-unilateral voltage response in odd-sized OBC circuits $N=41$, where the voltage profiles differ on either side of the current injection node. Specifically, the voltage amplitude remains non-zero but constant when the injection node is even (b3), and decreases linearly when the injection node is odd (b4). \textbf{(c1-c3)} The bilateral voltage response under PBC (red plots). In PBC circuits, three distinct voltage profiles emerge depending on the rank of the Laplacian matrix. Specifically, when the circuit size is a multiple of 4, the voltage on both sides of node $i$ decreases linearly, as shown in (c1). However, the voltage amplitudes on both sides of $i$ are constant, as illustrated in (c3). When the circuit size is odd, the voltage amplitude is uniform across all nodes (c2). In PBC circuits, we demonstrate a scenario with a single current injection at node 10 for all cases, as the voltage profile remains unaffected by the parity of node $i$. A consistent observation across both boundary configurations and different circuit sizes is that the voltage at nodes with the same parity as the current injection node $i$ is always zero, with the exception of odd-sized circuits under PBC. }
	\label{figVoltageOBCandPBC}
\end{figure*}

\subsubsection{Lateral voltage response (Unilateral and Semi-unilateral)}

Lateral voltage response refers to an asymmetrical voltage profile about the current injection node, occurring exclusively under OBC. While the parity of $N$ essentially determines whether the voltage profile is unilateral (Fig.~\ref{figVoltageOBCandPBC}a) or semi-unilateral (Fig.~\ref{figVoltageOBCandPBC}b), the parity of $i$ determines the side of the circuit where voltage amplitudes are present or larger, respectively. The erratic is when $N$ is even where the voltage profile exhibits unilateral behavior. While the node voltages are non-zero only on one side of the circuit with respect to the current injection node, the voltages are completely zero across all nodes on the other side. As illustrated in Fig.~\ref{figVoltageOBCandPBC}a, the injected current selectively induces a voltage on either the right or left side of the circuit depending on the parity of $i$. On the other hand, when $N$ is odd, the lateral voltage response persists but exhibits distinct profiles on both sides, which we dub the behavior as the semi-lateral voltage response. These peculiar lateral behaviors can be elucidated through the analysis of the eigenspectrum of the circuit Laplacian for both even and odd values of $N$.

According to \eqref{voltage2}, the voltage at node $x$ due to a current injection at node $i$ is determined by summing the $x$-th and $i$-th eigenvectors (i.e., $\Psi_{\alpha=x}$ and $\Psi_{\alpha=i}$, respectively) across all its components given by $\beta$, each weighted by its corresponding eigenvalue $\lambda_\beta$. Consider an even-sized OBC circuit and a current source attached to node $i$. As outlined in the previous section, for this circuit, the eigenvalues form pairs with opposite signs and the elements of each eigenvector have a pair with the same polarity when $\alpha$ is even and with the opposite polarity when $\alpha$ is odd, where $\alpha$ is the index of the eigenvectors. The voltage response is critically determined by this bipartite structure and polarity configuration of the eigenvector elements. Referring back to \eqref{voltage2}, we rewrite its summand as $\frac{1}{\lambda} \Psi_i \Psi_x$. When $i$ and $x$ share the same polarity, that is, either $\frac{1}{\lambda} \Psi_{i=\text{even}} \Psi_{x=\text{even}}$ or $\frac{1}{\lambda} \Psi_{i=\text{odd}} \Psi_{x=\text{odd}}$, the summation over all $\beta$s results in zero due to the bipartite structure inherent in all components. This phenomenon accounts for the observed zero voltage across all even or odd $x$ when the injection node $i$ shares the same parity as $x$, as illustrated in Fig.\ref{figVoltageOBCandPBC}. On the contrary, when $i$ and $x$ have different parities, namely $\frac{1}{\lambda} \Psi_{i=\text{even}} \Psi_{x=\text{odd}}$ or vice versa, a voltage appears at one side of the nodes whose parity is opposite to that of $i$. This occurs due to the disruption of the bipartite structure on either the left or right side of the circuit, depending on the parity of $i$. For instance, if $i$ is even, an odd number of elements will be present on the left side of the circuit, while an even number will be on the right side. The even number of elements on the right side results in mutual cancellation, leading to zero voltage across all nodes on this side. Conversely, a voltage will manifest at nodes with a polarity opposite to $i$'s due to the odd number of elements on the other side. This leads to a unilateral voltage response stemming from the circuit's fundamental symmetries. We will now turn our attention to OBC circuits of odd sizes, where a semi-lateral voltage response is observed.

In an OBC circuit with an odd $N$, the voltage at any node $x$ sharing the same parity as the injection node $i$ is always zero. This is due to $\frac{1}{\lambda} \Psi_{i=\text{odd;even}} \Psi_{x=\text{odd;even}}$ equating to zero, similar to the case where $N$ is even. However, when $i$ and $x$ possess opposite polarities, the voltage across all $x$ values varies owing to an unpaired component in the eigenvectors. As previously listed in the previous section, this unpaired element is non-zero for odd $\alpha$ and zero for even $\alpha$. Consequently, this leads to a distinctive voltage profile depending on whether the current is injected at an even or odd node. For instance, as illustrated in Fig.~\ref{figVoltageOBCandPBC}, when $i$ is even, the voltage at nodes with polarity opposite to $i$ remains constant on both sides. In contrast, when $i$ is odd, the voltage exhibits a linear decrease. The distinct voltage profiles observed on both sides of $i$ are directly linked to the presence of the unpaired element, however, the manifestation of constant or linearly decaying voltage characteristics is attributed to whether this unpaired element possesses a zero or non-zero value, respectively. The linear increase and decrease in voltage observed in cases of injection at odd nodes can be attributed to the non-zero unpaired element being divided by eigenvalues that range from $-\lambda$ to $+\lambda$, respectively. From the analysis of even and odd-sized OBC circuits, it is evident that linear algebraic properties, such as the rank and determinant of a square matrix, fundamentally influence the physical response of the circuit based on its size. With this understanding, we will now extend our examination to circuits with periodic boundary conditions.

\subsubsection{Bilateral voltage response}
Bilateral voltage response describes a voltage profile that is symmetrical about the current injection node occurring only in PBC circuits. In PBC circuits, three distinct voltage profiles emerge, influenced by only the parity of $N$. Focusing on odd-sized PBC circuits, we observe that while none of the eigenvalues is zero, there is consistently an unpaired eigenvalue, as detailed in the previous sections and illustrated in Fig.~\ref{FigEigenvalue}b3. Regarding the eigenvectors, each one has an unpaired component whose value equals the sum of its other components, with the exception of the eigenvector corresponding to the unpaired eigenvalue. The components of the eigenvector associated with the unpaired eigenvalue are all the same. While the components of other eigenvectors neutralize each other during summation, the unpaired component remains, consistently divided by the unpaired eigenvalue. Consequently, for each configuration of $i$ and $x$, $V(i,x)$ is determined by only the contribution form $\psi_N / \lambda_N$. This characteristic significantly impacts the voltage response in odd-sized PBC circuits, leading to a non-zero constant voltage measured at each node (refer to Fig.~\ref{figVoltageOBCandPBC}c2). Indeed, this circuit is the most stable circuit with a full rank and non-zero determinant among all other size configurations and boundary conditions. On the other hand, when the circuit size is even, the voltage response can be classified into two types based on the sizes of $4N$ and $4N-2$. As revealed by our spectral examination, the linearly decreasing voltage profile observed in circuits of size $4N$, as shown in Fig.~\ref{figVoltageOBCandPBC}c1, is attributed to an unpaired eigenstate. This eigenstate is divided continuously by all the eigenvalues arranged in ascending order, resulting in a linearly decreasing node voltage. In cases where the size is $4N-2$, an unpaired eigenstate is divided by the same eigenvalue in all summations. This leads to a constant voltage at the nodes with opposite parity to the current injection node, as illustrated in Fig.~\ref{figVoltageOBCandPBC}c3. Consequently, the fundamental structure of the eigenspectra, which varies depending on the circuit size, gives rise to three distinct voltage profiles extending to both laterals in PBC circuits. These voltage profiles are categorized as the bilateral voltage response, occurring exclusively in PBC circuits.

\subsection{Analytical approach}
To better understand the lateral behavior from an analytical perspective, we transform the real space Laplacian presented in \eqref{LaplacianRealSpace} into the k-space representation as $L(k)=2j\omega c(\cos(k)-\epsilon)$. For the sake of simplifying our discussion, we omit $j$ and assume that the driving AC frequency is set to the resonant frequency (where $\epsilon=0$), with $\omega c = 1\Omega^{-1}$, resulting in $L(k)=2\cos(k)$. This Laplacian represents an infinitely large periodic circuit, for which we define the Fourier basis as $\phi_s(k)$, along with the previously defined real space bases as $\psi_{\beta=i}(r)$ and $\psi_{\beta=x}(r)$. Now, let us introduce a notion of $\Phi(k|i,x) = \psi_{(i,x)}^\ast(r) \phi_s(k)$ in which we define the Fourier basis on the real space as
\begin{equation}
	\begin{aligned}
		&\Phi(k|i)=\sqrt{\frac{2}{N+1}} \sin(ki),\\
		&\Phi(k|x)=\sqrt{\frac{2}{N+1}} \sin(kx),
		\label{FourierbasisOBC}
	\end{aligned}
\end{equation}
where $\Phi^\ast \Phi = \delta$ and $k=\frac{s \pi}{N+1}$ where $s=\{1,2,\ldots,N\}$. Here, the wavevector vanishes beyond the boundaries of the circuit where $s=0$ and $s=N+1$, satisfying a 1D OBC circuit with $N$ nodes. Now, since we aim to obtain an analytical expression for the real space voltage at node $x$ in response to a current injection at node $i$, we utilize $G L^{-1}= -\delta$, in which the circuit Green's function $G$ is defined by the pseudoinverse of the circuit Laplacian $L$. Indeed, $G$ is equivalent to the voltage matrix, whose columns correspond to the node voltage when injecting a current at the column index. To reconstruct the real space circuit Laplacian, we apply the discrete Fourier transform and obtain the real space circuit Laplacian as
\begin{equation}
	L_{\text{OBC}}(i,x) = \sum_{s=1}^{N} \Phi^\ast(s|i) L(k) \Phi(s|x),
	\label{LanalyricalOBC}
\end{equation}
and because $V(k) = L^{-1}(k)$, the real space voltage 
\begin{equation}
	V_{\text{OBC}}(i,x) = \sum_{s=1}^{N} \Phi^\ast(s|i) L^{-1}(k) \Phi(s|x).
	\label{VanalyricalOBC}
\end{equation}
Now, using the Fourier bases on the real space in \eqref{FourierbasisOBC}, we explicitly obtain the real space voltage as
\begin{equation}
	V_{\text{OBC}}(i,x) = \frac{I_i}{N+1} \sum_{s=1}^{N} \frac{\sin(k i) \sin(k x)}{\cos(k)} ,
	\label{voltageanalyticalOBC}
\end{equation}
where $i$ indicates the current injection node, $x$ is the index of the node and $k$ is the momentum-space index defined as $k = s\pi/(N+1)$ where $s$ is varying from 1 to $N$. The expressions of \eqref{voltage2} and \eqref{voltageanalyticalOBC} equivalently provide the same node voltages of the 1D OBC chain circuit. To analyze the voltage responses of our 1D OBC circuit with odd and even sizes, we examine both the numerator and denominator of \eqref{voltageanalyticalOBC}. When $N$ is even, \eqref{voltageanalyticalOBC} yields non-zero voltage at nodes $x$ with the opposite parity of $i$, while the voltage at nodes $x$ are zero when the parity of $x$ and $i$ are the same.

We now derive an analytical expression for the voltage of the 1D PBC circuit. For this circuit, since the circuit periodicity is preserved due to the coupling between nodes 1 and $N$, we can define the Fourier basis by the Bloch exponent as $\Phi(s|i) =\phi_i(k) e^{jki}$ and $\Phi(s|x) =\phi_x(k) e^{jkx}$. Because $L(i,x)=\sum_{s=1}^{N} \Phi^\ast(s|i) L(k) \Phi(s|x)$, the real space Laplacian is thus explicitly given by 
\begin{equation}
	L_{\text{PBC}}(i,x) = \frac{1}{2N} \sum_{s=1}^{N} \cos(k) e^{j k(i-x)},
	\label{LanalyricalPBC}
\end{equation}
and the real space voltage as
\begin{equation}
	V_{\text{PBC}}(i,x) = \frac{I_i}{2N} \sum_{s=1}^{N} \frac{e^{j k(i-x)}}{\cos(k)},
	\label{voltageanalyticalPBC}
\end{equation}
where $j=\sqrt{-1}$ and $I_i$ is the magnitude of the injected current at node $i$. The wavevector is defined by $k=2\pi s/N$ where $s=\{1,2,\ldots,N\}$. The analytical expression above applies accurately to circuits with sizes of $2N-1$ and $4N-2$. To understand the constant amplitude voltage response when $N$ is odd and the existing and disappearing voltages at the odd or even $x$ depending on $i$ when $4N-2$, we analyze the numerator of \eqref{voltageanalyticalPBC}. The exponent in the numerator explicitly given as $\exp(j 2\pi(i-x)s/N)$ is always non-zero due to an odd $N$ and yields a constant magnitude voltage when divided by the denominator. However, when $N$ is even, specifically $4N-2$, the numerator becomes zero when summing over all $s$ depending on the parity of $i$ and $x$. When the parity of $i$ and $x$ is opposite, the $(i-x)$ becomes even. Because we consider sizes with $4N-2$, the exponent of the exponential term in the numerator has always opposite parity values in its numerator and denominator, i.e., $\exp(\pi s *\text{even}/ \text{odd})$. This guarantees the unity of the exponent when divided by the denominator and yields a constant magnitude voltage at the nodes where $i$ and $x$ have opposite parities. As can be seen in Table~\ref{tableRankDet}, the PBC circuits with sizes of $2N-1$ and $4N-2$ have full ranks and non-zero determinants which aligns with the observations above that there is always a solution to \eqref{voltageanalyticalPBC} when $2N-1$ and $4N-2$. However, \eqref{voltageanalyticalPBC} becomes indeterminate for circuits sized $4N$. This issue primarily arises from the denominator in \eqref{voltageanalyticalPBC}, which is $\cos(2\pi s/N)$. For circuits of size $4N$, this denominator turns into $\cos(2\pi s)$, leading to a value of zero for any integer value of $s$. As a result, the expression becomes indeterminate. This particular scenario typically occurs in an idealized circuit where $\epsilon$ introduced earlier is effectively zero. However, in physical circuits, the inevitable resistances and tolerance of the components, $\epsilon$ cannot be effectively zero. In the following sections, we will shift our focus to 1D circuits incorporating realistic effects such as parasitic resistances. However, before delving into these specifics, let us first explore the lateral voltage response in more generalized circuits such as the 1D topolectrical SSH circuit.

\subsection{Lateral voltage response in more general circuits: 1D dimer SSH, trimer SSH, and non-Hermitian Hatano-Nelson circuits}

\begin{figure*}[t!]
	\centering
	\includegraphics[width=13cm]{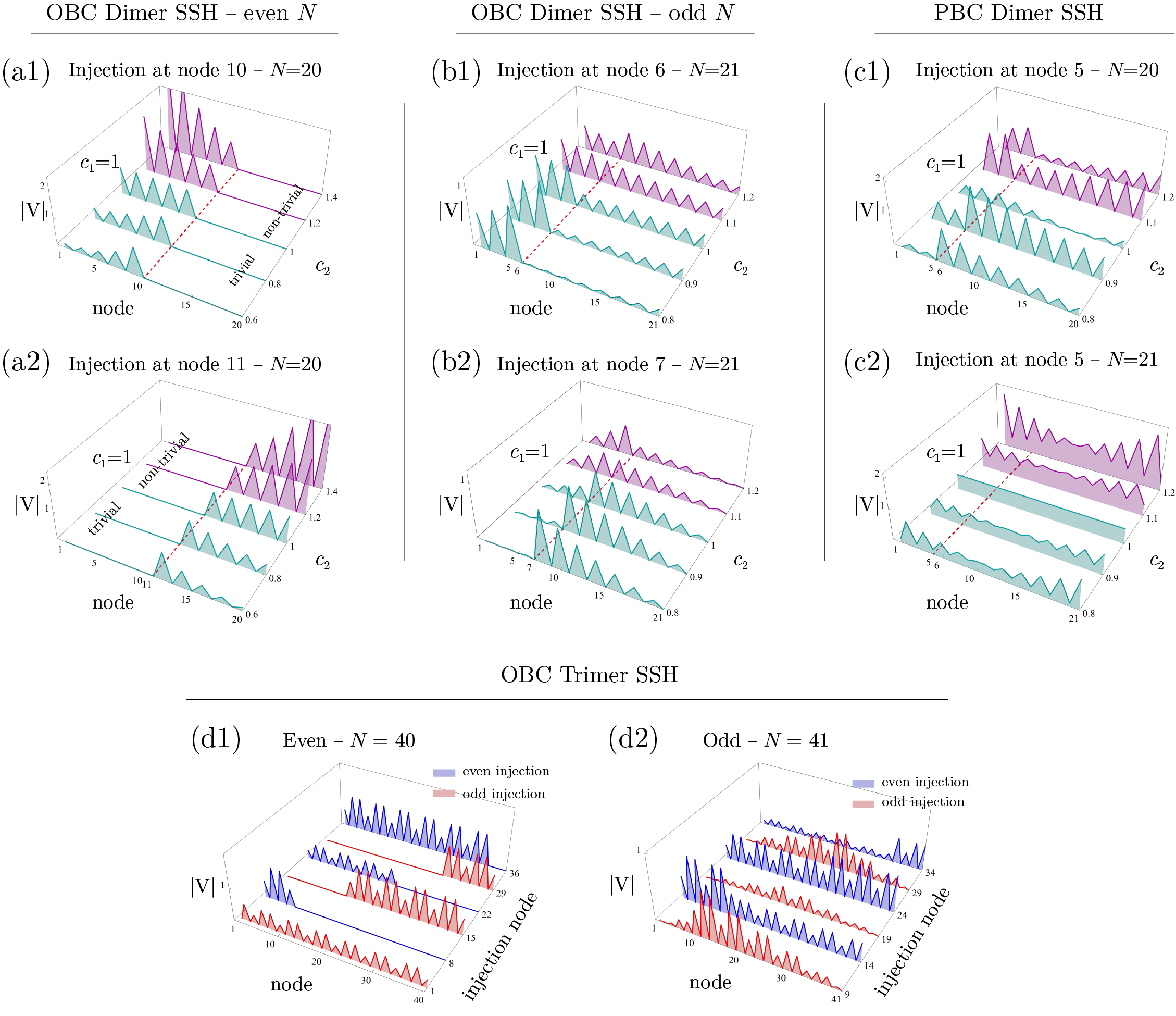}
	\caption{\textbf{Lateral voltage response in 1D dimer and trimer SSH circuits with even and odd sizes under both boundary conditions.} The 3D plots of the dimer SSH circuit illustrate the node voltages as a function of varying $c_2$ with $c_1 = 1$. The dark magenta-filled plots represent the topologically non-trivial phase, while the dark cyan-filled plots correspond to the trivial phase in dimer SSH circuits. The red dashed line in each plot highlights the current injection node. \textbf{(a1) and (a2)}: The unilateral voltage response of the dimer OBC SSH circuit with an even size ($N=20$), under current injection at even and odd nodes, respectively. Remarkably, the voltage amplitude is completely zero on one lateral side, while non-zero on the other. The z-axis is truncated where the voltage range exceeds $|V| = 2V$ to enhance the visual representation of nodes with smaller voltage amplitudes. \textbf{(b1) and (b2)}: The semi-unilateral voltage response of the dimer OBC SSH circuit with an odd size ($N=21$), under even and odd current injections, respectively. \textbf{(c1) and (c2)}: Illustrations of the bilateral voltage response in PBC circuits emerging in the dimer PBC SSH circuit with even ($N=20$) and odd ($N=21$) sizes. \textbf{(d1) and (d2)}: The lateral voltage response of the trimer OBC SSH circuit with even and odd sizes, under different parity current injections. The red-filled plots represent odd injections, while the blue-filled plots signify even injections. For $N=40$, a unilateral voltage response is apparent, whereas for $N=41$, a semi-unilateral response is evident. The parameters for the trimer SSH circuit are set at $c_1=2$, $c_2=1$, and $c_3=2$.}
	\label{FigSSHvoltage}
\end{figure*}

The lateral voltage response is not exclusive to the most basic circuit models; rather, it emerges as a fundamental behavior in more complex configurations, including dimer and trimer one-
 topolectrical circuits and non-Hermitian circuit models. In these cases, the parity dependence of the lateral behavior remains consistent. However, a notable distinction lies in the characteristics of the voltage profile. For instance, whereas the voltage amplitude remains constant in the simple single-node circuit previously discussed, it exhibits exponentially decaying or increasing characteristics in topological and non-Hermitian circuits. To discuss the lateral voltage response in generic models, we first consider the 1D dimer SSH circuit with two types of sublattice nodes in a unit cell, namely $A$ and $B$. In this bipartite circuit, $A$ type nodes correspond to odd-numbered nodes while $B$ nodes correspond to the even-numbered nodes. For consistency, we consider the total number of nodes instead of the total number of unit cells. For this circuit, the real space circuit Laplacian is defined as
\begin{equation}
		L_{\text{DSSH}} = j\omega \left( \sum_{x\in\text{mod2}}^{N-1} c_1 a_{x}^{\dagger}a_{x+1} + \sum_{x\in\text{mod2}}^{N-2} c_2 a_{x+1}^{\dagger}a_{x+2} \right) + \text{h.c.} - \sum_{x=1}^{N} \epsilon a_{x}^{\dagger}a_{x},
	 \label{LapDSSH}
\end{equation}
where `h.c.' refers to the Hermitian conjugate and $a_x^\dagger$ and $a_x$ are the creation and annihilation operators, respectively. $\epsilon$ denotes the total node conductance, which essentially vanishes when the driving signal frequency aligns with the resonant frequency. The summation is taken over the even $x$ indices for intracell ($c_1$) and intercell ($c_2$) couplings, varying from $1$ to $N-1$ and 1 to $N-2$, respectively. However, in a circuit of odd size, the summation for intracell hoppings with the coefficient $c_1$ ranges from $1$ to $N-2$. This Laplacian describes a PBC circuit when we set $a_1 = a_{N+1}$ and $a_2 = a_{N+2}$, indicating the circuit is wrapped around. Now, consider the SSH circuit with an even size under OBC, in which the uniform structure of the unit cells is preserved. In this case, since the translation symmetry of the circuit is preserved and the circuit is driven at the resonant frequency, there exists an exponentially localized voltage at one of the boundaries in the non-trivial phase, which is the case of $c_1<c_2$. In an ideal SSH circuit, the localization boundary is basically determined by the parity of the injection node. Explicitly, when $i$ is odd, there is an exponentially localized voltage at the right edge, whereas there is an exponentially localized voltage at the left edge when $i$ is even. By convention, the 1D SSH circuit possesses two topological zero eigenvalues whose eigenmodes show an exponential boundary localization in the non-trivial phase. The two eigenmodes, one corresponds to a zero eigenvalue and the other corresponds to the second zero eigenvalue, have the exactly same spatial distribution. This raises a crucial question: what causes left or right voltage localization when injecting a current at an even or odd node. The lateral voltage response is the answer. In Fig.~\ref{FigSSHvoltage}a1 and a2, we show the voltage response of the trivial (dark cyan) and non-trivial (dark magenta) SSH circuit with an even $N$ and when $c_1 =1$. In the non-trivial regime, while Fig.~\ref{FigSSHvoltage}a1 depicts the exponentially left-edge localized voltage when the parity of $i$ is even, Fig.~\ref{FigSSHvoltage}a2 shows a right localization when the parity of $i$ is odd. In both phases, the unilateral voltage response is evident, however, the voltage towards the boundary of the selected lateral has an exponential characteristic in both trivial and non-trivial phases, except for the special case of $c_1 = c_2 $ where the voltage amplitude is constant. In the trivial phase, while the maximum voltage amplitude attenuates and is bounded to the neighbor node of the injection node, in the non-trivial phase the node voltages are amplified towards a boundary and localized at the edge. In both cases, the direction of the exponentially decaying or increasing voltage aligns with our lateral voltage response definition. Therefore, the lateral voltage response is what determines the left or right localization in topolectrical circuits, which cannot be fully understood without considering the fundamental behavior of electrical circuits.

For further illustration, we now examine the SSH circuit with odd $N$. From the perspective of topology, an odd-sized circuit results in a broken chiral and translation symmetry due to the uncompleted unit cell. This scenario corresponds to an unstable circuit due to rank of $N-1$, hence, the circuit becomes topologically trivial in both parameter configurations $c_1<c_2$ (dark magenta) and $c_1>c_2$ (dark cyan), as illustrated in Fig.~\ref{FigSSHvoltage}b1 and b2. However, the exponential characteristic of the voltage response is preserved due to the dimer structure. Moreover, the semi-unilateral voltage response of the odd-sized OBC circuits is prominent for the parameter regimes where the attenuation of the exponential voltage is relatively small. For example, the semi-unilateral profile described in Fig.~\ref{figVoltageOBCandPBC}b1 and b2 is observed in the OBC SSH circuit with odd $N$ depicted in Fig.~\ref{FigSSHvoltage}b1 and b2, specifically when $c_2=1.4$. For the smaller $c_2$ values, since the decay length described as $\kappa=\log(c_2/c_1)$ is more predominant, the voltage amplitude on the nodes across the opposite lateral diminishes when combined with the attenuation nature of voltage in the trivial phase.

When it comes to the PBC SSH circuit, a crucial insight emerges about the lateral voltage response, particularly from a physical perspective. A PBC circuit is effectively formed by connecting the end nodes of an OBC circuit. From an electrical circuit engineering standpoint, a circuit without external excitation is deemed to be in a ground state, where the potential difference between any node and the ground is zero. Assuming the circuit is properly grounded without any open circuits, the current distribution follows the effective weight of the nodes on both sides of the current injection node.  As previously noted, PBC circuits typically exhibit a bilateral voltage profile, where the voltage on the left and right sides of the injection node is symmetric. This characteristic also applies to the PBC SSH circuit. However, contrasting the responses of even and odd PBC circuits yields significant insights. To illustrate this, consider a PBC SSH circuit with an even or odd number of nodes. In this circuit, excluding the injection node $i$ as it is the reference node, there are a total of $N-1$ remaining nodes in the circuit. A PBC circuit is accounted topologically trivial due to the absence of the boundaries. However, since the voltage is localized at one of the neighbor nodes of the injection node (depending on the parity of $i$), the localized voltage acts as a boundary in the voltage domain. This implication is evident when comparing the PBC SSH with even and odd $N$. For example, in Fig.~\ref{FigSSHvoltage}c1 where $N=20$ and $i=5$, the voltage profile in both regimes $c_1<c_2$ and $c_1>c_2$ has the same characteristics as the SSH circuit with an odd $N$, as in Fig.~\ref{FigSSHvoltage}b2. In the voltage domain, considering the injection node is the boundary, the remaining circuit with $N-1$ nodes corresponds to an odd-sized OBC circuit (in the voltage domain) where the voltage is exponentially localized at its right or left side. On the other hand, an odd-sized PBC SSH circuit has naturally a physical pseudo-boundary due to the uniformity in the end coupling. When $c_1 \neq c_2$, in any combination of the end coupling, the circuit no longer possesses the translational symmetry, leading to a defect localization. Considering the voltage response characteristic of the odd-sized PBC circuits, we expect a symmetrical voltage response with respect to the defect point, i.e., node 1 or node 21. In Fig.~\ref{FigSSHvoltage}c2, the voltage localizes at node 1 when $c_1<c_2$ (dark magenta) and at node 21 when $c_1>c_2$ (dark cyan). Physically both are the same since changing the rate of the couplings should give rise to a localization either node 1 or node 21. Here, the notable behavior is the symmetrical voltage profiles on both sides of the defect point, which aligns with the main characteristics of the odd-sized PBC circuits. 

We now discuss the lateral voltage response of a trimer SSH (TSSH) circuit whose real-space Laplacian is given by
\begin{equation}
	\begin{aligned}
		L_{\text{TSSH}} = &j\omega  \sum_{x\in\text{mod3}}^{N-1} c_1 a_{x}^{\dagger}a_{x+1}+ \text{h.c.} +  j\omega\sum_{x\in\text{mod3}}^{N-2} c_2 a_{x+1}^{\dagger}a_{x+2}  + \text{h.c.}\\
		& + j\omega \sum_{x\in\text{mod3}}^{N-3} c_3 a_{x+2}^{\dagger}a_{x+3}  + \text{h.c.}
		- \sum_{x=1}^{N} \epsilon a_{x}^{\dagger}a_{x}.
	\end{aligned}
	\label{LapTSSH}
\end{equation}
The Laplacian described above extends the $L_{\text{DSSH}}$ presented in \eqref{LapDSSH}, incorporating an additional summation that represents the third capacitor with a capacitance of $c_3$. It is important to note that this summation is calculated over the modulo of 3 for all $x$ values, with the exception of the onsite summation. Despite its complexity, the voltage response of the TSSH circuit aligns perfectly with the anticipated lateral voltage response behavior. This is clearly demonstrated in Fig.\ref{FigSSHvoltage}d1, where the node voltages on one side of an even-sized TSSH circuit are entirely zero (indicative of the unilateral response), dependent on the parity of the injection node $i$. Similarly, for an odd-sized TSSH, the voltage profile exhibits a semi-unilateral characteristic, as illustrated in Fig.\ref{FigSSHvoltage}d2. Specifically, injecting current at an odd node results in an asymmetric voltage profile with a linear decay. In contrast, injecting at an even node induces maximum voltage amplitudes that differ on either side of $i$. This behavior aligns seamlessly with the voltage responses observed in both the single-type-node circuit and the DSSH profile when $N$ is odd.

\begin{figure*}[ht!]
	\centering
	\includegraphics[width=\textwidth]{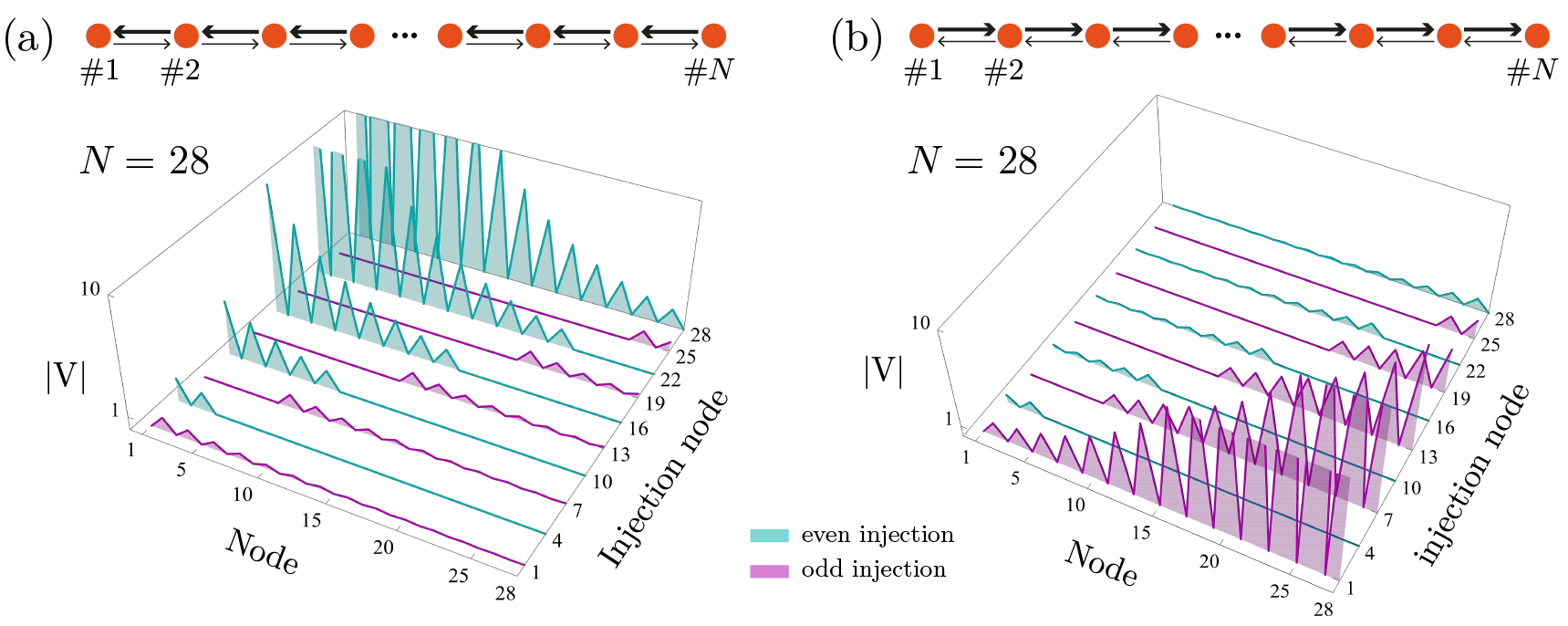}
	\caption{\textbf{Voltage response of the 1D non-reciprocal single-band circuit under OBC with an even size $(N=28)$.} For clarity in visual presentation, the voltage axis is truncated after $|V|=10V$ to better represent smaller amplitude voltage profiles. The dark cyan plots indicate even node injections, while dark magenta represents odd node injections. \textbf{(a)} The schematic depicts the 1D OBC HN circuit with stronger coupling towards the left edge. In this configuration, voltage exponentially localizes at the left edge for even node injections ($i$), and exponentially decays towards the right edge for odd $i$. \textbf{(b)} This panel shows the same HN circuit but with amplification towards the right edge. Unlike the circuit in (a), voltage localizes at the right edge for even node injections. For visual clarity, the x and y axes in the 3D plot are reversed. A common characteristic in both cases is the entirely zero voltage profile on one lateral side, while a non-zero voltage at nodes of opposite parity to $i$ signifies the unilateral voltage response in even-sized OBC circuits. NHSE occurs only when the amplification direction matches the side with non-zero voltage. The parameters used are $c=1$ and $\gamma=0.25$.}
	\label{FigHNNHSE}
\end{figure*}

Finally, we examine the lateral voltage response in the non-Hermitian version of the single-type-node circuit. This circuit indeed corresponds to the Hatano-Nelson (HN) circuit in which the coupling strength is stronger along one direction~\cite{hatano_localization_1996}. This asymmetry in the coupling strength leads to an accumulation of the eigenmodes of the circuit Laplacian, a phenomenon widely known as the non-Hermitian skin effect (NHSE)~\cite{lin_topological_2023,gong_topological_2018,rafi-ul-islam_unconventional_2022}. The circuit Laplacian is obtained by modifying the Hermitian counterpart of the HN model given in \eqref{LaplacianRealSpace} as
\begin{equation}
	L_{\text{HN}} = j\omega\sum_{x=1}^{N-1} \left( c a_x^\dagger a_{x+1} + (c \pm \gamma) a_{x+1}^\dagger a_x \right) - \sum_{x=1}^{N} \epsilon a_x^\dagger a_x,
	\label{LaplacianHN}
\end{equation}
where $\gamma$ is the non-reciprocal term and the sign of $t$ defines the accumulation direction is towards either right or left side. In this model, the eigenmodes of $L_{\text{HN}}$ exponentially localize with an inverse decay length described as 
\begin{equation}
	\kappa =  \log |\frac{c}{c \pm \gamma}|.
\end{equation}
Here, the scenario where $\gamma=0$ corresponds to a single-type-node circuit in which the exponential localization of the eigenmodes diminishes. In electrical circuits, the NHSE manifests as voltage localization at one of the boundaries. To achieve non-reciprocal coupling, we utilize operational amplifiers in negative impedance converters with current inversion (INIC) between each node, as discussed in Refs.~\cite{hofmann_chiral_2019,rafi-ul-islam_unconventional_2022}. This approach results in asymmetric coupling strength between nodes, leading to an amplified voltage towards the boundary in the direction of stronger coupling. Fig.~\ref{FigHNNHSE} examines the NHSE in an even-sized OBC HN circuit. We explore scenarios where the coupling strength is stronger either towards the left edge, as shown in Fig.~\ref{FigHNNHSE}a, or the right edge, as in Fig.~\ref{FigHNNHSE}b, by setting $c=1$ and $\gamma=0.25$. In both configurations, the exponential voltage localization occurs only when the parity of the injection node $i$ is even for the setup in Fig.\ref{FigHNNHSE}a, and odd for that in Fig.~\ref{FigHNNHSE}b. Interestingly, while one might intuitively expect boundary localization for both even and odd injections, voltage localization is observed only when the direction of skin localization aligns with the unilateral voltage response profile of the reciprocal counterpart of the HN circuit, as depicted in Figs.~\ref{figVoltageOBCandPBC}b1 and b2. For example, in our single-band circuit with an even size, as shown in Fig.~\ref{figVoltageOBCandPBC}a, we noted that non-zero voltage amplitudes appeared only on one lateral relative to the current injection node, determined by the parity of $i$. In the HN circuit, the unilateral voltage response means that when $i$ is odd, the voltage on the left side is entirely zero. Consequently, voltage localization at the left boundary does not occur when the circuit experiences left-edge amplification. Conversely, with even $i$, the voltage is non-zero on the left side, leading to the observation of the NHSE in the configuration depicted in Fig.~\ref{FigHNNHSE}a. Similarly, when the amplification direction is reversed, voltage localizes at the right edge for injections at odd nodes, as the voltage on the right side is non-zero for even $i$. This results in exponential voltage localization at the right edge, as illustrated in Fig.~\ref{FigHNNHSE}b. These examples demonstrate that understanding the parity-dependent localization in non-reciprocal circuits requires a fundamental grasp of the inherent features of circuit lattices, such as unilateral voltage response.

\subsection{Voltage characteristics of the realistic circuits with parasitic resistance}

\begin{figure*}[h!]
	\centering
	\includegraphics[width=13.7cm]{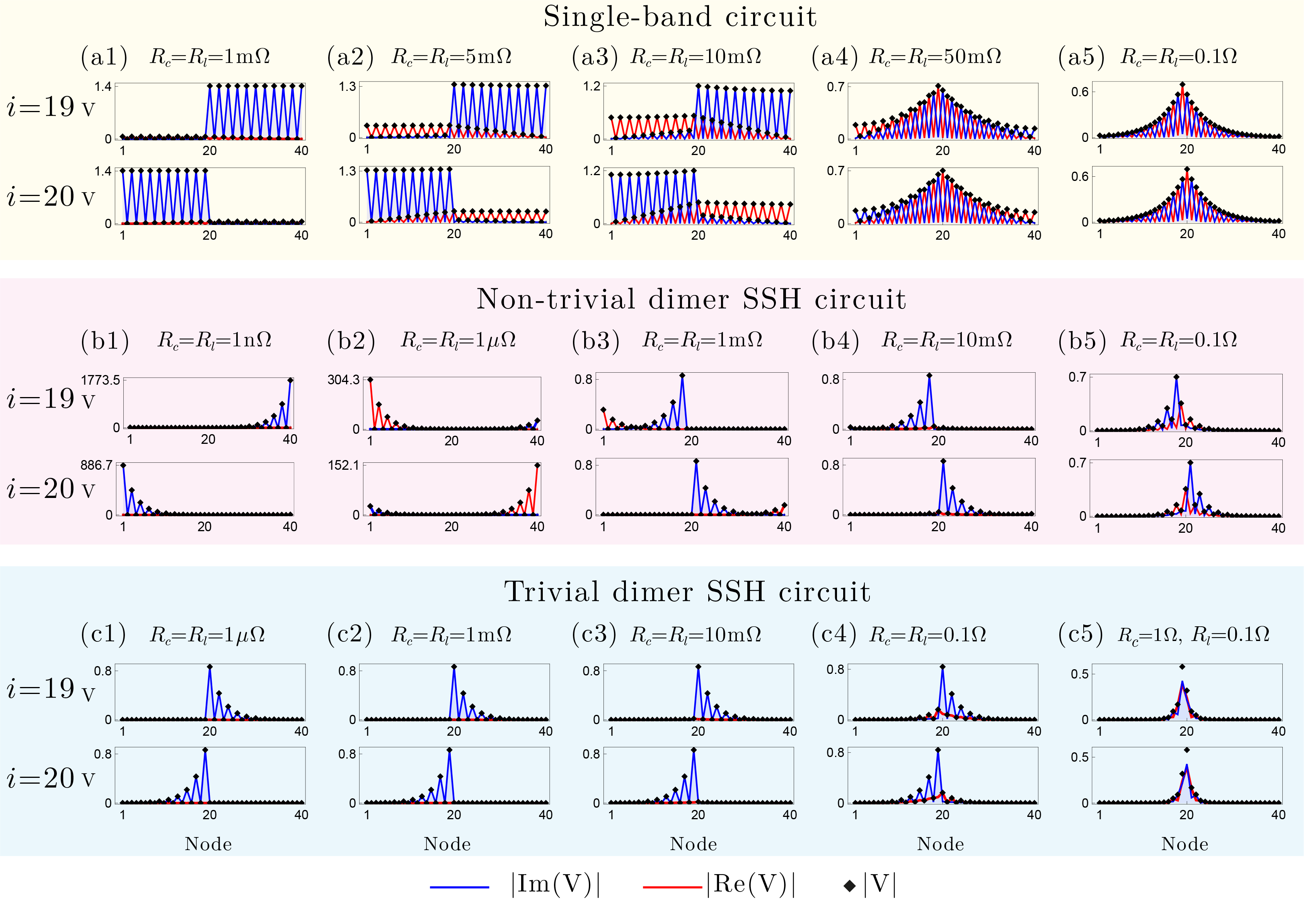}
	\caption{\textbf{The voltage profile of the single-band, topologically trivial and non-trivial dimer SSH circuits with the consideration of the ESRs.} In all panels, the blue (red) line represents the absolute value of the imaginary (real) part of the voltage. The black diamond represents the absolute value of the calculated voltage, which is indeed the RMS voltage measured in experimental implementations. For all circuits, we used $N=40$ and the current is injected at $i=19$ for the odd node injection examples and at $i=20$ for the even node injection examples. The parameters used are $c = 1\,\mu\text{F}$ and $L = 1\,\mu\text{H}$. \textbf{Panel a}: The voltage response of the single-band circuit with various ESRs is shown. The unilateral voltage response persists up to ESRs$~\sim 10\text{m}\Omega$, however, it diminishes for larger ESRs. \textbf{Panel b}: The non-trivial dimer SSH circuit is considered under various ESRs in both odd and even current injections. When the ESRs are relatively small, the imaginary voltage (i.e., the voltage having a 90-degree phase difference with current) localizes either edge of the circuit depending on the parity of $i$. However, as increasing the ESRs, the real part of the voltage experiences a boundary localization at the opposite edge with the imaginary voltage. For large ESRs, the injected signal is damped by the large resistivity of the circuit, hence, the voltage exponentially attenuates from the injection node. The parameters used are $c_1 = 1\mu$F, $c_2 = 2\mu$F and $L = 1\mu$H. \textbf{Panel c}: The voltage profile of the trivial SSH circuit demonstrates the unilateral voltage for even large ESRs. Additionally, the real part of the voltage is suppressed up to $\approx 0.1\Omega$. This is due to the decay characteristics of the trivial phase, hence, the real part of the voltage is dampened as opposite to the non-trivial part where the real part also grows. The parameters used are $c_1 = 2\mu$F, $c_2 = 1\mu$F and $L = 1\mu$H.}
	\label{FigESRcircuits}
\end{figure*}

Real-world applications of electrical circuits involve inevitable effects such as intrinsic parasitic resistance and tolerance of the components. In the previous sections, to depict the lateral voltage response, we have considered fully ideal circuits where the parasitic and real-world effects are omitted. We will now examine the lateral voltage response in circuits with inevitable realistic effects, which can be achieved by modifying the admittance of capacitors as $j\omega c \rightarrow \frac{j\omega c}{1 + j \omega c R_c}$ and admittance of the inductors as $\frac{1}{j\omega l} \rightarrow \frac{1}{R_l+j\omega l}$ in the circuit Laplacian. Involving series resistances into the circuit leads to the resonant frequency to be complex, implying that the injected signal experiences attenuation over time. This attenuation is directly proportional to the imaginary part of the frequency. In the Laplacian matrix, the presence of a complex frequency gives rise to non-zero real diagonal elements even though the circuit is driven at its resonant frequency. As a result, the voltage of the circuit is not purely imaginary but comprises both real and imaginary components. In practical implementations, these real and imaginary voltage components are observed with a phase difference. Intriguingly, in topolectrical circuits with the non-trivial phase, the voltage component that exhibits a 90-degree phase shift from the imaginary voltage also displays topological characteristics, such as exponential localization at the edge opposite to where the imaginary voltage localizes. 

To systematically examine the impact of parasitics, we start with our basic single-type-node circuit, which includes series resistances in the capacitors and onsite inductors, denoted as $R_c$ and $R_l$, respectively. In scenarios where the circuit is predominantly reactive, meaning the real part of the circuit impedance is negligible, the current and voltage (depicted in blue) are always out of phase, exhibiting a 90-degree phase difference. This is exemplified in Fig.~\ref{FigESRcircuits}a1, where $R_c = R_l = 1\text{m}\Omega$. However, larger equivalent series resistances (ESR) result in a voltage (shown in red) that is in phase with the current at nodes with the same parity as $i$. As illustrated in Figs.~\ref{FigESRcircuits}a2 and a3, the node voltages sharing parity with $i$ are no longer zero and align phase-wise with the current. While the unilateral voltage response persists up to $R_c = R_l \sim 10\text{m}\Omega$, as shown in Fig.~\ref{FigESRcircuits}a3, a larger ESR significantly alters the circuit's voltage response. In Figs.~\ref{FigESRcircuits}a4 and a5, where the ESR greatly exceeds that of standard surface mount components (SMD) currently available, the injected signal rapidly attenuates due to the high resistivity. In these cases, the real part of the circuit impedance becomes dominant, overshadowing the imaginary part. As a result, the circuit Laplacian is predominantly characterized by real couplings and real diagonal elements. Consequently, this circuit transforms into a resistive circuit, with each node effectively grounded through a resistor. Therefore, the circuit ceases to exhibit unique phenomena like the lateral voltage response. Consequently, a measurable voltage (indicated by black squares) appears at each node, with its magnitude characterized by a decay that is proportional to the damping factor associated with the complex frequency.

Similarly, the voltage profile of the dimer SSH TE circuit undergoes significant changes when Equivalent Series Resistances (ESRs) are introduced into capacitors ($R_c$) and inductors ($R_l$). For simplicity, we assume that the two capacitors with capacitances $c_1$ and $c_2$ have identical ESRs. Unlike the single-type-node circuit previously discussed, here the voltage exhibits exponential growth in the topologically non-trivial phase and exponential decay in the trivial phase. This behavior is demonstrated in Figs.~\ref{FigESRcircuits}b1 and c1, which depict both phases with negligible ESR. Interestingly, with increased ESR in both topological phases, the voltage phase aligned with the current (differing by 90 degrees) undergoes damping, while the in-phase voltage shows exponential localization at the opposite edge, as illustrated in Fig.~\ref{FigESRcircuits}b2. This exponentially localized in-phase voltage is observed only for relatively smaller ESRs; it too is dampened at larger ESRs, as shown in Figs.~\ref{FigESRcircuits}b3 and b4. However, exponential localization is still apparent with a larger ESR at the injection node $i$. Notably, the unilateral voltage response remains evident as in Figs.~\ref{FigESRcircuits}b3 and b4 (blue), with the existence of voltage on only one side determined by the parity of $i$. Conversely, a large ESR suppresses the boundary localization for voltages both in-phase and out-of-phase with the current, as depicted in Fig.~\ref{FigESRcircuits}b5. This voltage profile closely resembles that of the single-type-node circuit with a large ESR, leading to the circuit behaving in a trivial manner.

In the topologically trivial phase of the SSH circuit, the unilateral voltage response is maintained up to $R_c=R_l=0.1\text{m}\Omega$, as depicted in Figs.~\ref{FigESRcircuits}c1-c4. This persistence is attributed to the decay characteristics inherent to the circuit's trivial phase. Unlike the non-trivial phase, where the in-phase voltage can grow, the trivial phase sees attenuation of this voltage. Consequently, while a significant amplitude of in-phase voltage is noticeable in the non-trivial phase up to approximately $10\text{m}\Omega$, it becomes measurable in the trivial phase only when the ESR is exceptionally large. For instance, Fig.\ref{FigESRcircuits}c5 illustrates the voltage profile at a substantially high ESR of $R_c=1\Omega$, a value notably larger than that of standard SMD capacitors. Across all examples —single-type-node, non-trivial SSH, and trivial SSH circuits— the voltage profile converges to a similar characteristic, decaying in both directions. However, a distinctive observation in the trivial phase of the SSH circuit is that the real and imaginary components of the complex voltage consistently exhibit a 90-degree phase difference from the injected current.

\begin{figure}[ht!]
	\centering
	\includegraphics[width=12cm]{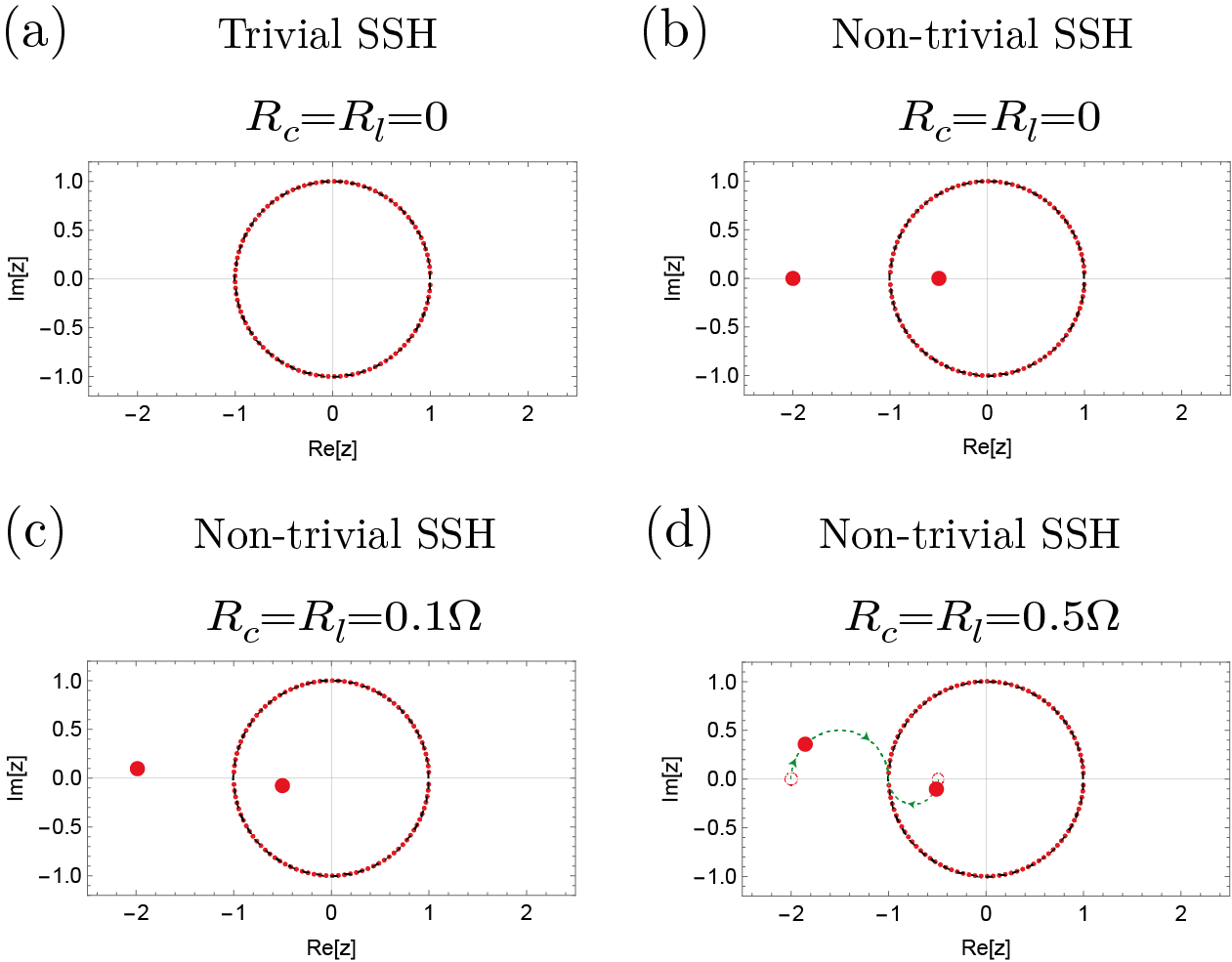}
	\caption{\textbf{The Brillouin zone of the SSH TE.} \textbf{(a)} This panel shows the BZ of the topologically trivial SSH with negligible ESRs. The dashed black circle represents the unit circle, and the energy-dependent Bloch momenta of the trivial SSH circuit align with this circle. \textbf{(b)} In contrast, the BZ of the non-trivial SSH circuit with negligible ESRs features two isolated states. These real states correspond to the non-trivial boundary momenta. \textbf{(c)} With $R_c = R_l = 0.1\Omega$, the two isolated states become complex. \textbf{(d)} At higher ESRs ($R_c = R_l = 0.5\Omega$), the isolated states further deviate from the real-valued boundary momenta. The dashed half-circle illustrates the trajectory of the isolated states as ESRs increase. At much higher ESRs, the two isolated momenta converge (i.e., $|z_l| = |z_r|$, where $z_l$ and $z_r$ represent the left and right Bloch momenta), indicating the evolution of the OBC spectrum towards the PBC spectrum. Common parameters are $N=40$, $c_1 = 1\mu$F, $L = 10\mu$H, $c_2 = 1\mu$F for the trivial phase, and $c_2 = 2\mu$F for the non-trivial phase. }
	\label{FigGBZ}
\end{figure}

We now focus on the voltage characteristics of the non-trivial realistic SSH TE circuit, examining its the Brillouin zone (BZ) profile. The BZ profile offers valuable insights, especially since the ESRs affect the energy spectrum due to the complex coupling strengths. To accomplish the BZ analysis, we modify the admittance of the capacitors to include the ESR and replace the Bloch momenta $e^{jk}$, where $k$ is the wavevector, with $z$ in the momentum space circuit Laplacian. With these modifications, the circuit Laplacian is represented as
\begin{equation}
   L_\text{SSH}(z) = 	\left( \begin{array}{cc}
		r & - \frac{ j \omega c_1}{1 + j \omega c_1 R_c} - \frac{ j \omega c_2}{1 + j \omega c_2 R_c}  z^{-1} \\
		- \frac{ j \omega c_1}{1 + j \omega c_1 R_c} - \frac{ j \omega c_2}{1 + j \omega c_2 R_c}  z & r
	\end{array} \right),
\end{equation}
where $r=\frac{j c_1 \omega }{1+j \omega c_1 R_c}+\frac{j \omega c_2 }{1+j \omega c_2 R_c}+\frac{1}{j \omega L}$. Here, to facilitate the discussion, we assume that the ESR of the inductors is negligible and consider the circuit to be driven at the resonant frequency. However, incorporating the ESR for the inductors also yields an equivalent analysis. While the phase transition occurs at $c_1=c_2$ in the ideal SSH circuit, the phase transition point in the realistic SSH TE circuit is defined as $c_1= (j c_2)/(-j+2 \omega c_2 R_c)$. The solutions for $z$ provide us with the OBC spectrum. These solutions are obtained by solving the characteristic equation $\det[L_\text{SSH}(z)-E\mathbb{1}_{2\times2}]=0$, where $E$ represents the OBC eigenenergy and $\mathbb{1}_{2\times2}$ denotes an identity matrix of the same dimension as $L_\text{SSH}(z)$. We then order the solutions of $z$ such that $|z_1|\le|z_2|\le|z_3|\le\ldots\le|z_N|$, with $N$ being the size of the OBC circuit. Here, the subscript of $z$ indicates that the solutions are sorted by their magnitudes. The BZ is determined by identifying the two roots with equal absolute values. In Figs.~\ref{FigGBZ}a and b, the BZ solutions of the trivial and non-trivial SSH circuits with zero ESRs are depicted. In the trivial phase's BZ profile, all solutions of the Bloch momenta $z$ form a unit circle capturing to the Bloch momenta profile varying from 0 to $2\pi$. However, in the non-trivial phase's BZ profile, there are two isolated momenta, as can be seen in Fig.~\ref{FigGBZ}b. These two isolated momenta are the topological boundary states associated with the value of the Bloch factor $e^{jk}$ at which the wave number is 0 and $2\pi$ corresponding to the boundary modes. In the BZ solution of the non-trivial SSH circuit, these two isolated boundary states are real when ESRs are zero, however, complex when ESRs are involved. For example, when the ESRs of the capacitors and inductor are relatively large, the two boundary states are visibly complex, as depicted in Figs.~\ref{FigGBZ}c and d. In the presence of smaller ESRs, this small complex part leads to a voltage localization having a 90-degree phase difference. This is shown in Fig.~\ref{FigESRcircuits}b2. This particular case provides us with insightful information. Introducing ESRs gives rise to an unequal deviation from the real-valued boundary states. As can be seen in Figs.~\ref{FigGBZ}c and d, the state falling outside of the unit circle experiences a larger deviation than that of the state falling inside the unit circle. This deviation leads to unbalanced amplification for the left and right boundary localized voltages since the imaginary parts of these states are unequal. For larger ESRs, the two boundary states further deviate from the real-valued Bloch momenta leading to the diminishing boundary states (refer to Fig.~\ref{FigGBZ}d). Therefore, we observe an exponential voltage localization at the injection node as in Fig.~\ref{FigESRcircuits}b5.

\subsection{Conclusion and remarks}
In conclusion, our study presents an analysis of lateral voltage response, beginning with a simple circuit model akin to the 1D free electron gas model. We identified five distinct voltage profiles based on the circuit size's parity; two in OBC and three in PBC. Voltage profiles in OBC exhibit unilateral characteristics, whereas in PBC, they show bilateral characteristics. The exact nature of these unilateral and bilateral responses also varies with the circuit size's parity in both boundary configurations. This variation is due to the inherent algebraic properties of the Laplacian matrix. We classified the lateral voltage response through rank and determinant analysis. Our examination extended to voltage profiles in more complex circuits, such as dimer and trimer SSH circuits and the non-Hermitian Hatano-Nelson circuit. Notably, features like exponential voltage localization at one edge are fundamentally related to the lateral voltage response. For example, in SSH circuits, the lateral voltage essentially determines the amplification direction in the non-trivial phase and attenuation in the trivial phase. Contrary to common belief, right and left edge states in an SSH circuit exhibit non-vanishing elements at $A$ or $B$ sublattice nodes only in an infinitely large circuit. In a finite SSH circuit, the edge states $|\psi_0^\text{left}|$ and $|\psi_0^\text{right}|$ possess identical spatial distributions. When a non-trivial SSH circuit is excited at any node, one might expect voltage localization at both edges. However, the ideal non-trivial SSH circuit shows voltage localization at only one edge, influenced by the parity of the injection node. This unilateral voltage response clarifies the inconsistency and offers a deeper understanding of the localization direction.

Furthermore, the Hatano-Nelson circuit demonstrates voltage localization only at odd or even node injections. In this non-Hermitian circuit, with each node having identical non-reciprocal couplings, it's intuitive to anticipate voltage localization at any injected node. Yet, we observe an exponentially localized voltage at the left edge for even parity node injections when the circuit is amplified towards the left edge and at the right edge for odd parity injections when the circuit is amplified towards the right edge. In other scenarios, the voltage exponentially decays towards the opposite edge. This phenomenon exemplifies the lateral voltage response in non-reciprocal, non-Hermitian circuit models, a behavior that would remain unexplained without considering the lateral voltage response.

The lateral voltage response is crucial for comprehending the voltage profiles in circuits affected by real-world factors such as Equivalent Series Resistances (ESRs). In practical implementations, although ESRs can be identified through post-processing, additional effects like contact resistance and parasitic resistances necessitate a phase analysis of the measured voltage profile. Our examinations highlight that the lateral voltage response offers an insightful tool for this analysis. This response not only aids in understanding inherent circuit characteristics but also in interpreting the impact of external factors, ensuring a more comprehensive view of circuit behavior under real-world conditions.

\chapter{Impedance responses and size dependent resonances in topolectrical circuits}
\label{ch:impedance}
\vspace{2em}

Our examination of the voltage response in electrical circuits, as presented in Chapter \ref{ch:lateralvoltage}, revealed a parity-dependent lateral voltage response in both classical and topolectrical circuits. In this chapter, we shift our focus to the corner-to-corner impedance characteristics of electrical circuits. Resonances in an electric circuit occur when capacitive and inductive components are present together. Such resonances appear in admittance measurements depending on the circuit's parameters and the driving AC frequency. In this chapter, we analyze the impedance characteristics of nontrivial topolectrical circuits such as one- and two-dimensional Su–Schrieffer–Heeger circuits and reveal that size-dependent anomalous impedance resonances inevitably arise in finite $LC$ circuits. Through the \textit{method of images}, we study how resonance modes in a multi-dimensional circuit array can be nontrivially modified by the reflection and interference of current from the structure and boundaries of the lattice. We derive analytic expressions for the impedance across two corner nodes of various lattice networks with homogeneous and heterogeneous circuit elements. We also derive the irregular dependency of the impedance resonance on the lattice size, and provide integral and dimensionally-reduced expressions for the impedance in three dimensions and above.

The main contribution of this chapter is revealing the size-dependent impedance responses in $LC$ circuits. Since various models, such as transmission lines, are studied from their equivalent circuit models, these size-dependent characteristics are important for analyzing any equivalent circuit model where the equivalent circuit consists of an array of inductive and capacitive components. Additionally, the size-dependent impedance resonances from the fractal scaling profiles allow us to distinguish other resonances induced by system properties, such as topological impedance resonances.

\section{Introduction}
Electric circuit networks are extremely versatile platforms for simulating a variety of condensed matter phenomena through their tight-binding representations~\cite{ningyuan_time-_2015,song_realization_2020,nakata_circuit_2012}. For instance, a uniform tiling of $LC$ oscillators composed of capacitor and inductor pairs gives rise to an AC signal propagating along an ideal transmission line, which simulates the lattice dynamics of a one-dimensional (1D) solid-state medium. Circuits that mimic condensed matter lattices, known as topolectrical (TE) circuits, have been extremely successful in demonstrating a wide range of topological and critical condensed matter phenomena~\cite{lee_topolectrical_2018,imhof_topolectrical-circuit_2018,helbig_band_2019,bao_topoelectrical_2019,rafi-ul-islam_topoelectrical_2020,rafi-ul-islam_realization_2020,wang_circuit_2020,wang_circuit_2020,li_emergence_2019,ezawa_electric_2019,lee_imaging_2020,helbig_generalized_2020,zhang_topolectrical-circuit_2020,olekhno_topological_2020, ni_robust_2020,hofmann_chiral_2019,rafi-ul-islam_system_2022,shang_2022_experimental}. In these circuits, topologically protected zero modes can be measured as impedance resonances at the resonant frequency. Beyond the simulation of linear condensed matter systems, the simulation of complex networks processes such as search algorithms has also been proposed through the use of non-linear circuit elements.~\cite{ezawa_electric_2020,ezawa_universal_2021,pan_electric-circuit_2021,quiroz-juarez_reconfigurable_2021,kotwal_active_2021,kengne_ginzburglandau_2022}. %Indeed, many physical phenomena can give rise to impedance resonances in their corresponding electrical circuit arrays. 

While tight-binding lattice models are usually faithful in representing their respective solid state systems~\cite{kane_z_2005,jin_solutions_2009,sun_nearly_2011,hasan_colloquium_2010,lee_lattice_2014,chen_impossibility_2014,gu_holographic_2016,rafi-ul-islam_unconventional_2022}, their intrinsic discreteness sometimes leads to additional anomalous contributions to the impedance with no analog in the continuum limit. In this chapter, we present an analytic formulation for computing the impedance across various finite circuit arrays. We also investigate the anomalous impedance behavior that emerges when components with different phase lags are simultaneously present in a finite electric circuit, which we call a heterogeneous circuit. We focus most specifically on heterogeneous circuits in which the nodes are coupled by a mixture of capacitors and inductors.

The electric potential distribution in electric circuit networks satisfies Kirchhoff's laws and can be described by the circuit Laplacian. The solutions of the circuit Laplacian contain all available information about the potential distribution. Therefore, any desired computation can be performed by solving the circuit Laplacian with appropriate boundary conditions. Although the two-point impedance in homogeneous circuits has been widely studied, most results pertaining to the condensed matter context are valid only for infinite circuit networks~\cite{kirkpatrick_percolation_1973,lavatelli_resistive_1972,bartis_lets_1967,zemanian_classical_1984,cserti_perturbation_2002,owaidat_two-point_2018,owaidat_perturbation_2014,koutschan_lattice_2013,izmailian_generalised_2014,izmailian_two-point_2014,tan_resistance_2017,cserti_uniform_2011,venezian_resistance_1994,owaidat_interstitial_2010,owaidat_resistance_2019,atkinson_infinite_1999,doyle_random_2000,tan_electrical_2020,tan_two-point_2016,asad_infinite_2005,essam_exact_2009,izmailian_asymptotic_2010,clerc_dielectric_1996,cserti_application_2000,morita_useful_1971,asad_perturbed_2014,owaidat_resistance_2016,owaidat_perturbation_2016,tan_electrical_2019,tan_recursion-transform_2015,zhang_resistance_2021,chen_equivalent_2021,tan_characteristic_2017,aitchison_resistance_1964,joyce_exact_2002,jeng_random_2000,chen_electrical_2019,chen_electrical_2020,chen_electrical_2020-1,fang_circuit_2022,cernanova_nonsymmetric_2014,joyce_exact_2017,asad_infinite_2014,asad_infinite_2013,guseinov_unified_2007,mamode_calculation_2019,zenine_lattice_2015,tan_recursion-transform_2015_1,tan_recursion-transform_2015_2,owaidat_resistance_2014,owaidat_electrical_2013,owaidat_substitutional_2010,jafarizadeh_calculating_2007,giordano_disordered_2005}. Two issues that present challenges in forming full analogies with condensed matter are the finite circuit boundaries (which differ from the usual open boundary conditions) and the homogeneity of the lattice array; while computations can certainly be performed numerically, a comprehensive analytical expression for the impedance in heterogeneous finite circuits has yet to be obtained. One way to implement boundary conditions in electrostatic theory is through the method of images, in which the boundaries are replaced by image charges located opposite the original charges~\cite{jackson_classical_1999,griffiths_introduction_2005,riley_mathematical_1999,yang_designing_2022,mamode_electrical_2017}. As a demonstration of this approach, we apply the method of images to periodic tilings of finite electric circuit networks to compute the two-point impedance of the finite circuit networks. 

While the impedance generally scales in a logarithmic manner with the circuit size in homogeneous circuits \cite{wu_theory_2004,tan_recursion-transform_2015,owaidat_regular_2014,cserti_uniform_2011,cserti_perturbation_2002,owaidat_resistance_2013,asad_infinite_2014,owaidat_substitutional_2010}, the same is not true in heterogeneous circuits, where the impedance can deviate very strongly from logarithmic scaling~\cite{abrahams_scaling_1979} at certain circuit sizes. The circuit size $N$ thus becomes a functional parameter alongside $LC$ and the driving AC frequency $\omega$. These independent parameters collectively affect the impedance resonances. This is in contrast to a waveguide or transmission line, in which the system size does not affect the behavior of the system in ideal cases. Moreover, the discreteness of $N$ results in fractal-like resonances when $N$ is varied at fixed $L$ and $C$ values.
\\
\\
\indent In this chapter, we study the size-dependent impedance resonances both numerically and analytically. We derive analytic expressions for the size-dependent impedance between two opposite corner nodes by utilizing the method of images. To reveal the origin of this anomalous impedance behavior, we examine circuits with a single node per unit cell, as well as those with nontrivial unit cells containing more than one node. As paradigmatic examples of the latter, we present detailed calculations for 1D and two-dimensional (2D) circuit lattices with Su-Schrieffer-Heeger (SSH) type dimerizations. As for homogeneous circuits with a single-type node per unit cell, we start from a 1D circuit and build higher-dimensional circuits by linking every node along the new direction with the same type of component (i.e., resistor, inductor, or capacitor). In the heterogeneous circuits section, we follow the same process but introduce at least two different types of components with different phase shifts. Finally, we discuss the emergent fractal-like structures that arise from the violated logarithmic scaling in $LC$ circuits.

\section{Formalism for the two-point impedance}\label{secII}
We first review the generic derivation of the expression for the two-point impedance in terms of the eigenvalues and eigenvectors of the circuit Laplacian~\cite{wu_theory_2004,tzeng_theory_2006,cserti_uniform_2011,izmailian_generalised_2014,izmailian_two-point_2014,cernanova_nonsymmetric_2014,mamode_revisiting_2021}. A $RLC$ circuit can be represented as a graph in which the vertices of the graph represent the voltage nodes and the edges represent the couplings between the nodes due to $R$, $L$, and $C$ components between the nodes. Under driving at a single AC frequency $\omega$, the circuit can be mathematically represented by its Laplacian matrix $J$, which relates the currents injected into the nodes with the voltages at each node via 
\begin{equation}
I = JV \label{LaplacianEq}
\end{equation}
where $I$ is a vector of the currents injected into each node and $V$ the corresponding vector of the node voltages. The Laplacian matrix for a circuit can be obtained simply by writing Kirchhoff’s current law at each voltage node. For example, consider a simple circuit consisting of two voltage nodes connected by a single capacitor with capacitance $C$. Applying Kirchhoff’s current law at the two nodes gives $I_1 = i\omega C (V_1-V_2)$ and $I_2 = i\omega C (V_2-V_1)$ where $I_a$ and $V_a$ are the injected current and voltage at node $a$, respectively. Using \eqref{LaplacianEq} and these relations between the node current and voltages, the Laplacian matrix $J$ of this simple circuit is then given by $J = i\omega C \begin{pmatrix} 1 & -1 \\ -1 & 1 \end{pmatrix}.$ By definition, the impedance between two nodes $i$ and $j$ is the factor of proportionality between the voltage difference $V_i-V_j$ that develops between the two nodes when a current of magnitude $I$ is injected into node $i$ and extracted at node $j$: 
\begin{equation}
    Z_{ij}=\frac{V_i - V_j}{I}.
    \label{ImpZ}
\end{equation}
\noindent
To determine the impedance between two nodes, the voltages $V_i$ and $V_j$ have to be determined. To achieve this, we employ the circuit Green's function $G$, which is defined as the pseudo-inverse of the circuit Laplacian $G=J^{-1}$. Using the Green's function, \eqref{LaplacianEq} can be rewritten as $V=GI$. The electrical voltage at node $i$ can thus be expressed as 
\begin{equation}
    V_i=\sum_{j}^{N} G_{ij} I_j,
    \label{Voltage_Greens}
\end{equation}
\noindent where $G_{ij}$ is the $(i,j)$th element of the pseudoinverse matrix $G$. By setting $I_i=I$ and $I_j=-I$ in \eqref{Voltage_Greens}, \eqref{ImpZ} can be rewritten as 
\begin{equation}
    Z_{ij}=\sum_{k=i,j} \frac{G_{i k} I_k - G_{j k} I_k}{I},
    \label{Z_G}
\end{equation}
which gives
\begin{equation}
    Z_{ij}=G_{ii}+G_{jj}-G_{ij}-G_{ji}.
    \label{ZGreensFunc}
\end{equation}
\noindent 
By resolving in the space of eigenstates, $J$ can be written in terms of its right eigenvectors $|\psi_k\rangle$, left eigenvectors $\langle \psi_k |$, and eigenvalues $\lambda_k$ as $J = \sum_k |\psi_k\rangle \lambda_k \langle \psi_k|$. In general, $|\psi_k\rangle \neq \langle \psi_k|^\dagger$, because $J$ is not Hermitian. However, $|\psi_k\rangle = \langle \psi_k|^\dagger$ holds in the following special cases: (i) in a purely $LC$ circuit, where $J$ is anti-Hermitian and the $\lambda_k$'s are imaginary, and (ii) in a purely resistive circuit, where $J$ is Hermitian and the $\lambda_k$ are real.  For a general $RLC$ circuit, the $\lambda_k$ are complex, and the general relation $G=\sum_k |\psi_k\rangle (\lambda_k)^{-1} \langle \psi_k|$ holds. Note that because $G$ is defined as the \textit{pseudo}inverse of $J$, any zero eigenvalues are excluded from the sum if they exist.  
Writing $|\psi_k\rangle$ as a column vector of the node voltages $|\psi_k\rangle=\{\psi_{k 1},\psi_{k 2},\dots,\psi_{k N}\}$ where $N$ is the total number of nodes, \eqref{ZGreensFunc} can rewritten as 
\begin{equation}
    Z_{ij}=\sum_{k, \lambda_k \neq 0} \frac{|\psi_{k i}-\psi_{k j}|^2}{\lambda_k},
    \label{Z_eig}
\end{equation}
\noindent where $\lambda_k$ is the corresponding eigenvalue of the Laplacian, and the $|...|$ norm is the biorthogonal norm. Any arbitrary two-point impedance can then be numerically calculated using \eqref{ImpZ}, \eqref{ZGreensFunc}, or \eqref{Z_eig}.
\\
\\
\begin{figure*}[htp]
	\centering
	\includegraphics[width=\textwidth]{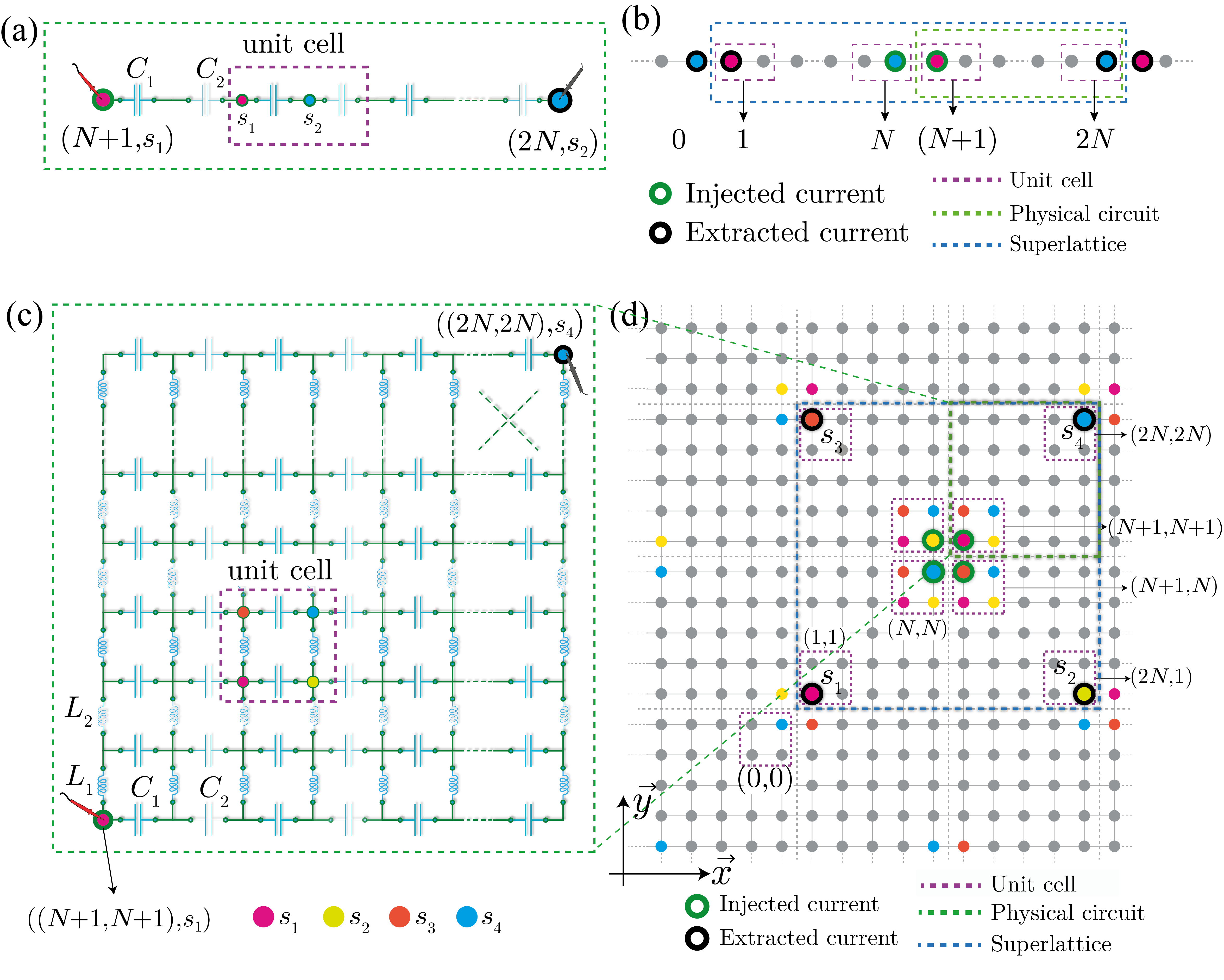}
	\caption{\textbf{Lattice structures of illustrative SSH circuits and their periodic image lattices for finite-lattice impedance computations}. (a) The 1D SSH circuit comprises two capacitors ($C_1$ and $C_2$) and two nodes ($s_1$ and $s_2$) in a unit cell. (b) The superlattice comprising the physical and image 1D SSH circuits is denoted by the blue dashed box, so constructed such that the impedance through a finite single block is recast into a problem with periodically placed injected and extracted currents. The green and purple dashed boxes denote the physical circuit and its unit cell, respectively. (c) The physical 2D SSH circuit consisting of two kinds of capacitors with capacitances $C_1$ (bold) and $C_2$ (thin) connecting the nodes along the horizontal direction and two kinds of inductors with inductances $L_1$ (bold) and $L_2$ (thin) linking the nodes along the vertical direction. The four distinct nodes in a unit cell (purple dashed square) $s_1$, $s_2$, $s_3$, and $s_4$ are represented as magenta, yellow, orange, and cyan circles, respectively. To measure the corner-to-corner impedance, current is injected at node $s_1$ of unit cell $(N+1,N+1)$ and extracted at node $s_4$ of unit cell $(2N, 2N)$ [also see the matrix representation in \eqref{2DcurrentMatrix}]. (d) The infinite periodic lattice tiling of the $(2N)^2$-unit cell superlattice consisting the image and physical circuits (blue dashed square), which is constructed such that the impedance across the finite lattice can be expressed in terms of translation-invariant momentum contributions. The nodes at which current is injected and extracted are denoted with green and black outlines, respectively.}
	\label{topo_circuits} 
\end{figure*}

\section{Method of images for analytic impedance formulas across bounded circuit lattices}
In this section, we review and derive the general analytical formalism for calculating the impedance between two edge or corner nodes in finite discrete circuit lattices based on \textit{the method of images} and discuss their impedance behaviors. The method of images is needed to put finite lattice impedance computations, which are not easily represented by exact analytic formulas due to the lack of translation symmetry, on equal footing with periodic lattices. We shall illustrate the approach with exemplary circuit arrays with nontrivial unit cells, such that the tiling of the periodic images is not trivial.

In the context of electrostatic theory, the electric potential distribution inside an area of interest in the vicinity of a grounded conducting plate can be obtained by placing an image charge reflected across the conducting plate~\cite{riley_mathematical_1999,griffiths_introduction_2005,jackson_classical_1999,feynman1965feynman,yang_designing_2022,li_non-hermitian_2022,mamode_electrical_2017,rafi-ul-islam_interfacial_2022,pantoja_greens_2021}. The image charge causes the surface of the plate to become equipotential (which is considered zero for a grounded conductor) and thus satisfy the boundary conditions of a grounded conducting plate. 

Inspired by this approach, we apply an analogous idea by placing image circuits symmetrically about our desired circuit to replicate the boundary conditions satisfied by a finite circuit under open boundary conditions~\cite{mamode_electrical_2017,mead_resistance_2009}. %Unlike the classical electrostatic picture in which the tangential component of the electrostatic potential distribution is zero at the boundaries
The translation and inversion symmetries obeyed by the original and image circuits in our method force the voltage potentials at the boundary nodes of the original and the image circuits to be the same[see Figure~\ref{VSpatial}(b)]. The equal potentials at the boundary nodes result in zero current flow across the boundaries between the physical and image circuits, and therefore replicates the boundary condition of zero current flow across the open physical boundaries of the finite bounded circuit. The potential distribution of the original circuit is then the same as that of the finite circuit with physical boundaries. This is the key insight that underlies our derivation of analytical expressions for the potential profiles of circuits with open boundaries via \textit{the method of images}. 

\subsection{Example: 1D SSH circuits}
To illustrate how \textit{the method of images} yields exact analytical expressions, we first consider sufficiently nontrivial circuit arrays in which each unit cell contains more than one type of node, such as the 1D and 2D topological Su-Schrieffer-Heeger (SSH) circuits. In such lattices, the transitions from the topologically trivial to non-trivial phases are accompanied by strong impedance resonances at the resonant frequency~\cite{lee_topolectrical_2018,helbig_band_2019,hofmann_chiral_2019,rafi-ul-islam_topoelectrical_2020,imhof_topolectrical-circuit_2018,2022arXiv220609931H,lenggenhager_simulating_2022,zhang2023electrical}. However, our focus here is on their edge-to-edge (1D SSH) and corner-to-corner (2D SSH) impedance profiles rather than the topological impedance characteristics.

The 1D SSH circuit consists of unit cells in which each unit cell contains two distinct nodes labeled as $s_1$ and $s_2$ where the nodes  are connected to each other by intra-cell capacitors of capacitance $C_1$ and inter-cell capacitors of capacitance $C_2$ [see Figure~\ref{topo_circuits}(a)]. The corresponding $2 \times 2$ circuit Laplacian of a periodic 1D SSH circuit in momentum space is therefore written as
\begin{equation}
\mathcal{L}_{\text{1D}}^{\text{SSH}}(k_1) = i \omega \begin{pmatrix}
     C_1+C_2 & - C_1 - C_2 e^{-i k_1}\\ - C_1 - C_2 e^{i k_1} & C_1+C_2
\end{pmatrix},
\label{lap_1d_ssh}
\end{equation}
\noindent where $\omega$ is the driving AC frequency and $k_1$ is the crystal momentum along the $x$ direction. We label the nodes in the SSH chain as $\mathbf{\tilde{r}}=(\mathbf{r},\mu)$ where $\mu \in (s_1, s_2)$ denotes the node within each unit cell, $\mathbf{r}=n_1 \mathbf{a_1}$ denotes the location of the unit cell with $\mathbf{a_1}$ being the unit vector along the length of the chain and $n_1$ the coordinate of the unit cell, and the tilde on $\mathbf{\tilde{r}}$ denotes a composite index consisting of both the location of the unit cell and the sub-lattice site.

To investigate the behavior of the impedance as the circuit size increases, we consider the impedance between node $s_1$ in the leftmost unit cell and node $s_2$ in the rightmost unit cell in the 1D SSH chain circuit. Note that the circuit size is increased unit cell-by-unit cell so as to preserve lattice uniformity. Therefore, $N$ refers to the number of unit cells throughout this study.
Henceforth, we shall call the actual circuit with open boundaries (the green dashed rectangles in Figure~\ref{topo_circuits}(b) and \ref{topo_circuits}(d)) \emph{the physical lattice} or \emph{the physical circuit}, the image(s) of the actual circuit \emph{the image lattice(s)} or \emph{the image circuit(s)}, and the block containing physical and image circuits (the blue dashed rectangles in Figure~\ref{topo_circuits}(b) and \ref{topo_circuits}(d)) \emph{the superlattice} or \emph{the supercircuit}. To obtain the voltages in a finite SSH chain containing $N$ unit cells that result from current injection at the left-most node and current extraction at the right-most node, we place an image chain of the same length to the left of physical chain circuit, but importantly reflected about the chain boundary so that no current flows across it due to reflection symmetry. We then inject current at the right-most node of the image circuit and the left-most node of the physical circuit,  and extract the injected current at the left-most node of the image circuit and the right-most node of the physical circuit, as shown in Figure~\ref{topo_circuits}(b).  The impedance between any two lattice points can then be obtained using by \eqref{ImpZ} once the voltage distribution is known. 

To determine the voltage distribution explicitly, we utilize Ohm's law, which states that the current distribution $\mathbf{\mathcal{J}} = \sigma \mathbf{E}$ where the electric field is given by $\mathbf{E}=-\grad V$. Therefore, the current distribution can be written as $\mathcal{J}=-z^{-1} \grad V$, where $z$, the uniform impedance between each node, is the inverse of the electrical conductivity $\sigma$. Owing to Kirchhoff's current law, the current density is also written as $\div{\mathcal{J}}= I \left( \delta_{(\mathbf{r'},\nu)\in\mathbf{\tilde{r}_{in}}}(\mathbf{r},\mu,\mathbf{r'},\nu)-\delta_{(\mathbf{r'},\nu)\in\mathbf{\tilde{r}}_{out}}(\mathbf{r},\mu,\mathbf{r'},\nu)\right)$ where $\delta$ denotes the Kronecker delta and $I$ represents the current magnitude. After performing the relevant substitutions, we arrive at a Poisson-type equation. Here, because the Green's function satisfies $\grad^2 G(\mathbf{r,\mu,r',\nu}) = -\delta(\mathbf{r,\mu,r',\nu})$ by definition \cite{katsura_lattice_1971,cserti_application_2000,guttmann_lattice_2010,joyce_exact_2002,joyce1973simple,montroll_random_1965} (note that we have rewritten the circuit Green's function $G$ of Sect.~\ref{secII} in the quasi-continuum picture to facilitate the discussion), the voltage at sub-lattice node $\mu$ of the unit cell at location $\mathbf{r}$ is found as
\begin{equation}
V(\mathbf{r},\mu) = I \Big(  \sum_{(\mathbf{r'},\nu
) \in \mathbf{\tilde{r}}_{in}} G(\mathbf{r},\mu,\mathbf{r'},\nu) - \sum_{(\mathbf{r'},\nu)\in \mathbf{\tilde{r}}_{out}} G(\mathbf{r},\mu,\mathbf{r'},\nu)  \Big),
    \label{volt_distr}
\end{equation}
\noindent where 
\begin{equation}
\begin{aligned}
	\mathbf{\tilde{r}}_{\text{in}} &\in \{(N\mathbf{a}_1,s_2),((N+1) \mathbf{a}_1,s_1)\},  \\
\mathbf{\tilde{r}}_{\text{out}} &\in \{(1\mathbf{a}_1 ,s_1),(2N \mathbf{a}_1,s_2)\} 
  \label{current_1dssh}
\end{aligned}
\end{equation}
denote the nodes where the currents are injected ($\mathbf{\tilde{r}}_{in}$) and extracted ($\mathbf{\tilde{r}}_{out}$). The spatial Green's function $G(\mathbf{r},\mu,\mathbf{r'},\nu)$ in \eqref{volt_distr} can be determined by applying the discrete Fourier transform given by
\begin{equation}
	G(\mathbf{r},\mu,\mathbf{r'},\nu) = \frac{1}{(2N)^D} \sum_\mathbf{k} G(\mathbf{k})_{[\mu,\nu]} e^{i \mathbf{k}\cdot(\mathbf{r} - \mathbf{r'})},
	\label{GreenFourier}
\end{equation}
\noindent where the momentum space index $\mathbf{k}=k_1 \mathbf{a_1}$ in which $k_1 = n_1 \pi/N$ and $n_1$ varies over a $2N$ period and $G(\mathbf{k})_{[\mu,\nu]}$ is the matrix element of the pseudoinverse of the momentum-space circuit Laplacian [for this example, it is $\mathcal{L}_{\text{1D}}^{\text{SSH}}(k_1)$ given in \eqref{lap_1d_ssh}]. $D$ represents the circuit dimension. We then tile the $2N$ unit cells comprising the physical and image circuits to form an infinite-sized lattice with a period of $2N$. Therefore, the Green’s function and the circuit Laplacian are constructed for the superlattice with a period of $2N$ such that the symmetric current injections and extractions in this periodic infinite lattice lead to a symmetric spatial voltage distribution with the period of $2N$; hence, the current entering the physical circuit cannot leak out through the boundaries of the physical circuit. Accordingly, the open boundary condition for the physical circuit with $N$ unit cells is satisfied. To realize this, we use the current distribution \eqref{current_1dssh} with \eqref{volt_distr} to determine the voltage at the leftmost node of the physical circuit as
\begin{equation}
    \begin{aligned}
    	 V(\mathbf{r}=N+1,\mu=s_1) =  I \big(& G((N+1),s_1, (N),s_2)+ G((N+1),s_1, (N+1),s_1)\\& - G((N+1),s_1, (1),s_1)  - G((N+1),s_1,(2N),s_2) \big),
    \end{aligned}
    \label{1Dssh_Vin}
\end{equation}
\noindent and that at the rightmost edge node as
\begin{equation}
    \begin{aligned}
    	V(\mathbf{r}=2N,\mu=s_2)  = I \big(& G((2N),s_2, (N),s_2) + G((2N),s_2, (N+1),s_1) \\&-G((2N),s_2, (1),s_1) -G((2N),s_2,(2N),s_2) \big).
    \end{aligned}
    \label{1Dssh_Vout}
\end{equation}
To find the voltages explicitly, we insert the momentum-space Green's function given in \eqref{GreenFourier} into \eqref{1Dssh_Vin} and \eqref{1Dssh_Vout} so that the impedance between the two edge nodes can be calculated as $Z_{1D}^{SSH}=V((N+1),s_1)-V((2N),s_2)$. At this point, we utilize the inversion and translation symmetries (i.e., $J(\mathbf{r})=J(\mathbf{-r})$ and $J(\mathbf{r})=J^{\intercal}(\mathbf{r})$ (where $(^\intercal)$ denotes the transpose operation), respectively) that our circuit possesses~\cite{venezian_resistance_1994,owaidat_resistance_2013,perrier_symmetries_2021,joyce1973simple,montroll_random_1965}. Because of the infinite periodic tiling, these symmetries imply that the voltage at the nodes where current is injected has the same magnitude as that at the nodes where current is extracted but with the opposite sign, i.e., $V(\mathbf{\tilde{r}} \in \mathbf{\tilde{r}}_{in}) = -V(\mathbf{\tilde{r}} \in \mathbf{\tilde{r}}_{out})$. Therefore, by using these symmetries and performing the relevant substitutions, the voltages at the edge nodes are found to be
\begin{equation}
\begin{aligned}
		V(&(N+1),s_1) = - V((2N),s_2) = \frac{I}{4N} \sum_{n_1=1}^{2N} \times  \\
		  & \frac{(e^{i n_1 \pi}-1)\left(C_2(1-e^{-i n_1 \pi})+C_1\left( 1-e^{-i n_1 \pi} e^{i n_1 \pi/N}\right) \right)}{i\omega C_1 C_2 \left(1-\cos (n_1 \pi /N)\right)},
\end{aligned}
\label{volt1DSSH}
\end{equation}
where we introduced $\mathbf{r}=n_1\mathbf{a_1}$ and $\mathbf{k}=k_1\mathbf{a_1}$ where $k_1 = n_1 \pi/N$. Here, because $e^{i n_1 \pi}-1=0$ when the integer $n_1$ is $even$, the summation is performed only for $odd$ $n_1$s. By considering $Z_{1D}^{SSH}=2V((N+1),s_1)/I=-2V((2N),s_2)/I$, and by means of trigonometric conversions [e.g., $1+e^{i n_1 \pi /N}=1+\cos (n_1\pi/N)+i \sin (n_1\pi/N)$ where the sine function can be neglected because of its zero contribution to the real part of the impedance], the edge-to-edge impedance as a function of the circuit size $N$ is given by
\begin{equation}
	Z_{1D}^{\text{SSH}}(N) = \frac{1}{N} \sideset{}{^*} \sum_{k_1} \frac{2C_2/C_1+(1+\cos k_1 )}{-i \omega C_2(1-\cos k_1 )}.
	\label{Imp1dsshCap}
\end{equation}
\noindent As before, $k_1 = n_1 \pi/N$ where $n_1$ is varied over the superlattice, i.e., $n_1\in\{1,2,\dots,2N\}$; however, the summation is restricted over $n_1 \in odd$ because of the above-mentioned summation rule, which the asterisk ($*$) on the summation operator indicates.
\\
\indent Using the same procedure, the edge-to-edge impedance in a circuit where the $C_1$ capacitors are replaced by capacitors with capacitance $C$ and the $C_2$ capacitors by inductors with inductance $L$ is obtained as
\begin{equation}
    Z_{1\text{D}}^{\text{SSH}_L}(N) = \frac{1}{N} \sideset{}{^*} \sum_{k_1} \frac{2- \omega^2 C L(1+\cos k_1)}{-i \omega C(1-\cos k_1)}.
	\label{Imp1dsshInd}
\end{equation}
%Note that, the same notation such as $k_1$ and the summation regulation in \eqref{Imp1dsshCap} apply to the above equation.
These formulas provide the impedance between nodes $s_1$ and $s_2$ in the leftmost and rightmost unit cells, respectively. Note that our sum includes a total of $N$ points, which corresponds to the $N$ unit cells.
\newline
\\
\subsection{Example: 2D SSH circuit}
We now proceed with a higher-dimensional circuit in which all the principal directions are non-trivial. For example, a 2D SSH circuit can be constructed by extending the 1D SSH circuit along the new $y$ direction. We first consider a capacitive 1D SSH circuit with intracell coupling $C_1$ and intercell coupling $C_2$. We then extend the circuit along the $y$ direction by using two inductances $L_1$ and $L_2$ to connect the nodes along the vertical direction, as shown in Figure~\ref{topo_circuits}(c). The resultant unit cell [dashed purple square in Figure~\ref{topo_circuits}(c)] has four distinct nodes denoted as $s_1$ to $s_4$ in which nodes $s_1$ and $s_4$ are located at opposite corners of the unit cell. Since preserving the uniformity of the unit cells requires the circuit size to be increased in multiples of the unit cells, the corner-to-corner impedance is measured between node $s_1$ in the first unit cell and node $s_4$ in the unit cell at the opposite corner of the 2D SSH circuit. The circuit Laplacian for a periodic 2D SSH circuit in momentum space is written as
\begin{equation}
    \mathcal{L}_{2\text{D}}^{\text{SSH}}(k_1,k_2) = i\omega \begin{pmatrix}
    \Sigma & \Gamma & \Delta & 0\\
    \Gamma^* & \Sigma & 0 & \Delta \\
    \Delta^* & 0 & \Sigma & \Gamma\\
    0 & \Delta^* & \Gamma^* & \Sigma
\end{pmatrix},
\label{2d_ssh_lap}
\end{equation}
\noindent where $\Gamma= -C_1 - C_2 e^{- i k_1} $, $\Delta= \frac{1}{\omega^2 L_1} + \frac{1}{\omega^2 L_2} e^{-i k_2}$, and $\Sigma = C_1 + C_2 - \frac{1}{\omega^2 L_1} - \frac{1}{\omega^2 L_2}$. Similar to the 1D SSH circuit, we evaluate the voltage distribution for the impedance measurement by introducing image circuits around the physical circuit [Figure~\ref{topo_circuits}(d)], injecting current at $\mathbf{\tilde{r}}_{\text{in}}$, and extracting current at $\mathbf{\tilde{r}}_{\text{out}}$ where 

\begin{equation}
\begin{aligned}
   \mathbf{\tilde{r}}_{\text{in}} &\in \{ ((N+1,N+1),s_1),((N,N+1),s_2),((N+1,N),s_3),((N,N),s_4)\}  \\
    \mathbf{\tilde{r}}_{\text{out}} &\in \{ ((1,1),s_1),((2N,1),s_2),((1,2N),s_3),((2N,2N),s_4)\}.
  \label{2dssh_current}
\end{aligned}
\end{equation}

\noindent Here, the spatial positions of the nodes at which current is injected and extracted are written in the form of $((n_1\mathbf{a_1},n_2 \mathbf{a_2}),s_\alpha)$ where $\alpha=(1,2,3,4)$ and $n_2$ and $\mathbf{a_2}$ are the coordinate and unit vector along the $y$ direction, respectively. 
The voltage $V\big(\mathbf{r}=(N+1,N+1),\mu=s_1\big)$ at the lower left corner of the physical circuit [see Figure~\ref{topo_circuits}(d)] is thus given by \\
\\

    \begin{equation}
		\begin{aligned}
			 V\Big(&\mathbf{r}=(N+1,N+1),\mu=s_1\Big) =\\
			 I \Big(&G\big((N+1,N+1),s_1,(N+1,N+1),s_1\big) + 
			 G\big((N+1,N+1),s_1,(N,N+1),s_2\big) \\
			 &+G\big((N+1,N+1),s_1,(N+1,N),s_3\big) + 
			 G\big((N+1,N+1),s_1,(N,N),s_4\big)\\ 
			 &-G\big((N+1,N+1),s_1,(1,1),s_1\big) 
			 -G\big((N+1,N+1),s_1,(2N,1),s_2\big) \\ 
			 -&G\big((N+1,N+1),s_1,(1,2N),s_3\big) - 
			 G\big((N+1,N+1),s_1,(2N,2N),s_4\big) \Big).
		\end{aligned}
		\label{2dssh_V1}
	\end{equation}
\noindent Similarly, the voltage at the upper right corner of the physical circuit is given by
	\begin{equation}
		\begin{aligned}
			 V\Big(&\mathbf{r}=(2N,2N),\mu=s_4\Big) = \\
			 I \Big(& G\big((2N,2N),s_4,(N+1,N+1),s_1\big) + 
			 G\big((2N,2N),s_4,(N,N+1),s_2\big) \\ 
			 &+G\big((2N,2N),s_4,(N+1,N),s_3\big) 
			 +G\big((2N,2N),s_4,(N,N),s_4\big) \\
			 &-G\big((2N,2N),s_4,(1,1),s_1\big) - 
			 G\big((2N,2N),s_4,(2N,1),s_2\big) \\
			&-G\big((2N,2N),s_4,(1,2N),s_3\big) -
			 G\big((2N,2N),s_4,(2N,2N),s_4\big) \Big).
		\end{aligned}
		\label{2dssh_V2}
	\end{equation}

\noindent We now employ the discrete Fourier transform given in \eqref{GreenFourier} to evaluate \eqref{2dssh_V1} and \eqref{2dssh_V2} explicitly. Because of the aforementioned circuit symmetries, the potential difference between the nodes at two opposite corners when the nodes are connected by a current source is $V((N+1,N+1),s_1) - V((2N,2N),s_4) = 2 V((N+1,N+1),s_1) = -2 V((2N,2N),s_4)$. By substituting \eqref{GreenFourier} into \eqref{2dssh_V1} (\eqref{2dssh_V2}), the voltage at the lower left (upper right) corner node can be obtained. The two-point impedance between opposite corner nodes is given by $Z_{2D}^{\text{SSH}}(N) = 2 V((N+1,N+1),s_1)$ or $Z_{2D}^{\text{SSH}}(N) = -2 V((2N,2N),s_4)$. After performing the substitutions, we arrive at
\begin{equation}
	Z_{2\text{D}}^{\text{SSH}}(N) = \sum_{k_1} \sum_{k_2} \frac{\text{Numerator($k_1, k_2$)}}{\text{Denominator($k_1, k_2$)}},
\end{equation}
where the numerator and denominator are explicitly given by
{\small
\begin{equation}
\begin{aligned}
\text{Numerator} = &\frac{1+(-1)^{1+n_1+n_2}}{2 i\omega^3 L_1^2 L_2^2} \\
& \times \Biggl[ L_1 \left[ 4  \omega^2 L_2 \left(2 C_1 +3 C_2\right)-4+e^{i(k_1-k_2)}\omega^2 C_1 L_2-2\omega^4 L_2^2\left(C_1^2+4 C_1 C_2+2 C_2^2\right)\right. \\
& \left. \quad + e^{-i k_1} \omega^4 C_1 C_2 L_2^2 - e^{-i(k_1-k_2)} \omega^4 C_1 C_2 L_2^2 + e^{-i k_2}\left(2-\omega^2 C_1 L_2 \right)\right. \\
& \left. \quad -e^{i k_1}\omega^2 C_1 L_2 \left(\omega^2 L_2 \left(2C_1+C_2\right)-4\right)- e^{i (k_1+k_2)} \omega^2 C_1 L_2\left( \omega^2 L_2 \left(2C_1+3C_2\right)-3\right)\right. \\
& \left. \quad +e^{i k_2} \left(2+ \omega^2 L_2 \left(C_1+4C_2-2\omega^2 L_2 \left(C_1^2+2C_1 C_2+ 2C_2^2\right)\right)\right)\right]\\
& -4 e^{i \frac{k_1}{2}} \omega^2 L_1^2 \left(2 \omega^2 L_2 \left(C_1+C_2\right)-2- \omega^4 C_1 C_2 L_2^2 \left(1- \cos(k_1)\right)\right) \\
& \quad \times \left( \left( C_1+C_2 \right) \cos(k_1/2)-iC_2 \sin(k_1/2)\right) \\
& +L_2 \left(2\omega^2L_2\left(2C_2\left(1+e^{i k_2}\right)+C_1\left(1+ e^{i k_1}+e^{i k_2}+e^{i (k_1+k_2)}\right)\right)\right. \\
& \left. \quad +\cos(2 k_2)+2i\sin(k_2)\left(\cos(k_2)-1\right)-1\right)\Biggr],\\
\text{Denominator} = & \frac{4 N^2}{\omega ^4 L_1^2 L_2^2} \Biggl[ - \omega^8 C_1^2 C_2^2 L_1^2 L_2^2 \left(\cos \left(2 k_1\right)+3\right) \\
& \quad +4 \omega ^4 C_1 C_2 \cos \left(k_1\right) \left(\omega ^4 C_1 C_2 L_1^2 L_2^2 -\omega ^2 L_1 L_2\left(C_1+C_2\right)  \left(L_1+L_2\right) \right. \\
& \left. \quad +L_1 L_2 \cos \left(k_2\right)+L_1^2+L_2^2+L_1 L_2\right) \\
& \quad +4 \cos \left(k_2\right) \left(\omega ^4L_1 L_2 \left(C_1^2+C_1 C_2+C_2^2\right) -\omega ^2\left(C_1+C_2\right) \left(L_1+L_2\right) +1\right) \\
& \quad +4 \omega ^6C_1 C_2 L_1 L_2 \left(C_1+C_2\right) \left(L_1+L_2\right) \\
& \quad -4 \omega ^4 \left(C_1 C_2 L_1^2+L_1 L_2\left(C_1^2+3 C_1 C_2+C_2^2\right) +C_1 C_2 L_2^2\right) \\
& \left. \quad +4\omega ^2 \left(C_1+C_2\right) \left(L_1+L_2\right)  - \cos \left(2 k_2\right) - 3 \right)\Biggr],
\end{aligned}
\end{equation}}
where $k_1$ and $k_2$ are the discrete momenta $k_1=n_1 \pi/N$ and $k_2=n_2 \pi/N$ where $(n_1,n_2) \in (1, 2, \cdots, 2N)$. Each $(k_1,k_2)$ contribution represents the impedance contribution from the length scale $(2\pi/k_1, 2\pi/k_2)$ in units of the lattice spacing. Notice that when taking the inverse of the circuit Laplacian given in \eqref{2d_ssh_lap}, any component that has zero eigenvalues should be omitted to avoid singularities. Although the analytical expression looks complicated, it demonstrates the utility of the method of images technique. The resulting voltage distribution over the superlattice reflects the inversion and translation symmetries of the circuit. 

Aside from the usual impedance peaks stemming from the $LC$ resonances, a topolectrical circuit exhibits an enormous impedance readout at the resonant frequency when the circuit is topologically non-trivial. It is well known that topological systems differ from the usual bulk systems owing to their special boundary modes. These boundary states are isolated from the bulk states and appear as mid-gap states in the admittance spectrum. Such mid-gap topological states lie on the zero-energy axis, and they are associated with very small eigenvalues. Therefore, any large impedance readout at the resonant frequency can be directly related to the topological mid-gap zero-modes when they exist~\cite{rafi-ul-islam_topoelectrical_2020,rafi-ul-islam_system_2022}. For example, the topological phase is defined by the ratio of two capacitors in the usual 1D SSH circuit in Ref.~\cite{lee_topolectrical_2018} and the circuit displays a non-trivial topological phase when $C_1 / C_2 < 1$ and a trivial phase when $C_1 / C_2 > 1$. However, even though the mid-gap states are protected by topological invariants such as a non-zero integer winding number, the unequal onsite energies lead to ill-defined invariants, hence, resulting in indistinguishable topological states~\cite{gong_topological_2018,kawabata_symmetry_2019,li_winding_2015,li_winding_2015}. In the TE context, the onsite energies are represented by the diagonal elements of the circuit Laplacian. Therefore, to unveil the topological boundary states, the circuit requires uniform grounding such that all the diagonal terms in the circuit Laplacian can vanish when the driving frequency is set to the resonant frequency. Otherwise, the boundary modes join the bulk modes due to the unequal potentials at different nodes and are no longer found as mid-gap states. To uniformly ground every node in the circuit, an artificial treatment is required, which would not be possible for analytical methods. Throughout this study, since we consider an infinite periodic lattice tiled with the original physical circuits and apply the method of images to obtain the exact analytical expression for the physical circuit, it may not be possible to introduce uniform grounding into the analytical two-point impedance formula. This is because the open boundary conditions are fulfilled as a direct consequence of the method of images, which results in the Neumann boundary condition in the circuit when the circuit Laplacian has non-uniform diagonal elements. Therefore, although the non-uniform diagonal elements representing the grounding mechanism cause the disappearance of the mid-gap topological states in most cases, in some examples, the topological states can remain isolated from the bulk states and appear in fractal-like diagrams as in Figure~\ref{fig:wr_N_MatrixPlot}d. 

We now present an example of the spatial voltage distribution in the superlattice containing the physical 2D SSH circuit and its image copies.

\subsubsection{Spatial voltage distribution of the 2D SSH circuit}

To show how the symmetric current injection and extraction over a periodic superlattice gives rise to equal potentials between the boundary nodes of the image and physical circuits, we present the spatial voltage distribution for a periodic 2D SSH circuit. The spatial voltage distribution can be calculated using the numerical Laplacian formalism [\eqref{LaplacianEq}]
\begin{equation}
	V=(J_{2\text{D}\text{-SSH}}^{\text{periodic}})^{-1} I_{\text{2D-SSH}},
\end{equation}
where the voltage matrix ($V$) over the periodic superlattice is obtained by performing the matrix multiplication of the inverse Laplacian matrix ($J_{2D\text{-SSH}}^{\text{periodic}}$) and the current matrix ($I_{\text{2D-SSH}}$). The current matrix corresponding to the 2D SSH circuit shown in Figure~\ref{topo_circuits}(d) is written as
\begin{equation}
	I_{2D\text{-SSH}}=I\left(\begin{NiceArray}{>{\strut}cccccccccc}[margin,extra-margin = 0pt]
		-1 & 0 & \cdots & 0 & 0 & 0 & 0 & \cdots & 0 & -1\\
		0 & 0 & \cdots & 0 & 0 & 0 & 0 & \cdots & 0 & 0\\
		\vdots & \vdots & \ddots & \vdots & \vdots  & \vdots & \vdots & \ddots & \vdots & \vdots  \\
		0 & 0 & \cdots & 0 & 0 & 0 & 0 & \cdots & 0 &0\\
		0 & 0 & \cdots & 0 & 1 & 1 & 0 & \cdots & 0 &0\\
		0 & 0 & \cdots & 0 & 1 & 1 & 0 & \cdots & 0 &0\\
		0 & 0 & \cdots & 0 & 0 & 0 & 0 & \cdots & 0 &0\\
		\vdots & \vdots & \ddots & \vdots & \vdots  & \vdots & \vdots & \ddots & \vdots & \vdots  \\
		0 & 0 & \cdots & 0 & 0 & 0 & 0 & \cdots & 0 &0\\
		-1 & 0 & \cdots & 0 & 0 & 0 & 0 & \cdots & 0 & -1\\
		\CodeAfter
		\begin{tikzpicture}
			\node [draw=red, rounded corners=3pt, inner ysep = 2pt,
			fit = (5-6) (1-10)] {} ;
		\end{tikzpicture}
	\end{NiceArray}\right)_{2N\times 2N}
	\label{2DcurrentMatrix}
\end{equation}

\begin{figure}[h!]
	\centering
	\includegraphics[width=12cm]{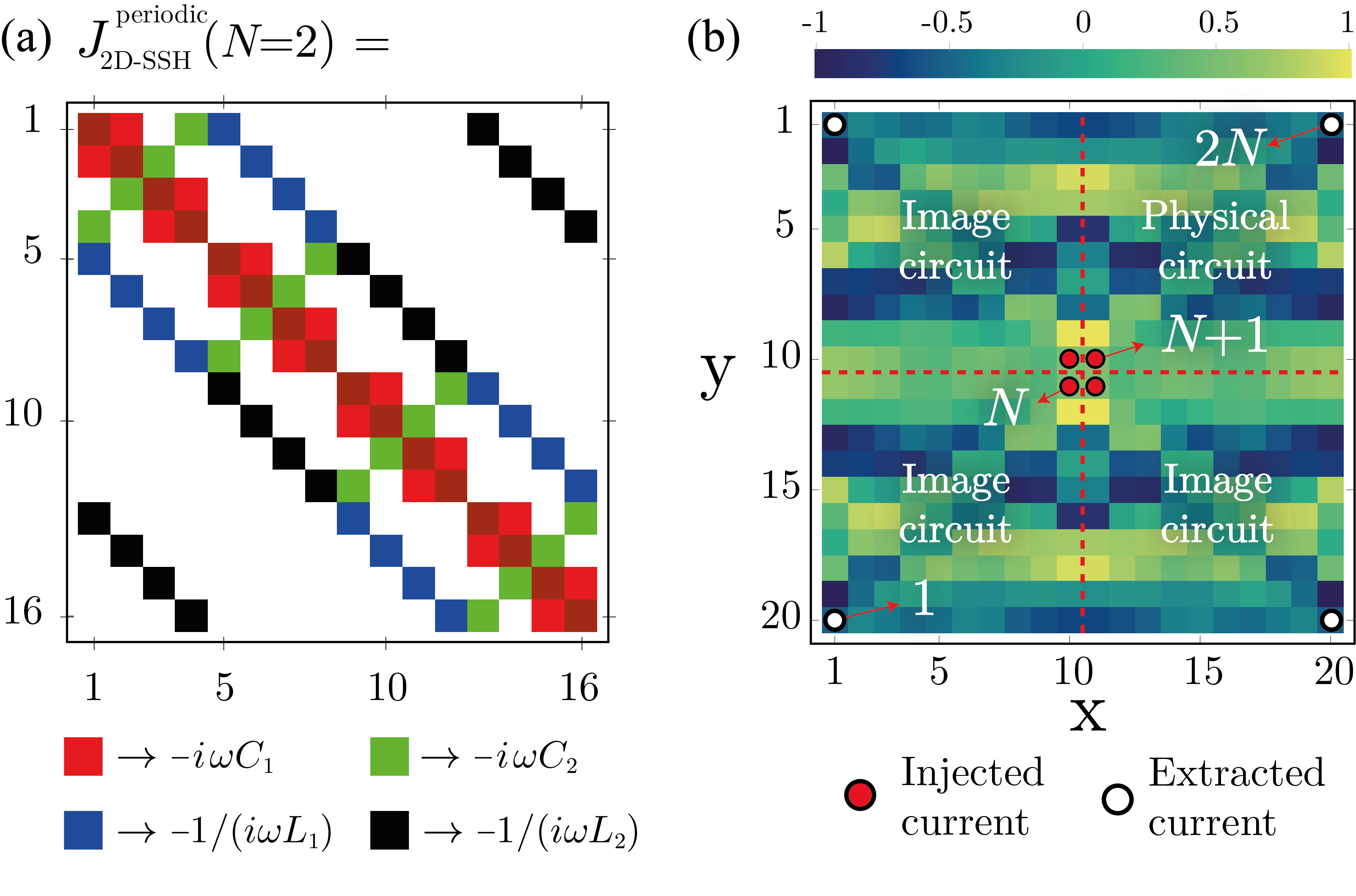}
	\caption{\textbf{The matrix representation of the periodic circuit Laplacian of the 2D SSH circuit for $N=2$ and its spatial voltage distribution for $N=10$.} (a) While the admittances between the node links are represented by the red, green, blue, and black squares in the off-diagonal elements, the darker red squares in the diagonal elements represent the total node conductance. %such that $\raisebox{.5ex}{\fcolorbox{white}{red}{\rule{0pt}{2pt}\rule{2pt}{0pt}}} \mathbf{\rightarrow} -i\omega C_1$, $\raisebox{.5ex}{\fcolorbox{white}{green}{\rule{0pt}{2pt}\rule{2pt}{0pt}}} \mathbf{\rightarrow} -i\omega C_2$, $\raisebox{.5ex}{\fcolorbox{white}{blue}{\rule{0pt}{2pt}\rule{2pt}{0pt}}} \mathbf{\rightarrow} -1/(i\omega L_1)$, $\raisebox{.5ex}{\fcolorbox{white}{black}{\rule{0pt}{2pt}\rule{2pt}{0pt}}} \mathbf{\rightarrow} -1/(i\omega L_2)$ 
	(b) Spatial voltage distribution matrix of the 2D SSH circuit for $N=10$ presented as a density plot. The red dashed lines separating the entire matrix into four blocks represent the boundaries between the physical and image circuits. The colors of the squares represent the magnitude of the voltage potential. Identical colors on both sides of the red dashed lines imply that there are zero potential differences between the boundary nodes. The red and white circles with black frames represent the nodes where the current is injected and extracted, respectively.}
	\label{VSpatial}
\end{figure}

The matrix elements framed by the red square in the upper right block corresponds to the current matrix of the physical circuit. Because we consider an infinite lattice tiling with a superlattice with a total period of $2N$ unit cells along each direction, we employ the periodic circuit Laplacian ($J_{2D\text{-SSH}}^{\text{periodic}}$). In Figure~\ref{VSpatial}(a), we display an example of the periodic circuit Laplacian of the 2D SSH circuit for $N=2$. Therefore, the matrix multiplication of the periodic Laplacian and the current matrix yields the spatial voltage distribution of the 2D SSH circuit. An example for the voltage distribution when $N=10$ is given in Figure~\ref{VSpatial}(b) where the voltage matrix is presented as a density plot. As can be seen from Figure~\ref{VSpatial}(b), the boundary nodes of the physical circuit have the same voltages as those of the boundary nodes of the image circuits. Due to the equal voltage potentials between the boundary nodes, the current injected at node $N+1$ cannot flow into the image circuits and is instead contained within the physical circuit. Therefore, the symmetrical current engineering as an analogy of the method of images leads to a perfectly symmetrical voltage distribution that therefore satisfies the boundary conditions. This makes it possible to obtain analytical expressions for finite-size circuits by applying the method of images to an infinite periodic lattice.

\section{Simplified analytic impedance formulas for RLC circuits with a single node per unit cell}\label{secRLC}

\begin{figure*}[t]
	\centering
	\includegraphics[width=\textwidth]{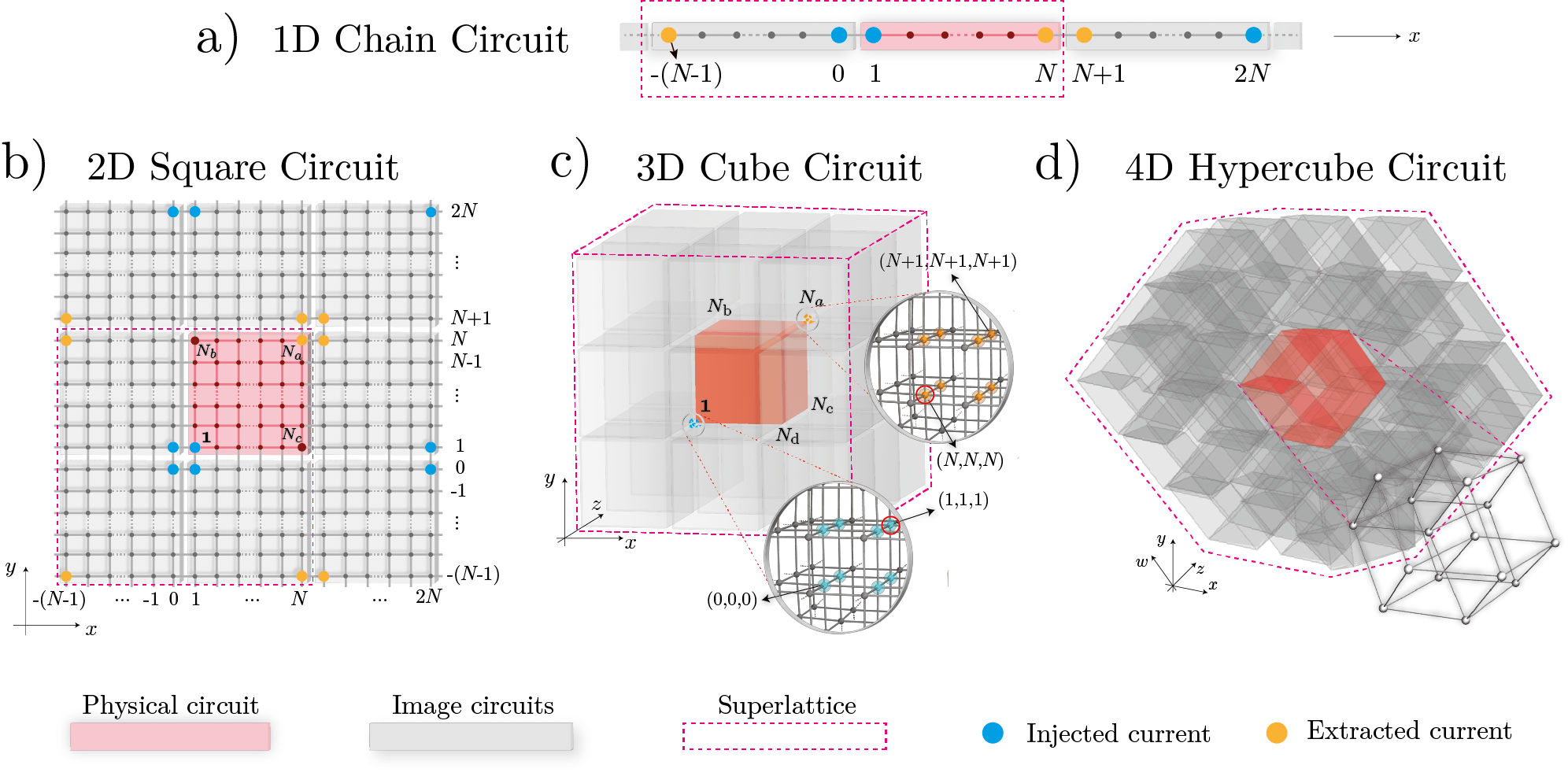}
	\caption{\textbf{Circuits lattices of different dimensionalities and their method of images implementations.} The impedance is measured (a) edge-to-edge in the 1D circuit and (b)-(d) corner to corner in the 2D and higher-dimensional circuits. Red and gray blocks represent the physical and image circuits constructing a periodic infinite lattice. Each superlattice consisting of a physical circuit and image circuits is marked by the magenta dashed lines in each illustration. The cyan and orange dots indicate the spatial positions of the nodes at which the current is injected and extracted, respectively. We label the corner nodes in the 2D square and 3D cube circuits as $\mathbf{N}_\alpha$ such that $\mathbf{N}_a$ is the diagonally opposite corner node to node $\mathbf{1}$ in every circuit. (b) The bigger red dots on the red plate denote the corner nodes $\mathbf{N}_c$ and $\mathbf{N}_b$ along the $x$ and $y$ directions, respectively. (c) The lower (upper) gray circle focuses on the node clusters where the current is injected (extracted). The red circles in the gray circles indicate the nodes belonging to the physical circuit i.e., nodes $\mathbf{1}$ and $\mathbf{N_a}$. (d) An illustrative representation of the 4D hypercube circuit. The input and output currents are not drawn to avoid excessive clutter. These circuits are homogeneous when the links are all resistive with admittances of $1/R_i$, all capacitive with admittances of $ z_{i} = i\omega C_{i} $, or all inductive with admittances $z_{i}= 1/(i\omega L_{i})$ [here $i=(1,2,\dots,D)$], but are heterogeneous when at least two distinct admittances with opposite phases such as $z_{1}=i\omega C$ and $z_{2}= 1/(i\omega L)$ are present.}
	\label{network_schematics}
\end{figure*}
 
Here, we derive a generalized analytical expression for the impedance between two nodes at the opposite edges in 1D and at opposite corners in higher dimensions in homogeneous and heterogeneous finite circuits that have a single node per unit cell. The nodes are connected by components with an admittance of $z_\alpha$ along the $\alpha$th direction where $\alpha \in  (1,2,\cdots, D)$ where $D$ is the dimension of the circuit. We define a homogeneous circuit as one in which the $z_\alpha$'s along all the directions have the same phase, i.e., they are either all capacitors or all inductors, and a heterogeneous circuit as one in which the $z_\alpha$'s along the different directions have different phases. To derive a generic expression, let us first consider a 2D infinite lattice made of image copies of a physical finite circuit [refer to Figure~\ref{network_schematics}(b)] where the current is injected at nodes $\mathbf{r}_{\text{in}}$ and extracted from nodes $\mathbf{r}_{\text{out}}$. Using the definition of the Green's function $G J = -\delta $ [\eqref{LaplacianEq}] and the translation invariance of the circuit [which implies that $G(\mathbf{r},\mathbf{r'}) = G (\mathbf{r}-\mathbf{r'})$], the voltage at any lattice point $\mathbf{r}$ is found to be
\begin{equation}
	V(\mathbf{r}) = I \left( \sum_{\mathbf{r'} \in \mathbf{r}_{\text{in}}} G(\mathbf{r - r'}) - \sum_{\mathbf{r'} \in \mathbf{r}_{\text{out}}} G(\mathbf{r - r'}) \right),
	\label{voltage_distribution}
\end{equation}
\noindent where $\mathbf{r} = n_1 \mathbf{a_1}+ n_2 \mathbf{a_2}$ with integers $(n_1,n_2) \in \{-(N-1),\cdots,N\}$, and $\mathbf{a_1}$ and $\mathbf{a_2}$ are unit vectors corresponding to the $x$ and $y$ directions, respectively. Note that we define the limits of the superlattice as $\{-(N-1),\cdots,N\}$ by choice unlike the limits of the superlattice that we define for the 1D and 2D SSH circuits. The period of the superlattice remains unchanged at $2N$. We now perform the discrete Fourier transformation [i.e., $G(\mathbf{r-r'})=1/(2N)^D \sum_{\mathbf{k}} G(\mathbf{k}) e^{i \mathbf{k}.(\mathbf{r-r'})}$] and recall $G(\mathbf{k})=\mathcal{L}^{-1}(\mathbf{k})$ to determine the spatial Green's function $G(\mathbf{r-r'})$ in \eqref{voltage_distribution} as
\begin{equation}
	G(\mathbf{r-r'})=\frac{1}{(2N)^2}\sum_{\mathbf{k}}{\frac{e^{i \mathbf{k \cdot (r-r')}}}{\mathcal{L(\mathbf{k})}}},
	\label{Green_r}
\end{equation}
\noindent where $\mathbf{k}=(k_1 \mathbf{a_1} + k_2 \mathbf{a_2})$ is the momentum space index where $k_i=n_i\pi/N$ where $i=(1,2)$ and $n_i\in\{-(N-1),\dots,N\}$. Notice that, because we derive an analytical expression for the circuit that has a trivial unit cell [i.e., the circuit is made of a single-type node], the momentum-space Laplacian can take the role of momentum-space Green's function since $G(\mathbf{k})$ is no longer a matrix but is instead just the reciprocal of $\mathcal{L}(\mathbf{k})$. To achieve a symmetric voltage distribution over the superlattice, the current is injected and extracted at the nodes as depicted in Figure~\ref{network_schematics}(b):
\begin{equation}
	\begin{aligned}
		&\mathbf{r}_{in} \in \{(0,0),(1,0),(0,1),(1,1)\}  \\
		&\mathbf{r}_{out} \in \{(N,N),(N+1,N),(N,N+1),(N+1,N+1)\},
	\end{aligned}
	\label{current_cases}
\end{equation}
where $(n_1,n_2)$ is a short-hand notation for $\mathbf{r}=(n_1\mathbf{a_1}+n_2\mathbf{a_2})$. From \eqref{current_cases} and \eqref{voltage_distribution}, the node voltage V($\mathbf{r}$) can be found through
\begin{equation}
	\begin{aligned}
		V(\mathbf{r}) = I  \Big( & G(\mathbf{r}) + G(\mathbf{r}+\mathbf{a_1}) + G(\mathbf{r}+\mathbf{a_2}) + G(\mathbf{r}+\mathbf{u}) \\
		& -G(\mathbf{r}+N\mathbf{u}) - G(\mathbf{r}+N\mathbf{u}+\mathbf{a_1}) \\ & -G(\mathbf{r}+N\mathbf{u}+\mathbf{a_2}) -
		G(\mathbf{r}+N\mathbf{u}+\mathbf{u}) \Big),
	\end{aligned}
	\label{voltage_final_with_green}
\end{equation}
where $\mathbf{u} = \mathbf{a_1}+\mathbf{a_2}$ denotes the unit vector for 2D lattices. To proceed, we insert the Green's function defined in \eqref{Green_r} into \eqref{voltage_final_with_green} to obtain the voltages at crosswise nodes $\mathbf{1}$ and $\mathbf{N}_a$ where $\mathbf{1}=1\mathbf{a_1}+1\mathbf{a_2}$ and $\mathbf{N}_a = N\mathbf{a_1}+N\mathbf{a_2}$. As mentioned, the translation symmetry in our circuit implies that $V(\mathbf{r})=-V(\mathbf{-r})$. Thus, the voltage difference between nodes $\mathbf{1}$ and $\mathbf{N}_a$ is $V(\mathbf{1})-V(\mathbf{N}_a) = 2 V(\mathbf{1}) = -2 V(\mathbf{N}_a)$. Therefore, we find the voltage at the corner nodes as
\begin{equation}
	\begin{aligned}
	V(\mathbf{1} ) = - V(\mathbf{N}_\alpha)=&  \frac{I}{4 N^2}\sum_{n_1} \sum_{n_2} \frac{1}{\mathcal{L}(\mathbf{k})} \times \\ & \left(1-e^{i\pi(n_1+n_2)}\right) \left(1+e^{i \pi n_1/N}+e^{i \pi n_2/N}+e^{i \pi (n_1+n_2)/N}\right).
		\end{aligned}
	\label{volt_final_exp}
\end{equation}
Notice that the term $(1-e^{i\pi (n_1+n_2)})$ in the numerator of \eqref{volt_final_exp} is zero when $(n_1+n_2)$ is $even$ and is 2 when $(n_1+n_2)$ is $odd$. Therefore, this term can be replaced by 2 provided that the summation is restricted to odd $(n_1+ n_2)$. After simplifying \eqref{volt_final_exp}, the impedance between the corner nodes $\mathbf{1}$ and $\mathbf{N}_a$ in a 2D square circuit, as a function of circuit size $N$, can be written as
\begin{equation}
		Z_{2\text{D}}^{\mathbf{1,N_a}}(N)= \frac{2}{N^2} \sideset{}{^*} \sum_{\mathbf{k}} 	\frac{\cos(k_{1}/2)\cos(k_2/2)\cos((k_1+k_2)/2)}{\mathcal{L}(\mathbf{k})},
	\label{final_imp}
\end{equation}
where the asterisk on the sum operator indicates that $\mathbf{k}=\sum_{i} k_i \mathbf{a}_i$ where $k_i=n_i \pi /N$ and $i=(1,2)$ and a restricted summation over odd $(n_1+n_2)$. $\mathcal{L}(\mathbf{k})$ represents the corresponding 2D circuit Laplacian. One can then calculate the two-point impedance for both 2D homogeneous and heterogeneous circuits by simply assigning the corresponding circuit Laplacian to \eqref{final_imp}. This equation is also valid for the impedance between any pair of corner nodes as long as the summation is taken over $n_2 \in \text{odd}$ for $Z_{2\text{D}}^{\mathbf{1,N_b}}$ and $n_1 \in \text{odd}$ for $Z_{2\text{D}}^{\mathbf{1,N_c}}$. This is because in the derivation of the expressions for $Z_{2\text{D}}^{\mathbf{1,N_b}}$ or $Z_{2\text{D}}^{\mathbf{1,N_c}}$, one arrives at \eqref{volt_final_exp} with the factor $(1-e^{i n_2\pi})$ in the numerator for  $Z_{2\text{D}}^{\mathbf{1,N_b}}$ and $(1-e^{i n_1\pi})$ for $Z_{2\text{D}}^{\mathbf{1,N_c}}$ because, for example, the current distribution when the current is injected and extracted at nodes $\mathbf{1}$ and $\mathbf{N_c}$ is written as $\mathbf{r}_{\text{in}} \in \{(0,0),(1,0),(0,1),(1,1)\}$ and $\mathbf{r}_{\text{out}} \in \{(N,0),(N+1,0),(N,1),(N+1,1)\}$, respectively. (Here, $\mathbf{N}_b = N\mathbf{a_1}$ and $\mathbf{N}_c =N\mathbf{a_2}$ are the vertical and horizontal opposite corner nodes to the lower left node $\mathbf{1} =1\mathbf{a_1}+1\mathbf{a_2}$, respectively [see Figure~\ref{network_schematics}(b)].) Therefore, there are contributions to the impedance only when $n_1\in \text{odd}$ for $Z_{2\text{D}}^{\mathbf{1,N_c}}$ or $n_2\in \text{odd}$ for $Z_{2\text{D}}^{\mathbf{1,N_b}}$. From here, we can deduce that the Fourier component(s) of the principal direction(s) corresponding to the corner node only contribute to the impedance when its (their) summation is odd.

Inspired by the derivation for the impedance formula for the 2D square circuit (\eqref{final_imp}) and taking into account the summation rule for the impedance between different corner nodes, we can obtain a general analytical expression for the corner-to-corner impedance of both $D$-dimensional homogeneous and heterogeneous circuits as
\begin{equation}
	Z(N)=\frac{2}{N^D} \sideset{}{^*} \sum_{\mathbf{k}} \frac{\left( \prod_{i=1}^D \cos(k_i /2)\right) \times \cos \left(\sum_{i=1}^D k_i /2 \right) }{\left(\sum_{i=1}^D \lambda_{i}(1-\cos(k_i)\right)+z_{gnd}/2},
	\label{general_Z}
\end{equation}
where $\lambda_i$ is the admittance of the coupling along each direction, $D$ represents the dimension of the circuit, and  $\mathbf{k}=\sum_{i=1}^{D}k_i \mathbf{a}_i$ where $k_{i}=\frac{n_i\pi}{N}$ where $n_i\in\{1,2,...,2N\}$ and $i=(1,2,\dots,D)$. The asterisk sign $(^*)$ on the summation operator implies that the impedance computation must be performed considering the summation rule. For example, since the diagonally opposite corner nodes can only be defined by considering all the spatial indices $n_i$s, one must take the summation over $(n_1+n_2+\dots+n_D)\in \text{odd}$. To uniformly attach a grounding component of a single type to every node in the circuit, the admittance $z_{gnd}$ can be assigned a non-zero admittance if desired, but it will be set to 0 if not. Note that because of the translation symmetry of the periodic infinite lattice, for simplicity, we can relabel the superlattice boundaries as $n_i\in \{1,2,\dots,2N\}$ instead of $\{-(N-1),\dots,N\}$. The corner-to-corner impedance between any pair of corners in a homogeneous or heterogeneous circuit can thus be calculated using \eqref{general_Z} by simply setting the $\lambda_i$'s in the denominator to the admittance values of the coupling along each principal direction. We will now apply this generalized expression to homogeneous and heterogeneous circuits in the following sections. We only consider passive circuit elements such as capacitors, inductors, and resistors, which can only store or dissipate energy~\footnote{But see the experiment in Ref.~\cite{zhang2022observation}, which demonstrates that such passive RLC elements can bring about non-local impedance responses in suitably designed circuits.} pumped into the circuit.

\section{Impedance results for homogeneous $RLC$ circuits with trivial unit cells in various dimensions }
In general, the corner-to-corner impedance of a homogeneous circuit constructed from passive components can be expected to increase uniformly with the circuit size because the components operate at the same phase (i.e., their admittances have the same sign). Since passive components cannot pump energy into the circuit, we intuitively expect the circuit to behave like a \textit{waveguide} as the circuit size increases. To illustrate this, we consider a 1D circuit constructed from a single type of circuit element with admittance $z_1$ and calculate the impedance between the two opposite edge nodes as new unit cells are added. Employing \eqref{general_Z} with $D=1$, $\lambda_{1} = z_1$, $z_{gnd}=0$, and $2\cos^2(k_1/2)=1+\cos(k_1)$, the edge-to-edge impedance in 1D homogeneous circuits is obtained as
\begin{equation}
    Z_{1D}^{\text{hom}}(N)=\frac{1}{N} \sideset{}{^*} \sum_{k_1} \frac{1+\cos k_{1}}{z_{1}(1- \cos k_{1})},
    \label{Z_imp_1D_homo}
\end{equation}

\begin{figure}[t]
    \centering
    \includegraphics[width=10cm]{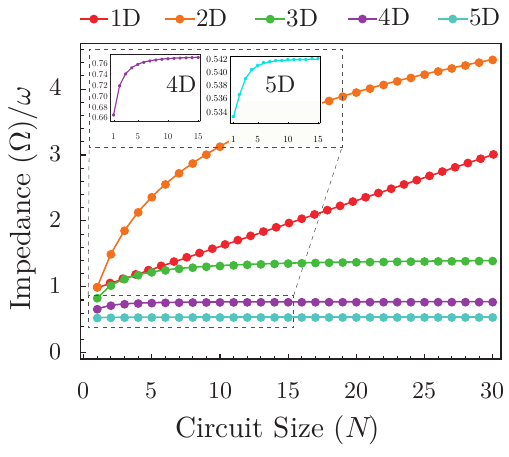}
    \caption{\textbf{Impedance across two opposite (edge) corner nodes of homogeneous (1D), 2D, 3D, 4D, and 5D circuits.} All the circuits are constructed from only a single type of capacitor $C$ and we set $C=1\ \mu \text{F}$ for all impedance computations. (The edge-to-edge impedance in the 1D chain circuit is normalized by $Z/10$ for illustration.) While the impedance between two edge nodes in a 1D chain circuit scales linearly with the size, the impedance between two opposite corner nodes scales logarithmically in 2D finite circuits and rapidly approaches a finite saturation value in three dimensions or higher, as further detailed in Section~\ref{appendixSaturation}.}
    \label{fig:uniform_circuits}
\end{figure}

\noindent where the asterisk means that $k_{1} = n_1 \pi/N$ and $n_1\in \{1,2,\dots,2N \}$ and $n_1\in odd$. Here, the impedance takes real values if the nodes are connected by resistors and takes imaginary values if $z_1$ corresponds to either a capacitor with an admittance of $i\omega C$ or an inductor with an admittance of $1/(i\omega L)$. Regardless of the component represented by $z_1$, the impedance increases linearly with the circuit size, as shown in Figure~\ref{fig:uniform_circuits}. Note that \eqref{Imp1dsshCap} and \eqref{Imp1dsshInd} provide the same edge-to-edge impedance as \eqref{Z_imp_1D_homo} when only one type of circuit element is present in the circuit and when $N$ is increased in steps of 2. This is because while a unit cell consists of two nodes for \eqref{Imp1dsshCap} and \eqref{Imp1dsshInd}, a unit cell comprises only a single node for \eqref{Z_imp_1D_homo}.

\begin{figure}[h!]
	\centering
	\includegraphics[width=10cm]{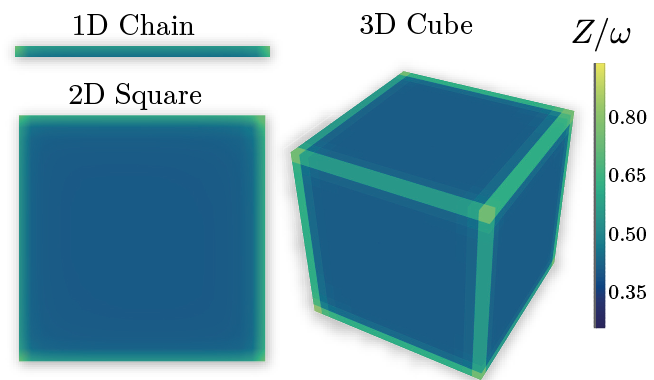}
	\caption{\textbf{Nearest-neighbor impedance \textit{distribution} in 1D, 2D, and 3D homogeneous bounded circuit arrays, with higher impedances near the boundaries.} The impedance is calculated between a node and its horizontal nearest neighbor (NN). Every cell on the density plot is colored according to the overall impedance between that node and its NN. The total impedance at a node decreases approaching the center, and increases approaching the boundaries. All the circuits are made of only a single type of capacitor with capacitance $C$ which is set as $C =1\,\mu\text{F}$.}
\label{fig:3d_cube_impedance_map}
\end{figure}

We now extend the 1D chain circuit to a 2D square circuit in which the nodes are connected by components with admittance $z_1$ along the horizontal direction and by components with admittance $z_2$ along the vertical direction. The corresponding circuit Laplacian becomes $\mathcal{L}_{2D}(k_1,k_2)=z_{1}(1- \cos k_{1}) + z_{2}(1- \cos k_{2})$. The two-point impedance for 2D circuits can be obtained by recalling \eqref{general_Z} with $k_{1} = n_1 \pi/N$ and $k_{2} = n_2 \pi/N$ and $z_1$ and $z_2$ set to $i\omega C$ for capacitors, $ \frac{1}{i\omega L} $ for inductors, and $1/R$ for resistors. Note that $z_1$ and $z_2$ must have the same phase (i.e., both are capacitors, inductors, or resistors; or capacitors + resistors; or inductors + resistors) in order to preserve the homogeneous structure of the circuit. We can intuitively expect the impedance of the 2D homogeneous circuits to scale logarithmically with the circuit size (Figure~\ref{fig:uniform_circuits}) because every addition of a new layer of unit cells results in a uniform increment in the overall impedance between two corner nodes with $Z=\int^N \rho \frac{d N'}{N^{'}} \sim z \log N$. Aside from the corner-to-corner impedance, the impedance between two first nearest-neighbor nodes (NNs) along the horizontal or vertical directions also scales uniformly with the size. Figure~\ref{fig:3d_cube_impedance_map} shows that the impedance between NNs is highest at the corner nodes and lower approaching the bulk nodes. This is because the current is pushed to the boundaries and accumulates there as it is reflected by the boundaries in a finite network. This trend also applies to higher dimensions such as 3D, 4D, 5D and beyond, for which interesting new phenomena can arise, and which can be feasibly implemented via circuits~\cite{bao_topoelectrical_2019,lee_electromagnetic_2018,li_emergence_2019,lee_imaging_2020,zhang_topolectrical-circuit_2020,wang_circuit_2020}.

Similar to the procedure with which we constructed the 2D homogeneous circuit, the 2D square circuit can be extended to a 3D cube circuit by simply linking every node with an additional component $z_3$ along the third direction, for which the Fourier component is written as $k_3$ where $k_3=n_3 \pi/N$. Figure~\ref{fig:uniform_circuits} shows the two-point impedance behavior for homogeneous circuits of different dimensions as the circuit size increases. What is most remarkable is the qualitatively different behavior in higher dimensionalities~\footnote{In higher-dimensional systems where non-reciprocity also accompanies resistive non-Hermiticity, the non-local response can alter the effective dimensionality of the entire system~\cite{jiang2022dimensional,li2021quantized}.}: for homogeneous circuits is, while 1D and 2D circuits exhibit unique scaling behaviors of linear and logarithmic impedance alternations, respectively, the circuits with the dimensionality of three and above tend to saturate at some finite constant values. For instance, the two-point impedance in the 3D cube circuit starts saturating after $N\sim25$, as can be seen in Figure~\ref{fig:uniform_circuits}. In general, for $D\geq3$, the impedance saturates to a finite value more quickly as the dimensionality of the network increases. The overall scaling in homogeneous circuits for $D\geq 3$ can be characterized by
\begin{equation}
    Z(N) \,\sim\, Z_{\text{D-dim}}^{\text{sat}} - \frac{(Z_{\text{D-dim}}^{\text{sat}})^D}{N^{D-1}},
    \label{zsaturation}
\end{equation}
\noindent where $Z_{\text{D-dim}}^{\text{sat}}$ is the saturation value, $D$ is the dimension of the circuit, and $N$ is the circuit size in terms of unit cells. To find the $Z_{\text{D-dim}}^{\text{sat}}$s, we take the limit of \eqref{general_Z} as $N \to \infty$. We present the full analytical expression for the saturation values in Section~\ref{appendixSaturation}. The saturation values obtained from \eqref{DDint} are: $Z_{\text{3-dim}}^{\text{sat}}=1.44015~\Omega$, $Z_{\text{4-dim}}^{\text{sat}}=0.774964~\Omega$, and $Z_{\text{5-dim}}^{\text{sat}}=0.542093~\Omega$, which confirm the impedance trends in the circuits for $D\geq 3$ in Figure~\ref{fig:uniform_circuits}. For $D\geq 3$, one can analytically show that in the continuum limit, the saturation impedances can be reduced to lower-dimensional integrals, thereby expressing lower-dimensional slices of the circuit in terms of lumped effective resistances.

\section{Impedance results for heterogeneous $RLC$ circuits with nontrivial unit cells}
\subsection{2D circuits}
We now turn to heterogeneous circuits, where the corner-to-corner impedance exhibits a peculiar scaling  behavior that differs significantly from that of homogeneous circuits. Heterogeneous circuits refer to circuits that have at least two types of components for which the admittances have different complex phases, such as capacitors and inductors. To illustrate the scaling behavior, we first consider a homogeneous 1D chain circuit consisting of $N-1$ capacitors with the admittance of $z_1 = i\omega C$ along the horizontal direction $x$. We then connect $N-1$ inductors with the admittance of $z_2=1/(i\omega L)$ along the vertical direction $y$ to each node in the 1D chain circuit. This results in a two-dimensional $LC$ circuit with $N \times N$ nodes. To calculate the impedance across the diagonal corner nodes, we recall \eqref{general_Z} and assign the admittances $z_1$ and $z_2$ to the $\lambda_i$'s such that $z_1$ and $z_2$ correspond to the principal directions. The corner-to-corner impedance of the 2D $LC$ circuit is thus given by
\begin{equation}
    Z_{2\text{D}}^{het}(N)=\frac{2}{N^2} \sideset{}{^*} \sum_{\mathbf{k}} \frac{\cos (k_1/2) \cos (k_2/2) \cos ( (k_1+k_2)/2)}{i\omega C (1- \cos k_1)+\frac{1}{i \omega L}(1- \cos k_2)},
    \label{Z_imp_2D_het}
\end{equation}
\begin{figure}[h!]
    \centering
    \includegraphics[width=10cm]{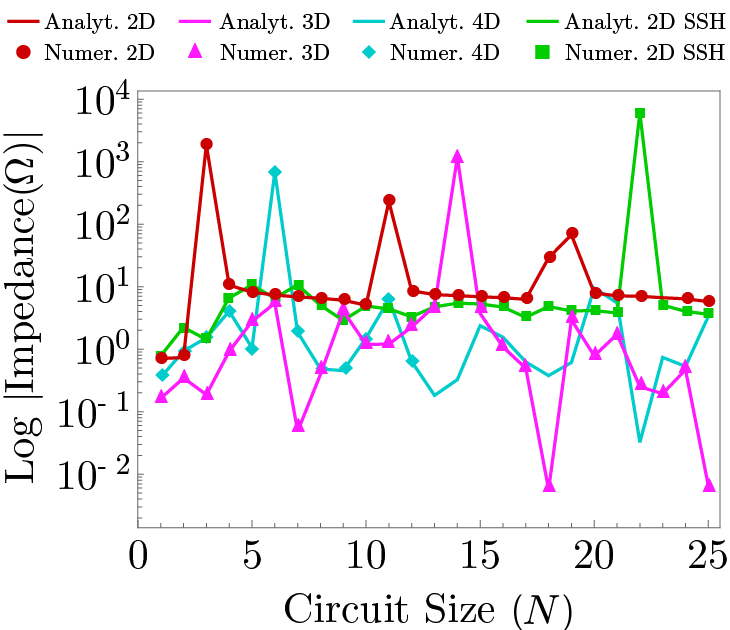}
    \caption{\textbf{Analytically and numerically calculated impedance between two opposite corner nodes in heterogeneous 2D (red), 3D (magenta), 4D (cyan), and 2D SSH (green) circuits.} Sharp peaks emerge at certain circuit sizes in the heterogeneous circuits, unlike the logarithmic-like impedance scaling in homogeneous circuits. The component values are $\omega (C, L)=(1.7~\Omega^{-1}, 2~\Omega$) for the 2D, $\omega (C_1, C_2, L)=(2.1~\Omega^{-1}, 1~\Omega^{-1}, 2.5~\Omega)$ for the 3D,  and $\omega (C_1, C_2, L_1, L_2)=(2.5~\Omega^{-1}, 1.8~\Omega^{-1}, 2.0~\Omega, 1.0~\Omega)$ for the 4D $LC$ circuits, and  $\omega (C_1, C_2, L_1, L_2)=(2.2~\Omega^{-1}, 1~\Omega^{-1}, 2~\Omega, 1~\Omega)$ for the 2D SSH circuit.}
    \label{fig:het_circuits_imp}
\end{figure}

\noindent where the asterisk on the summation operator indicates that $\mathbf{k}=k_1 \mathbf{a_1}+k_2\mathbf{a_2}$ where $k_1 = n_1 \pi/N$ and $k_2 = n_2 \pi/N$ and $n_1$ and $n_2$ are integers satisfying $(n_1,n_2)\in \{1,2,\cdots,2N \}$ where $(n_1+n_2)$ is $odd$. This equation can be used for any corner-to-corner impedance measurement in a 2D circuit by simply considering the following summation rules ($^*$): For instance, for the impedance across two opposite corner nodes, one must take the summation over $(n_1+n_2) \in \text{odd}$, while for the impedance between the two vertical corner nodes the summation should be taken over $n_2$ is odd. Apart from the driving AC frequency $\omega$ and component parameters $C$ and $L$, which are the independent parameters in the circuit, the circuit size $N$ becomes an additional independent parameter that affects the two-point impedance. It is well known that in $LC$ resonator circuits, resonances occur at the resonant frequency $\omega = 1/\sqrt{LC}$ regardless of the value of $N$. Here, we uncover a curious phenomenon in which strong resonances occur only at particular circuit sizes. Figure~\ref{fig:het_circuits_imp} shows the occurrence of impedance jumps with the variation of the circuit size in several dimensions. The origin of these impedance resonances can be explained by considering the denominator of \eqref{Z_imp_2D_het}. An  impedance resonance occurs when the values of $L$, $C$, and $N$ are such that there exist integers $n_1$ and $n_2$ at which the denominator of \eqref{Z_imp_2D_het} becomes nearly zero when the terms proportional to $iC$ and $1/iL$ cancel each other almost completely. Hence, strong resonances occur only at certain circuit sizes. Moreover, the impedance peaks stem not from the numerator but arise because of the almost-vanishing denominator at particular circuit sizes in $LC$ circuits. Therefore, heterogeneous circuits with $D>2$ can potentially exhibit more interesting circuit-size dependent impedance resonances since their Laplacians have more elements than that of lower-dimensional circuits leading to more possible combinations for the cancellation in the denominator.
 
\subsection{Heterogeneous circuits in 3D and higher dimensions}
We next investigate higher-dimensional $LC$ circuits starting with a 3D cube circuit array. We construct a cube circuit by extending the 2D $LC$ square circuit along the $z$ direction using capacitors with the admittance $z_3$. The cube circuit illustrated in Figure~\ref{network_schematics}(c) therefore comprises inductors with the admittance $1/(i\omega L)$ along the $y$ direction and capacitors with the admittances of $i \omega C_1$ and $i \omega C_2$ linking nodes along the $x$ and $z$ directions, respectively. We calculate the impedance across the corner nodes by using the $D=3$ analog of \eqref{general_Z} and employing the Laplacian $\mathcal{L}_{3D}(k_1,k_2,k_3)=i \omega C_1(1- \cos k_{1})+1/(i \omega L)(1- \cos k_{2})+i \omega C_2(1- \cos k_{3})$ where $k_{i} = n_i \pi/N$ where $n_i\in \{1,2,\dots,2N\}$ and $i=(1,2,3)$. Due to the oddness of the discrete momentum (i.e., $(1-(-1)^{\mathbf{k}})$, refer to \eqref{volt_final_exp}), the summation rule is determined by the corner nodes between which the impedance is to be measured. For example, to find the impedance between the diagonally opposite corner nodes $Z_{3D}^{\mathbf{1,N_a}}$, the summation is taken over $(n_1+n_2+n_3) \in odd$; while for $Z_{3D}^{\mathbf{1,N_b}}$ the rule is $(n_2+n_3) \in odd$, for $Z_{3D}^{\mathbf{1,N_c}}$ it is $(n_1+n_3) \in odd$ and finally for $Z_{3D}^{\mathbf{1,N_d}}$ it is $ n_1 \in odd$. Therefore, only the coordinate indices of the corner node opposite to the node $\mathbf{1}$ contribute to the impedance calculation and only when their sum is odd.

Similar to the 2D case, the denominator in \eqref{general_Z} with this Laplacian leads to an anomalously large impedance measurement when there exist integer values of $(n_1,n_2,n_3) \in \{1,\cdots,2N \}$ where it becomes almost zero. The procedure we have followed to build the cube circuit from the square circuit can be extended to construct hypercube circuits. For example, a 4D $LC$ circuit can be constructed by introducing an additional direction $w$, which is perfectly feasible in 3D space due to the versatile connectivity of electrical circuits, without resorting to synthetic (non-genuine) lattice dimensions. The momentum-space Laplacian of the 4D circuit in which the nodes along the $w$ direction are linked by inductors with the admittance of $z_4=1/(i \omega L_2)$ is  
\begin{equation}
\begin{aligned}
     \mathcal{L}_{4\text{D}}^{\text{het}}(k_1,k_2,k_3,k_4) 
    = i \omega \Big[ &C_1(1- \cos k_{1})-\frac{1}{\omega^2 L_1}(1- \cos k_{2}) \\ &+C_2(1- \cos k_{3})-\frac{1}{\omega^2 L_2}(1- \cos k_{4}) \Big],
    \label{Lap_4D}
\end{aligned}
\end{equation}
\noindent where $k_i = n_i\pi/N$  where $n_i \in \{1,2, \cdots, 2N\}$ and $i=(1,2,3,4)$. One can straightforwardly extend this circuit further to five dimensions (5D) or higher dimensions by invoking \eqref{general_Z}. Whereas the impedance between two opposite corner nodes in homogeneous circuits rapidly approaches a constant saturation value, it no longer exhibits a uniform trend in the heterogeneous cube and hyper-cube circuits. Instead, large impedance peaks are observed at certain values of $N$. Figure ~\ref{fig:het_circuits_imp} shows the variation of the corner-to-corner impedance of the 4D $LC$ circuit with the circuit size $N$. %We can conclude that the violation of uniform impedance scaling occurs when the circuit Laplacian is attenuated due to the polarity the Fourier components of each direction at a certain $N$. 
Interestingly, plotting the impedance peaks due to these size-dependent resonances against the driving frequency and circuit size gives a pattern that is reminiscent of fractals, as shown in Figure~\ref{fig:wr_N_MatrixPlot}. These fractal-like patterns are not due to the dimensionality of the circuits but rather depend on the homogeneity of circuits and, indeed, arise in the circuit size-versus-parameter diagrams of heterogeneous circuits of any dimensions, as we shall discuss in the following section.

\section{Emergent fractal-like resonances in impedance scaling behavior}
\begin{figure*}
	\centering
	\includegraphics[width=\textwidth]{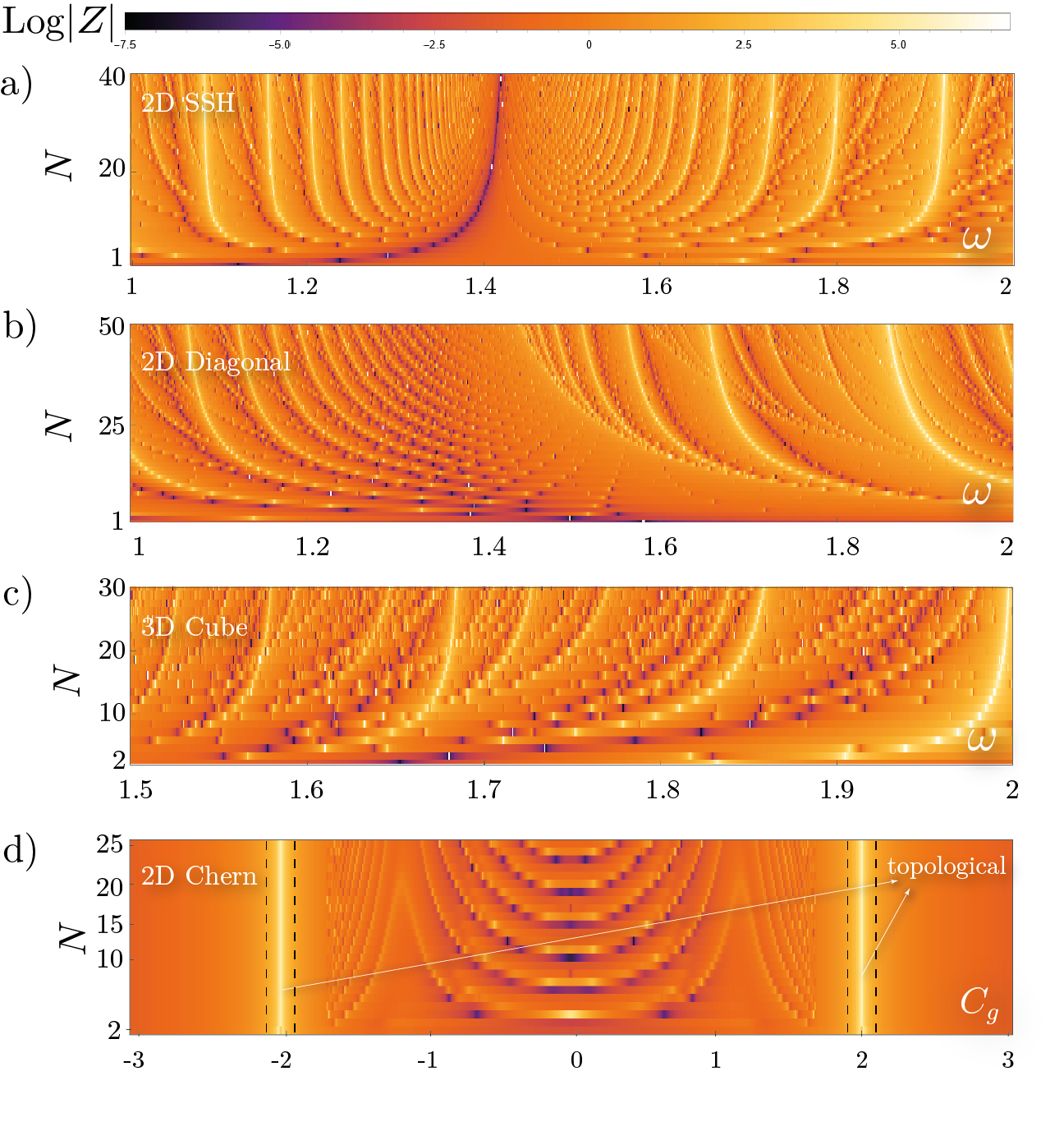}
	\caption{\textbf{Fractal-like structures emerging in the plots of the corner-to-corner impedance in heterogeneous circuits versus the circuit size $N$ and driving frequency $\omega$ [for (d), $N$ versus $C_g$]:} (a) 2D SSH circuit, (b) 2D square $LC$ circuit with diagonal cross links (obtained numerically), and (c) 3D cube circuit. The strong impedance resonances that appear as brighter branches are reminiscent of fractals. Despite the consideration of parasitic and intrinsic resistances that cause a variation of the component parameters, the structures  always survive robustly. (d) Impedance diagrams for the non-trivial topolectrical 2D Chern kink circuit. The brightest impedance branches arise in the non-trivial parameter regime and correspond to the topological zero edge modes. The circuit parameters are $(\omega C_1, \omega C_y, \omega L, R)=(1.2\ \Omega^{-1}, 0.6\ \Omega^{-1}, 0.6\ \Omega, 20\ \Omega)$.}
	\label{fig:wr_N_MatrixPlot}
\end{figure*}

Most physical systems do not change fundamentally as their system sizes are varied, except for special \textit{critical} systems exhibiting size-induced phase transitions~\cite{li_critical_2020,li_impurity_2021,rafi-ul-islam_critical_2022,siu_critical_2022}. However, in our heterogeneous circuits, we observe sharp impedance deviations from the overall trend. Whereas special resonances can be expected to occur because of specific modulations of the circuit parameters, the observed anomalous impedance resonances that arise due to the system length are unexpected. These anomalous impedance resonances can therefore be treated as violations of the logarithmic impedance scaling between two opposite corner nodes. \eqref{Z_eig} relates the two-point impedance to the eigensystem of the circuit Laplacian. The most significant difference between the eigenvalue spectrum of the Laplacians of homogeneous and heterogeneous circuits is that heterogeneous circuits typically have very small eigenvalues that can give rise to huge impedance readouts. Such a cancellation occurs in heterogeneous circuits between different components with admittance values of opposite signs, whereas no such cancellation occurs in homogeneous circuits as the admittance values have the same sign. Further, our analytical formulas show that at certain circuit sizes, there  exist combinations of the integers $n_1, n_2,\dots, n_D$ denoting the Fourier components at which the denominator of the impedance expression almost cancels out and results in strong impedance resonances. Most interestingly, the length-dependent impedance resonances seem self-organized and exhibit a fractal-like pattern in the impedance-length plots. As can be seen in Figure~\ref{fig:wr_N_MatrixPlot}, while the strongest impedance resonances (the brightest branches) occur along certain branches, the magnitude of the impedance varies between the branches. While one can obtain higher resolution diagrams (and spanning larger parameter spaces) with more computational resources, the overall trends do not change. Since the number of bulk states is defined by the total number of nodes in the circuit, the number of branches arising in these diagrams must be the same as the number of bulk states. Therefore, every row corresponding to a selected circuit size $N$ (i.e., the impedance peaks in a row corresponding to a fixed $N$) is indeed the projection of the zero energy axis of the admittance spectrum. Consequently, because the number of admittance states increases with the circuit size $N$, the number of states intersecting with the zero-axis increases, thus resulting in the formation of curly peak lines in the diagrams. In contrast, the topological impedance resonances, as seen in Figure~\ref{fig:wr_N_MatrixPlot}(d), form straight lines because the number of topological modes is not defined by the circuit size but rather defined by the topological invariants. As a result, the emergent fractal-like diagrams can be treated as a complete picture of the resonant properties of a circuit regardless of whether it is topological or not. We next examine two exemplary circuits to further discuss how these fractal-like diagrams differ depending on the circuit arrays.

\subsection{Two further examples for the emergent fractal-like resonances}
\subsubsection{Example: 2D square LC circuit with diagonal connections}
\begin{figure}[h!]
	\centering
	\includegraphics[width=10cm]{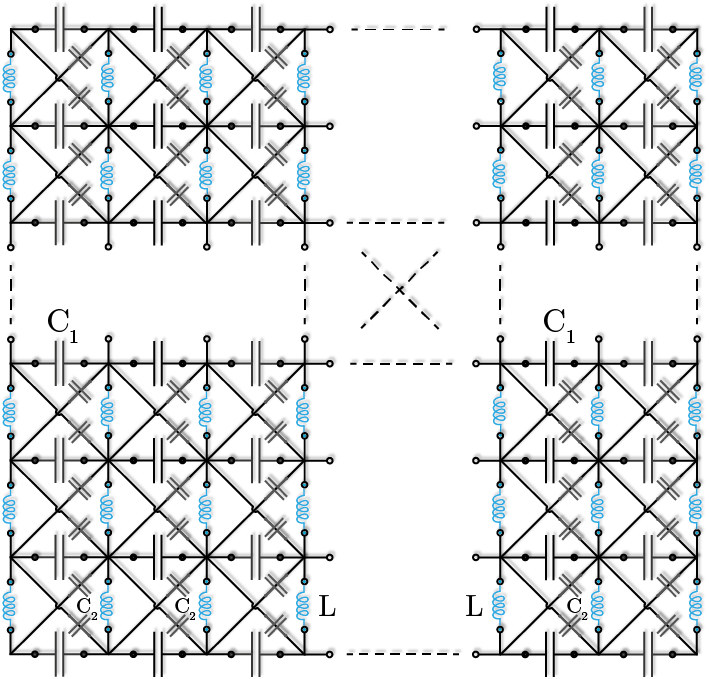}
	\caption{\textbf{2D LC circuit with diagonal cross connections with capacitors $C_2$.} The usual 2D $LC$ circuit in which the horizontal and vertical nodes are coupled with capacitors $C_1$ and inductors $L$, respectively, can be recovered by simply setting $C_2=0$. To obtain the fractal-like diagram [Figure~\ref{fig:wr_N_MatrixPlot}(b)], the impedance is numerically measured between two opposite corner nodes as the circuit size increases.}
	\label{fig:2ddiagonalcircuit}
\end{figure}
The heterogeneous 2D $LC$ circuit that we have studied so far (see Figure~\ref{network_schematics}(b)) can be made more complicated, which in turn leads to richer fractal-like diagrams. This circuit has capacitors $C_1$ along with the vertical direction, inductors $L$ along the horizontal direction and a second capacitor $C_2$ connecting every node with its diagonal neighbors [see Figure~\ref{fig:2ddiagonalcircuit}]. Because of these cross connections, the circuit Laplacian has an additional term representing the nearest neighbor links with the admittance $i\omega C_2$. Therefore, the circuit Laplacian [\eqref{Z_imp_2D_het}] becomes
\begin{equation}
		\mathcal{L}_{2\text{D}}^{\text{cross}}(k_1,k_2) = 2 i \omega C_1 (1-\cos k_1) + \frac{2}{i \omega L} (1-\cos k_2) + 4 i \omega C_2 (1- \cos k_1 \cos k_2)
\end{equation}
where $k_i = n_i \pi/N$ where $n_i \in \{1,2,\dots,2N\}$ and $i=(1,2)$. As we have discussed above, the size-dependent impedance resonances are induced by the attenuation of the Laplacian. Because $G=J^{-1}$ and $V=GI$, the attenuation of the Laplacian leads to voltage accumulation at the node where the current is injected. Thus, an enormous impedance read-out ($Z \propto V$) occurs. Since the circuit Laplacian for the usual 2D $LC$ circuit has a simpler form as given in the denominator of \eqref{Z_imp_2D_het}, it is expected that more complicated and fascinating branches may emerge in the fractal diagram of the 2D $LC$ diagonally connected circuit. As can be seen in Figure~\ref{fig:wr_N_MatrixPlot}(b), as the circuit size increases, many resonance branches that emerge shape a pattern. Although the diagonal connections result in a well-organized fractal-like diagram, we emphasize that the diagonal links in this circuit lead to a current leakage from the physical circuit to the image circuits. In the usual 2D $LC$ circuit [see Figure~\ref{VSpatial}(b)], the symmetrical current injection and extraction ensures that the same potential exists at the neighbor boundary nodes of the physical and image circuits so that no current leakage occurs at the boundaries. However, in a circuit with diagonal links, the diagonal links connect boundary nodes with different electrical potentials, resulting in current flow through the cross links. Therefore, it is not possible to obtain a simple analytical expression via the method of images for the diagonally connected 2D $LC$ circuit unless complicated current injection and extraction engineering is performed, which is not practical.

\subsubsection{Example: 2D Chern kink topolectrical circuit}
\begin{figure}[h!]
	\centering
	\includegraphics[width=9cm]{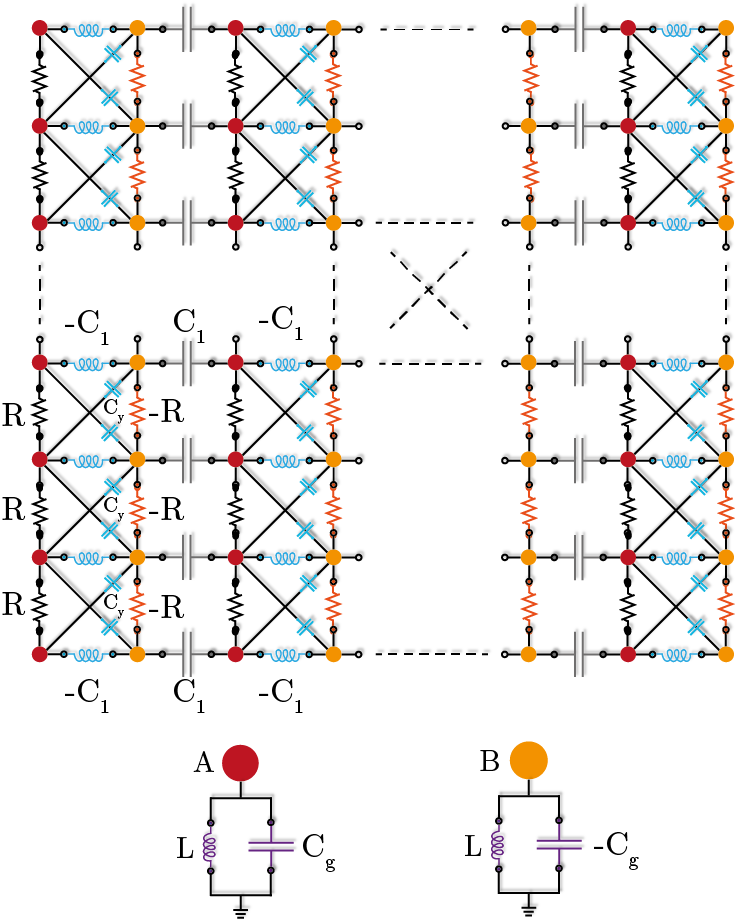}
	\caption{\textbf{The schematic of the 2D Chern kink circuit studied in Ref.~\cite{rafi2022valley}.} The circuit has two distinct nodes in a unit cell labeled as node $A$ (red circle) and node $B$ (orange circle). The two nodes within a unit cell are connected with intra-cell inductors, whose admittance is equal to that of the capacitor $-C_1$, i.e., $ L_1 = 1/ (\omega^2 C_1) = -C_1$. The horizontally adjacent unit cells are connected to each other via capacitors $C_1$. While the vertically adjacent $ A $ nodes are coupled via positive resistors $R$, two neighboring vertical $ B $ nodes are connected via negative resistors $ -R $ realized by means of negative resistance converters (NRCs). The circuit has also capacitors $C_y$ connecting $A-B$ and $B-A$ nodes along the $y$ direction. }
	\label{fig:2dvalleycircuit}
\end{figure}
Topolectrical circuits can exhibit sophisticated physical phases such as valley-dependent edge states~\cite{rafi2022valley}. For example, the valley-dependent corner or edge states can be modulated by setting the onsite capacitors $C_g$ connecting every node to the ground [refer to Figure~\ref{fig:2dvalleycircuit}]. The negative capacitance $-C_g$ and resistance $-R$ can be achieved by using negative impedance converters (NICs) and negative resistance converters (NRCs), respectively~\cite{hofmann_chiral_2019}. The circuit Laplacian at the resonant frequency is written as
\begin{equation}
	\begin{aligned}
		\mathcal{L}_{\text{TE}}^{\text{Chern}}(k_1,k_2) = &(-C_1 + C_1 \cos k_1 + 2 C_y \cos k_2) \sigma_x \\
		& + C_1 \sin k_1 \sigma_y + (C_g + 2 R \sin k_2) \sigma_z,
	\end{aligned}
\end{equation}
where the $\sigma_\alpha$;s ($\alpha=x,y,z$) represent the Pauli matrices and where $k_i=2n_i\pi/N$ where $i=(1,2)$. Although a uniform grounding mechanism can be employed to isolate the topological states from the bulk states, a Chern topolectrical circuit can be designed so that it exhibits clear chiral propagating boundary states in the admittance spectrum without a uniform grounding mechanism. This is due to nontrivial center-of-mass pumping through a Laughlin-style pumping argument~\cite{vanderbilt_chapter_2006,qi_topological_2008,soluyanov_wannier_2011,lee_free-fermion_2015,liu_quantum_2016,jurgensen_chern_2022,wang2023experimental}, which has been demonstrated in a variety of classical and quantum settings~\cite{chang_experimental_2013, nash_topological_2015,ding_experimental_2019,ni_robust_2020,motruk_detecting_2020,koh_simulation_2022}. Here, due to the nearest neighbor connections and the onsite capacitors represented by the factor $\sigma_z$ in the Laplacian, the topological states are separated despite the nonuniform grounding. Therefore, this circuit is an ideal candidate to demonstrate impedance resonances arising in the fractal-like diagrams due to the topological states. For instance, the impedance resonances stemming from the edge states clearly appear in the fractal-like diagram as seen in Figure~\ref{fig:wr_N_MatrixPlot}(d). The constant-magnitude impedance peaks can be treated as the evidence for zero-energy topological modes since the smallest eigenvalues always belong to these eigenstates. Aside from being the strongest resonances, the other remarkable property of these impedance resonances is that while the bulk states result in curved lines with the variation of the circuit size, the topological branches form distinctive straight lines. This is because even though the number of bulk states increases with the increase in circuit size, the number of topological states remains constant and the position of these states depends on the component parameters, which are fixed. Since we maintain the same component values except for the varying element ($C_g$), the topological impedance resonances appear at the same $ C_g $ value in the fractal-like diagram.

\section{Application of the method of images to general geometries}

The method of images can be applied to a general lattice model to derive an analytical expression. As discussed in the main text, the configuration of the current injection and extraction determines the spatial distribution of node voltages. To induce boundaries by utilizing the equipotential emerging on both sides of the symmetry axes, it is essential to inject and extract the current symmetrically. This results in a mirror-symmetric dispersion around the considered symmetry axes. (Note that such a symmetrical voltage distribution can only be achieved if inversion symmetry is present.) For instance, since the unit cell of the 2D SSH circuit (as well as the 2D square $LC$ circuit) exhibits translational symmetry along the $x$ and $y$ symmetry axes, the spatial potential profiles of the physical and image circuits are mirror symmetric about the $x$ and $y$ axes when there is symmetric current injection and extraction [refer to Figure~\ref{VSpatial}(b)]. On the other hand, the symmetry axes of a hexagonal lattice provide additional opportunities for creating unique boundary designs by strategically positioning current injection and extraction sites with respect to these axes. Consequently, we can achieve diverse boundary designs by configuring the current injection and extraction points relative to the symmetry axes in a honeycomb lattice.

\begin{figure*}[t]
	\centering
	\includegraphics[width=15cm]{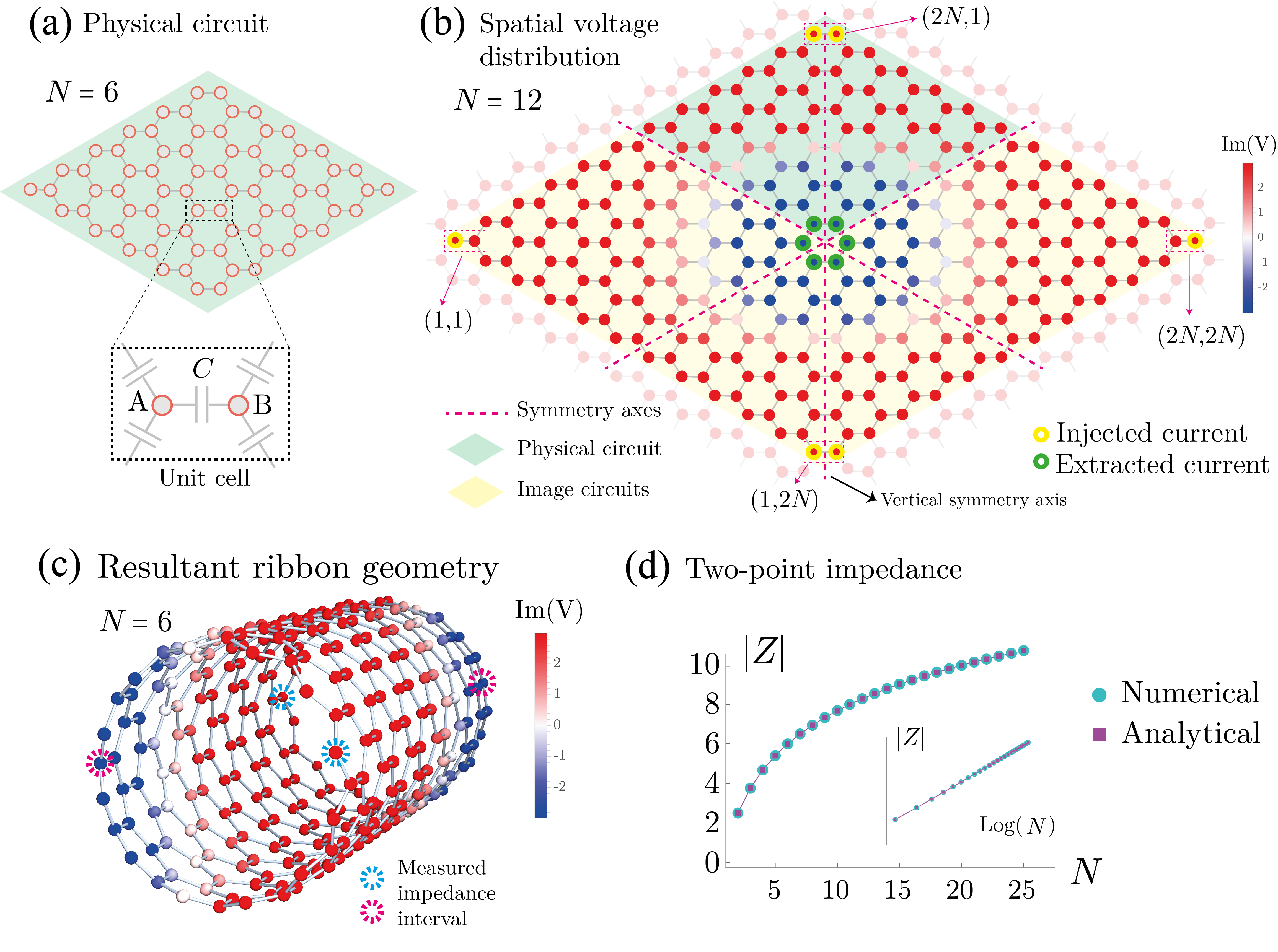}
	\caption{\textbf{An exemplary implementation of the method of images to achieve finite circuits in a general geometry using a 2D honeycomb lattice.} a) The rhombus-shaped finite physical circuit with $N=6$. A unit cell comprises two sublattice nodes $A$ and $B$, each connected by coupling capacitance $C$. b) The spatial voltage distribution response to the injected (yellow-outlined nodes) and extracted (green-outlined nodes) current configuration. The node colors represent the relative magnitude of the voltage at each node. A mirror-symmetrical potential distribution occurs about the vertical symmetry axis, ensuring the emergence of a boundary. c) The resultant distribution leads to a ribbon geometry with zigzag edges. The alternating node colors represent the voltage magnitude. d) The measured impedance between any two nodes outlined with magenta and cyan dashed circles is identical. The impedance increases logarithmically (inset) as the circuit size expands.}
	\label{fig:honeycomb}
\end{figure*}

To demonstrate how a symmetrical current injection and extraction configuration with respect to certain symmetry axes can be utilized to establish a boundary in general geometries, we examine a 2D honeycomb lattice composed of two sublattice nodes $A$ and $B$. Our objective is to derive an analytical expression for a ribbon geometry incorporating a honeycomb lattice structure and zigzag edges. We begin by considering a geometrically rhombic honeycomb lattice with a zigzag edge design. Its momentum space circuit Laplacian reads as
\begin{equation}
	\mathcal{L}_{\varhexagon}(\mathbf{k}) = i \omega C \begin{pmatrix}
		3& -(1+e^{-i k_1}+e^{-i k_2})\\ -(1+e^{i k_1}+e^{i k_2}) & 3
	\end{pmatrix},
	\label{lap_honey}
\end{equation}
where $C$ represents the coupling capacitance, $\omega$ is the driving frequency, and $k_1$ and $k_2$ are the momentum indices. We proceed to tile the space with rhombus-shaped circuits and apply currents at specific nodes to achieve a symmetrical voltage distribution about one of the symmetry axes. Since we initially assume a periodic lattice along each direction, it is sufficient to achieve a symmetrical voltage distribution around a single axis in order to obtain the desired ribbon geometry. In Figure~\ref{fig:honeycomb}(b), the magenta dashed lines represent the symmetry axes under consideration. Our current injection and extraction configuration results in a mirror-symmetric potential distribution about the vertical symmetry axis. This implies that the voltages of the sublattice nodes within a unit cell along the vertical axis have equal magnitudes, ensuring no current flows between these nodes. Since our circuit remains periodic in both directions, the equal voltages on either side of the vertical axis satisfy the boundary condition required to achieve a ribbon geometry.
We now move forward with deriving an analytical expression for the two-point impedance of the ribbon lattice. We recall \eqref{volt_distr} to determine the node voltages and express the spatial positions of the nodes where the current is injected and extracted, respectively, as
\begin{equation}
\begin{aligned}
	\mathbf{\tilde{r}}_{\text{in}} \in \{ &((1,1),1),((1,2N),1),((1,2N),2),((2N,1),1),\\&((2N,1),2),((2N,2N),2)\},  \\
	\mathbf{\tilde{r}}_{\text{out}} \in \{ &((N,N),2),((N+1,N),1),((N+1,N),2),\\&((N,N+1),1),((N,N+1),2),((N+1,N+1),1)\}.
	\label{current_honey}
\end{aligned}
\end{equation}
where $((n_1,n_2),\nu)$ is shorthand for $\mathbf{r}=(n_1\mathbf{a_1}+n_2\mathbf{a_2})$ where $\mathbf{a_1}$ and $\mathbf{a_2}$ are the unit vectors, $n_1$ and $n_2$ are the position indices and $\nu\in(1,2)$ represents the sublattice nodes A and B, respectively. We can calculate the voltage at the first node as
\begin{equation}
	\begin{aligned}
		&V(\mathbf{r}=1,\mu=1)\\
		&=I\Big(G((1,1),1,(1,1),1)+G((1,1),1,(1,2N),1)\\
		&+G((1,1),1,(1,2N),2)+G((1,1),1,(2N,1),1) \\ 
		&+G((1,1),1,(2N,1),2)+G((1,1),1,(2N,2N),2) \\ 
		&-G((1,1),1,(N,N),2)-G((1,1),1,(N+1,N),1) \\ 
		&-G((1,1),1,(N+1,N),2)-G((1,1),1,(N,N+1),1) \\ 
		&-G((1,1),1,(N,N+1),2)-G((1,1),1,(N+1,N+1),1)\Big).
	\end{aligned}
	\label{honeyVoltage}
\end{equation}
By employing the momentum space Green's function in \eqref{GreenFourier} and $\mathcal{L}_{\varhexagon}^{-1}(\mathbf{k})=G(\mathbf{k})$, the voltage at node $V(\mathbf{r}=1,\mu=1)$ is obtained as
\begin{equation}
	\begin{aligned}
		&V(\mathbf{r}=1,\mu=1) = \frac{1}{4i\omega C N^2(3-\cos(k_1) - \cos(k_2) -\cos (k_1-k_2))}\\
		&\times \left( 5(1+\cos(k_1)+\cos(k_2))+2\cos(k_1-k_2)+\cos(k_1+k_2)-\cos(\frac{N-1}{N}(n_1+n_2)\pi) \right.\\
		&\left. -2\cos(\frac{N-1}{N}(n_1-n_2)\pi)-4\cos(\frac{N-1}{N}n_1\pi-n_2\pi)-\cos(\frac{N-1}{N}n_1\pi+n_2\pi)\right.\\
        &\left.-\cos\left(\frac{N-1}{N}n_2\pi+n_1\pi\right)\right),
	\end{aligned}
\end{equation}
where $k_1=n_1\pi/N$ and $k_2=n_2\pi/N$. To simplify the above equation, we assume that $n_1\in(1,2,3,\cdots,2N)$ and $n_2\in(1,3,5,7,\cdots,2N-1)$. Owing to the translation symmetry in our circuit, the impedance between node 1 and its corresponding counterpart located at the opposite edge [i.e., $(N,N)$] is given by $Z_{\varhexagon}(N) = 2V(1, 1) = 2V(N, N)$. The impedance between any pair of nodes denoted by magenta and cyan dashed circles in Figure~\ref{fig:honeycomb}(c) is identical, a consequence of the symmetrical voltage distribution. In Figure~\ref{fig:honeycomb}(d), we plot our numerical and analytical impedance calculation results for the ribbon circuit depicted in Figure~\ref{fig:honeycomb}(c). There is an exact correspondence between the two sets of results. Our results also show that the logarithmic dependence of the circuit impedance on the circuit size $N$ applies for two-dimensional circuits regardless of the lattice geometry [see Figure~\ref{fig:honeycomb}(d)].

\subsection{Fractal-like nature of the anomalous impedance in the 2D honeycomb lattice with open edges}
\begin{figure}
	\includegraphics[width=12cm]{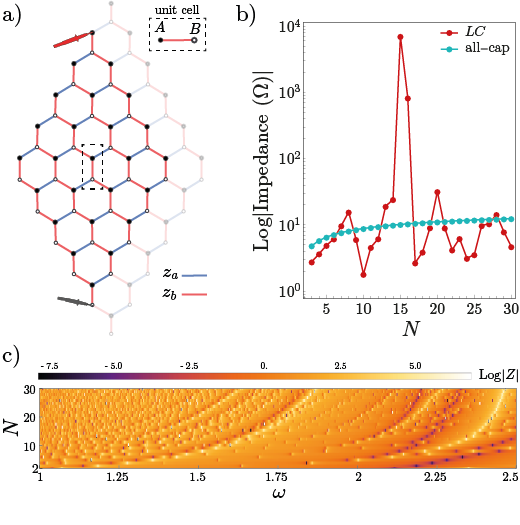}\centering
	\caption{\textbf{Honeycomb circuit lattice and its impedance results.} a) Illustrative honeycomb lattice with zigzag edge design when $N=5$. A unit cell consists of nodes belonging to two sublattices $A$ (black circles) and $B$ (black framed white circles). The blue and red lines represent the node links with the different admittances of $z_a$ and $z_b$, respectively. The faded cells indicate the extension of the circuit when $N=6$, as an example. b) The impedance response of the circuit in (a). The circuit is a resonant medium when $z_a=1(i\omega L)$ and $z_b=i\omega C$ and presents sharp impedance resonances as a function of the circuit size. The parameters used are $\omega C=1$ for the uniform circuit made of only capacitors and $\omega C=1$, $\omega L=2.21$ for the $LC$ honeycomb circuit. c) Fractal scaling of the 2D honeycomb lattice in the circuit size and driving frequency domain when $C=L=1$. The brightest and darkest branches represent the strong anomalous impedance resonances, which depend on the circuit size $N$. The legend located above the density plot indicates the logarithm of the absolute impedance.}
	\label{fig4}
\end{figure}

The presence of anomalous impedance in $LC$ circuits is a universal property that arises when two distinct components with opposing phases are present in the circuit lattice. Here, we investigate the impedance characteristics of the two most distant nodes in a honeycomb lattice with a zigzag edge design. We consider nodes belonging to two sub-lattices $A$ and $B$ that are connected by an admittance $z_a$ (blue lines in Fig.~\ref{fig4}a). The nearest neighbor nodes are also connected by an admittance $z_b$ (red lines in Fig.~\ref{fig4}a) in a unit cell. 
The resultant lattice is fully reactive and non-resonant when $z_a$ and $z_b$ have the same phases and resonant when they have opposite phase. To investigate the lattice size-dependent impedance characteristics, we examine the circuit in both cases and find that there are anomalous impedance resonances at specific circuit sizes when the circuit parameters are fixed. Fig.~\ref{fig4}b illustrates the impedance across circuit size behavior under two conditions: when the entire circuit is composed of only a single type of capacitor with a capacitance of $z_a=z_b=i\omega C$ (cyan), and when it is composed of two distinct admittances of $z_a=1/(i\omega L)$ and $z_b=i\omega C$ (red). The fractal scaling observed in the $LC$ circuits also arises in the 2D honeycomb lattice. Fig.~\ref{fig4}c illustrates the impedance resonances exhibiting fractal-like scaling with the variation of the circuit size and driving frequency. Furthermore, the edge design of the lattice affects only the form of the fractal-like branches, but the branches persist across different edge designs. This validates the fractal-like anomalous impedance scaling in $LC$ circuits with different lattice models.

\section{Saturation in homogeneous circuits}\label{appendixSaturation}
To find the finite saturation values in homogeneous circuits when $D>2$, let us first consider a homogeneous 3D cube circuit constructed by a typical resistor with the resistance $R$ in each principal direction. Now, it is essential to determine the electric potential at the opposite corner nodes whereby the corner-to-corner impedance as a function of the circuit size can be simply found by $Z=2|V|/I=2G$ due to the translation symmetry implying that $V(\mathbf{r}) = - V(-\mathbf{r})$. To impose the boundary conditions, symmetrical current injection and extraction [for example, from Figure~\ref{network_schematics}(c)] are performed such that the impedance between two opposite corner nodes in terms of the voltage distribution is written as
\begin{equation}
\begin{aligned}
	Z_{3\text{D}}=2\Big( & G(0,0,0)+G(1,0,0)+G(0,1,0)+G(0,0,1)\\
	&+G(1,1,0)+G(1,0,1)+G(0,1,1)+G(1,1,1) \\ 
	&-G(N,N,N)-G(N+1,N,N)\\ 
	&-G(N,N+1,N)-G(N,N,N+1) \\ 
	&-G(N+1,N+1,N)-G(N+1,N,N+1) \\ 
	&-G(N,N+1,N+1)- G(N+1,N+1,N+1)\Big),
\end{aligned}
\label{cubeV1}
\end{equation}
where $G(n_1,n_2,n_3)$ is the short notation of $G(\mathbf{r})$ [e.g., $G(\mathbf{r}) \rightarrow G(0,0,0)$ or $G(\mathbf{r}+N\mathbf{u}+\mathbf{a_1}) \rightarrow G(N+1,N,N)$], where $\mathbf{r} =\sum_i^{D=3} n_i \mathbf{a_i}$ where $n_i$s are the integers varying from 1 to $2N$ and where $\mathbf{u} = \mathbf{a_1} + \mathbf{a_2} + \mathbf{a_3}$ being the unit vector. As accomplished in Sec.~\ref{secRLC}, the circuit Green's function given by \eqref{Green_r} is inserted into the above equation. Because the circuit nodes are connected by resistors with the admittance $1/R$, the corresponding circuit Laplacian is written as $\mathcal{L}_{3D}^{\text{sat}}(\mathbf{k})= (2/R)(3-\cos k_1 - \cos k_2 - \cos k_3)$ where $\mathbf{k}=\sum_i^{D=3} k_i \mathbf{a_i} $ and where $k_{i}=n_i\pi/N$ where $n_i\in \{1,2,\dots,2N\}$ and $i=(1,2,3)$. However, for the sake of simplicity, we henceforth set $R=1\Omega$ in the derivation. By inserting the corresponding Laplacian substituted for $G(\mathbf{k})^{-1}$, we arrive
\begin{equation}
	\begin{aligned}
		Z_{3\text{D}}(N)=&\frac{2}{(2N)^3} \sum_{n_1=1}^{2N}\sum_{n_2=1}^{2N}\sum_{n_3=1}^{2N}\big( 1-e^{i\pi(n_1+n_2+n_3)}\big)  \\ 
		&\  \ \times \frac{8\cos\left(\frac{n_1\pi}{2N}\right) \cos\left(\frac{n_2\pi}{2N}\right) \cos\left(\frac{n_3\pi}{2N}\right) \cos\left(\frac{(n_1+n_2+n_3)\pi}{2N}\right)}{2\left(3-\cos\left(\frac{n_1\pi}{N}\right) - \cos\left(\frac{n_2\pi}{N}\right) - \cos\left(\frac{n_3\pi}{N}\right)\right)}.
	\end{aligned}
\label{3Dimp}
\end{equation}
Here, we obtain the analytical formula for the corner-to-corner impedance in the 3D homogeneous cube circuit. We can now generalize the corner-to-corner impedance in $D$ dimensions and write
\begin{equation}
	\begin{aligned}
		Z_\text{D-dim}(N)=&\frac{1}{N^D} \sum_{n_1=1}^{2N}...\sum_{n_D=1}^{2N}\big( 1-e^{i\pi\sum_{ i}^Dn_i}\big)  \\ 
		&\times \frac{\cos\left( \sum_{ i}^D\frac{n_i\pi}{2N}\right) \times \prod_{i}^D\cos\left(\frac{n_i\pi}{2N}\right) }{D-\sum_{ i}^D\cos\left( \frac{n_i\pi}{N}\right) }.
	\end{aligned}
	\label{DDimp}
\end{equation}
Here, $D$ refers to the circuit dimension, $n_i$s are the Fourier components varying between $1$ and $2N$ along each direction, and $N$ is the circuit size. Notice that because we have set $R=1\Omega$ earlier, there is no term representing the unit admittance. However, one can multiply the denominator of \eqref{DDimp} by a unit admittance if required because the admittance is a common factor in the Laplacian since the circuit nodes are linked by single-type components. The impedance between two opposite corner nodes calculated through the above expression saturates as $N$ approaches the continuum limit, i.e., $Z_{\text{D-dim}}^{\text{sat}}\approx\lim_{N \to\infty}Z_\text{D-dim}(N)$. The presence of saturation impedances in the large-$N$ limit in dimensions $D\geq 3$ suggests a convergent integral expression for the corner-to-corner impedances in this regime. In the large-$N$ limit, the evenness and oddness of the indices should have negligible effects if the sum can be approximated by an integral, since that entails shifts of $\pi/N$ to the momenta. Noting that the factor $1-e^{i\pi(\sum_{i}^D n_i)}$ effectively eliminates half of the terms, we have the approximation
\begin{equation}
	\begin{aligned}
		Z_\text{D-dim}(N\rightarrow\infty)\approx& \frac1{\pi^D}\int\displaylimits_{[0,2\pi]^D}\frac{\cos\left(\sum_i^D \frac{k_i}{2}\right) \times \prod_i^D\cos\frac{k_i}{2}}{D-\sum_i^D \cos k_i}\,d^D\mathbf{k}
	\end{aligned}
	\label{DDint}
\end{equation}
for $D\geq 3$. In practice, a very small positive term may have to be added to the dominator to make the integral converge; physically, this corresponds to inevitable parasitic resistances. For instance, we obtain $		Z_\text{3-dim}(N\rightarrow\infty)=1.44015~\Omega$, $Z_\text{4-dim}(N\rightarrow\infty)=0.774964~\Omega$ and $Z_\text{5-dim}(N\rightarrow\infty)=0.542093~\Omega$, which are very close to the values of $Z_{\text{3-dim}}(300)=1.43583~\Omega$, $Z_{\text{4-dim}}(50)=0.774739~\Omega$ and $Z_{\text{5-dim}}(40)=0.542011~\Omega$ computed from \eqref{DDimp}.

\subsubsection{Dimensional reduction of impedance formula}
Interestingly, the corner-to-corner impedance can be dimensionally reduced to a momentum integral in a lower number of dimensions, albeit with a seemingly more sophisticated integrand. To start, we separate out the last momentum coordinate $k_D=k$ by writing the integrand $\frac{\cos\left(\sum_i^D \frac{k_i}{2}\right)\prod_i^D\cos\frac{k_i}{2}}{D-\sum_i^D \cos k_i}$ of \eqref{DDint} as $\left(\prod\limits_i^{D-1}\cos\frac{k_i}{2}\right)\frac{\cos \frac{k}{2}\cos\left(a+\frac{k}{2}\right)}{D-b-\cos k}$ where $a=\sum\limits_i^{D-1} \frac{k_i}{2}$ and $b=\sum\limits_i^{D-1} \cos k_i$. The $k$-dependent fraction on the right can be integrated as follows: 
\begin{equation}
\int_0^{2\pi} \frac{\cos \frac{k}{2}\cos\left(a+\frac{k}{2}\right)}{D-b-\cos k}dk =\pi\left(\sqrt{\frac{D-b+1}{D-b-1}}-1\right)\cos a. 
\end{equation}
Since $\int_{[0,2\pi]^{D'}}\cos\left(\sum\limits_i^{D'} \frac{k_i}{2}\right)\prod\limits_i^{D'}\cos\frac{k_i}{2}\,d^D\bold{k}=\pi^{D'}$ for any $D'$ (Here and below, we denote $D'=D-1$ for brevity), we obtain
\begin{equation}
	\begin{aligned}
	Z_\text{D-dim}(N\rightarrow\infty)\approx 	-1+\int\displaylimits_{[0,2\pi]^{D'}} \frac{\cos\left(\sum\limits_i^{D'} \frac{k_i}{2}\right)\prod\limits_i^{D'}\cos\frac{k_i}{2}}{\pi^{D'}}\sqrt{\frac{D+1-\sum\limits_i^{D'} \cos k_i}{D-1-\sum\limits_i^{D'} \cos k_i}}\,d^{D'}\bold k.
	\end{aligned}
	\label{DDint2}
\end{equation}
In general, we can continue with this dimensional reduction procedure to obtain $	Z_\text{D-dim}(N\rightarrow\infty)\approx$
\begin{equation}
	\begin{aligned}
		\int\displaylimits_{[0,2\pi]^{D-d}} \frac{\cos\left(\sum\limits_i^{D-d} \frac{k_i}{2}\right)\prod\limits_i^{D-d}\cos\frac{k_i}{2}}{\pi^{D-d}}f_d\left(D-d-\sum\limits_i^{D-d}\cos k_i\right)\,d^{D-d}\bold k
	\end{aligned}
	\label{DDintred}
\end{equation}
where $d$ is the number of reduced dimensions and $f_d(x)$ is a weightage function that encapsulates the effects of dimensional reduction:
\begin{equation}
	\begin{aligned}
		f_0(x) &= \frac1{x}\\
		f_1(x) &= \sqrt{1+\frac{2}{x}}-1\\
		f_2(x) &= \frac{\left(x^2+5 x+8\right) K\left(-\frac{4}{x^2+4 x}\right)-x (x+4) E\left(-\frac{4}{x^2+4 x}\right)-2 i K\left(\frac{(x+2)^2}{x (x+4)}\right)}{\pi\sqrt{x (x+4)}}\\& \quad +\frac1{\pi}K\left(\frac{4}{(x+2)^2}\right)+\frac{2 i K\left(\frac{x (x+4)}{(x+2)^2}\right)}{\pi(x+2)}-1,
	\end{aligned}
\end{equation}
where $E(y)=\int_0^{\pi/2}\sqrt{1-y\sin^2\theta}\,d\theta$ and $K(y)=\int_0^{\pi/2}1/\sqrt{1-y\sin^2\theta}\,d\theta$ are elliptic integrals. Further expressions of $f_d$, $d>2$ exist in principle, although they would be much more complicated. Note that $f_0$ is just the inverse Laplacian spectrum for the unbounded circuit; the other terms in the integrand keeps track of the source/sink as well as boundary effects, as we have obtained from the method of images. As an illustration,
\begin{equation}
	Z_\text{3-dim}(N\rightarrow\infty)\approx \frac1{\pi}\int_0^{2\pi}\cos^2\frac{k}{2}\,f_2(1-\cos k)dk.
\end{equation}

\section{Conclusion}
In conclusion, we have revealed the circuit size-dependent anomalous impedance resonances in topolectrical circuits as well as various dimensional finite $LC$ circuits. We observed that the impedance scaling in heterogeneous circuits differs from the logarithmic-like scaling exhibited by homogeneous circuits as the circuit size $N$ increases. Conventionally, frequency-dependent elements such as capacitors and inductors are considered independent circuit parameters that can be set individually. However, we have demonstrated that the strong impedance resonances in our circuits are not due to variations in these parameters, but rather to specific circuit sizes. Therefore, the circuit size becomes an independent, albeit not widely known, parameter that affects the impedance behavior of the circuit. We invoked the method of images, inspired by free-space electrostatics, as a means to calculate the corner-to-corner impedance and provided a generic exact analytical expression homogeneous or heterogeneous circuits of any dimensions. This method naturally satisfies the open boundary conditions because it ensures that the potentials of the nodes at the boundaries of the physical and image circuits are equal [see Figure~\ref{VSpatial}(b)]. The size-dependent impedance jumps result in fractal-like patterns in the circuit size-resonant frequency plots. The existence of these patterns is common to all heterogeneous circuits but the details within the patterns are unique to each dimensionality and circuit structure. Therefore, this chapter establishes a framework for further investigation of anomalous impedance behaviors in more complex circuits, as well as their experimental realizations~\cite{zhang_anomalous_2023}.
\chapter{Anomalous impedance scaling in a two-dimensional electric network}
\label{ch:exp_impedance}
\vspace{2em}

\noindent 
In Chapter \ref{ch:impedance}, we theoretically explored the impedance responses of both classical $LC$ and topolectrical circuits. To delve deeper into the analysis of two-point impedance in electrical circuits, particularly in the context of scaling theory, this chapter focuses on a two-dimensional (2D) $LC$ circuit and its experimental observations~\footnote{The experiment on the 2D $LC$ circuit, which will be discussed in detail in this chapter, was conducted by Xiao Zhang and Boxue Zhang. I would like to express my gratitude for their significant contributions to our work, as reported in Ref.~\cite{zhang_anomalous_2023}.}. The combination of theoretical discussion and experimental findings on the anomalous size-dependent impedance scaling in a 2D $LC$ circuit broadens the insights gained in Chapter \ref{ch:impedance}. There, we observe a significant deviation from the expected logarithmic scaling in the impedance of a 2D $LC$ circuit network, contrary to continuum expectations. We find this anomalous impedance contribution to sensitively depend on the number of nodes $N$ in a curious erratic manner, and experimentally demonstrate its robustness against perturbations from the contact and parasitic impedance of individual components. This impedance anomaly is traced back to a generalized resonance condition reminiscent of the Harper's equation for electronic lattice transport in a magnetic field, even though our circuit network does not involve magnetic translation symmetry. It exhibits an emergent fractal parametric structure of anomalous impedance peaks for different $N$ that cannot be reconciled with continuum theory and does not correspond to regular waveguide resonant behavior. This anomalous fractal scaling extends to the transport properties of generic systems described by a network Laplacian whenever a resonance frequency scale is simultaneously present. Unlike the broader discussion of anomalous impedance scaling presented in Chapter \ref{ch:impedance}, this chapter will delve into the anomalous impedance scaling specifically within the framework of continuum scaling theory, and explore its breakdown attributable to the discrete nature of $LC$ circuit networks.

\section{Introduction}

From dimensional analysis to the universality of critical phase transitions, scaling theory provides a universal paradigm for the principal understanding of most physical phenomena~\cite{fisher_theory_1967, stanley_scaling_1999, hilfer_scaling_1992, cardy_scaling_1996, frohlich_scaling_1983, chen_scaling_2016, zuo_scaling_2021}. Particularly interesting are ``marginal'' scenarios, where observables exhibit great freedom in their functional dependency on the physical variables~\cite{leigh1995exactly, muller_marginal_2015, dresselhaus_numerical_2021, zirnbauer_marginal_2021}. A classic example is the electrical impedance $Z$ of a $D$-dimensional sample of characteristic length $N$, which scales as $Z\sim N^{2-D}$; in particular, for $D=2$, $Z$ must scale slower than any power of $N$, most commonly logarithmically.

Indeed, logarithmic scaling is ubiquitous in physics, appearing in a broad range of contexts as disparate as conformal field theory, disorder Green's functions, strongly coupled quantum fields, and graph complexity~\cite{mackinnon1983scaling, poland_conformal_2019, zamolodchikov_exact_1989, lee_disordered_1985, albert_statistical_2002}. It represents the paradigmatic slower-than-power-law behavior that appears naturally in various scale-free scenarios. Particularly, impedance scaling in electrical circuits, as a function of circuit size, dutifully displays this scaling behavior when the circuit is either entirely reactive or resistive, as discussed in Chapter \ref{ch:impedance}. For instance, the dimensionality of lattice models determines whether the impedance scales linearly, logarithmically or saturates to a constant impedance value. While 1D and 2D samples exhibit these scaling characteristics such as linear or logarithmic scaling,  the impedance in circuits characterized with dimensionality $D\geq3$ experiences rapid saturation proportional to $D$. Nevertheless, although the dimensionality of the lattice delineates the scaling characteristics, it ceases to be the predominant determinant in heterogeneous resonant medias. In fact, the scaling profile of $LC$ circuits with inductance $L$ and capacitance $C$ relies more on the form of the lattice array rather than the lattice dimension~\cite{chen_equivalent_2021,tan_electrical_2019}. This is because the impedance across two opposite farthest sites varies due to the parameter space irrespective of the lattice dimension. To explain this, we will conceptually elucidate and experimentally demonstrate the resonant conditions through a seemingly elementary physical 2D system by evaluating its corner-to-corner impedance behavior.

\begin{figure}
	\includegraphics[width=12cm]{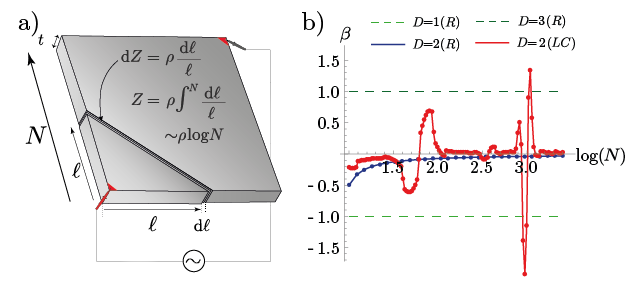}\centering
	\caption{\textbf{Origin of logarithmic impedance scaling in continuum media and its violation in lattices.} (a) In a continuous sample such as a square plate of length $N$ and resistivity $\rho$, the diagonal-to-diagonal impedance necessarily scales like $\rho\log N$. This is easily seen by slicing the sample into strips perpendicular to the diagonal and noticing that each strip approximately contributes a serial impedance that is inversely proportional to its width. This is because each successive ``shell'' in the sample scales with its linear dimension $l$ as $l^{D-1}=l$, such that the total impedance scales like $\int^N l^{-1}dl\sim \log N$. (b) Behavior of $\beta=-\frac{d\log |Z|}{d\log N}$, the fractional rate of change of impedance $Z$ diagonally with the system size $N$, across circuit lattices with and without a AC frequency scale. While a purely resistive 2D circuit (blue) exhibits a smoothly vanishing $\beta$ consistent with the continuum approximation in (a), our 2D $LC$ circuit (red) with an illustrative frequency scale of $\omega_r=1.95$ exhibits anomalous scaling behavior with pronounced and erratically located peaks. The dashed lines represent the constant or saturated scaling of other dimensions derived from $\beta$ of scaling theory applicable to non-resonant reactive media.
	}
\label{fig1}
\end{figure}

The two-point impedance and resistance problem has garnered significant attention~\cite{bartis_lets_1967,kirkpatrick_percolation_1973,venezian_resistance_1994,lavatelli_resistive_1972,zemanian_classical_1984,aitchison_resistance_1964,montroll_random_1965,morita_useful_1971} as it not only allows for the study of electrical conductivity, but also serves as a means of uncovering new physical phenomena related to lattice dimension, network model, lattice uniformity, and boundary design from the changes in the electrical characteristics in the presence of perturbations or disorder~\cite{asad_perturbed_2014,owaidat_perturbation_2016,giordano_disordered_2005,cserti_perturbation_2002,lee2021many}. The extensive research conducted in this expansive field has enhanced our fundamental understanding of electric circuits~\cite{izmailian_generalised_2014,tan_resistance_2017,cserti_uniform_2011,owaidat_interstitial_2010,asad_infinite_2005,doyle_random_2000,mamode_calculation_2019,pan_electric-circuit_2021,chen_electrical_2019,chen_electrical_2020,chen_electrical_2020-1,chen_electrical_2020-2,chen_electrical_2020-3,ammar_electrical_2022,tan_resistance_2017} and has practical applications in the design of various circuit systems, including topolectrical circuits~\cite{lee_topolectrical_2018,li2019emergence,wang_circuit_2020,wang2022observation,shang2022experimental,wu2023evidencing,zhang2023electrical}, non-linear systems~\cite{hohmann2023observation,tuloup2020nonlinearity,kotwal_active_2021,kengne_ginzburglandau_2022}, condensed matter counterparts~\cite{ningyuan2015time,rafi-ul-islam_topoelectrical_2020}, and metamaterials~\cite{kapitanova_photonic_2014}. In addition to numerical approaches such as the Laplacian formalism~\cite{wu_theory_2004,tzeng_theory_2006,cernanova_nonsymmetric_2014}, various analytical methods have been developed for determining the two-point impedance, including the recursion-transform method~\cite{tan_electrical_2019,tan_two-point_2016,tan_characteristic_2017,fang_circuit_2022,zhou_fractional-order_2017,tan_recursion-transform_2015,tan_recursion-transform_2015-2}, the lattice Green's function~\cite{joyce_exact_2017,cserti_application_2000,katsura_lattice_1971,mamode_revisiting_2021,joyce_exact_2002,joyce_1973}, asymptotic expansion~\cite{essam_exact_2009,izmailian_asymptotic_2010}, and the method of images~\cite{mamode_electrical_2017,sahin_impedance_2023}. While each of these methods employs a distinct approach to evaluate the impedance in both reactive and resonant media, they all require the circuit network to possess symmetries such as inversion and translation symmetries. The role of these symmetries has not been thoroughly explored in the literature, but their presence may result in anomalous behaviors that can be uncovered through the impedance scaling in electric circuits.

In this Chapter, we specifically examine a 2D square $LC$ circuit, wherein its reactive counterpart displays a notable logarithmic scaling. Although the same qualitative picture can be observed in circuits with different dimensions, the 2D $LC$ circuit allows us to investigate the origin of uniform scaling violations using a simpler yet richer example. Naively, one would expect from the impedance of a 2D circuit to vary smoothly with the number of unit cells from its continuum analogue since the circuit lattice can be construed as a discretization of 2D conducting plates. However, while this indeed holds for non-resonant circuits, such as those containing capacitors or resistors exclusively, the impedance behavior for resonant media i.e., $LC$ circuits, cannot be more different. Our theoretical and experimental investigations reveal curious impedance enhancements of up to a few orders at certain lattice sizes $N$, whose roots can be traced to a new commensurability criterion associated with a Hofstadter butterfly-like fractal structure. This challenges the applicability of a continuum description in even the simplest of resonant media.

\section{Results} 

\begin{figure*}[]
	\centering
	\includegraphics[width=\textwidth]{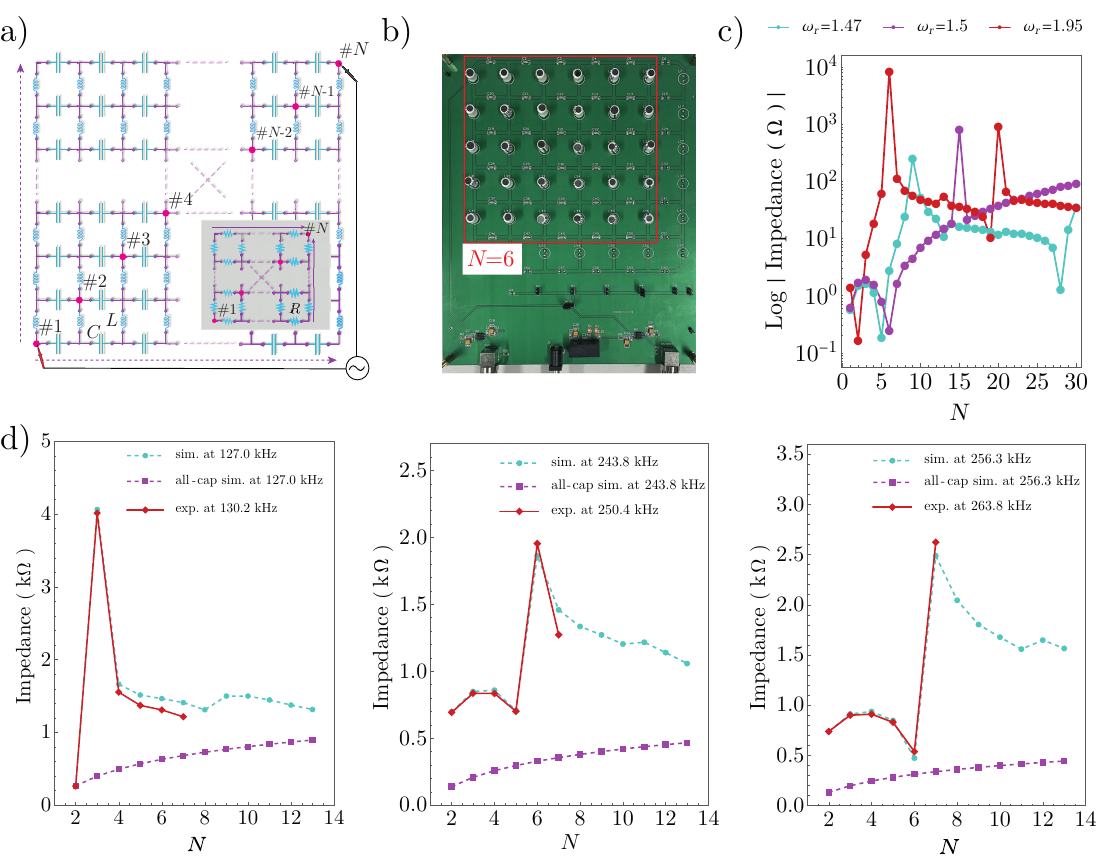}
	\caption{ \textbf{Circuit description and measured anomalous impedance scaling.} (a) Our circuit is a $N\times N$ square lattice array with the horizontal and vertical links being capacitors $C$ and inductors $L$ respectively. The corner-to-corner impedance $Z$ is measured by running a current between the lower left and upper right ($N$-th) nodes. The scaling behavior of $Z$ is revealed to contrast strongly with the logarithmic scaling of a similar but uniform circuit array consisting of only one type of element i.e., resistors (shown in the inset), or capacitors.	(b) We implement our $LC$ circuit arrays on circuit boards, and control the lattice size $N$ through switches. Shown here is the $6\times 6$ case - our board shown here admits up to the $N=7$ case. (c) In principle, with purely capacitive or inductive $LC$ components, the corner-to-corner impedance of our circuit becomes drastically higher by a few orders at particular lattice sizes $N$, and depends sensitively on $\omega_r=\omega\sqrt{LC}$ according to \eqref{Z}. (d) These anomalous corner-to-corner impedance peaks are attenuated in our experimental measurements but are still robustly prominent, as shown in these plots at three illustrative AC frequencies $\omega$. The measured (exp, red) data agrees well with the simulated values (sim, cyan) with estimated parasitic resistances (estimated to be R$_{pL}$ = 3.3 $\Omega$, R$_{pC}$ = 4.5 $\Omega$, R$_{pW}$ = 0.1 $\Omega$, see Methods: Analysis of uncertainties), and are captured by a complex effective $\omega_r$ with $\text{Im}(\omega_r)$ of the order of $10^{-2}$. This contrasts with uniformly capacitive circuits (all-cap, dark magenta), which exhibit logarithmic scaling with no non-monotonic peaks.
}
	\label{fig2}
\end{figure*}

\subsection{Violation of logarithmic impedance scaling}

To put our anomalous circuit impedance scaling behavior into perspective, we introduce the quantity $\beta = -\frac{d\log |Z|}{d\log N}$, which is the fractional rate of change of the impedance $Z$ with the system size $N$. It is closely related to the $\beta$-function in renormalization group analysis~\cite{callan_broken_1970, fisher_renormalization_1974, wilson_renormalization_1975}, and has also been famously employed in understanding the conductivity localization transition ~\cite{abrahams_scaling_1979, altshuler_interaction_1980, abrahams_quasiparticle_1981, lee_disordered_1985,garcia-garcia_anderson_2005,garcia-garcia_semi-poisson_2006} due to disorder scattering. 

In most conductors where $Z\sim N^{2-D}$, we have a constant $\beta = D-2$, which indicates that the impedance $Z$ increases (decreases) with the system size in a consistent qualitative manner for $D\leq2$ ($D>2$). This is the case for purely resistive media (such as the conducting plate depicted in Fig.~\ref{fig1}a) for which the impedance scales logarithmically viz. $Z\sim \log N$, giving rise to $\beta \sim -(\log N)^{-1}$ as sketched in Fig.~\ref{fig1}b (blue, green and dark green). However, we unveil that this crossover to the asymptotic limit can be far from smooth when a AC frequency scale exists in the circuit. As plotted in Fig.~\ref{fig1}b (red) for an illustrative 2D AC circuit lattice (detailed later), $\beta$ fluctuates erratically and dramatically as the system size $N$ increases. (Note that the irregular impedance scaling depicted in red in Fig.~\ref{fig1}b is not exclusive to a 2D sample but can occur in an $LC$ lattice, regardless of their dimensions.) In the following sections, this anomalous scaling behavior will be revealed to be part of an intricate fractal-like characteristic with slightly different reactance parameters often giving rise to unpredictably distinct anomalous impedance scaling.

\subsection{RLC Circuit with anomalous impedance scaling}

We investigate the discretization of the simplest 2D conducting sample, which is a $N\times N$ square lattice circuit array with fixed $RLC$ components connecting each node (Fig.~\ref{fig2}a). For consistency, we shall always measure the impedance across two diagonally opposite corner nodes, even though the subsequent results remain qualitatively valid for arbitrary impedance intervals. If every connection in the square lattice is composed of the same element $z$, it can be shown that the corner-to-corner impedance scales like $Z\sim z\log N$ (Fig~\ref{fig2}d). This is not surprising, since it is only natural to expect that the square circuit lattice inherits the same logarithmic scaling as its continuum counterpart. 

Yet, we find that this usual logarithmic impedance scaling becomes severely violated when the lattice connections $z$ are replaced by two different circuit components with impedance of opposite signs, such as $L$ and $C$ components, which define a frequency scale. Specifically, we built an $N\times N$ square lattice circuit array on circuit boards (Fig.~\ref{fig2}b) ($N=2,..,7$), such that each horizontal link contains a capacitor $C$ and each vertical link contains an inductor $L$. In momentum space, the circuit Laplacian $\mathcal{L}$, which relates the voltage and input current profiles via $\bold I=\mathcal{L}\bold V$, takes the form
\begin{align}
\mathcal{L}(k_x,k_y)&=2i\omega C(1-\cos k_x) + \frac{2}{i\omega L}(1-\cos k_y)\notag\\
&=2i\omega C\left[(1-\cos k_x)-\omega_r^{-2}(1-\cos k_y)\right],
\label{circuit_Laplacian}
\end{align}
where $\omega$ denotes the AC driving frequency. Barring the $2i\omega C$ overall prefactor, $\omega_r=\omega\sqrt{LC}$ is the only nontrivial parameter of our circuit besides the lattice size $N$, neither of which constitutes another competing length scale. 

For a fixed $\omega_r$, the measured corner-to-corner impedance does not follow a simple trend with the lattice size $N$, but varies erratically with abrupt and prominent peaks at certain $N$. As plotted in Fig.~\ref{fig2}c for an ideal $LC$ circuit without any dissipation, the impedance $Z$ fluctuates wildly as $N$ is increased, such that $|Z|$ can abruptly become a few orders of magnitude larger for particular values of $N$. Besides, the impedance behaves qualitatively differently for different $\omega_r$, even across small changes in $\omega_r$. It is noteworthy that such ``quasi-random'' behavior is robustly measurable in an actual experimental implementation with inevitable dissipation, as reflected in our measured data (Fig.~\ref{fig2}d), which agrees well with theory despite unavoidable parasite and contact resistances as well as component disorder.

\subsection{Emergent fractal resonances}

\begin{figure}
	\includegraphics[width=\linewidth]{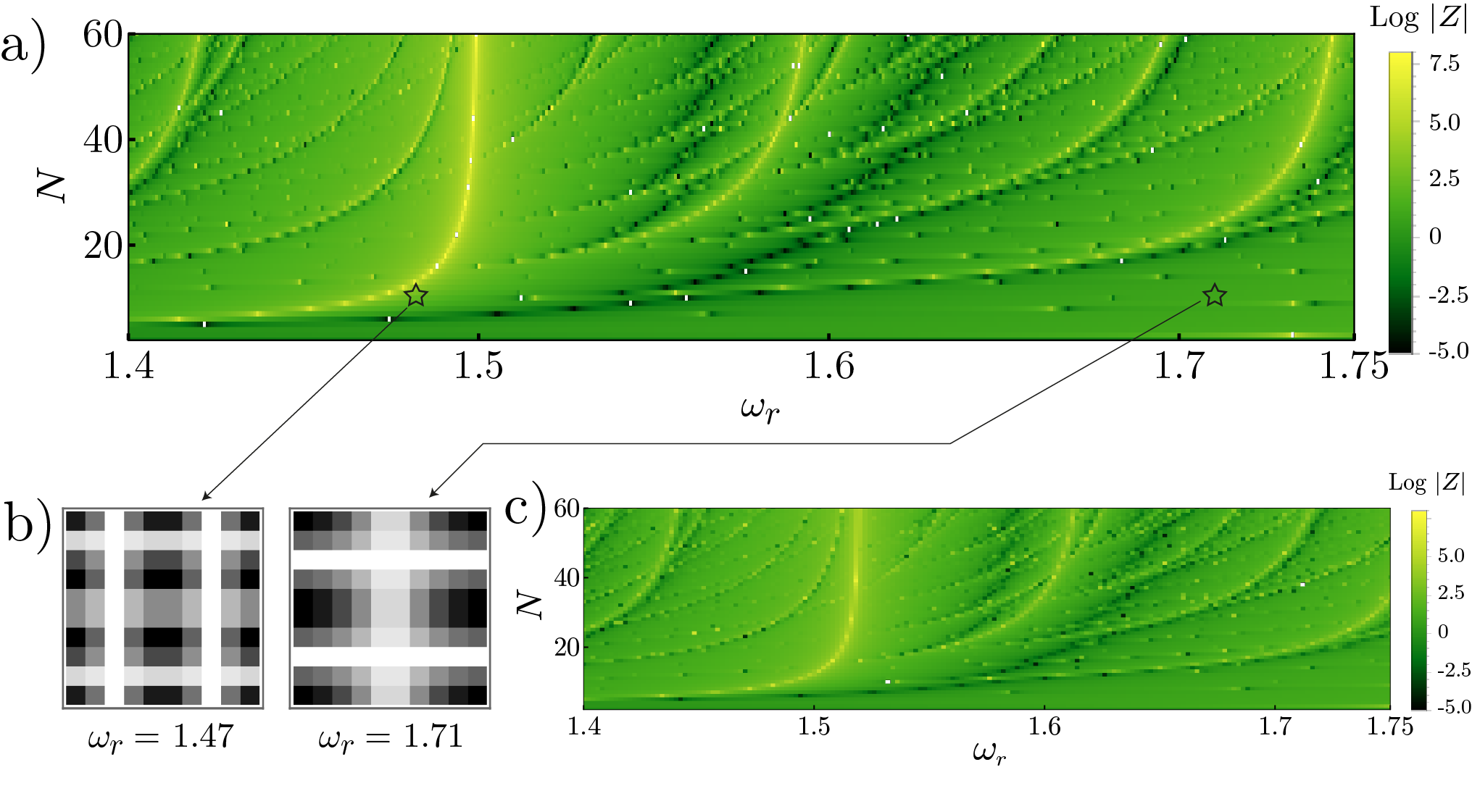}\centering
	\caption{\textbf{Fractal nature of anomalous impedance scaling.} (a) Log-density plot of the corner-to-corner impedance $Z$ in the parameter space of (real) $\omega_r$ and lattice size $N$ showing variations of $Z$ across several orders of magnitude in the form of fractal-like branches. The ``branch'' near $\omega_r=1.5$ is the strongest but, still, it contains strong impedance peaks only for certain system sizes $N$. (b) Representative minimal-eigenvalue eigenstates of the Laplacian $\mathcal{L}$ with $N=9$ at two illustrative $\omega_r$ (\eqref{resonance}) with very contrasting impedances $Z_{\omega_r=1.47}/Z_{\omega_r=1.71}=256/4.81\approx 53$. Unlike the case of waveguides, the markedly different impedances are not due to the spatial eigenstate distributions, which are qualitatively similar, but rather the ``vanishing energetics'' of $\omega_r$. (c) The impedance peak branches of $\log|Z|$ remain mostly robust in the presence of inevitable resistances, such as shown here for $\text{Im}(\omega_r)=-0.02$, which is of the same order as the parasitic resistances in our fabricated circuits. The plot legends in panels (a) and (c) indicate that the values represented are the logarithms of the absolute corner-to-corner impedance.}
	\label{fig3}
\end{figure}

The erratic, random-like behavior of the impedance across our $LC$ circuit suggests a hidden layer of emergent complexity in its resonance properties. Usually, one would expect a simple array of $LC$ components to behave as a waveguide with resonances that are simple enough to list down, for instance like the vibration modes on a stretched drumskin~\cite{kac1966can,tien1977integrated}. A complete impedance plot of our $LC$ circuit in $(N,\omega_r)$ parameter space, however, reveals a complicated fractal-like structure that bears resemblance to the energy bands in the Hofstadter butterfly~\cite{hofstadter_energy_1976, albrecht_evidence_2001, koshino2006hall, hunt2013massive}. In Fig.~\ref{fig3}a, we observe the following intricate hierarchy of impedance peaks: apart from some ``main'' branches, there exists a proliferation of less regular peaks that appear and disappear with the discreteness of $N$, akin to the cringes on the surface of a human palm. Additionally, these fractal patterns in our 2D $LC$ circuit are not confined to 2D instances, much like the Hofstadter butterfly, which is specific to 3D and quasi-1D systems~\cite{koshino_hofstadter_2001,koshino_phase_2002}.

To mathematically understand the origin of this fractal impedance behavior, we start from the formal expression for the impedance between two nodes $i$ and $j$~\cite{tzeng_theory_2006,lee_topolectrical_2018} 
\begin{align}
Z_{ij}&= \frac{V_i-V_j}{I}\notag\\
&=\frac{[\mathcal{L}^{-1}\bold I ]_i-[\mathcal{L}^{-1}\bold I]_j }{I}\notag\\
&=\sum_{\mu\neq 0} \frac{|\psi_\mu(i)-\psi_\mu(j)|^2}{\lambda_\mu},
\label{two-point-imp}
\end{align}
where $V_i$ and $V_j$ are respectively the voltage potentials at the current input sites $i$ and $j$ at which $I_l=I(\delta_{il}-\delta_{jl})$ is nonzero. Here $\psi_\mu$ and $\lambda_\mu$ are the corresponding eigenvectors and eigenvalues of the Laplacian, whose pseudoinverse is given by $\mathcal{L}^{-1}=\sum_{\mu\neq 0}\lambda_\mu^{-1}|\psi_\mu\rangle\langle\psi_\mu|$, where $\mu\neq 0$ indicates the omission of the uniform eigenvector corresponding to an overall voltage offset. 

Evidently, impedance peaks arise if there are eigenvalues $\lambda_\mu$ that are almost zero (not exactly zero, as they cannot perfectly vanish in a realistic circuit experiment). Such peaks have been featured as ``topolectrical'' resonances when the circuit band topology enforces topological zero modes~\cite{ningyuan2015time,imhof_topolectrical-circuit_2018, yang_observation_2020}. In our context, there is \emph{no} topological mechanism, and we proceed by deriving a compact albeit slightly complicated expression for the impedance $Z=Z_{ij}$ between the corner nodes $i$ and $j$, as detailed in the previous chapter. The idea is to first consider the circuit under a doubled system with periodic boundaries where $\mu$ in \eqref{two-point-imp} now labels the momentum eigenmodes $k_{x}=\frac{2\pi m}{2N}$, $k_{y}=\frac{2\pi n}{2N}$, and next employ the method of images to enforce the vanishing of currents across the $N\times N$ open boundaries. In doing so, we employ \eqref{general_Z} of Chapter 3 for $D=2$ and assign \eqref{circuit_Laplacian} to the denominator of \eqref{general_Z}. We then obtain the impedance as
\begin{equation}
Z(N)=\frac{2}{i\omega CN^2}\sum_{n+m\in\text{odd}}\frac{\cos \frac{n\pi}{2N}\cos \frac{m\pi}{2N}\cos \frac{(n+m)\pi}{2N}}{(1-\cos \frac{n\pi}{N})-\omega_r^{-2}(1-\cos \frac{m\pi}{N})}.
\label{Z}
\end{equation}
The denominator in \eqref{Z} resolves the origin of incommensurability leading to fractal-like behavior. Analogous to the Harper equation describing a Landau level due to a magnetic field~\cite{harper_single_1955, hofstadter_energy_1976, krasovsky_bethe_1999,poshakinskiy_quantum_2021}, we find the relation
\begin{equation}
\omega_r^2=\omega^2 LC = \frac{1-\cos\frac{m\pi}{N}}{1-\cos\frac{n\pi}{N}}
\label{resonance}
\end{equation}
describing a circuit resonance. Here, $\omega_r^2$ plays the analogous role to the energy in the Hofstadter butterfly, and $N$ plays the role of the denominator defining a fractional flux. In our case, however, all rational fractions with denominator $N$ simultaneously contribute to the impedance, and a strong resonance occurs if there exist integers $m,n$ of the same parity that accurately satisfy \eqref{resonance}.  

This relation explicitly expresses the resonance strength in terms of the commensurability properties of $\omega_r^2$ and $N$, even though the relation is hard to guess from intuitive reasoning. Unlike the Hofstadter butterfly problem, which is based on magnetic translation-symmetric Bloch states~\cite{hofstadter_energy_1976, koshino2006hall}, our circuit setup contains no such symmetries. While generic $LC$ (or likewise $RLC$) circuits do possess resonances, their resonance properties dimensionally depend on the frequency scale $LC$, and in general do not depend systematically on the system size. In our case, it is the mirror symmetry about the boundaries that fortuitously restores sufficient symmetry to give rise to an explicit, and hence also measurable, commensurability relation.

Stemming from the approximate solutions to \eqref{resonance}, the impedance peaks are primarily manifestations of commensurate ``energetics'' that lead to a vanishing $\omega_r$, rather than special spatial mode configurations. To illustrate this point, illustrative near-resonant and off-resonant eigenmodes of $\mathcal{L}$ are plotted in Fig.~\ref{fig3}b. Note that the near-resonant eigenmodes do not exhibit any spatial distribution particularities that distinguish them from ordinary eigenmodes contributing to far lower impedances.

\subsection{Robustness of fractal impedance peaks and crossover from logarithmic scaling}

In actual experiments, contact and parasitic resistances introduce inevitable dissipation and attenuate the impedance peaks, as evident in the comparison between Figs.~\ref{fig2}c and \ref{fig2}d. Yet, the key anomalous fractal scaling behavior of the impedance remains robust. Phenomenologically, we can represent these dissipative effects through modifying the capacitor and inductor impedances to $(i\omega C)^{-1} \rightarrow (i\omega C)^{-1} +R_C$ and $i\omega L \rightarrow i\omega L +R_L$, where $R_C$ and $R_L$ are real effective resistances. Incorporating the estimated $R_C$ and $R_L$ values from our experiments, which add an imaginary part of order $\mathcal{O}(10^{-2})$ to $\omega_r$, we find that the fractal parameter space profile of the impedance $|Z|$ becomes slightly smoothed out (Fig.~\ref{fig3}c), even though the main branches of the fractal structure remains qualitatively unchanged. This robustness stems from the strong impedance divergence due to commensurability effects on a discretized conducting medium, which holds for generic lattice discretizations, and not just for our square lattice (\eqref{Z}).

\section{Methods}

\subsection{Determination of the fractal dimension}
\begin{figure}[h]
	\centering
	\includegraphics[width=8.5cm]{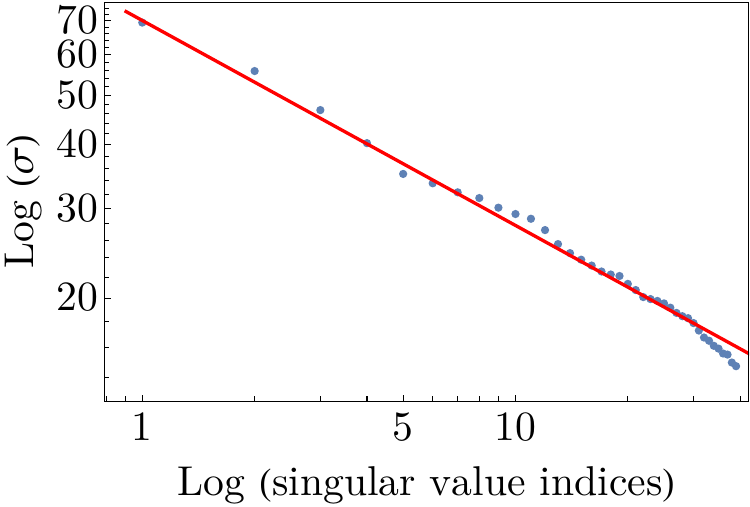}
	\caption{\textbf{Log-log scaling plot of the singular values extracted from the fractal diagram of our 2D circuit, obtained via Singular Value Decomposition (SVD).} The SVD is performed on the fractal matrix $\log |Z|$ displayed in Fig.~\ref{fig3}a. The blue dots represent the singular values given by \eqref{svd}, while the red solid line represents the best linear fit of the data. The fractal dimension is calculated as one minus the slope of the linear fit, resulting in $\mathcal{D}=1.4$ where $\mathcal{D}$ represents the fractal dimension.}
	\label{fig6}
\end{figure}
It is possible to estimate the fractal dimension $\mathcal{D}$ by using the Singular Value Decomposition (SVD) method~\cite{weng_singular_2022,alter_singular_2000,lee_exact_2015}. In this method, the self-similarities~\cite{mishra_effective_2021} or fractal dimension in a dataset, which is the fractal diagram in Fig.~\ref{fig3}a  (i.e., $\log|Z|$), is given by one minus the slope of the log-log plot of the singular values of the fractal matrix~\cite{malcai_scaling_1997,carr_practice_1991}. To determine the fractal dimension of our fractal structure, we performed SVD and write the decomposed matrices as
\begin{equation}
	\log|Z|=u~\sigma~v^\intercal
	\label{svd}
\end{equation}
where ($^\intercal$) denotes the transpose operation, $u$ and $v$ are the left and right singular matrices, respectively, and $\sigma$ is a diagonal matrix that comprises the singular values of the fractal. These diagonal values are nonnegative, and their squares give the eigenvalues of the $\log|Z|$ matrix~\cite{alter_singular_2000,weng_singular_2022}. We arrange the singular values in decreasing order, such that the largest value is $\sigma_1$, the second-largest is $\sigma_2$, and so on, i.e., $\sigma_1>\sigma_2>\sigma_3>\cdots$. This allows us to determine the scaling ratio, which is defined as the ratio of the largest singular value to the fractal matrix dimension. For example, in our case in Fig.~\ref{fig3}a, we find that $\sigma_1=276.71$ and $\text{dim}(\log|Z|)= 388$, and thus determine the scaling ratio as $\sim0.7$. The fractal dimension can be defined by utilizing the slope $(m_s=-0.4)$ of the best linear fit in the log-log scale plot (Fig.~\ref{fig6}) ~\cite{faloutsos_fast_2021}. Therefore, the fractal dimension is determined as $\mathcal{D}=1-m_s=1.4$.

\subsection{Effects of inevitable parasitic resistances}
\begin{figure}[h]
	\centering
	\includegraphics[width=11cm]{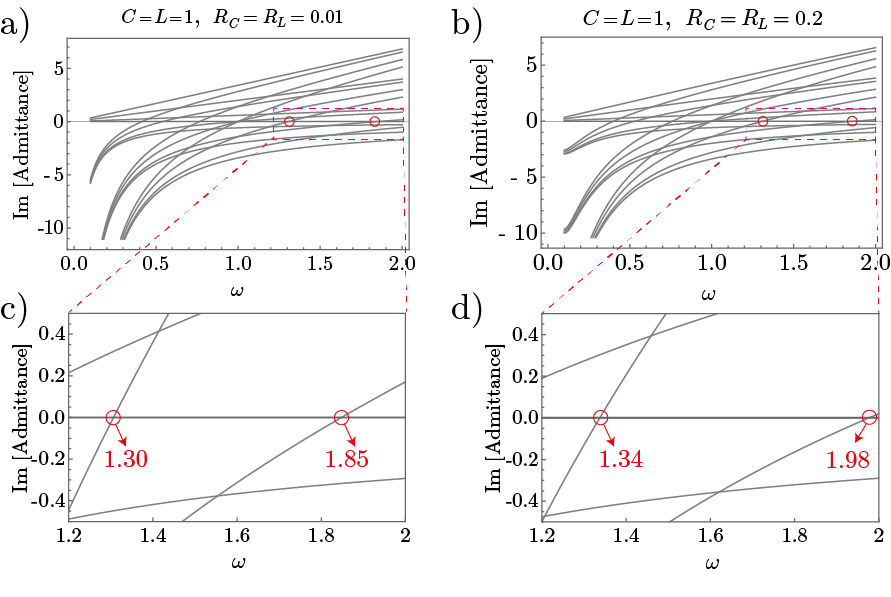}
	\caption{\textbf{Panels (a) and (b) show the admittance spectra of the 2D $LC$ square circuit when $C=L=1$ and $N=3$.} The upper row shows the admittance spectrum under the consideration of the parasitic series resistances when $R_C=R_L=0.01 \Omega$ and $R_C=R_L=0.2 \Omega$, respectively. Panels (c) and (d) display zoomed-in views of points where two randomly chosen admittance bands, depicted in (a) and (b) respectively, intersect the zero-admittance axis. An increase in parasitic resistances results in a shift towards higher frequencies at the band crossing points.}
	\label{fig7}
\end{figure}
Theoretically, the parasitic resistances can be incorporated by introducing additional the real effective resistances $R_C$ for the capacitors and $R_L$ for the inductors, as we discussed in the main text. In Fig.~\ref{fig3}c, we plot the fractal impedance peaks under the consideration of these parasitic resistances, which contribute to the imaginary part of the circuit resonance condition ($\omega_r$). As can be seen in comparison with Fig.~\ref{fig3}a, the realistic parasitic resistances only lead to a smooth shift of the impedance peak branches while the size-dependent resonances persist without further losses. This behavior can be understood from the admittance band structure. Since all the information for an ideal fully resonant media is contained within the imaginary part of its impedance, the presence of the parasitic resistances makes the impedance complex. This complexity implies that the stored energy represented by the imaginary part is dissipated due to the parasitic resistances. However, as long as the parasitic resistances do not become dominant, their presence results in a smooth shift in the frequency corresponding to zero admittance in the admittance band structure. To show this, we plot the admittance band structures in Fig.~\ref{fig7}a, and \ref{fig7}b for our 2D $LC$ circuit with $N=3$ by considering two different parasitic resistances. As evident from Figs.~\ref{fig7}c and ~\ref{fig7}d, which provide a zoomed-in view of the band crossing points in panels (a) and (b) respectively, the admittance bands themselves remain qualitatively unchanged, although there is a shift towards higher frequencies in the admittance band structure. This is significant because impedance resonances occur in the presence of nearly zero eigenvalues, which correspond to the band-crossing points in the Laplacian formalism. According to \eqref{two-point-imp}, a large impedance is obtained when at least one of eigenvalues ($\lambda_\mu$) becomes nearly zero provided that the wavefunction values at the measurement points of its corresponding eigenstate are not zero. Fig.~\ref{fig7} demonstrates that introducing small parasitic resistances results in a shift in the frequencies corresponding to the zero-energy eigenvalues. This explains the shift to the right in Fig.~\ref{fig3}c and demonstrates the robustness of our circuit against parasitic resistances despite the smooth shift in the resonant frequencies.

\subsection{Details of the experiment}

Our experiment consists of measuring the corner-to-corner impedance of a square lattice array of $LC$ elements, as pictured in Fig.~\ref{fig2}a, b. To fit our measured impedances with the theoretical predictions, we introduce serial resistances to the $L$ and $C$ components, such that the effective $\omega_r$ becomes complex. Below, we detail the procedures involved, as well as some of the subtleties.

\begin{figure}
	\centering
	\includegraphics[width=10cm]{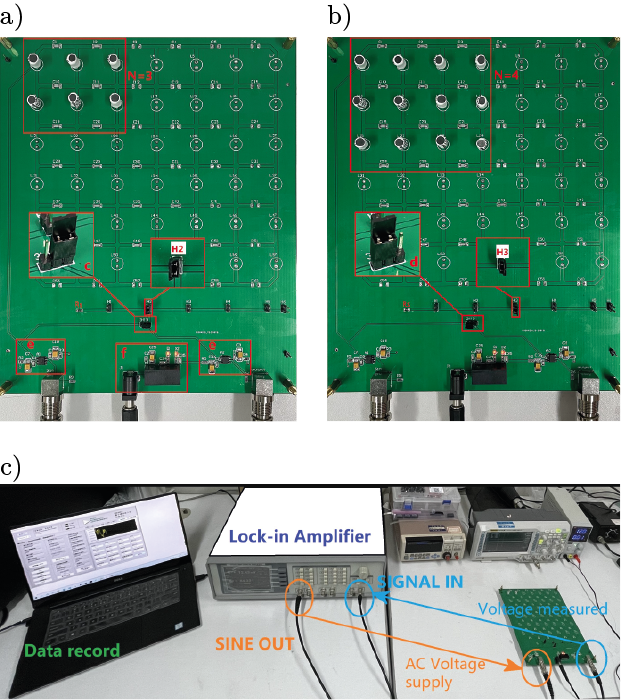}
	\caption{\textbf{Methodology of measuring and extending the $N\times N$ circuits. } (a) For the $N=3$ case measurement, H2 was connected through the jumper cap to connect the entire measurement circuit after the components were soldered. The voltage across a standard resistor of 110 $\Omega$ was then measured by connecting the right side of the switch 3H10 (c in Fig.(a)). The voltage across the entire circuit was next measured using lock-in amplifier by connecting the left side of the switch 3H10 with a jumper cap (d in Fig.(b)). After the measurement for $N=3$ was completed, all the switches were disconnected, and the $N=4$ circuit  (b) was extended from the $N=3$ circuit. H3 was subsequently connected and the above steps were repeated after the additional required components were soldered. In addition to the measured circuit, an operational amplifier (e in Fig.(a)) was also added at the input end and another operational amplifier at the output end of the signal as followers to ensure the stability of the lock-in amplifier signal. The power supply module (f in Fig.(a) ) supplies power to these two operational amplifiers. 
		(c) Lab setup. To effectively avoid interference to the weak signals, we used a lock-in amplifier for measurements. SINE OUT provides an AC voltage signal to the measurement circuit, and SIGNAL IN measures the voltage across either a standard resistor or the entire circuit (controlled by a switch). 
	}
	\label{figExp}
\end{figure}

\subsubsection{Measurement process}

Our measurements were performed on circuit lattices of different sizes corresponding to $N=2$ to $7$ (refer to Fig.~\ref{figExp}a and ~\ref{figExp}b). To mitigate the effects of component disorder, the larger lattices were built by extending the smaller lattices, i.e., measurements are carried out on a circuit of size $N$ before the circuit was extended by soldering additional circuit elements to form a circuit of size $N+1$.

Based on the fractal parameter space diagram, two  AC frequency ranges of interest were determined as 115-175 kHz and 215-290 kHz. For each lattice size $N$, we swept through both of these ranges with a sweep step size 200 Hz (an overly small step size will significantly increase the measurement time). The sweep time interval was set to 1000 ms, which was sufficient to ensure that the voltage reached a stable state each time the frequency $\omega$ was updated. For each $(\omega,N)$ point, the last three (stabilized) voltages were averaged and recorded. The configuration of the measurement setup is shown in Fig.~\ref{figExp}c.

\begin{figure}
	\centering
	\includegraphics[width=\linewidth]{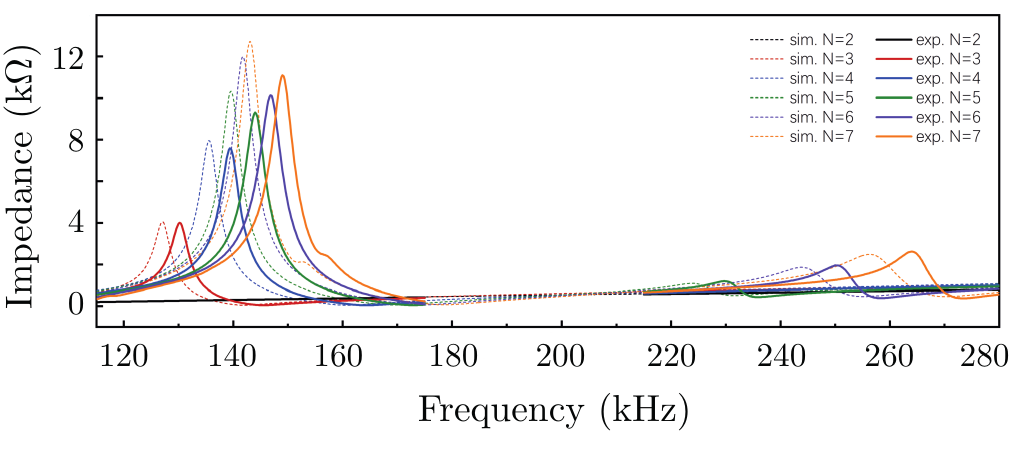}\\
	\caption{\textbf{Comparison of theoretically simulated and experimentally measured impedances}. The simulated and experimentally measured impedances differ slightly in both the peak positions and heights, but the discrepancies are fully accounted for by component uncertainly and tolerances as detailed in Table~\ref{table1}. }
	\label{fig9}
\end{figure}

\subsubsection{Analysis of uncertainties}
There are two main types of discrepancies between the theoretically predicted (\eqref{Z}) and experimentally measured impedances. The first is the discrepancy between the predicted and measured resonant frequencies \emph{f$_0$} where the impedance peaks, and the second is the discrepancy between the predicted and measured impedance values at \emph{f$_0$}. The first discrepancy can mainly be attributed to the uncertainties in the component values. The components we used are rated at $ C=4.7\ \mathrm{nF}\pm1\% $, $ L=1\ \mathrm{mH}\pm5\% $. Employing the frequency scale $\omega/\omega_r=(LC)^{-1/2}$ as a value estimator, we find that the discrepancy of \emph{f$_0$} presented in Fig.~\ref{fig9} is within a reasonable range, as further tabulated in Table~\ref{table1}.
\begin{sidewaystable}
\centering
	\begin{tabular}{|c|ll|ll|ll|ll|}
		\hline
		\multirow{3}{*}{\textbf{\begin{tabular}[c]{@{}c@{}}\end{tabular}}} & \multicolumn{2}{c|}{\multirow{2}{*}{\textbf{\begin{tabular}[c]{@{}c@{}}Sim. with   \\ C=4.7nF$\pm$ 1\%, L=1mH$\pm$ 5\%\end{tabular}}}} & \multicolumn{2}{c|}{\multirow{2}{*}{\textbf{\begin{tabular}[c]{@{}c@{}}Sim. with   \\ C=4.7nF, L=1mH\end{tabular}}}} & \multicolumn{2}{c|}{\multirow{2}{*}{\textbf{Exp.}}}                          & \multicolumn{2}{c|}{\multirow{2}{*}{\textbf{Error}}}               \\
		& \multicolumn{2}{c|}{}                                                                                                       & \multicolumn{2}{c|}{}                                                                                               & \multicolumn{2}{c|}{}                                                        & \multicolumn{2}{c|}{}                                              \\ \cline{2-9} 
		& \multicolumn{1}{l|}{\textbf{range of $f_0(kHz)$}}                           & \textbf{range of $Z(k\Omega)$}                          & \multicolumn{1}{c|}{\textbf{$f_0(kHz)$}}                     & \multicolumn{1}{c|}{\textbf{$Z(k\Omega)$}}                    & \multicolumn{1}{c|}{\textbf{$f_0(kHz)$}} & \multicolumn{1}{c|}{\textbf{$Z(k\Omega)$}} & \multicolumn{1}{c|}{\textbf{$f_0$}} & \multicolumn{1}{c|}{\textbf{$Z$}} \\ \hline
		\textbf{3}                                                                       & \multicolumn{1}{l|}{123.4$\sim$131.0}                                    & 3.9$\sim$4.2                                     & \multicolumn{1}{l|}{127.0}                                & 4.06                                                    & \multicolumn{1}{l|}{130.2}            & 4.01                                 & \multicolumn{1}{l|}{2.52\%}      & -1.23\%                         \\ \hline
		\textbf{4}                                                                       & \multicolumn{1}{l|}{131.7$\sim$139.8}                                    & 7.6$\sim$8.2                                     & \multicolumn{1}{l|}{135.6}                                & 7.96                                                    & \multicolumn{1}{l|}{139.4}            & 7.59                                 & \multicolumn{1}{l|}{2.80\%}      & -4.65\%                         \\ \hline
		\textbf{5}                                                                       & \multicolumn{1}{l|}{135.5$\sim$144.0}                                    & 9.9$\sim$10.7                                    & \multicolumn{1}{l|}{139.6}                                & 10.32                                                   & \multicolumn{1}{l|}{144.0}            & 9.31                                 & \multicolumn{1}{l|}{3.15\%}      & -9.79\%                         \\ \hline
		\textbf{6}                                                                       & \multicolumn{1}{l|}{236.8$\sim$251.4}                                    & 1.8$\sim$1.9                                     & \multicolumn{1}{l|}{243.8}                                & 1.86                                                    & \multicolumn{1}{l|}{250.4}            & 1.95                                 & \multicolumn{1}{l|}{2.71\%}      & 4.84\%                          \\ \hline
		\textbf{7}                                                                       & \multicolumn{1}{l|}{248.9$\sim$264.2}                                    & 2.4$\sim$2.6                                     & \multicolumn{1}{l|}{256.3}                                & 2.48                                                    & \multicolumn{1}{l|}{263.8}            & 2.62                                 & \multicolumn{1}{l|}{2.93\%}      & 5.65\%                          \\ \hline
	\end{tabular}
	\caption{Comparison of theoretical simulation results with given component error tolerances against experimental results. The given $LC$ components are rated at $L=1\ \mathrm{mH}\pm 5\%$ and $C=4.7\ \mathrm{nF}\pm 1\%$. $Z$ and $f_0$ are the peak impedance and the frequency at which it occurs. The experimentally measured (exp.) values are indeed within the theoretically predicted ranges (sim.) corresponding to the error tolerances. The full exp. and sim. impedance curves given in Fig.~\ref{fig9}. }
	\label{table1}
\end{sidewaystable}

The second type of discrepancy, i.e., the impedance peak heights, is greatly affected by the parasitic resistance in addition to the component uncertainties. The parasitic resistance effectively suppress the peak of the measured impedance. This is reflected in the impedance-frequency curve in which the decrease in the peak value is accompanied by an increase in the FWHM (full width at half maximum), which makes it difficult to distinguish between the curves of different system sizes if the parasitic resistances were too large (fortunately, they were not). There may be several sources that contribute to parasitic effects, such as parasitic resistance, capacitance, and inductance. However, through numerous simulation studies, we found that the parasitic resistances are the most significant contributors that affect the measured impedance resonances. Using the estimated serial parasitic resistances of $R_{pL} = 3.3 \Omega$, $R_{pC} = 4.5 \Omega$, $R_{pW} = 0.1 \Omega$ for the inductors, capacitors, and solder contacts respectively, we find that the experimental and simulation results match reasonably, as shown in Tables~\ref{table1} and \ref{table2}, and plotted in Fig.~\ref{fig2}d of the main text.

\subsubsection{Reducing the influence of parasitic resistances}
Parasitic resistance has a strong impact on the experiment. The most direct way to reduce its impact is to increase the inductances $L$ while decreasing the capacitances $C$, since doing so does not necessitate a proportional increase in the parasitic resistances. However, the inductance value should not be too large in order to keep R$_{pL}$ within a reasonable range. At the same time, if the capacitance value is too small, the equivalent series resistance of the capacitors becomes dominant and the frequency $f_0$ increases, which may increase the uncertainty in the measurement. In order to strike a balance, we chose \emph{C=4.7} nF, \emph{L=1} mH.

\subsection{Determination of $\omega_r$ for experimental setups}

Here, we provide details on how resistive contributions from $L$ and $C$ components (not necessarily parasitic) affect $\omega_r$, which is the most important dimensionless parameter in our setup. The addition of serial resistances to each capacitor and inductor modifies their admittance contributions to the circuit Laplacian as follows:
\begin{align}
i\omega C&\rightarrow \frac{i\omega C}{1+i\omega CR_C}\\
\frac{1}{i\omega L}&\rightarrow \frac{1}{R_L+i\omega L}
\end{align}

The Laplacian from Eq.~1 of the main text is hence modified to
\begin{align}
	\mathcal{L}(k_x,k_y)&=\frac{2i\omega C}{1+i\omega CR_C}(1-\cos k_x ) + \frac{2}{R_L+i\omega L}( 1-\cos k_y )\notag\\
	&=\frac{2i\omega C}{1+i\omega CR_C} \left[ (1-\cos k_x)-\frac{\frac{L}{C}-R_CR_L+i\omega LR_C + \frac{i R_L}{\omega C}}{R_L^2+\omega^2 L^2}(1-\cos k_y)\right]\notag\\
			&=\frac{2i\omega C}{1+i\omega CR_C} \left[ (1-\cos k_x)-\omega_r^{-2}(1-\cos k_y)\right]
\end{align}
with the important parameter $\omega_r^{-2}$ modified to
\begin{equation}
\omega_r^{-2}=\frac{\frac{L}{C}-R_C R_L}{R_L^2+\omega ^2 L^2}+\frac{\omega LR_C+\frac{R_L}{\omega C}}{R_L^2+\omega ^2 L^2}i.
\end{equation}
Substituting the measured parasitic resistances for our fabricated circuits via $R_C=R_{pC}+2R_{pW}, R_L=R_{pL}+2R_{pW}$, and $\omega_r$ into the simulations, we find an excellent fit to the measured circuit impedances and their peaks (Fig.~\ref{fig2}d of the main text). Their corresponding $\omega_r$ are given in Table~\ref{table2}. Note that an imaginary part $\text{Im}\omega_r$ of $\sim 0.02$ to $\sim 0.06$ was acquired due to these resistances. Since the components used were not of particularly high quality, $\text{Im}(\omega_r)$ can potentially be reduced by one or more orders if necessary - in our case, they already suffice for demonstrating the anomalous impedance scaling.

\begin{table*}
\centering
	\begin{tabular}{cccccllll}
		\cline{1-5}
		\multicolumn{1}{|c|}{\multirow{2}{*}{\begin{tabular}[c]{@{}c@{}}Z max at\\  N=\end{tabular}}} & \multicolumn{2}{c|}{sim.}                                  & \multicolumn{2}{c|}{exp.}                                  &  &  &  &  \\ \cline{2-5}
		\multicolumn{1}{|c|}{}                                                                        & $f_0(kHz)$                 & \multicolumn{1}{c|}{$\omega_r$}           & $f_0(kHz)$                 & \multicolumn{1}{c|}{$\omega_r$}           &  &  &  &  \\ \cline{1-5}
		\multicolumn{1}{|c|}{3}                                                                       & 127.0                  & \multicolumn{1}{c|}{$1.7297-0.0190i$} & 130.2                & \multicolumn{1}{c|}{$1.7733-0.0198i$} &  &  &  &  \\
		\multicolumn{1}{|c|}{4}                                                                       & 135.6                & \multicolumn{1}{c|}{$1.8468-0.0211i$} & 139.4                & \multicolumn{1}{c|}{$1.8986-0.0222i$} &  &  &  &  \\
		\multicolumn{1}{|c|}{5}                                                                       & 139.6                & \multicolumn{1}{c|}{$1.9013-0.0222i$} & 144.0                  & \multicolumn{1}{c|}{$1.9612-0.0234i$} &  &  &  &  \\
		\multicolumn{1}{|c|}{6}                                                                       & 243.8                & \multicolumn{1}{c|}{$3.3195-0.0600i$} & 250.4                & \multicolumn{1}{c|}{$3.4092-0.0630i$} &  &  &  &  \\
		\multicolumn{1}{|c|}{7}                                                                       & 256.3                & \multicolumn{1}{c|}{$3.4895-0.0658i$} & 263.8                & \multicolumn{1}{c|}{$3.9515-0.0695i$} &  &  &  &  \\ \cline{1-5}
		\multicolumn{1}{l}{}                                                                          & \multicolumn{1}{l}{} & \multicolumn{1}{l}{}                & \multicolumn{1}{l}{} & \multicolumn{1}{l}{}                &  &  &  & 
	\end{tabular}
	\caption{\textbf{The effective $\omega_r$ for various experimental data points and their simulated values (from Fig.~\ref{fig2}d of the main text)}. Due to parasitic resistances, $\omega_r$ acquires a small imaginary part on the order $10^{-2}$. }
	\label{table2}
\end{table*}

\section{Discussion}

In this chapter, we theoretically discussed and experimentally investigated the pronounced yet seemingly random impedance scaling behavior of $RLC$ circuit lattices arrays. This anomalous impedance scaling contrasts greatly with the usual logarithmic scaling expected in 2 dimensions, and is rooted in the commensurability properties of the circuit Green's function eigenvalues, reminiscent of the commensurability conditions pertaining to a Hofstadter lattice with magnetic flux. This results in a curious fractal-like impedance behavior in the parameter space of dimensionless frequency $\omega_r$ and lattice size $N$, whose complexity and structure elude any simplistic waveguide analysis.

In generic circuit lattices with more complex connections, unit cells, and feedback elements, more sophisticated fractal impedance fringes would be expected due to the more complicated commensurability conditions for the vanishing of the circuit Laplacian eigenvalues. This points towards the hitherto unnoticed general breakdown of a continuum description of resonant conducting media, which highlights the need for more careful analysis in the discretization of device geometries in electrostatics simulations. The discretization of continuous media involves dividing the medium into discrete units or elements that can be represented using discrete variables. In the context of electrical circuits, this can involve dividing continuous electrical fields or currents into discrete components such as resistors, capacitors, and inductors, which can be connected in various ways to create a circuit. Discretization allows for the use of mathematical tools and techniques to analyze physical phenomena in a continuum media~\cite{webman_theory_1977,kirkpatrick_percolation_1973}, as in this study.

More generally, the fractal anomalous scaling behavior extends to the steady state behavior of systems governed by network Laplacians where a resonance frequency also enters the dynamics. This includes, for instance, directed information networks, which are physically unrelated to electrical circuits. While we have focused on a very regular square lattice network that should have possessed simple logarithmic impedance scaling naively, such fractal scaling also exists in more generic network structures, albeit in possibly more concealed manners. 
\SetPicSubDir{ch-nonlinear}
\SetExpSubDir{ch-nonlinear}

\chapter{Non-linear Topological Chaos}
\label{ch:Topochaos}
\vspace{2em}
Building on our investigations into the fundamental properties of electrical circuits, including voltage and impedance responses discussed in the previous chapters, we now present an implementation of these circuits in a non-linear topological system. In this chapter, we explore the intricate interplay between topological and chaotic phases within a one-dimensional topological Su-Schrieffer-Heeger (SSH) circuit combined with onsite chaotic Chua's circuits. Although chaos and the topological phases of matter are seemingly distinct phenomena, we investigate how topological phases influence chaotic dynamics in both self-driven and externally stimulated circuit configurations. The presence of a non-trivial topological phase results in distinct chaotic phase portraits at the edge nodes, attributable to the exponential topological edge localization in both self-driven and externally stimulated arrangements of chaotic SSH circuit (cSSH). While an inserted signal perturbs the chaotic dynamics of our circuit to some extent, we observe that non-trivial topology protects chaotic oscillations from perturbations induced by the injected signal in the externally stimulated scenario. The topological and chaotic characteristics of our nonlinear chaotic SSH circuit are governed by the Lorenz equations, involving the dimensionless intracell and intercell couplings. Our study reveals that topological and chaotic dynamics can be found individually but collectively contribute to the dynamics of the circuit. This work presents the realization of topological phenomena in nonlinear systems, with implications for the design of robust and adaptable electronic nonlinear devices.

\section{Introduction}
Linear systems serve as the foundation for modern technologies, thanks to their predictable and controllable dynamics. However, the vast majority of physical systems in the real world often exhibit nonlinear characteristics. As we push the boundaries of linear systems, it becomes necessary to explore nonlinear dynamical systems. For example, topological phenomena have been observed in a range of linear systems, such as photonics~\cite{ozawa_topological_2019}, mechanics~\cite{zheng_progress_2022}, acoustics~\cite{xue_topological_2022}, electrical circuits~\cite{lee_topolectrical_2018}, and even in a dance performance~\cite{du_chiral_2023}. These examples, though, are predominantly linear realizations, despite some precedent examples~\cite{hohmann_observation_2023,kotwal_active_2021,hadad_self-induced_2018,ezawa_topological_2022,ezawa_nonlinearity-induced_2021,marquie_nonlinear_1995}. Meanwhile, chaotic dynamics of nonlinear systems have long been recognized and studied~\cite{watts_collective_1998,hofstrand_discrete_2022,chaunsali_stability_2021,chaunsali_dirac_2022,chen_circuit_2014} since the first discovery of chaotic dynamics by Lorenz~\cite{lorenz1963deterministic}. The renowned realization of the Lorenz system is the Chua's circuit—a simple electrical circuit that models Lorenz's nonlinear system—serving as a prime example in electrical circuits~\cite{kennedy_three_1993,lu_network_2022,nekorkin_homoclinic_1995}. Despite extensive research on both topological phenomena and chaotic dynamics, their interplay within nonlinear circuits remains underexplored.

Here, we introduce the concept of topological chaos, which arises from the interaction between chaotic oscillations and topologically non-trivial behaviors in a one-dimensional chaotic Su-Schrieffer-Heeger (cSSH) circuit. The circuit consists of inductive intracell and intercell couplings and identical onsite Chua's circuits. We observe that the characteristics of chaotic oscillations at the edge Chua's circuits differ from bulk oscillations in the non-trivial topological phase due to edge localization. Distinct chaotic phases exist in the topologically non-trivial phase, whereas in the trivial phase, chaotic oscillations remain uniform throughout the circuit. To investigate signal evolution across both topological phases, we introduce a sinusoidal signal into the cSSH circuit. External excitations disrupt the circuit's chaotic dynamics, leading to disturbed chaotic oscillations in the trivial phase. However, we observe that the topologically non-trivial phase preserves the chaotic oscillations, owing to topological protection. This protection is attributed to the presence of topological boundary localization. 

To investigate the interplay between chaos and topology, we develop a novel method based on node conductance. This approach originates from our observation that chaotic dynamics in Chua's circuits are not solely determined by individual component values, but rather by the cumulative effective conductance at the nodes, particularly under varying coupling conditions. By focusing on the effective inductance at the coupling nodes of Chua's circuits, we explore both chaotic and topological behaviors within our circuit array. Our methodology simplifies the complex interdependencies of various elements by fixing the values of capacitors and the IV characteristics of non-linear resistors, allowing us to modify only the inductance values. This reduction to a primarily inductive circuit aids in isolating the effects of inductance on the system's dynamics, enabling a clearer examination of the relationship between topological arrangements and chaotic phenomena. Therefore, our circuit and the effective inductance method demonstrate how topological and chaotic dynamics interact individually and collectively, constructing a comprehensive picture of topological chaos.

\section{Results}
\begin{figure*}[ht!]
	\centering
	\includegraphics[width=\linewidth]{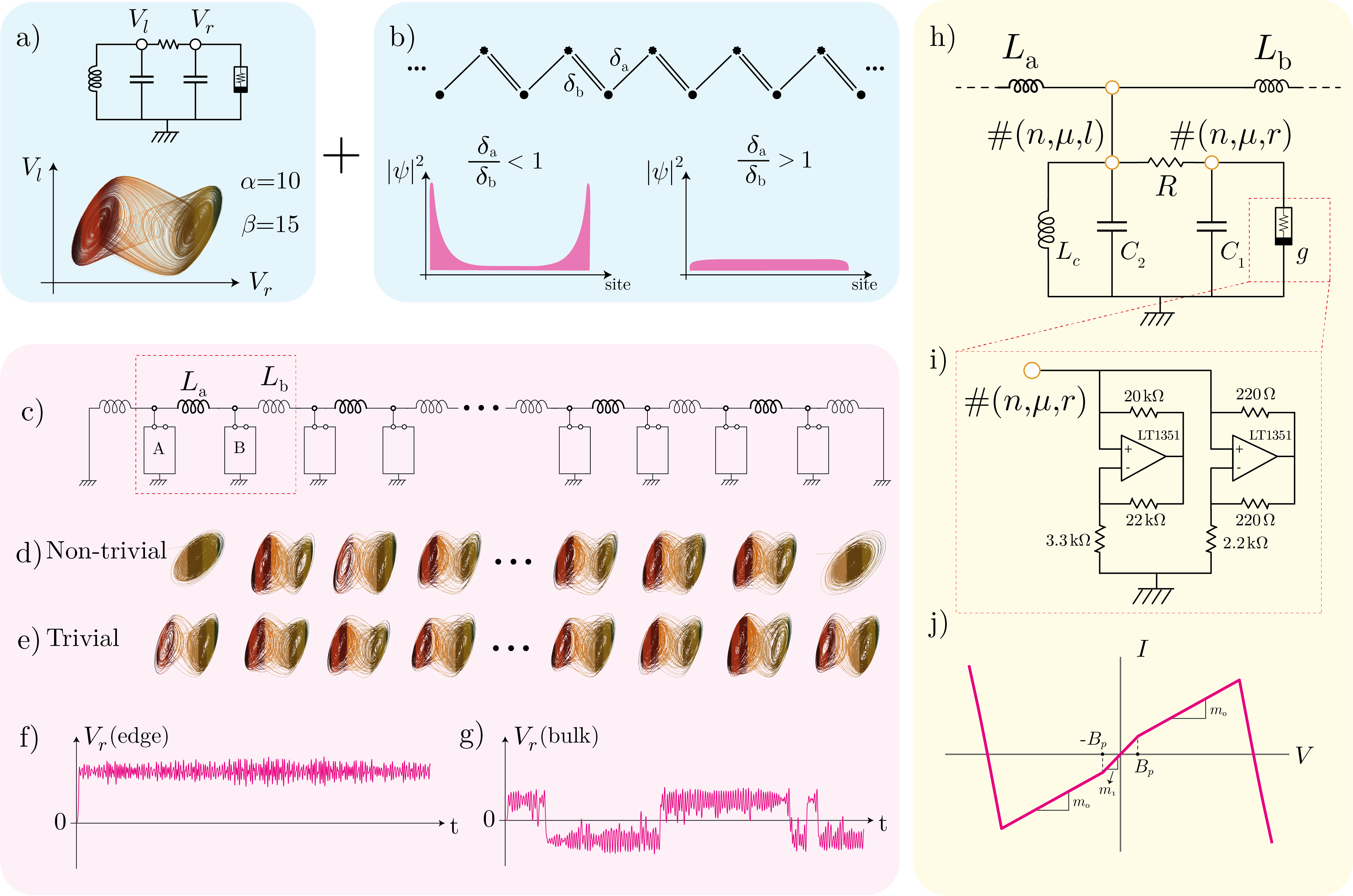}
	\caption{\textbf{The summary of topological chaos and the generic illustrative picture.} \textbf{a} A single Chua's circuit and its exemplary chaotic double scroll phase portrait when $\alpha=10$ and $\beta=15$. \textbf{b} The conventional 1D SSH lattice with intra-cell and inter-cell couplings $\delta_a$ and $\delta_a$, respectively, and its eigenmode profile in the topologically trivial and non-trivial phases. \textbf{c} The schematic of our chaotic SSH circuit with identical Chua's circuits attached between each node and ground. A unit cell comprising of intra-cell inductor with inductance $L_a$ and inter-cell inductor with inductance of $L_b$ is indicated with a dashed magenta rectangular. \textbf{d} The chaotic phase portraits of each Chua's circuit attached to every node of the chaotic SSH circuit in the non-trivial ($L_a=100$mH, $L_b=20$mH) and \textbf{e} trivial phases ($L_a=20$mH, $L_b=100$mH). The topological non-trivial phase leads to a transition in the chaotic phase of the edge Chua's circuits due to the boundary localization. \textbf{f} The time variation of the voltage oscillations at the right node of the Chua's circuits of the edge nodes and \textbf{g} of the bulk nodes in the non-trivial phase in (\textbf{d}). \textbf{h} A detailed illustration of a Chua's circuit employed at each node of our chaotic SSH circuit. We denote the two nodes found on the left and right side of the conventional resistor $R$ as $(n,l)$ and $(n,r)$, respectively. \textbf{i} The non-linear Chua's diode is realized with two op-amps and their feedback configurations. For our LTspice simulations, we employ op-amps with Analog catalog numbers LT1351. \textbf{j} The five segments piece-wise current-voltage characteristics of the non-linear diode depicted in \textbf{(i)}. $B_p$ represents the breakpoint voltage, $m_0$ and $m_1$ are the slope of the corresponding linear parts.}
	\label{fig1}
\end{figure*}
In our one-dimensional chaotic SSH circuit, the chaotic phase portraits of the edge and bulk nodes differ depending on the topological phase. While the chaotic dynamics are uniform and identical across all nodes in the topologically trivial phase, the oscillations at the edge nodes become distinct from those of the bulk in the topologically non-trivial regime, as depicted in Figs.~\ref{fig1}d,e. This difference in the chaotic phase of the edge nodes is a result of the bulk topology, known as bulk-boundary correspondence (BBC)\cite{helbig_generalized_2020}. 

Note that, unlike linear topological models where the topological invariant can be calculated to define the topological phase through established classifications, the topological phases in our circuit model cannot be defined using linear methods due to the inclusion of non-linear components. For instance, in an SSH circuit, the boundary between topologically trivial and non-trivial phases is defined by the ratio of intra-cell to inter-cell couplings, with the two phases distinctly separated along a linear axis. However, given that our model is non-linear, we anticipate a non-linear phase boundary, where chaotic scrolls at the edge and bulk of the circuit scale distinctly in different parameter regimes. Consequently, the occurrence of topology in this context will be referred to as non-linear topological phases, which we will define through the effective inductance approach and the chaotic scroll amplitude differences of the bulk and edge scrolls in the subsequent sections.

To understand the underlying mechanism, we examine $N$ fully identical, uncoupled onsite Chua's circuits arranged along the $x$ direction. Each Chua's circuit displays identical oscillatory behavior due to the same initial conditions and a consistent parameter setting for the four passive elements: inductance $L_c$, capacitances $C_1$ and $C_2$, and resistance $R$ (refer to Fig.~\ref{fig1}h). Without the nonlinear resistor $g(V)$, also known as Chua's diode (detailed in Fig.~\ref{fig1}i), a Chua's circuit is simply a trivial linear RLC circuit. However, with the nonlinear Chua's diode connected in parallel to the three energy storage elements, the circuit becomes nonlinear. Under specific parameters such as $L_c=24\text{mH}$, $C_1=10\text{nF}$, $C_2=100\text{nF}$, and $R=1.85\text{k}\Omega$, and following the non-linear current-voltage ($I$-$V$) characteristic shown in Fig.~\ref{fig1}j, chaotic double-scroll oscillations emerge between the $(n,l)$ and $(n,r)$ nodes in each individual Chua's circuit, where $(n,l)$ and $(n,r)$ denote the left and right nodes of the regular resistor $R$, respectively. (Note that we consistently use the same parameters for these four passive components throughout this study.)

A Chua's circuit~\cite{leon_chua_ten_years,kennedy_three_1993_2} is basically the realization of the Lorenz system~\cite{lorenz1963deterministic} described by the set of equations as
\begin{equation}
	\begin{aligned}
		\dot{x} &= \alpha(y-x-f(x))\\
		\dot{y} &= x-y+z\\
		\dot{z} &= -\beta y
	\end{aligned}
	\label{eqLorenz}
\end{equation}
where $\dot{x}$, $\dot{y}$ and $\dot{z}$ are the first-order time derivative of the dimensionless variables $(x,y,z)$ representing the dimensionless form of the circuit's states and $f(x)$ is the dimensionless form of the non-linear function. Remarkably, the Lorenz system exhibits \textit{chaos} for a narrow range of values of empirical parameters $\alpha$ and $\beta$ such as the Lorenz attractor when $\alpha\approx10$, $\beta\approx15$ as shown in Fig.~\ref{fig1}a.

To achieve topological chaos, we now incorporate the Lorenz equations, which describe a single chaotic Chua's circuit, into a one-dimensional chain circuit. First, we group every two neighboring sites, starting from the leftmost site, into two sublattice nodes labeled $A$ and $B$, such that each group forms a unit cell. Next, we introduce a coupling inductance, denoted as $L_a$, which links the left nodes (i.e., $(n,l)$) of each Chua's circuit in the groups, as illustrated in Fig.~\ref{fig1}h. This coupling leads to energy exchanges between the circuits and influences their individual chaotic dynamics. When two Chua's circuits are coupled, they eventually become synchronized, a phenomenon known as \textit{chaos synchronization}, and it has been extensively studied through various coupling components and combinations\cite{chaos_sync_leo,yao_synchronization_2019,kapitaniak_experimental_1997,gamez-guzman_synchronization_2009,feki_adaptive_2003,kolumban_role_1998}. As expected, our coupled Chua's circuits in each unit cell also present synchronized phase portraits over a short period of time. This is because the information exchange leads to the merging of the two attractor points of the chaotic systems~\cite{nekorkin_travelling_1996,nekorkin_chaotic_1996,anishchenko_dynamics_1995,xie_hybrid_2002}. The time required for synchronization depends on the coupling strength, which essentially determines the amount of shared information. For instance, with a relatively strong $L_a$ (e.g., around a few tens of millihenries), our circuit exhibits the behavior of a fully dimerized SSH circuit. In this case, two coupled Chua's circuits in the unit cells display identical and synchronized chaotic phase portraits.

Now, by introducing a second coupling inductance, $L_b$, between unit cells for inter-cell coupling, our circuit transforms into a complete coupled 1D chain circuit. To accurately achieve an SSH model, which involves both intracell and inter-cell coupling as well as an onsite Chua's circuit at each site, we add an additional $L_a$ between the left edge node and the ground and directly ground the right edge, as shown in Fig.~\ref{fig1}c. This configuration ensures that each site in our circuit experiences the same potential. With this proper setup, the Lorenz equations for the coupled chaotic SSH circuit are given as
\begin{equation}
	\begin{aligned}
		& \dv{x(n,A)}{\tau} = \alpha(y(n,A) - x(n,A) -  f(x(n,A))),\\
		& \dv{y(n,A)}{\tau} =  x(n,A) - y(n,A) + z(n,A) + u(n) - v(n-1) ,\\
		& \dv{z(n,A)}{\tau} = - \beta y(n,A),\\
		& \dv{u(n)}{\tau} = \delta_a (y(n,B) - y(n,A)),\\
		& \dv{v(n)}{\tau} = \delta_b (y(n+1,A) - y(n,B)),\\
		& \dv{x(n,B)}{\tau} = \alpha(y(n,B) - x(n,B) -  f(x(n,B))),\\
		& \dv{y(n,B)}{\tau} = x(n,B) - y(n,B) + z(n,B) -u(n) + v(n),\\
		& \dv{z(n,B)}{\tau} = - \beta y(n,B),
	\end{aligned}
	\label{fullequations}
\end{equation}
where, $x,y,z$ are the state variables, $\alpha, \beta$ are positive constant parameters, $\tau$ is the dimensionless time variable and $f(x)$ represents the non-linear piece-wise function. $n$ labels the unit cells. The open boundary conditions are imposed as follow
\begin{equation}
	\dv{v_0}{\tau} = \delta_b y(1,A) \quad \text{and} \quad y(n+1,A)=0.
	\label{iL24}
\end{equation}
Unlike \eqref{eqLorenz}, our coupled system incorporates two new dimensionless system variables, $\delta_a$ and $\delta_b$ in \eqref{fullequations} (for details, refer to `Methods: The Lorenz equations for the chaotic SSH'). These variables represent the dimensionless forms of the intracell and intercell couplings, respectively. The ratio of these variables is crucial in defining the topological phases of the system. The influence of these couplings on the system's behavior will be discussed in the following sections.

\subsection{Self-driven chaotic SSH}
Each Chua's circuit is equipped with direct current (DC) sources to power the operational amplifiers (op-amps) found in the non-linear resistors, as shown in Fig.~\ref{fig1}i. Consequently, it is feasible to simulate the circuit without additional sources, although we will address this scenario in a later section. Given appropriate initial conditions and the same above-stated parameter settings, the Chua's circuits begin to oscillate. In contrast to the synchronization behavior observed in the case of fully dimerized unit cells, introducing couplings ($L_b$) between the unit cells does not guarantee instantaneous and simultaneous synchronization. By incorporating the chaotic dynamics of the Chua's circuit with the topological dimerization results that a distinct chaotic phase portrait (single scroll) is observed at the edge nodes while the remaining nodes exhibit the same portraits (double scroll) for the topologically non-trivial phase with $L_a=100\,\text{mH}$ and $L_b=20\,\text{mH}$, as shown in Fig.~\ref{fig1}d. On the other hand, in the topologically trivial phase (where $L_a=20\,\text{mH}$ and $L_b=100\,\text{mH}$), both bulk and edge nodes display the same type of chaotic portraits (double scroll), as in Fig.~\ref{fig1}e. This unique behavior emerges depending on the ratio of the intracell and intercell couplings and absent if there is no topological mechanism. 
\begin{figure}[t]
	\centering
	\includegraphics[width=\linewidth]{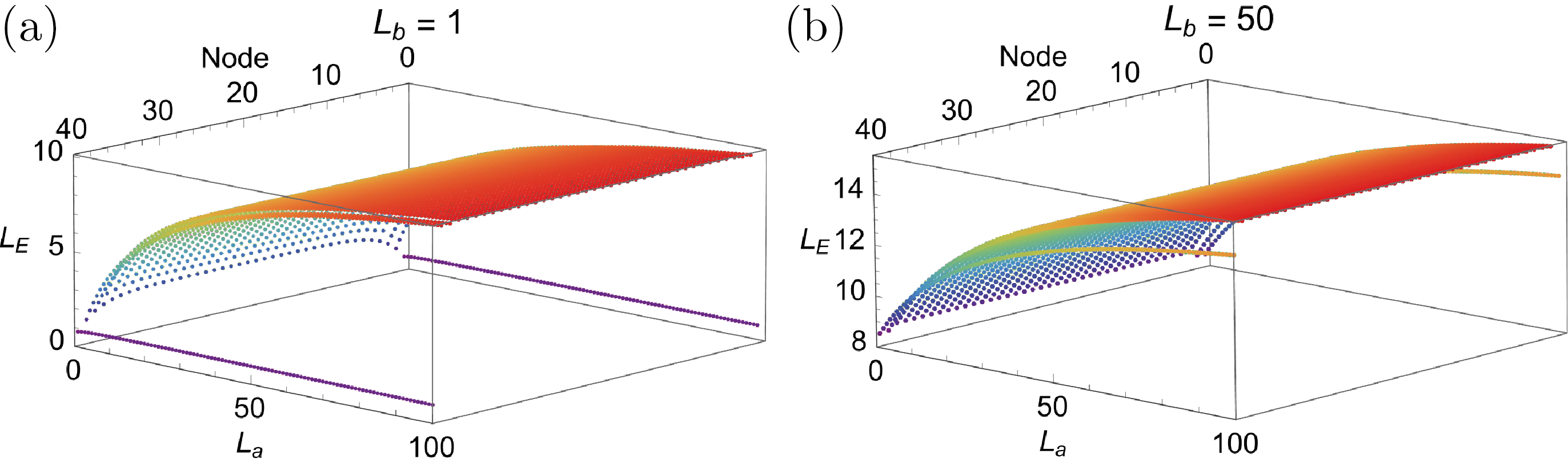}
	\caption{\textbf{The effective inductance profiles of open boundary cSSH from the numerical recursive equations in 3D plots for two different fixed $L_b$ values.} (a) The effective inductance profile when $L_b = 1\,\text{mH}$ (b) when $L_b = 50\,\text{mH}$. The edge inductances are very distinct for a small $L_b$ while the distinction decreases for $L_b$s. This behavior is unique to the open boundary circuit, whereas the effective inductance of all nodes remains constant in a periodic boundary circuit. For each case, we set $N=20$, $L_c = 24\,\text{mH}$.}
	\label{figinductance}
\end{figure}
To better understand how the alternation of $L_a$ and $L_b$ affects the chaotic phase profiles of each node, we calculate the total effective inductances of all nodes analogous to the lumped-element method. As the only variation in the circuit is the intracell and intercell couplings, altering the ratio of the coupling inductances results in a change in the effective inductance of the edge nodes due to the open boundaries of our circuit. The recursion relation for the effective inductance at a specific node is given by
\begin{equation}
	L_E(i)=\left(\frac{1}{L_c}+\frac{1}{L_E^\text{left}}+\frac{1}{L_E^\text{right}}\right)^{-1},
	\label{eqtotalindMain}
\end{equation}
where $L_E(i)$ represents the total effective inductance at the $i$th node, with $L_c$ denoting the inductance of the inductors in each Chua's circuit. $L_E^\text{left}$ and $L_E^\text{right}$ are the effective lumped inductances on the left and right sides of node $i$, respectively (for a detailed derivation, refer to `Methods: Effective inductance calculation'). As shown in Fig.~\ref{figinductance}a, the total effective inductances at the edge nodes (indicated by purple dots) are noticeably different from those of the bulk nodes in the non-trivial topological phase, where $L_b=1\,\text{mH}$ and $L_b < L_a$ consistently. In contrast, the effective inductances of the edge nodes align more closely with those of the bulk nodes when the circuit is in a topologically trivial phase, as illustrated in Fig.~\ref{figinductance}b. Figure~\ref{figinductance}b demonstrates both trivial and non-trivial topological regimes. In our circuit, the topological phases are defined by the ratio of $L_a/L_b$, with the circuit being topologically non-trivial when $L_a < L_b$ and trivial when $L_a > L_b$, due to the inverse relation between the admittance of the inductors and inductance. As seen in Fig.~\ref{figinductance}b, while the edge effective inductance aligns with the bulk inductances in the trivial regime, a distinct edge effective inductance emerges in the non-trivial regime. This characteristic of the edge nodes in the topologically non-trivial phase results in distinct chaotic phase portraits at the edges and manifests the interplay of topological and chaotic dynamics in the cSSH. 

The interplay primarily arises from the effective inductance profile of the topological SSH circuit. The chaotic scrolls are obtained through the voltage oscillations at nodes $(n,l)$ and $(n,r)$. Consequently, the scroll amplitudes are defined by the voltage amplitude at the node $(n,l)$, which is the node where each Chua's circuit is connected. Since our effective inductance method calculates the effective node inductances specifically at the $(n,l)$ nodes, the chaotic scroll amplitudes can be inferred from the effective inductance at each respective $(n,l)$ node. This inference is based on the relationship $V = L \dv{I}{t}$, where the inductance is proportional to the voltage. Therefore, a lower effective inductance at the edge nodes leads to the emergence of distinct edge chaotic scrolls. Since the effective inductance profile is influenced by the topological setting, the chaotic profile of the cSSH circuit also depends on the topological phase through effective inductance. Consequently, the total inductance profile elucidates how the topological phase contributes to a distinct chaotic phase. We will now proceed to further examine our circuit's response to additional signal injections in both topological phases.

\subsection{cSSH with an external signal injection}

\begin{figure}[h!]
	\centering
	\includegraphics[width=10cm]{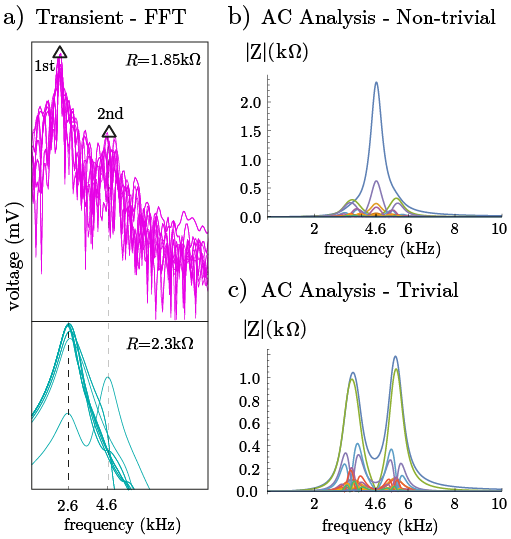}
	\caption{\textbf{Transient and AC responses of the cSSH.} \textbf{a} The Fast Fourier Transform obtained from LTspice simulations when $R=1.85$k$\Omega$ (magenta) and $R=2.3$k$\Omega$ (cyan) without any external signal excitation. The oscillations are chaotic as the magenta FFT profile due to the chaotic dynamics of the circuit when $R=1.85$k$\Omega$. However, the oscillations become regular when, for example, $R=2.3$k$\Omega$ resulting a regular frequency domain profile. There are two distinct peaks indicated with two rectangular shapes in the FFT profile. The second harmonic corresponds to the calculated resonant frequency by \eqref{resfreq}. \textbf{b} and \textbf{c} show the AC response of our cSSH circuit. The emergence and disappearance of the impedance peaks at $f\sim4.6$kHz depending on the topological phase is the manifestation of the topology of our circuit.}
	\label{figacresponse}
\end{figure}

The voltage profile of a purely linear reactive SSH circuit exhibits exponential localization due to its non-trivial bulk topology. To examine our circuit in this context, we introduce an external current source that generates sinusoidal waves, denoted as $I(t)=I_0 \sin(2\pi f_r t)$, where $I_0$ is the constant current magnitude, and $f_r$ is the resonant frequency given by
\begin{equation}
	f_r=\frac{1 \pm \sqrt{1-4 R^2 C_2(L_c^{-1} + L_a^{-1} + L_b^{-1})}}{4 \pi i R C_2}.
	\label{resfreq}
\end{equation}
We determine the resonant frequency, $|\text{Re}(f_r)|=4.62$ kHz, using common parameters: $L_c=24\,\text{mH}$, $L_a=100\,\text{mH}$, $L_b=30\,\text{mH}$, $C_2=100\,\text{nF}$, and $R=1.85\,\text{k}\Omega$. We assume that Chua's diode acts as an open circuit, since the current passing through it is negligible compared to the current through the regular resistor and capacitor $C_1$. We also validate the calculated steady-state $f_r$ of our circuit by comparing it with the AC response and the Fast Fourier Transform (FFT) profile shown in Fig.~\ref{figacresponse}, obtained from LTspice simulations. The second harmonics in the FFT profile, depicted in Fig.~\ref{figacresponse}a, correspond to the calculated resonant frequency. In addition to the FFT results, the two-point impedance measured between the two edge nodes demonstrates a high impedance at the calculated frequency in the non-trivial phase, as seen in Fig.~\ref{figacresponse}b. Conversely, this impedance peak is absent in the trivial phase, as depicted in Fig.~\ref{figacresponse}c. This distinction serves as one of the main hallmarks for distinguishing the topological phases of TEs. Since the topological response of topolectrical circuits manifests when the signal frequency aligns with the resonant frequency~\cite{lee_topolectrical_2018}, we consistently maintain the signal frequency at the resonant frequency of $4.62\,\text{kHz}$, unless specified otherwise. However, we do examine our circuit under various values of the signal magnitude $I_0$.

\begin{figure}[t!]
	\centering
	\includegraphics[width=11cm]{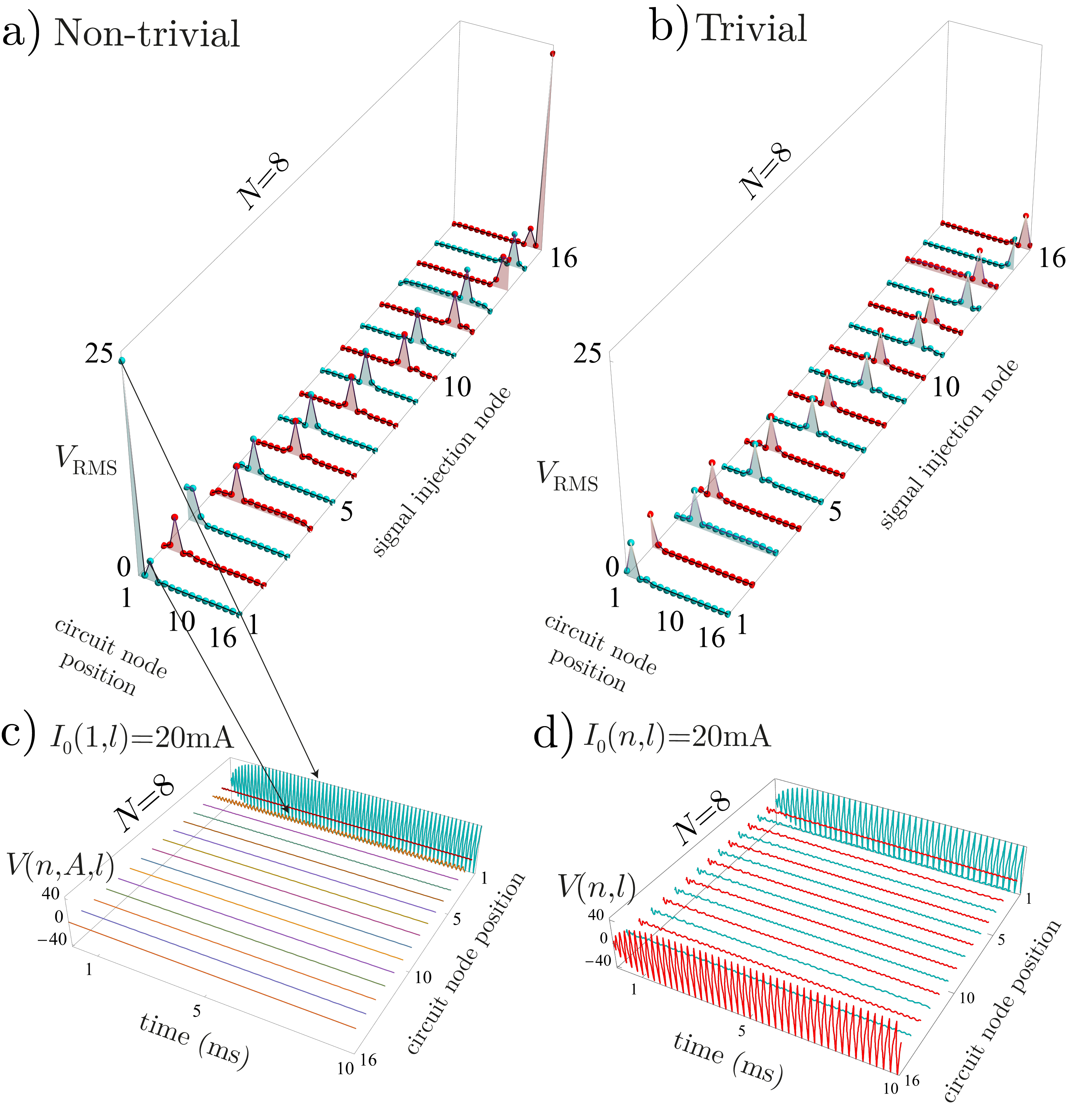}
	\caption{\textbf{Transient LTspice analysis and voltage responses of the cSSH with external signal injection.} \textbf{a} and \textbf{b} show the peak voltages of the circuit nodes with respect to the signal injection node in the topologically non-trivial and trivial phases, respectively. The peak voltages are obtained by measuring the root-mean-square (RMS) voltage values of the signal oscillations at the corresponding node. The voltage response of the edge nodes to the injected signal is significantly distinct from that of the bulk nodes, manifesting the topological non-trivial edge localization. The dark red and cyan bottom-filled nodes correspond to $A$-type and $B$-type nodes, respectively. \textbf{c} The node voltage oscillations at $A$-type nodes when the signal is injected at the leftmost edge node in the topologically non-trivial phase. The magnitude of the oscillations decays exponentially towards the bulk nodes. \textbf{d} The voltage oscillations over time only at the nodes where the signal is injected in the topologically non-trivial phase. Red and cyan color oscillations represent the $A$- and $B$-type nodes. The voltage response is highest at the edge nodes. To obtain plots \textbf{a} and \textbf{b}, we measure the RMS voltages and plot the voltage profiles of cSSH for all nodes and for each injection. The LTspice simulations are performed with total 8 unit cells with non-trivial parameters $L_a=80\,\text{mH}$ and $L_b=8\,\text{mH}$. The magnitude and frequency of the injected sinusoidal signal are $I_0=20$mA and $f=6.72$kHz.
	}
	\label{figexternal}
\end{figure}

Given that our cSSH circuit inherently experiences damping due to the regular resistors in Chua's circuits, accurately observing genuine topological behaviors becomes infeasible. This is attributed to the frequency becoming complex, with its imaginary part representing the damping effect, leading to signal attenuation. Consequently, signals introduced at bulk nodes fail to undergo topological amplification towards edge nodes. Nevertheless, the node voltages exhibit an exponential decay from edge nodes towards bulk nodes in the non-trivial phase, as seen in Fig.~\ref{figexternal}d. Moreover, the voltage response is highest when the signal is applied at the edge nodes, decreasing towards the bulk nodes. 
To demonstrate this, we measure the voltage peak at the node where the signal is injected in our transient analyses and display the injected signal response of our cSSH circuit in the topologically non-trivial phase in Fig.~\ref{figexternal}a and in the trivial phase in Fig.~\ref{figexternal}b. A higher voltage response at the edge nodes compared to the bulk nodes in the non-trivial phase indicates the presence of topological boundary resonances. These findings further elucidate the topological properties and their subsequent implications on signal evolution within our cSSH circuit. However, introducing a substantial external signal results in the regularization of chaotic oscillations. Thus, it is crucial to keep the injected signal amplitude below approximately $5\,\text{mA}$ to preserve the chaotic properties in the presence of an external signal. In the following section, we will delve deeper into the investigation of our circuit's chaotic characteristics.

\subsection{Topology protects chaos}

\begin{figure*}[ht!]
	\centering
	\includegraphics[width=15cm]{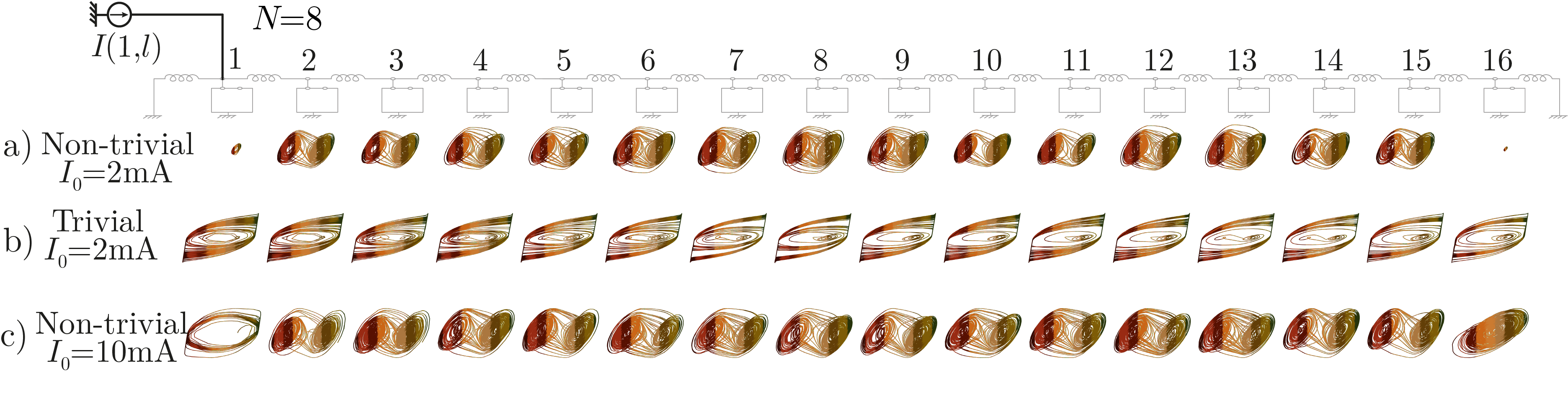}
	\caption{\textbf{Topological protection of chaotic phase profiles in topologically non-trivial and trivial phases under small and large signal perturbations.} A sinusoidal signal is injected at the left edge node, denoted as 1, in all cases. \textbf{a} and \textbf{b} display the evolution of the chaotic phase portraits under a small signal with a magnitude of $2\,\text{mA}$ and frequency of $6.72\,\text{kHz}$. In the topologically non-trivial phase (\textbf{a}), edge localization preserves the chaotic dynamics of the circuit while the phase portraits evolves from single to the double scroll. In the topologically trivial phase (\textbf{b}), the injected signal disrupts the previously existing chaotic profile (double-scroll) and all nodes exhibit limit cycle oscillations, even though the signal is relatively small. \textbf{c} Even a large signal with a magnitude of $10\,\text{mA}$ and frequency of $6.72\,\text{kHz}$ cannot perturb the chaotic phase profile in the topologically non-trivial phase due to the perturbation effect decaying exponentially from edge nodes to bulk nodes, resulting in a perturbed chaotic profile of the left edge node. The circuit parameters used are $L_a=80\,\text{mH}$, $L_b=8\,\text{mH}$ for the non-trivial phase and $L_a=8\,\text{mH}$, $L_b=80\,\text{mH}$ for the trivial phase.}
	\label{figprotect}
\end{figure*}

Introducing an external signal can indeed disrupt the chaotic dynamics inherent to each Chua's circuit. However, in the topological non-trivial phase, the signal injected at the edge nodes undergoes exponential decay, allowing the topology to protect the chaotic dynamics of the cSSH. This protection occurs because the injected current tends to localize at the circuit's boundaries, without spreading throughout the entire circuit. The reason for this localization is the proportional relation between inductance and voltage. In the non-trivial setting, since the edge effective inductances are always much smaller than those of the bulk nodes, the injected current at the edges flows to the ground due to the relationship $V(t) = L \dv{I}{t}$. To demonstrate this and examine the cSSH's response to a small perturbation, we initially introduce a sinusoidal signal with an amplitude of $I_0=1\,\text{mA}$ at the leftmost edge. Figure ~\ref{figprotect}a shows that the injected signal fails to disturb the bulk oscillations due to voltage localization and the exponentially decaying signal in the topologically non-trivial phase. Conversely, in the topologically trivial phase, the injected signal spreads into the circuit's bulk, leading to perturbations in chaotic oscillations, as illustrated in Fig.~\ref{figprotect}b.

When a signal with a large magnitude, such as $I_0=10\,\text{mA}$, is introduced, the chaotic oscillations at the injection node undergo alterations. However, oscillations farther from the signal insertion node retain their original phase portraits, due to the high degree of non-trivial topological localization. As detailed in the `Methods: Chua's Circuit' section, the voltage at the left node (i.e., $(n,l)$) of a Chua's circuit can be manipulated to alter the chaotic phase profile of the circuits. Typically, chaotic phase portraits evolve in a particular sequence: starting as a single scroll, they transition to a double scroll, then to a double+limit cycle portrait, and finally to a full limit cycle as the inductance of the left node increases. In the non-trivial topological phase of our cSSH circuit, even when a strong signal is introduced, the chaotic phase portraits still evolve in this sequence. For instance, in Fig.~\ref{figprotect}c, a larger signal with an amplitude of $10\,\text{mA}$ and a frequency of $6.72\,\text{kHz}$ is injected at node 1. However, in a topologically trivial configuration, the same signal completely disrupts the chaotic dynamics. Thus, the circuit topology effectively protects the chaotic dynamics of the cSSH due to the non-trivial localization. This characteristic is significant for preserving the chaotic circuits used in encryption and secure communications~\cite{kolumban_role_1998,feki_adaptive_2003,moon_chaos_2021,guan_chaos-based_2005,xiong_simplest_2021}, as even minor variations in the initial conditions of the circuits can lead to substantial changes.

\section{Methods}
\subsection{Chua's circuit}
\begin{figure}[b!]
	\centering
	\includegraphics[width=13cm]{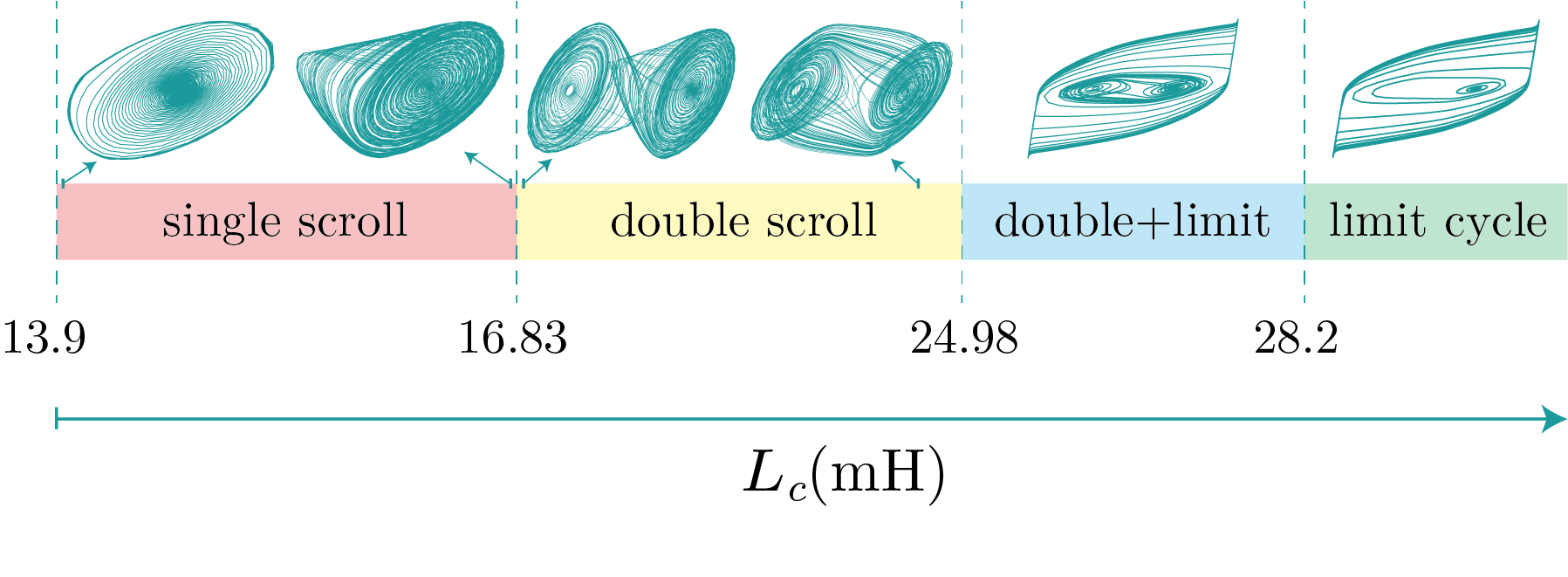}
	\caption{\textbf{Phase portrait evolution of a single Chua's circuit with respect to the inductor's inductance.} The chaotic phase portrait transitions from a single scroll to a limit cycle as the inductance increases. In a single Chua's circuit, the phase transition values are well-defined, such as $16.83\,\text{mH}$ for the transition from single scroll to double scroll when $C_1=10\,\text{nF}$, $C_2=100\,\text{nF}$, and $R=1.85\,\text{k}\Omega$. No oscillations occur below $L=13.9$mH, and a full limit cycle, i.e., regular oscillations, emerges above $L_c=28.2\,\text{mH}$.}
	\label{figportraits}
\end{figure}
The Chua's circuit is recognized as the first and one of the simplest electrical circuits capable of exhibiting chaos~\cite{muthuswamy_simplest_2010}. A single Chua's circuit comprises four passive components: one inductor ($L_c$), two capacitors ($C_1$ and $C_2$), and one conventional resistor ($R$). The nodes located on the left and right sides of the resistor $R$ are denoted as $(n,l)$ and $(n,r)$, respectively, as shown in Fig.~\ref{fig1}h. The chaotic dynamics of the Chua's circuit depend on a non-linear resistor, also known as Chua's diode, in conjunction with the four passive components, satisfying $\alpha=C_2/C_1=10$ and $\beta=R^2 C_2/L_c \approxeq 14.2$. In this study, we consistently set their values at $L_c=24\,\text{mH}$, $C_1=10\,\text{nF}$, $C_2=100\,\text{nF}$, and $R=1.85\,\text{k}\Omega$, under which the Chua's circuit exhibits a double-scroll chaotic portrait. The Chua's diode introduces non-linearity through its five-segment piecewise $I$-$V$ characteristics, as given in Fig.~\ref{fig1}i. Practically, the non-linear $I$-$V$ characteristic is achieved by employing two op-amps~\cite{michael_peter_kennedy_robust_1992}, and in our LTspice simulations, we used op-amps with the catalog number LT1351. These op-amps constitute a negative resistor in some regime, pumping energy into the circuit. Conversely, the standard resistor $R$ consumes energy and stabilizes oscillations between the nodes $(n,l)$ and $(n,r)$. 

Each component plays a coherent role in constructing oscillations between nodes $(n,l)$ and $(n,r)$, where the energy flow is unstable due to the chaotic regime. These chaotic oscillations are primarily classified based on stable points known as attractors~\cite{tsonis_chaos_1989,kuznetsov_hidden_2023}. For instance, by varying $L_c$ while keeping $C_1$, $C_2$, and $R$ constant, we can define intervals of $L_c$ to ascertain the characteristics of the chaotic portrait. A single scroll portrait emerges when $13.9\,\text{mH} < L_c < 16.83\,\text{mH}$, and a double scroll appears when $16.83\,\text{mH} < L_c < 24.98\,\text{mH}$, as illustrated in Fig.~\ref{figportraits}. These intervals remain consistent in a single Chua's circuit as long as the total inductance is unchanged, regardless of whether additional inductors are added. Consequently, it can be asserted that the characteristics of chaotic portraits depend on the total effective inductance at node $(n,l)$ (and also at $(n,r)$, although our focus is on node $(n,l)$ since we couple Chua's circuits at every $(n,l)$ node). For instance, as the effective inductance at $(n,l)$ increases, the phase portrait of the Chua's circuit evolves from a single scroll to a double scroll, initially double then transitioning to a limit cycle, and finally to a complete limit cycle. Oscillations do not occur outside these intervals. However, while the chaotic dynamics of a single circuit rely on four passive and one active component, introducing couplings between Chua's circuits introduces additional energetic freedoms, as observed in the chaotic SSH circuit.

\subsection{Chaotic SSH circuit}

A unit cell of the cSSH circuit consists of two sub-lattice nodes. Each of these nodes is grounded with identical onsite Chua's circuits, and they are connected by the inductances of the intracell coupling inductor $L_a$ and the intercell coupling inductor $L_b$. Additionally, the left edge node is grounded with an extra $L_b$. In the linear limit, the condition $L_a > L_b$ corresponds to the topologically non-trivial phase, while $L_a < L_b$ indicates the trivial phase. It is important to note that since $L_a$ and $L_b$ represent the coupling strength, and the admittance of an inductor is inversely proportional to this strength (i.e., $(i\omega L_{a,b})^{-1}$), the topological phase is defined by this inverse relationship.

In our cSSH implementation, we utilize inductors to connect Chua's circuits at their $(n,l)$ nodes. However, various coupling components and configurations can be employed to realize cSSH. An open question for future research is the exploration of cSSH when Chua's circuits are coupled at their $(n,r)$ nodes using alternative components, such as capacitors. Our study highlights the potential for achieving topological features using diverse components and connection methods, provided that the values of the dimensionless parameters fall within the regimes where the system retains non-linearity. This indicates that the chaotic dynamics of Chua's circuit depend on the effective impedances at each node. In our circuit design, for instance, linking Chua's circuits at their $(n,l)$ nodes with inductors proves to be an effective approach for studying the circuit's dynamics. This is because all components are linear at the $(n,l)$ node of a Chua's circuit. The linearity of the components at this node simplifies the determination of the total effective impedances.

It is important to note that while topological features can be observed across a wide range of parameters (e.g., nano, milli, or mega) in linear circuits, topological chaos only occurs within a specific parameter regime. This is because excessively extreme coupling parameters can compromise the chaotic dynamics in Chua's circuits. As discussed in the `Methods: Chua's Circuit' section, there are distinct regimes where Chua's circuit exhibits chaotic oscillations. Our cSSH circuit demonstrates topological chaos under various parameter settings, provided they are approximately within the millihenry range. The determination of total inductance will be discussed in the subsequent section.

\subsection{The Lorenz equations for the chaotic SSH}
\begin{figure}[h!]
	\centering
	\includegraphics[width=14cm]{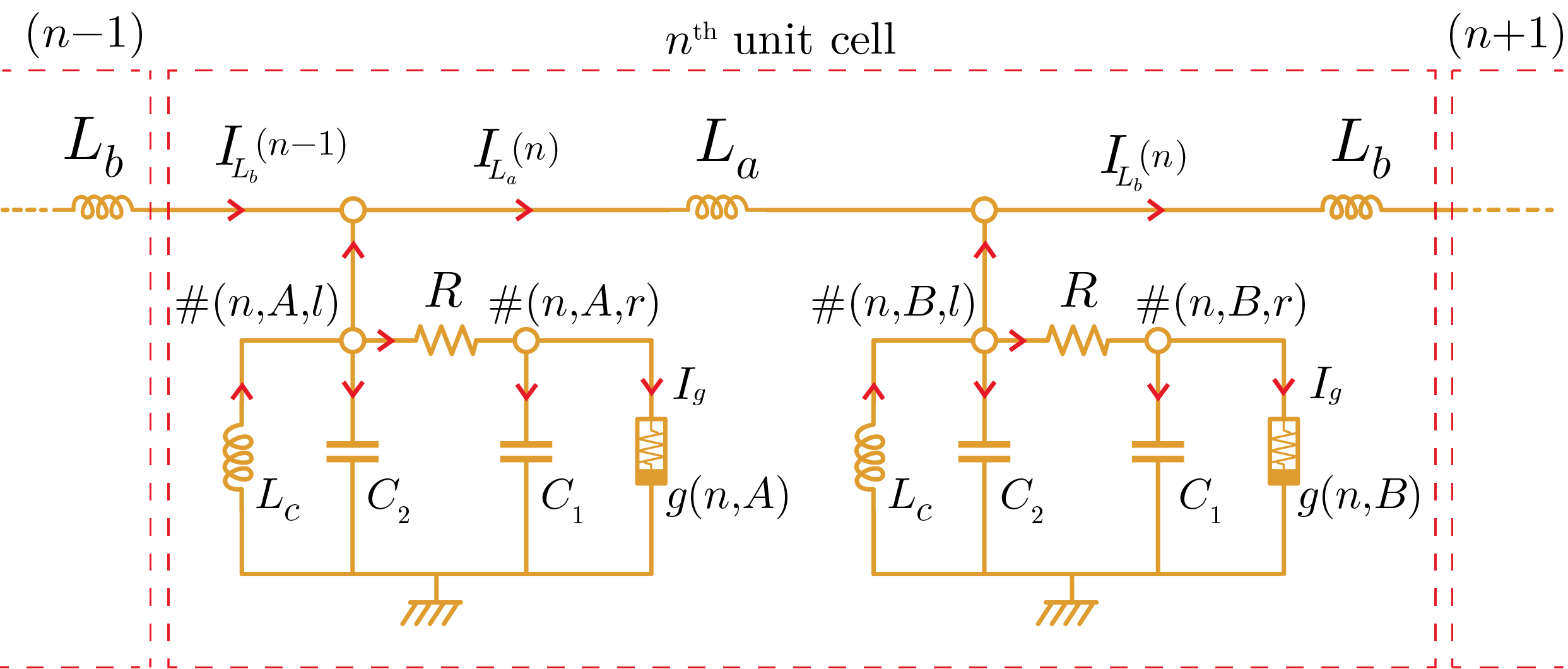}
	\caption{\textbf{The schematic of a unit cell of the chaotic SSH circuit.} Two sublattice nodes within a unit cell are connected by a coupling inductor with inductance $L_a$, and the nearest neighboring unit cells are connected with an inductor of inductance $L_b$. The two nodes found on the left and right side of the conventional resistor $R$ in each Chua's circuit are denoted as $(n,\mu,l)$ and $(n,\mu,r)$, respectively, where $n$ represents the unit cell number and $\mu$ denotes the sublattice nodes, $A$ and $B$. The red arrows indicate the direction of the current flows. The non-linear Chua's diodes are denoted as $g(n,\mu)$.}
	\label{figunit}
\end{figure}

The equations of motion for our cSSH circuit are defined by a set of eight dimensionless equations for each unit cell, as illustrated in Fig.~\ref{figunit}. These equations govern the temporal variation of the current passing through each energy storage component. They are expressed as a function of the unit cell index $n$, as shown in Eq.\eqref{fullequations}. Our circuit's equation set includes two new dimensionless parameters, $\delta_a$ and $\delta_b$. These parameters represent the energy exchange between the nearest neighboring circuits and cells, respectively. The introduction of these parameters allows us to adjust the circuit's topological phase, owing to their dependence on the intracell and intercell coupling inductances. To derive the system's dimensionless Lorenz equations, which describe our coupled Chua's circuits, we apply Kirchhoff's current and voltage laws (KCL and KVL). The currents passing through the components connected to the $(n,A,l)$ and $(n,B,l)$ sublattice nodes are given by
\begin{equation}
	\begin{aligned}
		&  I_{C_2}(n,\mu) = I_{L_c}(n,\mu) - I_{R}(n,\mu) - I_{L_b}(n-1) + I_{L_a}(n),\\
		&  I_{C_1}(n,\mu) = I_{R}(n,\mu) - I_{g}(n,\mu) ,\\
		&  L_c(n,\mu) \dv{I_{L_c}(n,\mu)}{t} = - V(n,\mu,l),
	\end{aligned}
	\label{currentsA}
\end{equation}
where $n$ and $\mu$ denote the unit cell number and the sublattice nodes $A$ and $B$, respectively, that is, $\mu \in {A,B}$. The terms $I_{C_1}(n,\mu)$, $I_{C_2}(n,\mu)$, $I_{R}(n,\mu)$, and $I_{L_c}(n,\mu)$ represent the currents across the capacitors, the resistor between nodes $(n,l)$ and $(n,r)$, and the inductor in each of Chua's circuits, respectively. Since we couple two of Chua's circuits using $L_a$ and the neighboring unit cells with $L_b$, we apply KCL to these two additional inductors as follows
\begin{equation}
	\begin{aligned}
		& L_a(n) \dv{I_{L_a}(n)}{t} = V(n,A,l) - V(n,B,l),\\
		& L_b(n) \dv{I_{L_b}(n)}{t} = V(n,B,l) - V(n+1,A,l),
	\end{aligned}
	\label{iLs}
\end{equation}
where $L_a$ and $L_b$ represent the coupling inductances of the intra-cell and inter-cell couplings, respectively, and $V$ denotes the voltage at the node indicated by its subscript. We proceed by employing the relationship $I_a = C_a \frac{\mathrm{d}V_a}{\mathrm{d}t}$ for capacitors. For inductors, the voltage across each inductor is described by the temporal evolution of the current passing through it, as given by $V_a = L_a \frac{\mathrm{d}I_a}{\mathrm{d}t}$. To transform the aforementioned KCL and KVL equations into Lorenz equations, we introduce dimensionless variables as
\begin{equation}
	\begin{aligned}
		& x(n,\mu) = \frac{V(n,\mu,r)}{B_p}, \qquad  y(n,\mu) = \frac{V(n,\mu,l)}{B_p}, \qquad  z(n,\mu) = I_{L_c}(n,\mu) \frac{R(n,\mu)}{B_p},\\
		& \qquad \qquad  u(n) = I_{L_a}(n) \frac{R(n,\mu)}{B_p}, \qquad  v(n) = I_{L_b}(n) \frac{R(n,\mu)}{B_p},
	\end{aligned}
	\label{xyzvariables}
\end{equation}
where, in addition to the dimensionless variables $x, y, z$, we introduce $u$ and $v$. The term $B_p$ represents the breakpoint voltage that defines the characteristics of the non-linear Chua's diode. We further introduce dimensionless parameters as
\begin{equation}
	\begin{aligned}
		& \alpha = \frac{C_2(n,\mu)}{C_1(n,\mu)}, \qquad  \beta= R(n,\mu)^2 \frac{C_2(n,\mu)}{L_c(n,\mu)}, \qquad \tau = \frac{t}{R(n,\mu) C_2(n,\mu)},\\
		& \qquad \qquad \qquad \delta_a= \frac{C_2(n,\mu)}{L_a(n)} R(n,\mu)^2, \qquad \delta_b= \frac{C_2(n,\mu)}{L_b(n)} R(n,\mu)^2 ,
		\label{dimensvar}
	\end{aligned}
\end{equation}
and finally the non-linear resistors, we introduce
\begin{equation}
	f(x(n,\mu)) = I_{g}(n,\mu) \frac{R(n,\mu)}{B_p}, \quad a=m_1 R(n,\mu) , \quad b=m_0 R(n,\mu),
	\label{nonlinfunc}
\end{equation}
where $I_{g}(n,\mu)$ represents the non-linear current across the Chua diode, with the slopes of the three segments being defined by $m_0$ and $m_1$.

\subsubsection{Dimensionless non-linear function}
The non-linear current-voltage characteristic of Chua's diode is expressed as a piece-wise function comprising three segments, given by
\begin{equation}
	\begin{aligned}
	I_{g}(n,\mu) =&\frac{1}{2}(m_{1}-m_{0}) \left(|V(n,\mu,r)+B_p|-|V(n,\mu,r)-B_p| \right) + m_{0} V(n,\mu,r),
	\end{aligned}
	\label{g1}
\end{equation}
where $B_p$ denotes the breakpoint voltage and $V(n,\mu,r)$ is the voltage at the node $r$ of the $\mu$ sublattice node of the $n$th unitcell. Utilizing the dimensionless slope parameters specified in Eq.~\eqref{nonlinfunc}, the dimensionless non-linear function is expressed as
\begin{equation}
	 f(x(n,\mu)) = b x(n,\mu) +\frac{1}{2}(a-b)(|x(n,\mu)+B_p|-|x(n,\mu)-B_p|).
	\label{diodef}
\end{equation}
This non-linear function becomes linear when $B_p = 0$. In our simulations, we specifically set $B_p = 3\,\mathrm{V}$, $a = -8/7$, and $b = -5/7$ to realize the non-linear function described above.
\subsubsection{Detailed derivations of dimensionless equations}
We express the KCL for the left node (i.e., $(n,\mu,l)$) of each Chua's circuit in our chaotic SSH circuit as a function of the unit cell $n$ and sublattice node $\mu$, as follows
\begin{equation}
	I_{C_2}(n,\mu) = I_{L_c}(n,\mu) - I_{R}(n,\mu) \pm I_{L_a}(n) \mp I_{L_b}(n-1) .
	\label{derc2_1}
\end{equation}
Here, the sign of $I_{L_a}(n)$ and $I_{L_b}(n-1)$ varies depending on the sublattice node. For example, when $\mu=A$, $I_{L_a}(n)$ and $-I_{L_b}(n-1)$, but when $\mu=B$, $-I_{L_a}(n)$ and $I_{L_b}(n-1)$. The above equation can be reformulated to express it in terms of the voltage across the capacitors $C_2$ and the resistors $R$ as
\begin{equation}
	\begin{aligned}
	C_2(n,\mu) \dv{V(n,\mu,l)}{t} =&  I_{L_c}(n,\mu) \pm I_{L_a}(n) \mp I_{L_b}(n-1) \\
	& - \frac{1}{R(n,\mu)} \left(V(n,\mu,l)-V(n,\mu,r)\right).
	\end{aligned}
	\label{derc2_2}
\end{equation}
Now, we substitute the dimensionless variables specified in Eq.\eqref{xyzvariables} and Eq.\eqref{dimensvar} into the above equation and rewrite it as
\begin{equation}
	\begin{aligned}
	\dv{y(n,\mu)}{\tau} =& x(n,\mu) - y(n,\mu) + \frac{R(n,\mu)}{B_p} \left( I_{L_c}(n,\mu) \pm I_{L_a}(n,\mu) \mp I_{L_b}(n-1) \right).
	\end{aligned}
	\label{derc2_3}
\end{equation}
We then rearrange Eq.~\eqref{xyzvariables} and derive one of the dimensionless system equations as
\begin{empheq}[box=\fbox]{equation}
	\dv{y(n,\mu)}{\tau} = x(n,\mu) - y(n,\mu) + z(n,\mu) \pm u(n) \mp v(n-1) .
	\label{derc2_4}
\end{empheq}
Notice that the sign of the dimensionless variables $u(n)$ and $v(n-1)$ changes such that $\dot{y}(n,A)=\dots + u(n)-v(n-1)$ and $\dot{y}(n,B)=\dots - u(n)+v(n-1)$. Similarly, we express KCL for the right node (i.e., $(n,\mu,r)$) of each Chua's circuit as
\begin{equation}
	I_{C_1}(n,\mu) = I_{R}(n,\mu) - I_{g}(n,\mu),
	\label{derc1_1}
\end{equation}
and
\begin{equation}
	\begin{aligned}
	C_1(n,\mu) \dv{V(n,\mu,r)}{t} = &\frac{1}{R(n,\mu)} \left(V(n,\mu,l)-V(n,\mu,r)\right) - I_{g}(n,\mu).
	\end{aligned}
	\label{derc1_2}
\end{equation}
We apply the dimensionless variables from Eq.~\eqref{xyzvariables} and Eq.~\eqref{dimensvar}, and rewrite the above equation as
\begin{equation}
	\frac{C_1(n,\mu) }{C_2(n,\mu)} \dv{x(n,\mu)}{\tau} = \left(y(n,\mu) - x(n,\mu)\right) - I_{g}(n,\mu) \frac{R(n,\mu)}{B_p},
	\label{derc1_3}
\end{equation}
and finally we have
\begin{empheq}[box=\fbox]{equation}
	\dv{x(n,\mu)}{\tau} = \alpha \big( y(n,\mu) - x(n,\mu) - f(x(n,\mu))\big).
	\label{derc1_4}
\end{empheq}

To derive the third equation in the Lorenz system, we apply KVL as
\begin{equation}
	L_c(n,\mu) \dv{I_{L_c}(n,\mu)}{t} = - V(n,\mu,l),
	\label{iLc1}
\end{equation}
and explicitly write as
\begin{equation}
	\dv{I_{L_c}(n)}{\tau} \frac{L_c}{R(n,\mu) C_2(n,\mu)} = - B_p y(n,\mu),
	\label{iLc2}
\end{equation}
and with $\beta = R(n,\mu)^2 C_2(n,\mu) / L_c(n,\mu)$, we obtain
\begin{empheq}[box=\fbox]{equation}
	\dv{z(n,\mu)}{\tau} = - \beta y(n,\mu).
	\label{iLc4}
\end{empheq}

We now derive the dimensionless form of the intra-cell coupling, denoted as $L_a(n)$, within a unit cell. The KVL equation for the inductors $L_a(n)$ is expressed as
\begin{equation}
	L_a \dv{I_{L_a}(n)}{t} = V(n,B,l)-V(n,A,l).
	\label{iL11}
\end{equation}
We incorporate the dimensionless parameters from Eq.~\eqref{xyzvariables} and Eq.~\eqref{dimensvar} and reformulate the above equation as
\begin{equation}
	\dv{I_{L_a}(n)}{\tau} \frac{L_a}{R(n,\mu) C_2(n,\mu)} = B_p \left(y(n,B)-y(n,A)\right),
	\label{iL12}
\end{equation}
and
\begin{equation}
	\frac{R(n,\mu)}{B_p} \dv{I_{L_a}(n)}{\tau}\frac{L_a(n)}{R(n,\mu)^2 C_2(n,A)} =y(n,B) - y(n,A).
	\label{iL13}
\end{equation}
We now employ the dimensionless new variables $u(n)$ and $\delta_a$, as explicitly defined in Eq.\eqref{xyzvariables} and Eq.\eqref{dimensvar}, to further simplify the above equation as
\begin{empheq}[box=\fbox]{equation}
	\dv{u(n)}{\tau} = \delta_a \big( y(n,B)-y(n,A) \big).
	\label{iL14}
\end{empheq}
Similarly, we can now derive the dimensionless form of the inter-cell coupling, denoted as $L_b(n)$, within a unit cell. The KVL for the inductors $L_b(n)$ is expressed as
\begin{equation}
	L_b(n) \dv{I_{L_b}(n)}{t} = V(n+1,A,l) - V(n,B,l).
	\label{iL21}
\end{equation}
By following similar steps and utilizing the new dimensionless variables $v(n)$ and $\delta_b$, we obtain
\begin{empheq}[box=\fbox]{equation}
	\dv{v(n)}{\tau} = \delta_b \big(y(n+1,A) - y(n,B) \big).
	\label{iL22}
\end{empheq}

We have derived the dimensionless system equations in terms of the unit cell number $n$ and the sublattice node $\mu$, which can be either $A$ or $B$. In these equations, while $n$ can represent any desired size, different components can be assigned to each unit cell. However, for the implementation of topology in our cSSH, we set all components of the same type are identical across all unit cells. Consequently, we set $L_a(n)=L_a$, $L_b(n)=L_b$, $L_c(n,\mu)=L_c$, $C_1(n,\mu)=C_1$, $C_2(n,\mu)=C_2$, and $R(n,\mu)=R$.

\subsection{Chaotic scroll space across $L_a$ and $L_b$}
\begin{figure*}[ht!]
	\centering
	\includegraphics[width=\textwidth]{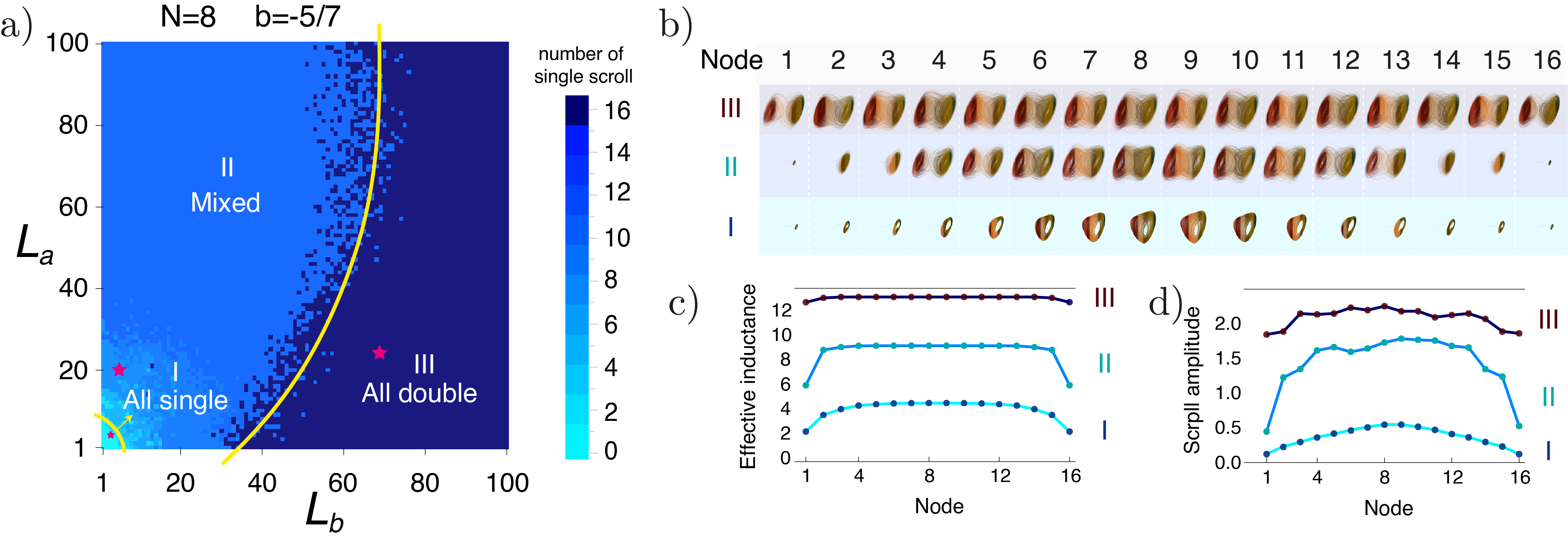}
	\caption{\textbf{Simulated chaotic scroll distribution across all the nodes of the cSSH circuit depending on the intra- and inter-cell coupling inductances $L_a$ and $L_b$.} \textbf{(a)} To evaluate the scroll types at each node in our cSSH circuit, we simulated our circuit using \eqref{fullequations} with $N=8$ and a range from $1\,\mathrm{mH}$ to $100\,\mathrm{mH}$ for $L_a$ and $L_b$. For given $L_a$ and $L_b$ parameters, we identify the scroll type, either single or double scroll. The colored points in the diagram represent different single scroll distributions classified by the total number of single scrolls. \textbf{(b)} The spatial distribution of the scrolls based on the parameters selected from the diagram in (a) is illustrated. \textbf{(c)} The effective inductance profiles of the three example values of $L_a$ and $L_b$ indicated by the magenta stars in (a). \textbf{(d)} The scroll amplitudes are plotted for the three examples in (b). The correlation between (c) and (d) allows us to study the scroll dynamics based on the effective inductance approach. The parameters on the right of the scroll examples are all in units of mH. }
	\label{figscrollspace}
\end{figure*}
Until now, our focus has primarily been on the transition from double to single scroll behavior at the edge nodes. For instance, the transition from double scroll to single scroll at only the edge nodes occurs within the parameter space of the dark blue region (region III) shown in Fig.~\ref{figscrollspace}a. However, different parameter settings affect the effective inductance of the bulk nodes, leading to variations in scroll amplitudes. When the inductances $L_a$ and $L_b$ are smaller, the corresponding dimensionless coupling variables $\delta_a$ and $\delta_b$ become larger, due to the inverse relationship between $L_{a,b}$ and $\delta_{a,b}$, i.e., $\delta_{a,b} = 1/L_{a,b}$. Therefore, smaller values of $L_a$ and $L_b$ imply stronger coupling strength. Importantly, as shown in Fig.~\ref{figscrollspace}a, the scroll profile is predominantly influenced by the inter-cell coupling $L_b$. For example, when $L_b$ exceeds $60\,\text{mH}$, the scroll distribution in the cSSH circuit is consistently double scroll, regardless of the value of $L_a$. The magnitude of the edge scroll for smaller $L_b$s is always smaller compared to those for larger $L_b$ settings, as illustrated in Fig.~\ref{figscrollspace}b. The scroll profiles are fully consistent with our effective inductance profile, as depicted in Figs.~\ref{figinductance}c and d. For instance, when $L_b$ is below $8\,\text{mH}$ (the magenta stars in regions I and II), the effective inductance across all nodes is smaller, leading to reduced scroll amplitudes. This smaller scroll amplitude results in a change in the scroll type. Therefore, as we move towards smaller parameter regimes, more single scrolls appear in the spatial distribution.

\subsection{Effective inductance calculation}
Each Chua's circuit consists of a non-linear resistor, two capacitors, and one inductor. Chaos emerges due to the presence of the non-linear resistor but also depends on the linear components. However, the chaotic dynamics of a Chua's circuit do not solely rely on the values of these components individually; rather, they indeed rely on the total effective inductance, capacitance, and resistance at its nodes, especially in the presence of additional linear components attached. This means that the total of each type of conductance essentially determines the oscillation dynamics, whether chaotic or regular.

In our study, we couple Chua's circuits at their left nodes via inductors with inductances $L_a$ and $L_b$, where $L_a$ and $L_b$ represent the intercell and intracell couplings in a unit cell, respectively, forming an open SSH circuit with onsite Chua's circuits. These onsite Chua's circuits are identical in terms of the IV characteristics of non-linear resistors, the capacitance of capacitors $C_1$ and $C_2$, and the inductance of inductors $L_c$. Coupling the identical Chua's circuits results in a breakdown of the identity of Chua's circuits because each Chua's circuit experiences effectively different conductances at the coupling nodes due to the open boundary condition of the circuit array. Although the coupling nodes comprise the total of $L_c + L_a + L_b + C_2$, the effective node conductance varies depending on the position of the node relative to the circuit boundary. This dependency allows us to examine the interplay between the chaotic Chua's circuits and the topological circuit array. For this purpose, we fix the values of the capacitors and the IV characteristics of the non-linear resistors while incorporating $L_c$ into our examinations as a variable element. This simplification is necessary due to the complexities involving various independent variables. 

\begin{figure}[ht!]
	\centering
	\includegraphics[width=12cm]{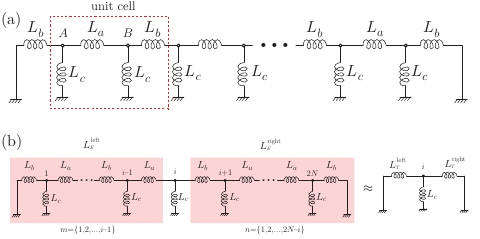}
	\caption{\textbf{The reduced inductive SSH circuit model with open boundary condition.} \textbf{(a)} Because chaotic Chua's circuits are coupled by inductors with inductances $L_a$ and $L_b$, the chaotic dynamic of the cSSH circuit can be examined through the examination of the effective inductance at each node. For this purpose, we consider an equivalent SSH circuit with onsite inductors with inductance $L_c$ corresponding to the cSSH circuit network with only the inductors. \textbf{(b)} The application of the lumped-element method, which considers the conductance of each direction as a single equivalent conductance. This method requires lumping all the elements along each direction and allows us to determine the conductance at a node (inductance for our circuit) by considering each side as a single entity. The conductance of the lumped elements can be determined starting from the edge elements. $m$ and $n$ represent the indices of each iteration. }
	\label{circuitschematic}
\end{figure}

The result when fixing the component values other than the inductors is a reduced inductive circuit comprising only inductors $L_a$, $L_b$, and $L_c$, where $L_c$ represents the onsite inductors, as shown in Fig.~\ref{circuitschematic}a. Reducing the degree of freedom to only the inductance of the inductors is particularly useful since we couple each Chua's circuit with two types of inductors with inductances $L_a$ and $L_b$. By examining the total effective inductance at the left nodes of the Chua's circuit, where the Chua's circuits are coupled, we can investigate both topological and chaotic dynamics in our circuit array. Since each Chua's circuit is connected via inductive couplings, and all other parameters remain consistent, we can assess the circuit's chaotic and topological dynamics by examining the overall effective node inductances. This feasibility is due to the fact that our circuit reduces to a linear topological SSH circuit without involving non-linearity. Additionally, in terms of chaos, the chaotic dynamics solely depend on the effective inductance when fixing the other degrees of freedom. This behavior is evident when connecting various inductors, whether in parallel, series, or both, at a left node of a single Chua's circuit. Therefore, we utilize this effective inductance dependency to investigate the interplay between topology and chaos.

\subsubsection{Effective node inductance derivation}
For the sake of generality in our discussion, we will begin by deriving the effective inductance of bulk nodes. Consider a full inductive open boundary circuit with $N$ unit cells, corresponding to the reduced model of the chaotic SSH circuit (see Fig.~\ref{circuitschematic}a). The effective inductance at a node is not simply the sum of the inductances of the connected inductors (i.e., $L_a + L_b + L_c$). Instead, it is determined by the contributions of all the inductors on both sides of the node. To calculate the effective node inductance, we consider a total of three nodes, with the node under examination sandwiched between the effective total inductances of both sides. This concept corresponds to the lumped-element model, where the circuit elements are concentrated at a single node, as illustrated in Fig.~\ref{circuitschematic}b. As a result, we recall \eqref{eqtotalindMain} in which the effective inductance of three parallel inductors can be calculated as
\begin{equation}
	L_E(i)=\left(\frac{1}{L_c}+\frac{1}{L_E^\text{left}}+\frac{1}{L_E^\text{right}}\right)^{-1},
	\label{eqtotalind}
\end{equation}
where $L_E(i)$ represents the effective inductance of $i$th node, $L_c$ is the inductance of the onsite inductor, $L_E^\text{left}$ and $L_E^\text{right}$ are the effective lumped inductances of the left and right sides of node $i$, respectively. As shown in Fig.~\ref{circuitschematic}b, the effective inductance of any arbitrary bulk node is analogous to three inductors in parallel, taking into account that the left and right lumps are equivalent to a single inductor with inductances $L_E^\text{left}$ and $L_E^\text{right}$, respectively.  We will now discuss how to obtain the effective inductance of the two lumps on both sides of each node.

\subsubsection{Recursive effective inductances of left and right lumps}
\begin{figure}[ht!]
	\centering
	\includegraphics[width=13cm]{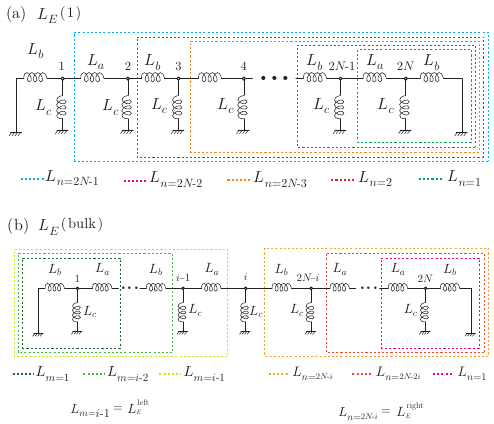}
	\caption{\textbf{Examples for the application of recursive effective inductance for edge and bulk nodes.} \textbf{(a)} This is the case where $n=0$. $L_E(1)$ is obtained by starting to lump the inductors from the right side of the circuit. As we proceed to the opposite edge, we label each summation with the recursion index $m$. $n=2N-1$ gives us the effective inductance of the right side of the circuit. \textbf{(b)} This is the case where $m,n \neq 0$ in which the effective inductance of a bulk node is obtained by lumping both sides of the circuit. The effective inductances of the left and right lumps are obtained when $L_E^{\text{left}} = L_{m=i-1}$ and $L_E^{\text{right}} = L_{n=2N-i}$ where $N$ is the total number of unit cells. Once $L_E^{\text{left}}$ and $L_E^{\text{right}}$ are obtained, $L_E(i)$ is given by \eqref{eqtotalind}. }
	\label{recursivemethod}
\end{figure}
Due to the open boundary conditions in our circuit, the effective node inductance varies depending on the node's distance from the boundaries. To calculate $L_E^\text{left}$ and $L_E^\text{right}$, we begin from both boundaries and sum the inductances towards node $i$. We label the summation of the left and right sides as $L_m$ and $L_n$, where $m$ and $n$ are the indices that range from $1$ to $i-1$ and $2N-i$, respectively. Here, $N$ represents the total number of unit cells. For example, at both edges (refer to Fig.~\ref{circuitschematic}b), inductors $L_b$ and $L_c$ are in parallel, and they are in series with $L_a$ in the first iteration. This leads to $L_{m,n=1}=(\frac{1}{L_c} + \frac{1}{L_b})^{-1} + L_a$ for the first iterations. The total inductance of these three inductors is effectively reduced to a single equivalent inductor connected to the ground. Continuing from the effective inductance of the first iteration as $L_{m,n=1}$, the second iteration (i.e., $\{m,n\}=2$) results in $L_{m=2}=(\frac{1}{L_c} + \frac{1}{L_{m=1}})^{-1} + L_b$ and $L_{n=2}=(\frac{1}{L_c} + \frac{1}{L_{n=1}})^{-1} + L_b$. As we progress to node $i$, a recursive relation arises, expressed as
\begin{equation}
	\begin{aligned}
		L_{m} =& \left(\frac{1}{L_c}+\frac{1}{L_{m-1}}\right)^{-1}+L_{a,b} ,\\
		L_{n} =& \left(\frac{1}{L_c}+\frac{1}{L_{n-1}}\right)^{-1}+L_{a,b} ,
		\label{recursiveRelations}
	\end{aligned}
\end{equation}
where $L_0=L_b$ and $L_{a,b}$ alternates between $L_a$ and $L_b$ depending on the parity of the indices $m$ and $n$. Specifically, for the odd values of $m$ and $n$, $L_{a,b}$ assumes the value of $L_a$, while for even values, it takes on the value of $L_b$. The above recursive relations provide us with the effective inductance of each lump when $m=i-1$ and $n=2N-i$ implying that $L_{m=i-1} = L_E^\text{left}$ and $L_{n=2N-i} = L_E^\text{right}$. In the case of edge node effective inductances (i.e., $L_E(1)$ and $L_E(2N)$), where $i=1$ corresponds to the left edge and $i=2N$ to the right edge, the recursive indices become $m=0$ and $n=2N-1$, or $m=2N-1$ and $n=0$, respectively. This implies that we have a single lump covering the remaining nodes corresponding to $m=0$ or $n=0$. Once the effective inductances on both sides of the $i$th node are determined, the total inductance at the $i$th node  is given by \eqref{eqtotalind}. 

\subsection{Interpretation of the effective inductance through the recursive relations}
The effective inductance profile is highly dependent on the boundary conditions of the circuit. To illustrate this, let us consider a semi-infinite circuit where $N$ is reasonably large. We recall \eqref{eqtotalind}, which is equivalent to the lumped-element model that consolidates all the circuit components found along both the $-x$ and $x$ directions into a single equivalent conductance. This approach simplifies the calculation of effective conductance into a single-node circuit. In our case, since we are dealing only with inductors, the conductance corresponds to the inductance. \eqref{eqtotalind} implies that the effective inductance of the bulk nodes in the depths of the circuit is approximately equivalent when $m \gtreqqless n$. This is because $L_E^\text{left}$ and $L_E^\text{right}$ have nearly equal weight due to the long-depth of the fractions of $L_E^\text{left}$ and $L_E^\text{right}$. This can be exemplified by
\begin{equation}
	\scriptstyle
	L_E^{\text{left,right}} =  L_{a,b} + \cfrac{1}{\frac{1}{L_c} + \cfrac{1}{L_{a,b} + \cfrac{1}{\frac{1}{L_c} + \cfrac{1}{L_{a,b}  + \cfrac{1}{\frac{1}{L_c} + \cfrac{1}{L_{a,b} + \cfrac{1}{\ddots}}}}}}},
	\label{recursionEquation}
\end{equation}
where $L_{a,b}$ takes on value of $L_a$ or $L_b$ depending on the parity of the node number $i$. (Note that the above continued fraction aims to present the general structure, not the explicit values of $L_E^\text{left}$ and $L_E^\text{right}$.) As we approach one of the edges, the depth of one of the recursions of $L_E^\text{left}$ or $L_E^\text{right}$ decreases while the other increases, corresponding to $n\ll m$ or $m \ll n$. A decrease (increase) in the depth of recursion leads to an overall increase (decrease) in $L_E^\text{left}$ or $L_E^\text{right}$. Because $m \ll n$ or $n \ll m$ as we approach the boundaries, the weights of $L_E^\text{left}$ and $L_E^\text{right}$ must be different. This results in a rapid change in the effective inductance of the boundary nodes because the weights of $L_E^\text{left}$ or $L_E^\text{right}$ change non-linearly due to the depth of the fraction. From \eqref{eqtotalind}, because the effective inductances are in the denominator, an increase in the inductance of the lumps as we approach the edges leads to a decrease in the overall node inductance, i.e., $L_E(i \rightarrow 1)$ or $L_E(i \rightarrow 2N)$. In the case of a periodic boundary condition, because $m$ and $n$ are approximately equal to $N$, the weights of $L_E^\text{left}$ and $L_E^\text{right}$ are approximately the same, implying that $L_E(i)$ remains constant for all $i$, regardless of the parity of $i$.

\subsection{Analytical effective inductance formulae}
We begin by deriving the closed-form analytical expressions for the edge nodes, specifically for $L_E(i=1)$ and $L_E(i=2N)$, which involve continued fractions due to recursive relations. Given that our circuit exhibits reflectional symmetry about node $N$, the effective inductance at both edge nodes is equivalent, i.e., $L_E(1) = L_E(2N)$. Therefore, we initially focus on deriving a closed-form expression for the left edge node, under the assumption that the right side of the circuit extends semi-infinitely. The effective inductance for the left edge node, where $i=1$, can be ascertained by substituting \eqref{recursiveRelations} into \eqref{eqtotalind}. This yields
\begin{equation}
	\scriptstyle
	L_E(1) =  \cfrac{1}{\frac{1}{L_c} + \frac{1}{L_b}+ \cfrac{1}{\fcolorbox{red}{white}{$\displaystyle L_{a} + \cfrac{1}{\frac{1}{L_c} + \cfrac{1}{L_{b}  + \cfrac{1}{\frac{1}{L_c} +  \cfrac{1}{L_{a} + \cfrac{1}{\ddots}}}}} $} }}.
	\label{LE1closedform}
\end{equation}
The above equation realizes \eqref{eqtotalind} when $m = 0$ and $n = \propto$. The red box represents $L_E^\text{right}$. Since the red box involves a continued fraction ($L_E^\text{right}$) and we are dealing with a nested structure, we can express the two repeated sequences as
\begin{equation}
	A = L_a + \left( \frac{1}{L_c}+\frac{1}{B}\right)^{-1},
\end{equation}
where $A$ is defined in terms of $B$, which means that to solve for $A$, we express $B$ in a form that eventually leads back to $A$ itself as
\begin{equation}
	B = L_b + \left( \frac{1}{L_c}+\frac{1}{A}\right)^{-1}.
\end{equation}
The continued fraction represented by $A$ is, in essence, an infinite nested structure that, due to its recursive nature, can be described by a finite formula. Breaking down the recursive dependency typically involves substituting one expression into the other and solving for the variable of interest. Therefore, we solve the above two relations simultaneously and find them as
\begin{equation}
	A = \frac{1}{2} \left(L_a+\sqrt{\frac{\left(L_a+2 L_c\right) \left(2 L_c \left(L_a+L_b\right)+L_a L_b\right)}{L_b+2 L_c}}\right),
	\label{Aexplicit}
\end{equation}
and 
\begin{equation}
	B = \frac{1}{2} \left(L_b + \sqrt{\frac{\left(L_b+2 L_c\right) \left(2 L_c \left(L_a+L_b\right)+L_a L_b\right)}{L_a+2 L_c}}\right).
	\label{Bexplicit}
\end{equation}
Now, by rewriting \eqref{LE1closedform} in terms of $A$, 
\begin{equation}
	L_E(1) =  \left( \frac{1}{L_c} + \frac{1}{L_b}+ \frac{1}{A} \right)^{-1},
	\label{LE1closedform2}
\end{equation}
and substituting $A$ into the above equation, we obtain the closed-form expression for the left edge node as
\begin{equation}
	L_E(1) = \left( \frac{1}{L_c}+\frac{1}{L_b} + \frac{2}{L_a + \sqrt{\frac{\left(L_a+2 L_c\right) \left(2 L_c \left(L_a+L_b\right)+L_a L_b\right)}{L_b+2 L_c}}}\right)^{-1}.
\end{equation}
The above closed-form expression provides an effective inductance for the left edge node, assuming the right side of the circuit is semi-infinite. This expression aligns well with the numerically derived $L_E(1)$, obtained from recursive relations. It is important to note that $L_E(1) = L_E(2N)$, reflecting the translational symmetry of our circuit. Following a similar approach, we can derive an analytical expression for the second left-edge node (where $i=2$) by considering the right side of node 2 as a semi-infinite circuit. Applying \eqref{eqtotalind} with $m=1$ and $n=\infty$ leads to the following numerical recursive relation
\begin{equation}
	L_E(2) =  \cfrac{1}{\frac{1}{L_c} + \cfrac{1}{L_a + \cfrac{1}{\frac{1}{L_c}+\frac{1}{L_b}}}+ \cfrac{1}{L_{b} + \cfrac{1}{\frac{1}{L_c} + \cfrac{1}{L_{a}  + \cfrac{1}{\frac{1}{L_c} + \cfrac{1}{\ddots}}}}}}.
	\label{LE2closedform}
\end{equation}
In the equation above, the right fraction has a continued form due to the semi-infinite approximation, through which we can rewrite it as
\begin{equation}
	L_E(2) =  \cfrac{1}{\frac{1}{L_c} + \cfrac{1}{L_a + \cfrac{1}{\frac{1}{L_c}+\frac{1}{L_b}}}+ \cfrac{1}{B}}.
	\label{LE2closedform2}
\end{equation}
We can explicitly obtain $L_E(2)$ by substituting $B$ in the above equation as
\begin{equation}
		L_E(2) = \left( \frac{1}{L_c}+\frac{1}{L_a+\frac{1}{\frac{1}{L_c}+\frac{1}{L_b}}} +\frac{2}{L_b+\sqrt{\frac{\left(L_b+2 L_c\right) \left(2 L_c \left(L_a+L_b\right)+L_a L_b\right)}{L_a+2 L_c}}} \right)^{-1}
	\label{LE2closedform3}
\end{equation}
Using the same approximation methods, it is possible to obtain an analytical expression for $L_E(3)$, which is given by
\begin{equation}
	L_E(3) =  \cfrac{1}{\frac{1}{L_c} + \cfrac{1}{L_b+\frac{1}{\frac{1}{L_c}+\frac{1}{L_a+\frac{1}{\frac{1}{L_c}+\frac{1}{L_b}}}}}+ \cfrac{1}{L_{a} + \cfrac{1}{\frac{1}{L_c} + \cfrac{1}{L_{b}  + \cfrac{1}{\frac{1}{L_c} + \cfrac{1}{\ddots}}}}}}.
	\label{LE3closedform}
\end{equation}
We now replace the continued fraction with $A$ and rewrite the above recursive relation as
\begin{equation}
	L_E(3) =  \cfrac{1}{\frac{1}{L_c} + \cfrac{1}{L_b+\frac{1}{\frac{1}{L_c}+\frac{1}{L_a+\frac{1}{\frac{1}{L_c}+\frac{1}{L_b}}}}}+ \cfrac{1}{A}}.
	\label{LE3closedform2}
\end{equation}
The explicit analytical expression for $L_E(3)$ is obtained when $A$ given in \eqref{Aexplicit} is substituted in \eqref{LE3closedform2} as

\begin{equation}
	L_E(3) =  \left(\frac{1}{L_c} + \cfrac{1}{L_b+\frac{1}{\frac{1}{L_c}+\frac{1}{L_a+\frac{1}{\frac{1}{L_c}+\frac{1}{L_b}}}}}+\frac{2}{L_a+\sqrt{\frac{\left(L_a+2 L_c\right) \left(2 L_c \left(L_a+L_b\right)+L_a L_b\right)}{L_b+2 L_c}}} \right)^{-1}.
	\label{LE3closedform3}
\end{equation}
We have derived the closed-form analytical expressions for the first three edge nodes. Since the effective inductance rapidly converges to the same value as the bulk nodes, as evidenced in Fig.~\ref{figinductance}, this is sufficient to obtain the behaviors of the effective inductance at the boundaries.

We now proceed to derive analytical closed-form expressions for the bulk nodes, employing the same methodology. In contrast to the edge nodes, we are considering a bulk node where both sides are considered to be semi-infinitely large. The recursive relation for such a bulk node is given as
\begin{equation}
	L_E(\text{bulk}) =  \cfrac{1}{\frac{1}{L_c} + \cfrac{1}{L_a + \cfrac{1}{\frac{1}{L_c}+\frac{1}{L_b+\frac{1}{\ddots}}}}+ \cfrac{1}{L_b + \cfrac{1}{\frac{1}{L_c}+\frac{1}{L_a + \frac{1}{\ddots}}}}}.
	\label{LEBulkclosedform}
\end{equation}
Here, since both sides of the bulk node are semi-infinitely large, there are two continued fractions in the denominator of $L_E(\text{bulk})$. Therefore, we can rewrite the above recursive relation as
\begin{equation}
	L_E(\text{bulk}) =  \left( \frac{1}{L_c} + \frac{1}{A}+ \frac{1}{B} \right)^{-1}.
	\label{LEBulkclosedform2}
\end{equation}
We obtain the explicit analytical expression for the bulk nodes by substituting $A$ and $B$ into \eqref{LEBulkclosedform2} and simplifying the result as
\begin{equation}
	L_E(\text{bulk}) = \frac{L_a L_c \left(L_b+L_c\right)+L_b L_c^2}{\sqrt{\left(L_a+2 L_c\right) \left(L_b+2 L_c\right) \left(2 L_c \left(L_a+L_b\right)+L_a L_b\right)}}.
	\label{LEBulkclosedform3}
\end{equation}

\begin{figure}[ht!]
	\centering
	\includegraphics[width=8.5cm]{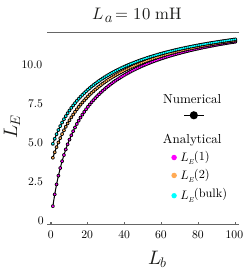}
	\caption{\textbf{Comparison between numerical and analytical expressions of effective inductance.} The black dots denote the numerically calculated effective inductance, while the magenta, orange, and cyan colored dots represent the effective inductance calculated from analytical expressions for the first two edge nodes and a bulk node, respectively. The parameters used are $L_a = 10$ and $L_c = 24$.}
	\label{LEcompNumAnalytical}
\end{figure}

\subsection{Comparison of the phase diagram from the numerical and analytical relations}
\begin{figure}[ht!]
	\centering
	\includegraphics[width=12cm]{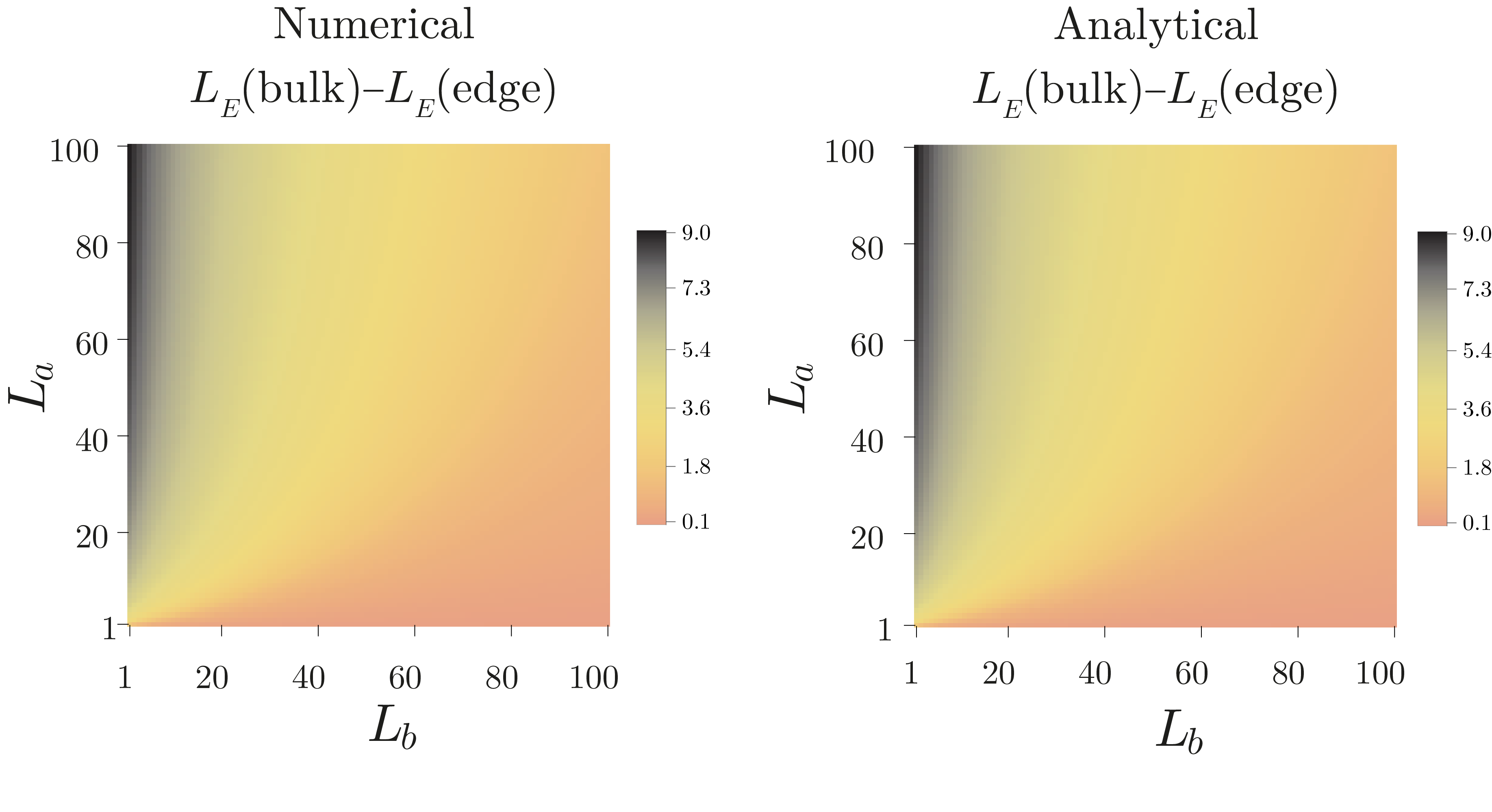}
	\caption{\textbf{The difference between the edge and bulk nodes from the numerical and analytical expressions.} The distinct black region represents the non-trivial non-linear topological phase. A relatively rapid change indicates that our circuit possesses two distinct phases. The black region represents distinct edge effective inductances from that of the bulk and is mostly independent of $L_a$. For small $L_b$ values ($L_b \lesssim 25$), the edge effective inductances are always distinct as long as $L_a$ is not too small. The parameters used are $N=200$ and $L_c = 24$.}
	\label{phasedensityplots}
\end{figure}
In the previous sections, we discussed how to obtain the effective inductance of the reduced SSH circuit both numerically and analytically. Now, we present phase diagrams obtained from both the numerical and analytical effective inductance formulas.

Unlike an inductive periodic boundary SSH circuit, the open boundary SSH exhibits a peculiar effective inductance profile. In an open circuit, the effective inductance towards the edge nodes decreases due to the grounded edges, while $L_E(i)$ remains consistent for all $i$ in a periodic circuit. This is because the node effective inductance is always given by \eqref{LEBulkclosedform2}, in which both continued fractions have nearly the same weight. Also, $L_E(i)$ is the same regardless of the parity of $i$ since $L_E^\text{left} \approxeq L_E^\text{right}$ due to $m \approxeq n$. In addition to the boundary dependency, the exact effective inductance profile also depends on the inductances of the inductors, as shown in Fig.~\ref{figinductance} of the main text. For example, while the edge effective inductances (i.e., $L_E(1)$ and $L_E(2N)$) are particularly distinct for nearly all $L_a$ values when $L_b$ is small, the gap between the edge and bulk effective inductances decreases for larger $L_b$ values. This distinction can be utilized to determine the non-linear topological phases because the edge inductances quickly converge to the level of bulk inductances as $L_b$ increases. 

To understand the relative behavior of the edge and bulk inductances in the parameter space of $L_a$ and $L_b$, we define a quantity $L_E^\text{diff} = L_E(\text{bulk}) - L_E(1)$, which evaluates the convergence behavior of the effective inductance profile. $L_E^\text{diff}$ provides us with insight regarding the variance of the bulk and edge $L_E$ values in the parameter space of $L_a$ and $L_b$. In Fig.~\ref{phasedensityplots}, the density plots present $L_E^\text{diff}$ profiles from the numerical recursive relations and analytical closed-form expressions. The two distinct black and white regions represent the non-trivial and trivial non-linear topological phases, respectively. While the numerical density plot shows the exact behavior, there is a deviation in the magnitude of $L_E^\text{diff}$ in the analytically calculated one. This deviation is attributed to the approximations assumed for the analytical expression derivations, such as the semi-infinite circuit approximation and the order of truncation in the continued fractions. However, the similar profile in both density plots enables us to determine the distinct phases. Although the phase boundary is not directly apparent from the density plots, it can be precisely determined by employing image processing algorithms, which we will discuss in a separate section later.

\subsection{Correspondence between the effective inductance and chaotic scroll amplitudes}
\begin{figure}[ht!]
	\centering
	\includegraphics[width=12cm]{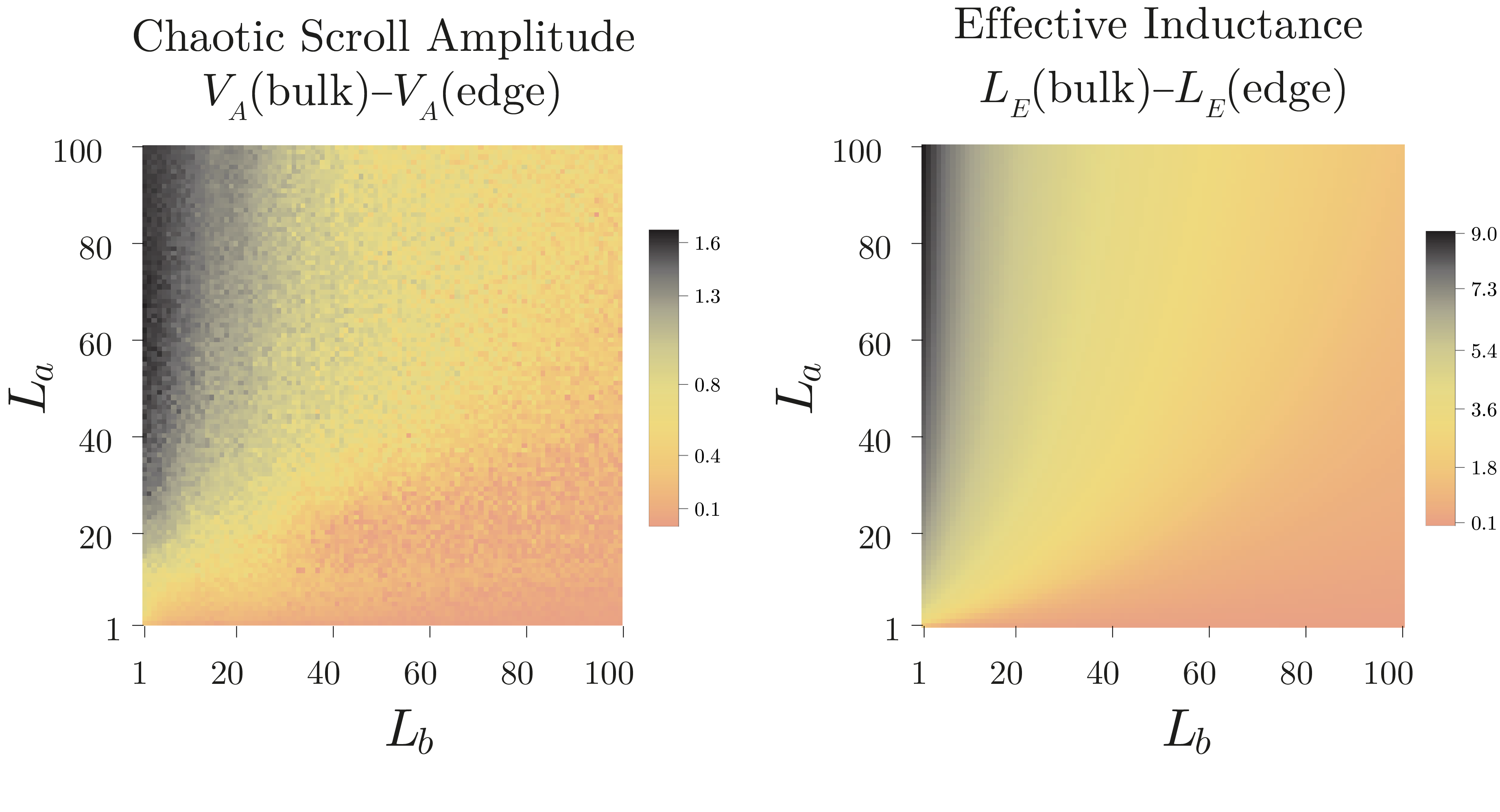}
	\caption{\textbf{The correspondence between the chaotic scroll amplitudes (measured as voltage) and the effective inductance profiles.} The chaotic scroll amplitudes were obtained through numerical simulations of our circuit with $N=8$ and $L_c = 24$. To evaluate our circuit, we take the difference of the maximum voltage amplitudes between an edge node and a bulk node. Two distinct regions appear in the density plots of $V_A^{\text{diff}}$. The black region in density plots indicates that the difference between the edge and bulk voltage amplitudes is largest due to the distinct edge inductance for smaller $L_b$ values. The slightly noisy appearance of the density plot is due to the dynamical characteristics of the chaotic scrolls.}
	\label{VAdiffLEdiffdensityplots}
\end{figure}
In this section, we examine the previously mentioned argument that emphasizes the direct relationship between the effective inductance of each node and the chaotic scroll amplitudes of Chua's circuits connected to those nodes. In our circuit, where we have fixed the values of components other than the inductors, the voltage at the left node of a Chua's circuit is determined solely by the effective inductance at the nodes. While the left-node voltage of a single uncoupled Chua's circuit is given by $V = L \frac{dI}{dt}$, the oscillation amplitudes of Chua's circuits in our circuit are given by $V(i) = L_E(i) \frac{dI}{dt}$, where $I$ represents current. We can examine this argument by comparing the $L_E^\text{diff}$ and $V_A^{\text{diff}}$ profiles. Similar to $L_E^{\text{diff}}$, we define the voltage amplitude of the chaotic scrolls as $V_A^{\text{diff}} = V_A(\text{bulk}) - V_A(\text{edge})$.As shown in Fig.~\ref{VAdiffLEdiffdensityplots}, the amplitude of the chaotic oscillations aligns with the effective inductance of the bulk and edge nodes.

\section{Conclusion}
In summary, we have demonstrated topological chaos in a one-dimensional circuit array of inductively coupled identical Chua's circuits. The topologically non-trivial phase of the circuit results in a chaotic profile where edge nodes exhibit distinct chaotic phases due to the presence of topological boundary modes. In our proposed circuit setup, single-scroll phase portraits are observed at edge nodes, while the bulk nodes retain their original double-scroll portraits. This boundary effect is absent in the topologically trivial phase; hence, the circuit maintains uniform chaotic phase portraits throughout, characterized by double scroll oscillations. This distinction stems from the lower effective inductance at the edge nodes in the trivial and non-trivial phases. When an external signal is applied, the current localizes at the edge nodes to preserve the relation between the edge voltage and effective inductance of the edge nodes. Notably, this localization leads to topological protection for the chaotic dynamics of each Chua's circuit. We have shown that the topological non-trivial phase can protect the chaotic oscillations to a certain extent. In the event of substantial perturbation, chaotic oscillations are disrupted around the affected sites, but the chaotic profile is preserved a few unit cells away, owing to the exponential localization.

Our study provides an exemplary exploration of the coexistence of topological phenomena and chaotic dynamics. For example, this research lays the groundwork for subsequent studies that investigate the interaction between topology and multi-scroll chaotic systems~\cite{liu_generation_2016,chen_circuit_2014}. It also prompts further observations on the effects of using intrinsic components as intra-cell and inter-cell couplings or combining topological features with non-oscillatory couplings, such as resistors. Consequently, our study may open a new avenue for exploring topological features in chaotic systems, which could be realized using various electrical components, including non-linear diodes, transistors, or memristors~\cite{zheng_analysis_2018,yu_dynamic_2015,yu_new_2017}.
\SetPicSubDir{ch-nonhermitian}
\SetExpSubDir{ch-nonhermitian}

\chapter{PT-sensing through topological state morphing in a topolectrical circuit}
\label{ch:nonhermitian}
\vspace{2em}

Following the discussion of the non-linear chaotic topolectrical circuit realization in Chapter \ref{ch:Topochaos}, this chapter presents another implementation of electrical circuits in a system characterized by topological, non-Hermitian (NH), and parity-time (PT) symmetry phenomena. The concept of topological, non-Hermitian (NH), and parity-time (PT) symmetric systems presents individually intriguing features and has led to various applications in sensing implementations. Two ring-shaped, one-dimensional (1D) Su-Schrieffer-Heeger (SSH) topolectrical circuits, corresponding to SSH circuits under periodic boundary conditions (PBC), exhibit a sensitive voltage response. This sensitive response arises from a single coupling between the two SSH circuits, one equipped with uniform onsite gain and the other with uniform loss. The uniform gain and loss terms introduce an offset energy for the bulk, while the four modes corresponding to the coupling node are isolated from the bulk modes, depending on the coupling strength. Depending on the gain and loss strength, we observe PT and anti-PT (APT) transitions over the four isolated states. At the phase transition, exceptional points (EPs) appear, leading to a sensitive and high voltage at the coupling nodes. The modulation of these four isolated modes through a single coupling makes our system an excellent candidate for sensing applications. Electrical simulation validation of our circuit's implementation is provided, advancing our understanding of deformed topological states in PT symmetric NH systems. This study offers critical insights into the dynamics of topological defect states and could have broad implications for the development and control of PT-symmetric sensing applications.

Overall, this chapter presents an example of applying profound physical phenomena in a simple electrical circuit design. By adapting the coupling module to a measurand, our circuit proposal can be harnessed as a sensitive sensor. Our circuit offers an easily applicable design for integration into real-world applications. The discussion in this chapter uniquely contributes a roadmap for how a physical concept can be implemented in electrical circuits.

\section{Introduction}
Symmetries are the underlying mechanism in many physical phenomena, and their presence or absence gives rise to substantial changes in the properties of the system. One of the best examples of the role of symmetries is the bulk-boundary correspondence in topological lattices. In these systems, the bulk symmetries induce special boundary states, known as edge states and surface states. These special topological states have been extensively studied and realized in systems such as acoustic~\cite{yan_acoustic_2020,ferdous_observation_2023,ma_topological_2019,xue_topological_2022}; metamaterials~\cite{paulose_topological_2015,ni_topological_2023}; photonics~\cite{takata_photonic_2018,meng_exceptional_2023,kim_recent_2020}; and topolectrical circuits (TEs) ~\cite{imhof_topolectrical-circuit_2018,lee_topolectrical_2018,hofmann_chiral_2019,yang_realization_2023,rafi-ul-islam_chiral_2024}, serving as examples of translational, inversional, chiral, mirror, and time-reversal symmetries. Yet, the PT symmetry is particularly intriguing due to its effects on the eigenspectrum of NH systems~\cite{lin_topological_2023,jana_emerging_2023}, where the energy spectrum is typically complex, unlike in Hermitian systems~\cite{wu_topology_2021,cao_fully_2022,wang_interconversion_2022,choi_observation_2018,schindler_experimental_2011,jin_solutions_2009,wang_2_2013,li_band_2023,guo_exceptional_2022,krasnok_parity-time_2021,heiss_physics_2012,miri_exceptional_2019,itable_entanglement_2022,meden_mathcalpt-symmetric_2023,xia_nonlinear_2021}. Achieving PT symmetry in an electrical circuit usually involves introducing alternating onsite gain and loss components. Electrical circuits with alternating gain and loss components may exhibit the PT, broken PT, and APT phases depending on the magnitudes of the gain and loss terms, and present real, complex, and imaginary eigenenergy spectra, respectively. In a defective lattice, while the broken PT phase can suppress the defect state, the topological edge states experience a shift through the imaginary plane~\cite{stegmaier_topological_2021}. However, we introduce a novel type of defect engineering in a coupled SSH circuit, enabling us to modulate the topological defect states rather than the bulk spectra.
\begin{figure}[h]
	\includegraphics[width=\textwidth]{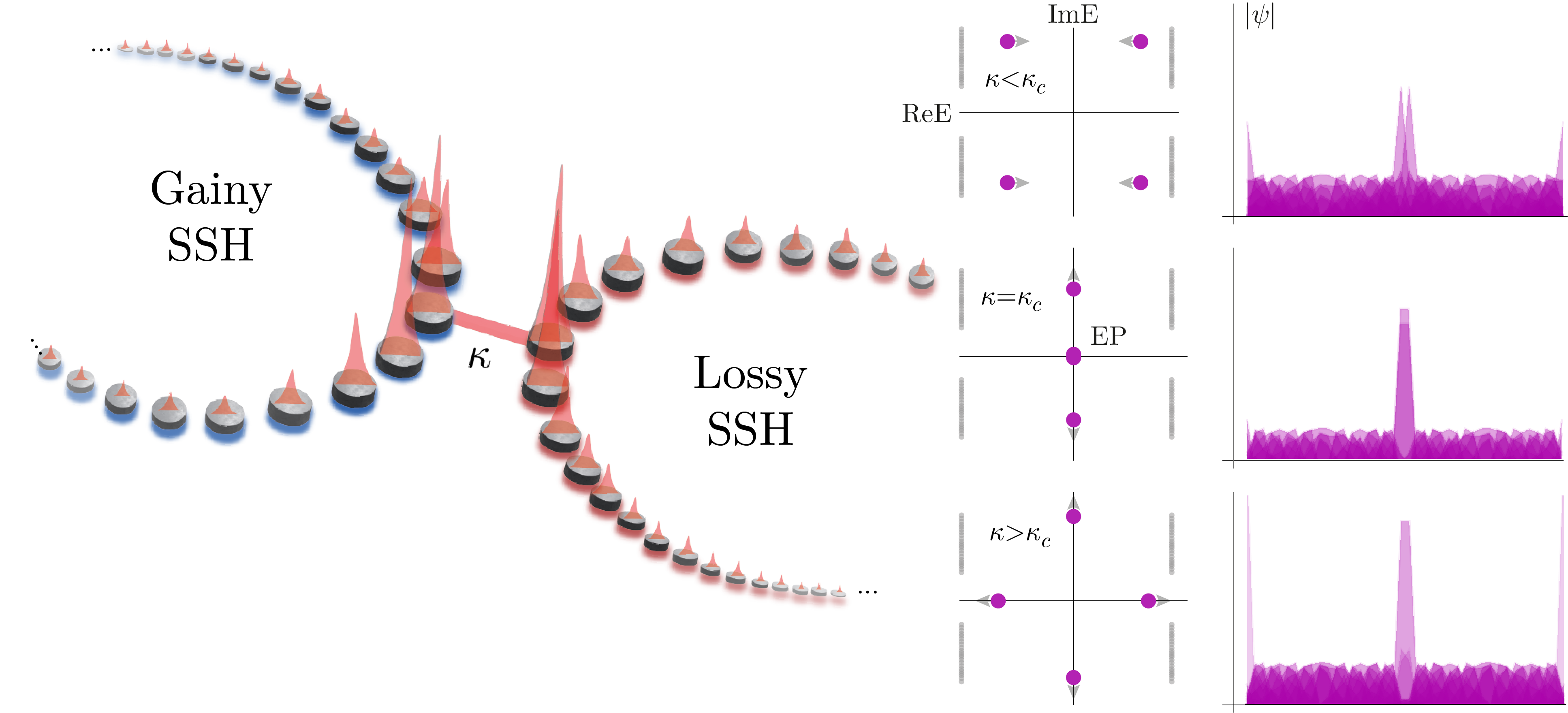}\centering
	\caption{\textbf{An illustrative representation of our PT-sensing mechanism.} Two non-trivial SSH circuits under PBC, one with uniform gain and the other with uniform loss groundings, are coupled at a single site. The coupling strength, denoted as $\kappa$, initiates four isolated modes. At the critical phase transition point, an EP emerges. In the broken PT phase, two out of the four isolated modes become real, leading to the emergence of topological edge states. }
	\label{figheader}
\end{figure}

A topological defect state in a topological lattice emerges where the periodic structure of the bulk is broken due to a defect~\cite{liu_topological_2019,lang_effects_2018,ferdous_observation_2023}, which is generally undesirable. In this chapter, we present a topolectrical circuit that can indeed take advantage of the defect states for a sensing application~\cite{de_carlo_non-hermitian_2022}. Our circuit involves two 1D SSH circuits under PBC. Each ring-shaped SSH circuit (corresponding to the PBC) hosts non-trivial topological zero modes under open boundary conditions (OBC). Unlike conventional PT symmetric TEs, we incorporate a uniform gain to each node in one of the SSH rings, while applying a uniform loss to the second SSH ring~\cite{hashemi_uniform_2022,wu_flux-controlled_2022}. These uniform gain and loss terms in each ring effectively introduce an offset energy, shifting all the bulk modes to a constant negative or positive real energy, proportional to the magnitudes of the gain and loss, respectively. (It is noteworthy that the eigenvalues of each SSH ring are fully imaginary due to the reactive components, in the absence of gain and loss.) When the two isolated SSH rings are coupled at a single node, the entire system adheres to PT symmetry. The significance of coupling the two SSH rings through a single connection is its ability to modulate the coupling modes rather than the entire bulk modes~\cite{budich_non-hermitian_2020,stegmaier_topological_2021,ye_controlling_2022,yuan_nonhermitian_2023,konye_non-hermitian_2023}. Since both SSH circuits are under PBC and possess a periodic structure, their coupling results in broken translational symmetry, leading to the emergence of a pseudo-boundary~\cite{rafi-ul-islam_interfacial_2022,guo_accumulation_2023,ramezani_anomalous_2022,zhu_delocalization_2021,li_topological_nodate,poli_selective_2015,poli_selective_2015-1}. The four modes, isolated from the bulk modes, correspond to the coupling nodes and arise due to the varying strength of the coupling component. 

In our system, we observe distinct PT phases through the modulation of the coupling strength, which specifically alters the modes of the four nodes. In the PT-symmetric phase, the four states are purely imaginary, while in the broken PT-symmetric phase, two of the four isolated modes become purely real. The two zero admittance modes arise at the critical phase transition point as the coupling strength varies (see Fig.~\ref{figheader}). The eigenstates of these four isolated eigenvalues are interestingly localized at the coupling nodes. In the broken PT-symmetric phase, where two of the isolated modes become purely real, they relocalize at the pseudo-edge nodes, manifesting the emergence of topological edge states.

\begin{figure}[ht!]
	\includegraphics[width=7cm]{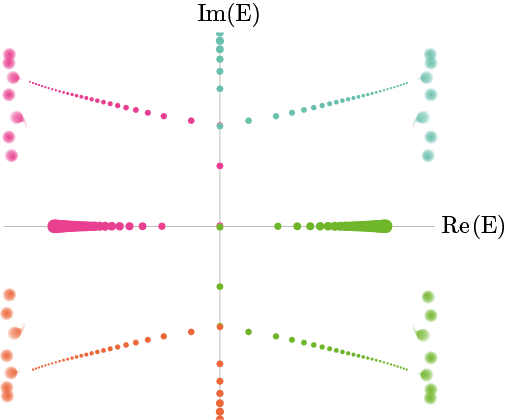}\centering
	\caption{\textbf{The eigenvalue evolution of the PT-sensing system as $\kappa$ varies.} The increasing size of points represent different $\kappa$ values ranging from $0.4\,\text{nF}$ to $0.65\,\text{nF}$ in ascending order, from smaller to larger. The isolated modes converge at the imaginary axis when $\kappa=0.5\,\text{nF}$. Two out of the four modes become zero at $\kappa=0.51\,\text{nF}$, while the remaining two continue to be imaginary and diverge further from zero. The larger spacing around the critical value (i.e., $\kappa=0.51\,\text{nF}$) highlights the exponentially sensitive response of the eigenvalues to small perturbations. The other parameters used are $c_1=0.1\,\text{nF}$, $c_2=0.22\,\text{nF}$, $l=10\,\text{mH}$, $R=5\,\text{k}\Omega$ and $n_1=n_2=8$. }
	\label{FigKappavaries}
\end{figure}

\begin{figure*}[ht!]
	\includegraphics[width=\textwidth]{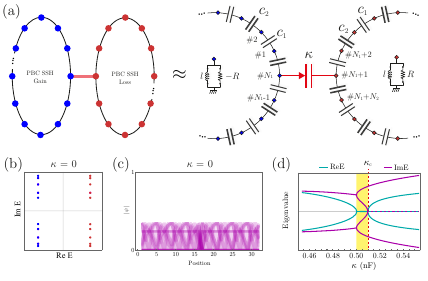}
	\caption{\textbf{The schematic illustration of the PT-sensing circuit and its eigenspectra.} \textbf{(a)} Our circuit consists of two main segments; the gainy PBC SSH (left), and the lossy PBC SSH (right). The gainSSH and lossSSH circuits are constructed on topologically non-trivial settings with 8 unit cells per segment, denoted as $n_1$ and $n_2$ for the gainSSH and lossSSH, respectively. The coupling capacitor $\kappa$ connects the two segments at the last node ($N_1$) of the gainSSH and the first node ($N_1+1$) of the lossSSH. While the negative resistors (gain) in the gainSSH are realized using INICs, the loss effect in the lossSSH is realized by attaching a regular resistor at each node. \textbf{(b)} The eigenvalue spectrum when $\kappa=0$ corresponds to the eigenvalues of the gainSSH and lossSSH under PBC with an equal offset energy but with opposite polarity due to the negative and regular resistors. \textbf{(c)} The eigenstate spectrum is trivial due to the absence of boundaries when $\kappa=0$. We will see how a single parameter can modify the eigenstates in our system when $\kappa\neq0$. \textbf{(d)} The band structure of the isolated modes in the coupled system with respect to $\kappa$. While cyan represents the real energy bands, magenta represents the imaginary energy bands. The yellow-shaded region signifies the sensing regime. The critical phase transition point is shown with a red dashed line where $\kappa=0.51\,\text{nF}$. The parameters used are $c_1=0.1\,\text{nF}$, $c_2=0.22\,\text{nF}$, $l=10\,\text{mH}$, $R=5\,\text{k}\Omega$, and $n_1=n_2=8$. }
	\label{FigCircuitSchematic}
\end{figure*}

From an electrical circuit perspective, the modulation of these four states leads to a high voltage response at the critical phase transition point, where the circuit hosts an EP. Furthermore, this high voltage response at the coupling nodes demonstrates extraordinary sensitivity to changes in the coupling strength. In our system, we harness this sensitive property for sensor applications. Despite the multitude of tight-binding topological lattice models proposed for sensing applications, such as non-Hermitian topological sensors~\cite{guo_sensitivity_2021,budich_non-hermitian_2020}, the voltage response in our model is unique as it relies on a single parameter, rather than the entire bulk. This sensitivity to a single coupling parameter is reliant on the coexistence of topological, non-Hermitian, and PT-symmetric phenomena in our circuit model.

\section{Results}

To realize the sensitive voltage response at the EP, where the PT phase transition occurs, we utilize LTspice simulation software with realistic components and $n_1=n_2=8$ unit cells, as detailed in the `Methods: PT-sensing through LTspice simulations' section. The two PBC SSH circuits depicted in Fig.~\ref{FigCircuitSchematic}a, one with onsite uniform gain and the other with loss, are henceforth referred to as 'gainSSH' for the circuit with negative resistance, and 'lossSSH' for the circuit with regular resistance, respectively. For the coupled system, the circuit Laplacian at the resonant frequency is written as
\begin{equation}
	L = L^{\mathrm{SSH}}_{\mathrm{gain}}  + L^{\mathrm{SSH}}_{\mathrm{loss}} + L_\mathrm{coup}, 
	\label{totalL}
\end{equation}
where
\begin{equation}
	L_\mathrm{coup} = i\omega\kappa \left( |n_1\rangle \begin{pmatrix} 0 & 0 \\ 1 & 0 \end{pmatrix} \langle n_1+1| + | n_1+1\rangle  \begin{pmatrix} 0 & 1 \\ 0 & 0 \end{pmatrix} \langle n_1| \right).
	\label{LcoupTB}
\end{equation}

Above, $L^{\mathrm{SSH}}_{\mathrm{gain}}$ represents the SSH circuit with negative resistors and $L^{\mathrm{SSH}}_{\mathrm{loss}}$ represents the SSH with regular onsite resistors, explicitly given in ~\eqref{LgainTB} and \eqref{LlossTB}, respectively. For convenience, we chose to couple the two SSH rings with a capacitor, whose capacitance is denoted by $\kappa$. The specific symmetry regime, whether PT or APT, is determined by the strength of the onsite resistors where $r=1/R$. For instance, to transition from the PT to broken PT phases, we set $R=5\,\text{k}\Omega$. Conversely, for the transition from APT to broken PT phase, we could set $R=1.04\,\text{k}\Omega$. On the other hand, the variation in the strength of this coupling capacitor leads to either preserved or broken PT symmetry. Following this, we identify the critical phase transitions at $\kappa_c = 0.51\,\text{nF}$ for the PT phase (refer to Fig.~\ref{FigKappavaries} and Fig.~\ref{FigCircuitSchematic}d) and $\kappa_c = 1.82\,\text{nF}$ for the APT phase. While it is equally feasible to utilize both phase transitions for our sensing purpose, as an EP occurs at both values of $\kappa_c$, we have specifically designed our circuit to undergo the transition from PT to broken PT phase where $R=5\,\text{k}\Omega$.

\begin{figure*}[ht!]
	\includegraphics[width=15cm]{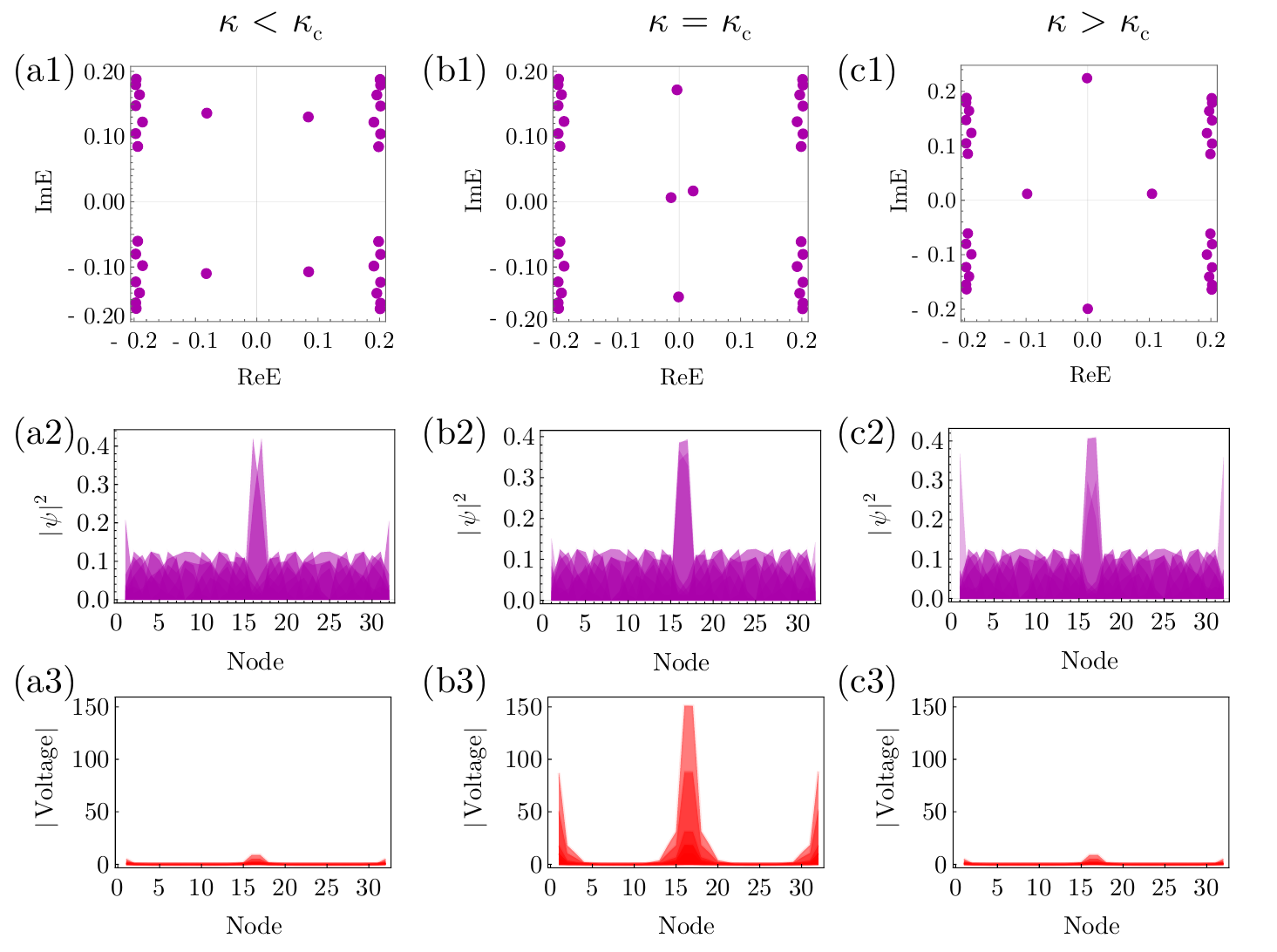}\centering
	\caption{\textbf{Comparative simulation analysis of PT-sensing circuit behavior across different PT phases with realistic components.} The figure presents LTspice simulation results demonstrating eigenspectra and voltage profiles for a PT-sensing circuit in the PT phase ($\kappa < \kappa_c$), at the phase transition point ($\kappa = \kappa_c$), and in the broken PT phase ($\kappa > \kappa_c$). The transition from PT to broken PT phase is achieved by varying the capacitance of the coupling capacitor, emphasizing the pivotal role of the coupling parameter $\kappa$ in the phase characteristics of the system and its sensitive voltage response at the phase transition point. \textbf{(a1)-(a3)} The four defect modes are isolated from the bulk modes when $\kappa=0.495\,\text{nF}$. The emergence of the isolated modes gives rise to localized eigenmodes at nodes 16 and 17. \textbf{(b1)-(b3)} Initially, the four isolated modes fall on the imaginary axis and two of the four modes reach the zero energy at the critical phase transition where $\kappa=0.51\,\text{nF}$. At this critical value, the eigenstates of the four isolated modes become highly localized at the junction nodes, leading to a significant voltage measurement at these nodes. \textbf{(c1)-(c3)} In the broken PT phase where $\kappa=0.57\,\text{nF}$, while two of the four modes remain imaginary, the other two become fully real. As the value of $\kappa$ is further increased, these two real eigenvalues approach the band gap of the gainSSH and lossSSH, resulting in the emergence of topological edge states. The common parameters used in the simulations are $c_1=0.1\,\text{nF}$, $c_2=0.22\,\text{nF}$, $l=10\,\text{mH}$, $R=5\,\text{k}\Omega$, and $f_r=88.97\,\text{kHz}$, where $f_r$ denotes the resonant frequency. }
	\label{FigLTspiceResults}
\end{figure*}

Our analysis begins with an investigation of the eigenspectra of gainSSH and lossSSH when $\kappa=0$. At this point, $L_\mathrm{coup}$ in \eqref{totalL} vanishes, resulting in two uncoupled SSH PBC circuits. Although both circuits exhibit topological zero edge modes under OBC due to the topologically non-trivial setting with $c_1=0.1\,\text{nF}$ and $c_2=0.22\,\text{nF}$, the midgap zero modes are absent under PBC, as depicted in Fig.\ref{FigCircuitSchematic}b and c. While the energies remain fully imaginary in the absence of onsite resistors, the situation changes when gainSSH is equipped with negative resistors, denoted as $-r$, and lossSSH with regular resistors, represented by $r$. In this configuration, the energies of both gainSSH and lossSSH acquire a real part of the same magnitude, but with opposite polarity, as illustrated in Fig.\ref{FigCircuitSchematic}b. As we gradually increase $\kappa$, which connects the last node of gainSSH and the first node of lossSSH, localized defect states emerge due to broken translational symmetry at the nodes of the B sublattice in the last unit cell of gainSSH and the A sublattice in the first unit cell of lossSSH. Up until $\kappa$ reaches $0.5\,\text{nF}$, the four isolated states remain complex and their corresponding eigenstates' spatial distribution is non-topological, as shown in Fig.~\ref{FigCircuitSchematic}d. They exhibit a spatial distribution characteristic of a defect state, similar to those in isolated PBC SSH circuits, decaying exponentially across all nodes in both directions. However, in the parameter regime from $\kappa=0.5\,\text{nF}$ to $\kappa_c$, which is $0.51\,\text{nF}$, the four isolated modes become fully imaginary and their states localize strongly at the coupling nodes, as shown in Fig.~\ref{FigLTspiceResults}b2. The amplitude of these localized states reaches its maximum at $\kappa_c$. At the critical point, where $\kappa=0.51\,\text{nF}$, the PT phase transition occurs, and two of the four fully imaginary eigenvalues shift to the real axis (see Fig.~\ref{FigKappavaries} and in Fig.~\ref{FigLTspiceResults}b1). Importantly, at this phase transition point, two of the four eigenvalues become zero modes, resulting in a high voltage at the junction nodes, owing to the inverse proportionality between voltage and eigenvalues, i.e., $V\approx1/\lambda$ where $\lambda$ is the eigenvalue~\cite{sahin_impedance_2023,zhang_anomalous_2023}. As illustrated in panel b of Fig.~\ref{FigLTspiceResults}, this change in the eigenstate localization profile, and correspondingly the voltage response, occurs with a variation in the coupling capacitance of only $0.1\,\text{nF}$. Compared to the changes in voltage response under other parameter regimes of $\kappa$, the pronounced voltage response to a mere $0.1\,\text{nF}$ variation highlights the sensitivity of our system. In this regime, the response is indeed exponential, as deduced from Fig.~\ref{FigKappavaries}, where the scaling of the two isolated modes towards the EP is much faster than in other parameter regimes. In circuits with significant parasitic elements, although the two eigenvalues may not be exactly zero, the voltage response remains comparably high at the critical value of $\kappa$, especially when contrasted with other regimes.

In our PT sensing circuit, we exploit this exponential response in the parameter regime where the PT phase transition occurs at $\kappa=\kappa_c$. This exponential characteristic can be translated into a highly sensitive sensor, as the entire response hinges on a variation in a single parameter, namely $\kappa$. In our LTspice simulations, we employ a variable capacitor (varactor) whose capacitance is dependent on the bias voltage at its terminals. The capacitance of this varactor can range from $100\,\text{pF}$ to $200\,\text{pF}$, as detailed in the `Methods: PT-sensing through LTspice simulations' section. For practical sensor applications, this varactor could be replaced with a material whose capacitance varies in response to the measurand. An alternative method to utilize the sensitive response of this system involves using an inductive material that alters inductance based on changes in an external magnetic field. The sensitivity of the system, rooted in its bulk design, allows for flexibility in the choice of the coupling component. Adapting this system to different platforms is possible, as long as the tight-binding model remains under PBC and includes a uniform gain and loss mechanism. 
%At the critical value of $\kappa$, the eigenstates of the isolated modes localize entirely at the junction nodes, leading to a significant accumulation of voltage at these nodes, as simulated in panel b of Fig.\ref{FigLTspiceResults}. When compared to the voltage amplitudes shown in Figs.\ref{FigLTspiceResults}a and c, a substantial increase in voltage is observed when the coupling capacitance undergoes a change of only $0.1,\text{nF}$. This heightened voltage response primarily results from the emergence of an EP, where two of the four isolated modes are nearly zero. Since voltage is inversely proportional to the relative magnitude of the eigenvalues, the near-zero modes contribute to a high voltage response. 

We now turn our attention to the regime where $\kappa > \kappa_c$, in which two of the four isolated eigenvalues are real, while the remaining two are imaginary. This scenario aligns with the broken PT phase, as each twin pair of the four eigenvalues forms real and imaginary pairs. Due to the non-trivial topological setting of our system, a band gap exists between the bulk bands. As depicted in Fig.~\ref{FigLTspiceResults}c1, the bulk eigenvalues of gainSSH exhibit a negative offset, whereas those of lossSSH show a positive offset, yet both are of the same magnitude owing to the equal magnitude of $r$ in both circuits. Intriguingly, as $\kappa$ increases, the two real eigenvalues approach the band gap of the bulk modes; one associated with gainSSH moves towards the spectrum of gainSSH, and the other, related to lossSSH, moves towards the spectrum of lossSSH. This behavior results in the emergence of topological edge states at the first node of gainSSH and the last node of lossSSH, shown in Fig.~\ref{FigLTspiceResults}c2. The spatial distribution of this eigenstates are topological, i.e., the amplitude of the states decays exponentially and present at only A sublattice nodes in the gainSSH and at the B sublattice nodes in the lossSSH circuits. On the other hand, the eigenstates corresponding to the imaginary pair exhibits a defect localization at the junction nodes.

\section{Methods}

\subsection{PT-sensing circuit model} 
The PT-sensing circuit consists of two 1D SSH TE circuits, each containing $N_1$ and $N_2$ nodes, respectively, as depicted in Fig.~\ref{FigCircuitSchematic}a. These SSH circuits share an identical coupling structure: a unit cell comprising two sublattice nodes, namely $A$ and $B$, with the capacitances of the intracell and intercell capacitors being $c_1$ and $c_2$, respectively. Additionally, each node is grounded with a common inductor of inductance $l$. To facilitate PT-symmetric phase transitions, we introduce uniform resistors of equal magnitude but opposite signs (i.e., $-r$ and $r$) to each node in the gainy SSH and lossy SSH circuits, respectively. Both SSH circuits adopt a ring design, implying that gainSSH and lossSSH operate under PBC. The uniform gain and loss mechanism in each SSH circuit, as opposed to the conventional alternating pattern in the circuit array, and the ring structure of the circuits are key aspects of our circuit. Consider a scenario where the gain SSH chain has an even number of nodes, $N_1$, numbered from 1 to $N_1$, and the loss SSH chain also has an even number of nodes, $N_2$, numbered from $N_1+1$ to $N_1+N_2$. By introducing $n_1 \equiv N_1/2$ and $n_2 \equiv N_2/2$ to represent the number of \textit{unit cells} in each ring, we number the unit cells in the gain SSH ring from 1 to $n_1$, and those in the loss SSH ring from $n_1+1$ to $n_1+n_2$. The Laplacian for each isolated circuit, operating at the resonant frequency where the reactive diagonal terms vanish, is given as
\begin{align}
	L^{\mathrm{SSH}}_{\mathrm{gain}} =& \sum_{n=1}^{n_1} |n\rangle \left(i\omega c_1\sigma_x - r\mathrm{I}_2\right) \langle n| \\&\quad+ i \omega c_2 \sum_n^{n_1-1} \left( |n\rangle \begin{pmatrix}  0 & 0 \\ 1 & 0 \end{pmatrix}  \langle n + 1| 	+  |n+1\rangle \begin{pmatrix}  0 & 1 \\ 0 & 0 \end{pmatrix}  \langle n| \right) \nonumber  \\
	&\quad+ i \omega c_2 \left( |n_1\rangle \begin{pmatrix}  0 & 0 \\ 1 & 0 \end{pmatrix}  \langle 1| + |1\rangle \begin{pmatrix}  0 & 1 \\ 0 & 0 \end{pmatrix}  \langle n_1| \right), \label{LgainTB} \\
	L^{\mathrm{SSH}}_{\mathrm{loss}} =& \sum_{m=n_1}^{n_1+n_2} |m\rangle \left(i \omega c_1 \sigma_x + r\mathrm{I}_2\right) \langle m|  \\&\quad+i \omega c_2 \sum_{m=n_1}^{n_1+n_2-1} \left( |m\rangle \begin{pmatrix}  0 & 0 \\ 1 & 0 \end{pmatrix}  \langle m + 1| 	+ |m+1\rangle \begin{pmatrix}  0 & 1 \\ 0 & 0 \end{pmatrix}  \langle m| \right) \nonumber \\
	&\quad + i \omega c_2 \left( |n_1+n_2\rangle \begin{pmatrix}  0 & 0 \\ 1 & 0 \end{pmatrix}  \langle n_1+1| +  |n_1+1\rangle \begin{pmatrix}  0 & 1 \\ 0 & 0 \end{pmatrix}  \langle n_1+n_2| \right) \label{LlossTB}.
\end{align} 
where $r=1/R$ and $\omega$ is the angular frequency. The $|n\rangle$s to the basis states for the $n$th unit cell, and the 2 by 2 matrices are in the A / B sublattice basis. To form a PT-symmetric circuit system, we introduce a single coupling capacitor between the two SSH rings and connect them at the last node of gainSSH and the first node of lossSSH, which is given in \eqref{LcoupTB}. The non-Bloch form of the corresponding surrogate Laplacians for the gainy and lossy circuits can be written as 
\begin{align}
	L^{\mathrm{SSH}}_{\mathrm{gain}}(\beta) &= \begin{pmatrix} -r & i\omega c_1 +i\omega c_2/\beta  \\ i\omega c_1+i\omega c_2\beta & - r \end{pmatrix}, \\
	L^{\mathrm{SSH}}_{\mathrm{lossy}}(\beta) &= \begin{pmatrix} r & i\omega c_1 + i\omega c_2/\beta  \\ i\omega c_1 + i\omega c_2\beta & r \end{pmatrix}.
\end{align}
Afterwards, for the simplicity of our analysis, we omit the term $i\omega$, considering it as a common factor. The wavefunctions within the gain and loss portions of the circuit can be expressed in generic forms as 
\begin{align}
	|\psi_{\mathrm{g}}\rangle(n) &= a_1 \beta_{1,\mathrm{g}}^n \begin{pmatrix} 1 \\ \chi_{1,\mathrm{g}} \end{pmatrix} + a_2 \beta_{2,\mathrm{g}}^n  \begin{pmatrix} 1 \\ \chi_{2,\mathrm{g}} \end{pmatrix} \label{psilg} \\
	|\psi_{\mathrm{l}}\rangle(m) &= b_1 \beta_{1,\mathrm{l}}^{(m-n_1)} \begin{pmatrix} 1 \\ \chi_{1,\mathrm{l}} \end{pmatrix} + b_2 \beta_{2,\mathrm{l}}^{(m-n_1)}  \begin{pmatrix} 1 \\ \chi_{2,\mathrm{l}} \end{pmatrix}  \label{psiln}
\end{align}
where $( 1 , \chi_{i,\mathrm{g}})^{\mathrm{T}}$, $i \in (1,2)$, is an eigenvector of $L^{\mathrm{SSH}}_{\mathrm{gain}}(\beta_{i, \mathrm{g}})$ with the given eigenvalue $E$, i.e., 
\begin{align}
	L^{\mathrm{SSH}}_{\mathrm{gain}}(\beta_{i,\mathrm{g}}) \begin{pmatrix} 1 \\ \chi_{i,\mathrm{g}} \end{pmatrix}  = E\begin{pmatrix} 1 \\ \chi_{i,\mathrm{g}} \end{pmatrix}, \\
	E  = ir + (-1)^i \sqrt{ (c_1 + c_2 / \beta_{i,\mathrm{g}})(c_1 + c_2\beta_{i,\mathrm{g}}) },.
\end{align} 
Given that our primary interest lies in the modulation of the topological modes through the tuning of the coupling capacitor's strength $\kappa$, our analysis is particularly focused on the coupling nodes. Consequently, substituting \eqref{psiln} and \eqref{psilg} into \eqref{totalL} gives, at the $N_1$ and $N_1+1$ th nodes, 
\begin{equation}
\begin{aligned}
	\langle\psi|N_1\rangle=& (-ir - E) (a_1 \beta_{1,\mathrm{g}}^{n_1} \chi_{1,\mathrm{g}} +  a_2 \beta_{2,\mathrm{g}}^{n_1} \chi_{2,\mathrm{g}}) + c_1 ( a_1 \beta_{1,\mathrm{g}}^{n_1} + a_2 \beta_{2,\mathrm{g}}^{n_1})\\ &+  c_2 ( a_1 \beta_{1,\mathrm{g}} + a_2 \beta_{2,\mathrm{g}}) + \kappa ( b_1 \beta_{1,\mathrm{l}}+ b_2 \beta_{2,\mathrm{l}} ), \\ 
	\langle\psi|N_1+1\rangle=&(ir - E)( b_1 \beta_{1, \mathrm{l}} + b_2 \beta_{2, \mathrm{l}} ) + \kappa (a_1 \beta_{1,\mathrm{g}}^{n_1} \chi_{1,\mathrm{g}} +  a_2 \beta_{2,\mathrm{g}}^{n_1} \chi_{2,\mathrm{g}}) \\&+  c_1 ( b_1 \beta_{1, \mathrm{l}} \chi_{1,\mathrm{l}}  + b_2 \beta_{2, \mathrm{l}} \chi_{2,\mathrm{l}} ) + c_2 ( b_1 \beta_{1, \mathrm{l}}^{n_2} \chi_{1,\mathrm{l}}  + b_2 \beta_{2, \mathrm{l}}^{n_2} \chi_{2,\mathrm{l}} ).
\end{aligned}
\end{equation}
where the variables $a_1$ and $a_2$ correspond to the amplitude coefficients of gainSSH, while $b_1$ and $b_2$ correspond to the amplitude coefficients of lossSSH. The state of the junction nodes reveals that the contributions from the neighboring ring's states diminish when $\kappa = 0$. However, when $\kappa > 0$, the states of the two junction nodes begin to contribute to each other simultaneously.

\subsection{PT-sensing through LTspice simulations}
The proposed circuit model can be effectively implemented using electrical circuit simulation software such as LTspice. This software facilitates the realization of realistic conditions, including equivalent series resistance (ESR), parasitic resistance, and tolerance analysis. We have designed the circuit depicted in Fig.~\ref{FigCircuitSchematic}a in LTspice, utilizing real-world electrical components. Let us initially discuss the data processing obtained from the simulations to obtain the circuit eigenspectrum, which requires to construct the circuit Laplacian. The Laplacian, denoted as $L$, links the voltage $V$ to the current $I$, following $I=LV$, where $I$ and $V$ are matrices of current and voltage, respectively, matching the dimension of $L$. To ease the construction of the Laplacian and to better understand the voltage distribution, we opt to use a current source wherever feasible. Employing a current source is advantageous because it ensures the amplitude of the current supplied to each node during the analysis remains constant. This constant amplitude allows us to eliminate the current matrix in the construction of $L$. In this scenario, $L$ is obtained by inverting the voltage matrix, as the current matrix effectively becomes an identity matrix due to Kirchhoff's Current Law (KCL). However, it is important to normalize the inverted voltage matrix by the current amplitude to accurately represent the true energies. In our LTspice simulations, we connect a grounded current source to a node and record the node voltages, resulting in a total of $N_1+N_2$ voltage values. By repeating this procedure $N_1+N_2$ times (i.e., injecting current at each node and extracting the corresponding node voltages), we compile the voltage matrix. Each column of this matrix is filled with the voltages extracted from the respective node. Upon completing the construction of the voltage matrix, we determine the circuit Laplacian using $L=V^{-1}I$. This process enables us to ascertain the eigenspectrum of the circuit.

For our realistic PT-sensing implementation, we use a total of 8 unit cells totaling of 16 nodes in each SSH circuit. The gainSSH and lossSSH share the same bulk structure in terms of their capacitors and inductors. In each circuit, we selected $0.1\,\text{nF}$ capacitors for the intracell couplings (Murata Electronics GRM21A5C2E101FW01D) and $0.22\,\text{nF}$ capacitors for the intercell couplings (Murata Electronics GRM21A5C2E221FW01D). To keep the resonant frequency as small as possible, we used $10\,\text{mH}$ inductors with the library number of W\"urth Elektronik 7447221103. Selecting capacitors with smaller capacitance while inductors larger inductance ensures a lower effective ESRs since the admittance of capacitors $i\omega c \rightarrow (i\omega c )/(1+ i \omega c R_c)$ and admittance of inductors $  1/(i\omega l ) \rightarrow 1/(R_l + i \omega l)$ with ESRs, where $R_c$ and $R_l$ represent the effective average ESRs of capacitors and inductors, respectively. With these components, the resonant frequency corresponds to $f_r =88.97\,\text{kHz}$. Each SSH circuit with these component selections is topologically non-trivial and hosts two topological zero eigenvalues under OBC when the frequency of the driving AC signal is set to the resonant frequency. Since these pure reactive circuits are Hermitian, their eigenvalues are fully imaginary. 

We now introduce a negative resistor between the nodes and ground in the gainSSH and a regular resistor in the lossSSH. Our selection of resistance of resistors is $5\,\text{k}\Omega$ (Bourns CRT0805-BY-5001EAS) for the feasibility to realize PT-symmetric phase transitions and the coupling module. The circuit implementation of negative resistors in the gainSSH circuit requires to employ active components alongside to the passive components introduced above. The negative resistance can be achieved by using operational amplifier (opamp) with negative impedance converters with current inversion (INIC), as will be detailed in following `Methods: Implementation of INICs' section. Considering the criteria discussed in `\textit{Implementation of INICs}', we choose AD8047 opamp from Analog Devices for our INIC implementation. The INICs serve as a negative resistance converter when the two resistors found in the feedback configuration have the same resistance. While this feedback configuration ensures a coupling resistance with the same magnitude but with opposite phase, we also include parallel capacitors to these resistors to filter the noises in the opamp outputs. These auxiliary capacitors do not affect the circuit behavior, instead help to stabilize the opamp's response. Through our DC response tests with our negative feedback configuration of opamps, we determine the capacitance of the auxillary capacitors as $33\,\text{pF}$ (Murata Electronics GCM1885C2A330FA16D) and the resistance of the feedback resistors as $6.19\,\text{k}\Omega$ (TE Connectivity RN73C2A6K19BTDF).

Upon introducing the onsite resistors, our two isolated SSH circuits become reciprocal non-Hermitian SSH circuits. The resistors with negative and positive resistors lead to a complex eigenvalue spectrum with $\pm$ same magnitude of real parts.

\subsubsection{Coupling module}
\begin{figure}[h!]
	\centering
	\includegraphics[width=7cm]{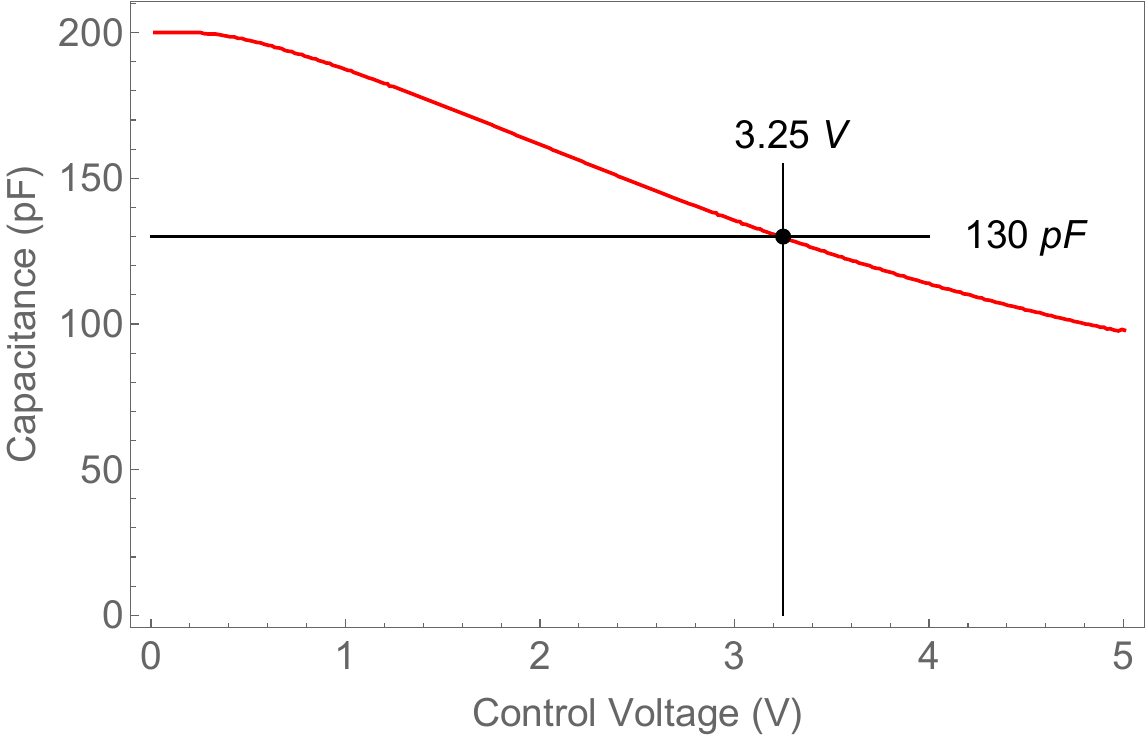}
	\caption{\textbf{The voltage-dependent variation of the capacitance of Murata LXRW19V201-058.} The red solid line represents the capacitance of the varactor as a function of the control voltage. The critical phase transition point is identified at a control voltage of $3.25\,\text{V}$.}
	\label{fig:VaractorV}
\end{figure}

The gainSSH and lossSSH circuits are coupled by a single component. In our circuit simulations, we use a voltage-dependent capacitor (varactor), which varies in the range of 100pF to 200pF (Murata LXRW19V201-058). For the transition from the PT phase to the broken PT phase, where the critical phase transition occurs at $\kappa_c = 0.51\,\text{nF}$, we connect the varactor in parallel with a fixed capacitor of $380\,\text{pF}$ capacitance (KYOCERA AVX 1210GA391JAT1A). This configuration ensures that the capacitance value at the critical phase transition falls within the range of the varactor. With this coupling configuration, we observe a high voltage at the junction nodes (nodes 16 and 17) when the effective capacitance of the varactor reaches $130\,\text{pF}$. This capacitance value corresponds to a control voltage of $3.25\,\text{V}$, as illustrated in Fig.~\ref{fig:VaractorV}. 

In addition to the capacitive coupling between the two nodes, we also utilize two varactors to eliminate the additional terms in the circuit Laplacian corresponding to the junction nodes. This elimination requires grounding nodes 16 and 17 with $-\kappa$, which can be achieved using INICs. As the capacitance of the coupling varactor varies with the control voltage, we include two varactors in two INICs and connect them between the junction nodes and ground. By implementing this setup, we effectively neutralize the additional terms introduced by the coupling varactor at the 16th and 17th diagonal elements of the circuit Laplacian.

\subsection{Implementation of INICs}
\begin{figure}[h!]
	\centering
	\includegraphics[width=7cm]{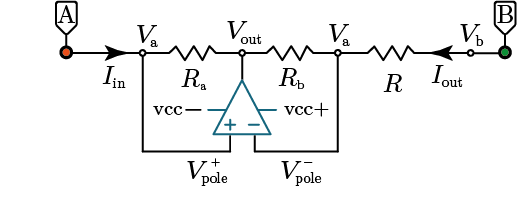}
	\caption{The INIC implementation of an opamp with the negative feedback configuration between node A and B. Each node is labeled by its respective voltage $V$ except the pole indicators $V_\text{pole}^{+}$ and $V_\text{pole}^{-}$. The initial input and output current directions are indicated with an arrow and labeled as $I_\text{in}$ and $I_\text{out}$, respectively.}
	\label{fig:INIC}
\end{figure}
In our circuit, the negative resistances found in the gainSSH are realized by using INICs in which an opamp is employed in the inversion configuration. Such a configuration as shown in Fig.~\ref{fig:INIC} results in a directional current flow in which the currents along the two directions have the same magnitudes but opposite signs, i.e., $I_{AB} = I_{BA}$~\cite{hofmann_chiral_2019}. This implies that the currents flowing in the components between nodes A and B are no longer reciprocal and the coupling symmetry is broken. Denoting the input and output currents as $I_\mathrm{in}$ and $I_\mathrm{out}$, we can, respectively, define the current flowing in the components $R_a$ and $R$, as
\begin{equation}
	I_{\mathrm{in}} = \frac{V_a - V_\mathrm{out}}{R_a}
	\label{I_in}
\end{equation}
\begin{equation}
	I_{\mathrm{out}}= \frac{V_b - V_a}{R}
	\label{I_out}
\end{equation}
An ideal opamp outputs a voltage that is proportional to the product of its amplification (gain) and the difference between the potentials at its poles $V_\mathrm{pole}^{+}$ and $V_\mathrm{pole}^{-}$. However, because our main interest is to obtain a current flow with opposite phase with the input current instead of gain, we configure the poles of the opamp such that its inputs are connected to its output ($V_\mathrm{out}$). Creating a loop between the input and output poles pushes the opamp to make the potentials at its poles equal. For instance, as the potential at the pole $V_\mathrm{pole}^{+}$ increases, according to the above-mentioned definition, the output pole $V_\mathrm{out}$ experiences a corresponding increment with the amplification factor $A$. However, because the negative feedback is correlated with the output as well, the increment in the output also leads to an increment in the negative pole $V_\mathrm{pole}^{-}$. This negative feedback process persists until the opamp reaches the stable condition where the potential difference between the poles is zero. Therefore, the opamp with negative feedback in the INIC configuration ensures that the node to which the negative pole is connected has the same potential as the node $V_a$. Hence, it is also possible to define the input current as $ I_\mathrm{in} = (V_a - V_\mathrm{out}) / R_b$. By interpreting this relation with ~\eqref{I_in} and \eqref{I_out}, an INIC satisfying $I_\mathrm{in} = I_\mathrm{out}$ can be represented in the Laplacian form as
\begin{equation}
	\begin{pmatrix}
		I_\text{in}\\ I_{\text{out}}
	\end{pmatrix}
	= \frac{1}{R} 
	\begin{pmatrix}
		-\eta & \eta \\
		-1 & 1
	\end{pmatrix}
	\begin{pmatrix}
		V_{a}\\V_{b}
	\end{pmatrix}
	\label{INIC_Lap}
\end{equation}
where $\eta$ represents the ratio of the capacitors $ R_b/R_a $. The value of $\eta$ defines the correlation between the $I_\mathrm{in}$ and $I_\mathrm{out}$ currents. However, because we aim to obtain a current with the same magnitude but with the opposite sign, we set $ R_a=6.19\mathrm{k}\Omega$ and $ R_b=6.19\mathrm{k}\Omega$ in our LTspice simulations so that $\eta = 1$.

At this point, one should be careful about the placement of the resistor $R$. As the Laplacian in \eqref{INIC_Lap} is written for the configuration in Fig.~\ref{fig:INIC}, if the resistor $R$ is placed at the left-hand side of the opamp, the Laplacian will differ from \eqref{INIC_Lap} because of the asymmetry between the poles of the opamp employed in the INIC. While the positive pole is known as the non-inverting pole, the negative pole is known as an inverting pole. When $R$ is placed at the non-inverting side, the INIC will not act as a resistance converter. The INIC functionality can be guaranteed by simply placing the component at which the inverting pole is connected to. Such a placement ensures that the A-B coupling is $ -R $ because the current propagating in the forward direction leaves the INIC from the inverting pole ($ V_\mathrm{pole}^{-} $), which causes the admittance phase of the coupling resistor $R$ to change by $ 180^\circ$. This phase shift results in a change in the sign of $R$ and hence the A-B coupling effectively becomes $-R$. Conversely, the B-A coupling is $R$ because the current flowing along the backward direction leaves the INIC from the non-inverting pole. The directional forward and backward admittance due to the INIC enables the realization of negative resistance in our TE circuit. Apart from the INIC architecture, the instability issue that may occur when opamps are present must be carefully considered because opamps are active circuit elements.

\subsubsection{Instability Issue}
\label{subsubsec:OpampInstability}
Unlike passive circuit elements such as capacitors, inductors, and resistors, an active circuit element such as an opamp can drain or pump energy into the circuit. An ideal opamp has infinite impedance, which results in no current flowing into it. However, in realistic cases, the opamp in the INIC may be disturbed before reaching the stable condition  $V_\mathrm{pole}^{+} = V_\mathrm{pole}^{-}$ by parasitic effects that may arise from the initial oscillations or the circuit itself. Once instability emerges in the circuit, the opamps operate nonlinearly, which means that the voltage may accumulate at a certain point. In order to overcome this problem, one can firstly choose \textit{high speed precision opamps} to eliminate the initial oscillations because their quick response will not allow the accumulation of voltage. One can secondly use \textit{low noise opamps} because of their superior ability to suppress parasitic effects. In our LTspice simulations, we chose an opamp from the Analog library with the model number AD8047, which meets these two conditions. Apart from the choice of the opamp, an extensive discussion for the experimental realization of INICs was given by Helbig \textit{et al.} in Ref.~\cite{helbig_generalized_2020}. It was shown there that using additional resistors can help to avoid undesired gain or loss in energy due to the opamps in the experimental realisations.

\section{Conclusion}
In summary, we presented the interplay of topological, non-Hermitian, and PT-symmetric phenomena in a topolectrical circuit through the circuit simulations. By integrating these concepts into a topolectrical circuit, we have demonstrated a novel mechanism for modulating topological defect states through a single coupling in a coupled SSH circuit configuration. This approach leads to the emergence of sensitive and high voltage responses at critical phase transition points, specifically at the EPs, which are of great interest for sensing applications.

In all three PT regimes, namely when $\kappa < \kappa_c$, $\kappa = \kappa_c$, and $\kappa > \kappa_c$, we observed modulation solely in the modes of the junction nodes, while the bulk modes remained unchanged. The complex isolated modes lead to defect localization in the PT phase, while the two real eigenvalues result in the emergence of topological states in the broken PT phase. At the critical phase transition point, all eigenstates localize at the junction nodes. Depending on the distinct PT phases, our system demonstrates how defect states can transform into topological states. The integration of topological, non-Hermitian, and PT-symmetric phenomena in a single TE circuit unveils various profound features that can be harnessed for a sensitive TE sensor.

Our work stands out by focusing on the modulation of isolated topological modes at the coupling nodes, rather than altering the entire bulk modes. This precise control, enabled by the unique interplay of uniform gain and loss in the coupled PBC SSH rings, showcases the potential of topological defect engineering in non-Hermitian PT-symmetric TEs. The exceptional sensitivity of our system to changes in coupling strength underlines its viability as a robust and highly responsive sensor due to the topological nature of the circuit's bulk. This is a notable departure from traditional non-Hermitian topological sensors, which typically depend on the modulation of entire bulk properties. Future work could extend the application of this model to other fields where precise control of topological states is crucial. The integration of our findings into more complex systems may enhance sensor functionality and open new possibilities for utilizing topological phenomena in practical scenarios. The insights from this research contribute to the broader understanding of PT-symmetric non-Hermitian systems and their practical applications.
\chapter{Conclusion and Future Work}
\label{ch:concl}

\section{Conclusion}

This thesis has intricately navigated the realm of topolectrical circuits, presenting a nuanced understanding of their behavior and potential applications in condensed matter systems and electrical engineering. Each chapter has contributed a unique perspective, collectively enhancing our comprehension of these complex systems.

In Chapter 2, we unveil the parity-dependent voltage response in topolectrical circuits. Our analysis, grounded in the simplest one-dimensional electrical circuit analogous to the 1D free electron gas in condensed matter physics, sheds light on the more intricate voltage behaviors observed in topolectrical circuits, including topological and non-Hermitian circuits. The lateral voltage response, a fundamental characteristic of electrical circuits, elucidates the exponential voltage localization characteristics inherent in both topolectrical and non-Hermitian circuits. Particularly notable is the voltage localization in the non-Hermitian Hatano-Nelson model, which exhibits a parity dependence based on the current injection node and cannot be fully comprehended without considering the fundamental characteristics we have delineated in this chapter. This analytical framework is crucial for interpreting circuit behaviors in real-world applications, especially those affected by inevitable factors such as parasitic resistances. Therefore, understanding the parity dependency that arises from both the circuit size and the current injection node is pivotal for the experimental implementation of topolectrical circuits.

We further analyze the fundamental characteristics of electrical circuits through their impedance responses in Chapter 3. Our research challenges the traditional belief that LC resonances only occur based on the parameter space of an LC circuit network. We have discovered that circuit size can also create a significant impedance resonance. Our analysis is based on the two-point impedance measurements in various dimensional LC classical and topolectrical circuits. Remarkably, the size-dependent impedance responses lead to the emergence of fractal structures within the circuit size and parameter space domain. To analytically examine the two-point impedance, we have developed the method of images technique, which facilitates the implementation of open boundary conditions. This approach has yielded a generalized analytical expression applicable to any dimensional homogeneous and heterogeneous circuits. Such a generalized expression for OBC circuits represents a significant contribution to fundamental circuit theory. Consequently, this chapter advances our comprehension of the resonance conditions and analytical analysis in classical and topolectrical circuits.

In Chapter 4, we present the experimental verification of size-dependent impedance scaling in a 2D LC circuit. Our experimental findings substantiate the robust nature of the size-dependent impedance resonances and their associated fractal structures. The combined analytical and experimental analyses of the LC circuits provide profound validation for the method of images, which we developed in the previous chapter. Observations of anomalous impedance scaling, along with its robustness against perturbations, offer valuable insights into the practical applications and limitations of these circuits in real-world scenarios.

In the preceding sections, after detailing the fundamental characteristics and features of both classical and topolectrical circuits, Chapter 5 ventures into the implementation of nonlinear topolectrical circuits. This chapter elucidates the interaction between chaos and topology. The combination of the 1D SSH circuit with the chaotic Chua's circuit results in an extraordinary occurrence: topological chaos. Although the majority of TE implementations are based on linear models, our study paves the way for exploring the dynamics of both topological and nonlinear circuits. The effective conductance method introduced in this chapter offers a novel approach to investigating the interplay between topology and chaos. Additional examples, such as the topological protection in chaotic systems and the analytical derivation of dynamical equations for nonlinear topological systems, provide insightful perspectives into these complex systems.

Finally, in Chapter 6, we propose a real-world application for topolectrical circuits by combining topological, non-Hermitian, and PT-symmetry elements within a single circuit model. The proposed circuit demonstrates a highly sensitive response at the PT-symmetric phase transition point. Simulations of our circuit, conducted using an electrical circuit simulation software under realistic conditions, reveal the robustness and applicability of our model. This chapter serves as an exemplary demonstration of how topolectrical circuits can encapsulate and manifest various complex phenomena, such as non-Hermitian and PT-symmetry, and how they can be effectively harnessed for real-world applications based on our fundamental investigations in electrical circuits discussed in the first three chapters.

Together, these chapters paint a comprehensive picture of topolectrical circuits as versatile platforms for simulating and understanding a wide range of phenomena in condensed matter physics. The insights gained from this thesis not only advance the theoretical framework of topolectrical circuits but also provide practical guidelines for experimental implementations. The interplay between theoretical predictions and experimental observations underscored throughout this thesis reinforces the potential of topolectrical circuits in future technological applications, particularly in the fields of sensing, nonlinear devices, and electrical engineering.

\section{Future Research Directions}
As explored in this thesis, topolectrical circuits provide an outstanding platform for realizing and investigating various phenomena in condensed matter physics, including intriguing aspects of topology, non-Hermitian behavior, and non-linearity~\cite{sahin_topolectrical_2025,yang_circuit_2024,chen_engineering_2025}. Based on our findings and identified potential research gaps in this field, I propose the following topics for future investigation.

\subsection{Floquet circuit realization}
Floquet physics is a promising avenue that involves new dimensions in the field~\cite{traversa_generalized_2013} and provides a powerful framework for the analysis of time-periodic Hamiltonians, which are ubiquitous in systems subjected to periodic perturbations~\cite{rechtsman_photonic_2013,yuce_pt_2015}, such as lattices influenced by external factors like electric fields~\cite{liu_symmetry_2022}. This framework has profound implications in various domains, including electrical circuits. Notably, recent research has demonstrated the potential of analog multipliers to introduce periodic variations in inductance, effectively creating Floquet-like behavior in electrical circuits. This promising approach was exemplified in the preprint titled "Realizing efficient topological temporal pumping in electrical circuits."~\cite{stegmaier2023realizing}.

This research endeavors to investigate alternative techniques for the realization of Floquet circuits, complementing the analog multiplier approach. We aim to identify novel circuit elements and configurations that enhance the efficiency and versatility of Floquet circuit implementations. Our research proposal encompasses a comprehensive blend of theoretical analysis, circuit design, simulations, and experimental validation. Advanced modeling and simulation tools will be employed to investigate diverse circuit configurations and their Floquet-like characteristics. Furthermore, prototypes will be constructed and rigorously tested to confirm the practical feasibility of the proposed Floquet circuit realization techniques. The advancement of Floquet circuit realization techniques holds tremendous potential for the field of topological signal processing. By exploring alternative approaches and practical applications, this research aims to contribute significantly to the development of circuits capable of manipulating signals in a topologically protected manner. 

\subsection{Non-linear memristive circuits}
In the realm of non-linear topology, a promising avenue of exploration lies in the integration of memristors within topological circuits. Memristors, known for their ability to retain resistance values and store information, represent a significant advancement in non-linear components. This becomes particularly relevant when examining topological phase transitions in two-dimensional systems, such as the SSH model.

The intriguing aspect of incorporating memristors into such topological circuits revolves around their potential to influence and preserve topological characteristics. For instance, within a 2D SSH circuit, the presence of memristors could potentially retain non-trivial edge properties even when the system undergoes a phase transition into a trivial topological phase. This hypothesis is rooted in the unique memristive properties initially described by Chua in 1971~\cite{chua_memristor-missing_1971} and later experimentally realized by Strukov et al. in 2008~\cite{strukov_missing_2008}.

The capacity of memristors to `store' nontrivial topology proposed may open up new avenues for research. It suggests potential applications in quantum computing and advanced memory technologies~\cite{kozma_advances_2012}, where the stability and manipulation of topological states are of paramount importance. This concept aligns with recent studies on memristive systems~\cite{waser2016introduction}, and their application in neuromorphic computing~\cite{prezioso_training_2015}. By extending these principles to topological phases, we can explore the convergence of memristive properties and topological phases, an area that promises to unveil novel physical phenomena and technological applications. In summary, the proposed study aims to bridge the gap between the physics of non-linear components and topological insulators, with a particular focus on the unique interplay between memristors and topological phase transitions in 1D and 2D SSH circuits. This research has the potential to deepen our understanding of non-linear topology.

\subsection{The fractal spectral behavior due to the mode interference in critical non-Hermitian skin effect chain with a tail}

The emergence of fractal patterns in the spectral profiles of physical systems is a subject of enduring fascination. A particularly interesting instance of this is observed in the eigenvalue spectra of a critical non-Hermitian skin effect (cNHSE) chain~\cite{li_critical_2020}. The cNHSE chain is characterized by a spectral flow that converges towards the PBC spectra on the complex plane~\cite{qin_universal_2023,siu_terminal-coupling_2023}. This convergence is a result of coupling two oppositely amplified non-reciprocal Hatano-Nelson chains~\cite{rafi-ul-islam_critical_2022}.

In such a system, the OBC spectrum expands into the complex plane at a critical system size, offering new insights into the nature of non-Hermitian (NH) systems. A further intriguing phenomenon emerges when a `tail', comprising a single HN chain, is attached to the cNHSE system. In this extended system, the energy spectra undergo significant alterations with the inclusion or exclusion of just a single site in the chain. This minor modification can lead to the spectrum switching between real and complex values, depending on the specific sizes of the cNHSE and the tail.

Our preliminary research suggests that these drastic spectral changes are attributable to the interference of oscillating states. In our future research, we aim to delve deeper into this phenomenon. Specifically, we will explore how oscillatory behaviors in NH systems can induce such profound spectral transformations. This investigation promises to shed new light on the dynamics of NH systems and their spectral characteristics.

\subsection{Exceptional Bound states}
In the domain of last decade, topological boundary states and non-Hermitian skin states have emerged as two of the most prominent research areas in physics. These boundary states are known for their robustness, which is protected by the bulk properties of the systems in which they are found. More recently, a novel category of boundary states, termed `exceptional bound states' (EB states), was introduced by Lee in 2022~\cite{lee_exceptional_2022}. These EB states have subsequently been observed in electrical circuits, where they manifest as unique circuit resonances (as discussed in Zou et al., 2023~\cite{zou_experimental_2023}).

What sets EB states apart is their origin in the distinctive nature of non-Hermitian exceptional points~\cite{rafi-ul-islam_knots_2024}, which are characterized by their defectiveness. This leads to intriguing phenomena such as the emergence of negative entanglement entropy within the realm of quantum mechanics. Experimental investigations into EB states have revealed a new kind of robustness in these states. Unlike the robustness found in topological or non-Hermitian skin states, the robustness of EB states relies on the geometric defectiveness of the exceptional points.

As a burgeoning class of robust bound states, EB states represent a significant breakthrough. They hold the potential to fundamentally expand our understanding of physical phenomena, paving the way for groundbreaking conceptual developments in the field.

%\bookmarksetup{startatroot}
%\printbibliography[heading=bibintoc]

\appendix
\bibliographystyle{ieeetr}
\bibliography{main}

\end{document}